%% file: RMP_main.tex
\documentclass[rmp,aps,preprint,showpacs,floatfix]{revtex4}
\usepackage{epsf}
\usepackage{pslatex}
\usepackage{graphicx}
\usepackage{bbm}
\usepackage{amsmath}
\usepackage{latexsym}
\usepackage{color}

\begin{document}
\bibliographystyle{apsrmp-arxiv}
\epsfclipon

%%%%%%%%%%%%%%%%%%%%%%%%%%%%%% MACROS %%%%%%%%%%%%%%%%%%%%%%%%%%

\def\MeV{{\rm Me\!V}}
\def\GeV{{\rm Ge\!V}}
\newcommand{\msbar}{\text{$\overline{\text{MS}}$}}

\def\gtwid{{\,\raise.3ex\hbox{$>$\kern-.75em\lower1ex\hbox{$\sim$}}\,}}
\def\ltwid{{\,\raise.3ex\hbox{$<$\kern-.75em\lower1ex\hbox{$\sim$}}\,}}

\def\ask#1{$\langle\!\langle${\small #1}$\rangle\!\rangle$}

\def\ie{{\it i.e.},\ }
\def\eg{{\it e.g.},\ }
\def\et{{\it et al.}}
\def\etc{{\it etc.}\ }
\def\ala{{\it \`a la}}
\def\via{{\it via}\ }
\def\vs{{\it vs.}\ }
\def\cf{{\it c.f.}\ }

\def\cA{{\cal A}}
\def\cB{{\cal B}}
\def\cC{{\cal C}}
\def\cD{{\cal D}}
\def\cE{{\cal E}}
\def\cF{{\cal F}}
\def\cG{{\cal G}}
\def\cH{{\cal H}}
\def\cI{{\cal I}}
\def\cJ{{\cal J}}
\def\cK{{\cal K}}
\def\cL{{\cal L}}
\def\cM{{\cal M}}
\def\cN{{\cal N}}
\def\cO{{\cal O}}
\def\cP{{\cal P}}
\def\cQ{{\cal Q}}
\def\cR{{\cal R}}
\def\cS{{\cal S}}
\def\cT{{\cal T}}
\def\cU{{\cal U}}
\def\cV{{\cal V}}
\def\cW{{\cal W}}
\def\cX{{\cal X}}
\def\cY{{\cal Y}}
\def\cZ{{\cal Z}}
 
\def\chpt{\raise0.4ex\hbox{$\chi$}PT}
\def\schpt{S\raise0.4ex\hbox{$\chi$}PT}
\def\rschpt{rS\raise0.4ex\hbox{$\chi$}PT}
\def\g{\gamma}
\newcommand*{\Tr}{\textrm{Tr}}
\newcommand*{\tr}{\textrm{tr}}

\def\opg{{\scriptstyle \frac{(1+\gamma_5)}{2}}}
\def\omg{{\scriptstyle \frac{(1-\gamma_5)}{2}}}

\def\eqn#1{\label{eq:#1}}
\def\Equation#1{Equation~(\ref{eq:#1})}
\def\Equations#1#2{Equations~(\ref{eq:#1}) and (\ref{eq:#2})}
\def\eq#1{Eq.~(\ref{eq:#1})}
\def\eqsthru#1#2{Eqs.~\ref{eq:#1} through \ref{eq:#2}}
\def\eqs#1#2{Eqs.~(\ref{eq:#1}) and (\ref{eq:#2})}
\def\eqsthree#1#2#3{Eqs.~(\ref{eq:#1}), (\ref{eq:#2}) and (\ref{eq:#3})}
\def\eqsfour#1#2#3#4{Eqs.~(\ref{eq:#1}), (\ref{eq:#2}), (\ref{eq:#3}) and (\ref{eq:#4})}
\def\eqsfive#1#2#3#4#5{Eqs.~(\ref{eq:#1}), (\ref{eq:#2}), (\ref{eq:#3}), (\ref{eq:#4}) and (\ref{eq:#5})}
\def\figref#1{Fig.~\ref{fig:#1}}
\def\Figref#1{Figure~\ref{fig:#1}}
\def\figrefs#1#2{Figs.~\ref{fig:#1} and \ref{fig:#2}}
\def\Figrefs#1#2{Figures~\ref{fig:#1} and \ref{fig:#2}}
\def\figrefthree#1#2#3{Figs.~\ref{fig:#1}, \ref{fig:#2}, and \ref{fig:#3}}
\def\Figrefthree#1#2#3{Figures~\ref{fig:#1}, \ref{fig:#2}, and \ref{fig:#3}}
\def\figreffour#1#2#3#4{Figs.~\ref{fig:#1}, \ref{fig:#2}, \ref{fig:#3}, and \ref{fig:#4}}
\def\Figreffour#1#2#3#4{Figures~\ref{fig:#1}, \ref{fig:#2}, \ref{fig:#3}, and \ref{fig:#4}}
\def\figreffive#1#2#3#4#5{Figs.~\ref{fig:#1}, \ref{fig:#2}, \ref{fig:#3}, \ref{fig:#4}, and \ref{fig:#5}}
\def\Figreffive#1#2#3#4#5{Figures~\ref{fig:#1}, \ref{fig:#2}, \ref{fig:#3}, \ref{fig:#4}, and \ref{fig:#5}}
\def\figrefthru#1#2{Figs.~\ref{fig:#1} -- \ref{fig:#2}}
\def\Figrefthru#1#2{Figures~\ref{fig:#1} -- \ref{fig:#2}}
\def\secref#1{Sec.~\ref{sec:#1}}
\def\secrefs#1#2{Secs.~\ref{sec:#1} and \ref{sec:#2}}
\def\secrefthree#1#2#3{Secs.~\ref{sec:#1}, \ref{sec:#2}, and \ref{sec:#3}}
\def\Secref#1{Section~\ref{sec:#1}}
\def\tabref#1{Table~\ref{tab:#1}}

% DT added
\newcommand{\eps}{\epsilon}
\newcommand{\pbp}{\bar\psi\psi}
\newcommand{\epbp}{\langle\bar\psi\psi\rangle}
\newcommand{\qbq}{\bar q q}
\newcommand{\psibar }{\bar \psi}
\newcommand{\chibar}{\bar\chi}
\newcommand{\etal}{{\it et al.}}
\newcommand{\Plaq}{\Box}
\newcommand{\eplaq}{\langle\Box\rangle}
\newcommand{\LL}{\left\langle}   % left angle bracket
\newcommand{\RR}{\right\rangle}  % right angle bracket
\newcommand{\LP}{\left(}         % left parenthesis
\newcommand{\RP}{\right)}        % right parenthesis
\newcommand{\LB}{\left\{}        % left curly bracket
\newcommand{\RB}{\right\}}       % right curly bracket
\def\PAR#1#2{ {{\partial #1}\over{\partial #2}} }
\def\PARTWO#1#2{ {{\partial^2 #1}\over{\partial #2}^2} }
\def\PARTWOMIX#1#2#3{ {{\partial^2 #1}\over{\partial #2 \partial #3}} }
\newcommand{\BE}{\begin{equation}}
\newcommand{\EE}{\end{equation}}
\newcommand{\BEA}{\begin{eqnarray}}
\newcommand{\EEA}{\end{eqnarray}}
\newcommand{\EL}{\nonumber\\}

% Jim added
\newcommand{\Chi} {\raisebox{0.4ex}{$\chi$}}

% Jack added
\def\bea{\begin{eqnarray}}                                                      
\def\eea{\end{eqnarray}}                                                        

% Urs added
\def\babar{\mbox{\sl B\hspace{-0.4em} {\small\sl A}\hspace{-0.37em}
  \sl B\hspace{-0.4em} {\small\sl A\hspace{-0.02em}R}}}

%%%%%%%%%%%%%%%%%%%%%%%%%%%%%% TITLEPAGE %%%%%%%%%%%%%%%%%%%%%%%%%%

% \draft

\title{Full nonperturbative QCD simulations with 2+1 flavors of improved  
staggered quarks}

\author{A. Bazavov and D. Toussaint}
\affiliation{Department of Physics, University of Arizona, Tucson, AZ  
85721, USA}

\author{C. Bernard and J. Laiho}
\affiliation{Department of Physics, Washington University, St.~Louis,  
MO 63130, USA}

\author{C. DeTar, L. Levkova and M.B. Oktay}
\affiliation{Physics Department, University of Utah,
Salt Lake City, UT 84112, USA}

\author{Steven Gottlieb}
\affiliation{Department of Physics, Indiana University, Bloomington,  
IN 47405, USA}

\author{U.M. Heller}
\affiliation{American Physical Society, One Research Road,
Ridge, NY 11961, USA}

\author{J.E. Hetrick}
\affiliation{Physics Department, University of the Pacific, Stockton, CA 95211, 
USA}

\author{P.B. Mackenzie}
\affiliation{Theoretical Physics Department, MS 106, Fermilab, PO Box 500,
Batavia, IL 60510-0500, USA}

\author{R. Sugar}
\affiliation{Department of Physics, University of California, Santa  
Barbara, CA 93106, USA}

\author{R.S. Van de Water}
\affiliation{Department of Physics, Brookhaven National Laboratory, Upton,
NY 11973, USA}
\date{\today}

\begin{abstract}
Dramatic progress has been made over the last decade in the
numerical study of quantum chromodynamics (QCD) through the use
of improved formulations of QCD on the lattice (improved actions), the
development of new algorithms and the rapid increase in computing power
available to lattice gauge theorists. In this article we describe
simulations of full QCD using the improved staggered quark formalism,
``asqtad'' fermions. These simulations were carried out
with two degenerate flavors of light quarks
(up and down) and with one heavier flavor, the strange quark. Several
light quark masses, down to about 3 times the physical light quark
mass, and six
lattice spacings have been used. These enable
controlled continuum and chiral extrapolations of many low energy QCD
observables.
We review the improved
staggered formalism, emphasizing both advantages and drawbacks. In
particular, we review the procedure
for removing unwanted staggered species in the continuum limit.
We then describe the asqtad lattice ensembles created
by the MILC Collaboration. All MILC lattice ensembles
are publicly available, and they have been used extensively
by a number of lattice gauge theory groups. We review physics
results obtained with them, and discuss the
impact of these results on phenomenology.
Topics include the heavy quark potential,
spectrum of light hadrons, quark masses, decay constant of light and
heavy-light pseudoscalar mesons, semileptonic form factors, nucleon
structure, scattering lengths and more. We conclude with a brief
look at highly promising future prospects.
\end{abstract}


\maketitle

\newpage
\tableofcontents
\newpage

%%%%%%%%%%%%%%%%%%%%%%%%%%%% SECTIONS %%%%%%%%%%%%%%%%%%%%%%%%%%%%

\input RMP_sec1.tex

\input RMP_sec2.tex

\input RMP_sec3.tex

\input RMP_sec4.tex

\input RMP_sec5.tex

\input RMP_sec6.tex

\input RMP_sec7.tex

\input RMP_sec8.tex

\input RMP_sec9.tex

\input RMP_sec10.tex

\input RMP_sec11.tex

%%%%%%%%%%%%%%%%%%%%%%%%%%% ACKNOWLEDGMENTS %%%%%%%%%%%%%%%%%%%%%%%%%%% 

\section*{Acknowledgments}

We thank Maarten Golterman for careful reading of several sections of
this manuscript, and for his helpful suggestions. We are also grateful
to Yigal Shamir for suggested clarifications to \secref{Rooting}.
We thank Heechang Na for help with the heavy baryon section and
Subhasish Basak for help with the section on electromagnetic effects.
This work was supported in part by the United States Department of Energy
grants DE-FG02-91ER-40628, DE-FG02-91ER-40661, DE-FG02-04ER-41298,
DE-FC02-06ER-41443, DE-FC-06ER-41446, DE-AC-02-98CH10886, and by the
National Science Foundation grants PHY05-55234, PHY05-55235, PHY05-55243,
PHY05-55397, PHY07-03296, PHY07-04171, PHY07-57035, and PHY07-57333.
Fermilab is operated by Fermi Research Alliance, LLC, under Contract
No. DE-AC02-07CH11359 with the United States Department of Energy.
Computations for this work were carried out in part on facilities of
the NSF Teragrid under allocation TG-MCA93S002, facilities of the USQCD
Collaboration, which are funded by the Office of Science of the United
States Department of Energy, at the Argonne Leadership Class Computing
Facility under an Incite grant to the USQCD Collaboration, at the National
Energy Research Scientific Computing Center, at Los Alamos National Lab,
and at the University of Arizona, the CHPC at the University of Utah,
Indiana University, and the University of California, Santa Barbara.

%%%%%%%%%%%%%%%%%%%%%%%%%%%%% REFERENCES %%%%%%%%%%%%%%%%%%%%%%%%%%%%% 

%\begin{thebibliography}{999}
\bibliography{RMP_refs}{}
%\end{thebibliography}

\end{document}

%% file: RMP_sec1.tex
% File for section 1 for RMP article
%
\section{Introduction}
\label{sec:Intro}

The standard model of high energy physics encompasses our current
knowledge of the fundamental interactions of subatomic physics. It
consists of two quantum field theories: the Weinberg-Salam theory of
electromagnetic and weak interactions, and
quantum chromodynamics (QCD), the theory of the
strong interactions.  The standard model has been enormously
successful in explaining a wealth of data produced in accelerator and
cosmic ray experiments over the past thirty years. Our knowledge of it
is incomplete, however, because it has been difficult to extract many of
the most interesting predictions of QCD: those that depend on the strong
coupling regime of the theory and therefore require nonperturbative
calculations.

At present, the only means of carrying out nonperturbative QCD
calculations from first principles and with controlled errors is
through large-scale numerical simulations within the framework of
lattice gauge theory. These simulations are needed to obtain a
quantitative understanding of the physical phenomena controlled by the
strong interactions, such as the masses, widths, and scattering lengths
of the light hadrons, and to make possible the determination of the weak
interaction Cabibbo-Kobayashi-Maskawa (CKM) matrix elements from
experiment.
A central objective of the experimental program in
high-energy physics, and of lattice QCD simulations, is to determine the
range of validity of the standard model, and to search for new physics
beyond it. Thus, QCD simulations play an important role in efforts to
obtain a deeper understanding of the fundamental laws of physics.

Major progress has been made in the numerical study of QCD over the 
last decade through the use of improved formulations of QCD on the lattice,
the development of new algorithms, and the increase in computing power
available to lattice gauge theorists.
The lattice formulation of QCD is not merely a numerical approximation
to the continuum formulation.  The lattice regularization is
every bit as valid as any of the popular continuum regularizations,
and has the distinct advantage of being nonperturbative.
The lattice spacing $a$ establishes a momentum cutoff $\pi/a$ that
removes ultraviolet divergences.  Standard renormalization methods
apply, and in the perturbative regime they allow a straightforward conversion of
lattice results to any of the standard continuum regularization
schemes.

There are several formulations of the lattice QCD Lagrangian in
current widespread use.  The gauge field action can be constructed
with varying degrees of improvement that are designed to reduce cutoff
effects at nonzero lattice spacing.  The quark action can be
formulated using Wilson's original method \cite{Wilson:1974sk} with
modern improvements \cite{Sheikholeslami:1985ij} or with the twisted
mass \cite{Frezzotti:1999vv, Frezzotti:2000nk, Frezzotti:2003ni} or
other variants \cite{Zanotti:2001yb,Morningstar:2003gk}, with the
Kogut-Susskind or staggered fermion formulation \cite{Kogut:1974ag,
Banks:1975gq, Banks:1976ia, Susskind:1976jm} with improvements, and
with the more recently implemented chiral methods that include
domain-wall fermions \cite{Kaplan:1992bt,Shamir:1993zy,Furman:1994ky} and
overlap fermions \cite{Narayanan:1994gw, Neuberger:1997fp}.  Other
improvements also in production use are Wilson quarks with HYP smearing
to reduce lattice artifacts \cite{Hasenfratz:2007rf,Schaefer:2007dc},
or to approximate good chiral behavior \cite{Gattringer:2000js}.

In this article, we review a ten-year research program founded on a
particular improvement of staggered fermions called ``asqtad''
\cite{Blum:1996uf, Lepage:1997id, Orginos:1998ue, Lagae:1998pe, Orginos:1999cr, Bernard:1999xx}
(named for its ${\mathcal O}(a^2)$ level of improvement and its inclusion
of a ``tadpole'' renormalization).  Over this time,
the MILC Collaboration has created
significant library of gauge field configuration ensembles with the
full complement of the light sea quarks $u$, $d$, and $s$. 
The masses of the $u$ and $d$ quarks have been taken to be equal,
which has a negligible effect ($< 1\%$) on isospin-averaged quantities.
In planning the parameters of these ensembles, an attempt has been made
to address the three primary sources of systematic errors in lattice QCD
calculations: the chiral and continuum extrapolations and finite size
effects. It is straightforward to perform simulations with the mass of the
$s$ quark close to its physical value, and in most of the ensembles
that has been done. However, up to now it has been too computationally 
expensive to perform simulations at the physical mass of the $u$ and $d$ quarks.
Instead, ensembles have been generated with a range of light quark masses
in order to perform extrapolations to the chiral (physical value of the
$u$ and $d$ quark mass) limit guided by chiral perturbation theory.
Simulations have been performed with six values of the lattice spacing in 
order to enable controlled extrapolations to the continuum (zero lattice 
spacing) limit, and in almost all cases the physical size of the box in
which the simulations have been carried out  has been taken to be more than 
four times the Compton wavelength of the pion in order to minimize finite 
size effects.  Finally, because
SU(3) chiral perturbation theory converges rather slowly for the $s$
quark mass close to its physical value, a number of ensembles have been
generated with lighter than physical $s$ quark masses to improve the
chiral extrapolation.
These ensembles are publicly available, and have
been used by a number of research groups,  including our own, to
calculate a wide variety of hadronic quantities ranging from chiral
properties of light mesons to hadronic parton distributions to
semileptonic decays of mesons with a charm or bottom quark to the
spectroscopy of heavy quarkonium.

\begin{figure}
\begin{center}
\begin{tabular}{c c}
\epsfxsize=3.5in
\epsfbox[0 0 4096 4096]{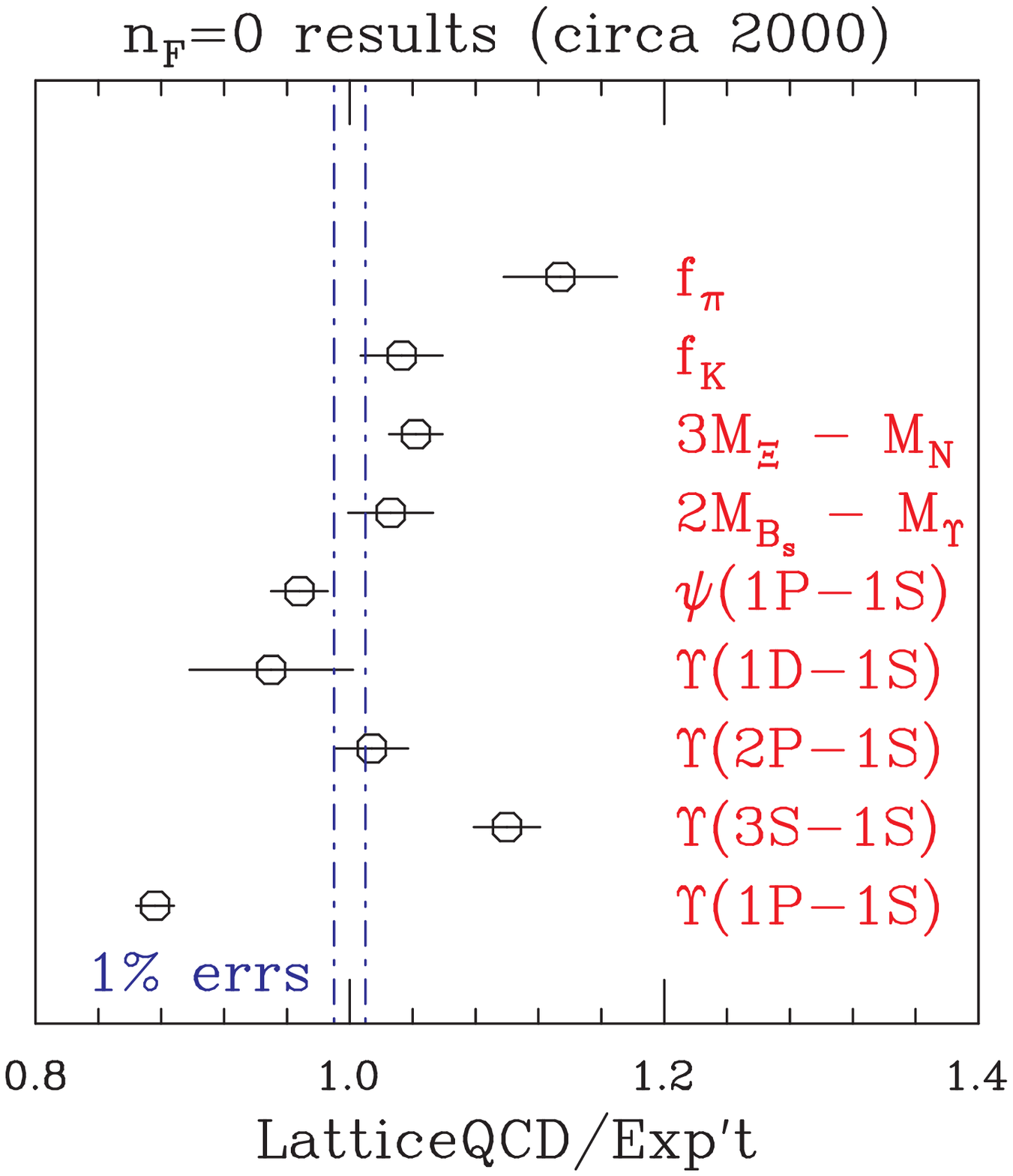}
&
\hspace{-0.3in}
\epsfxsize=3.65in
\epsfbox[0 0 4096 4096]{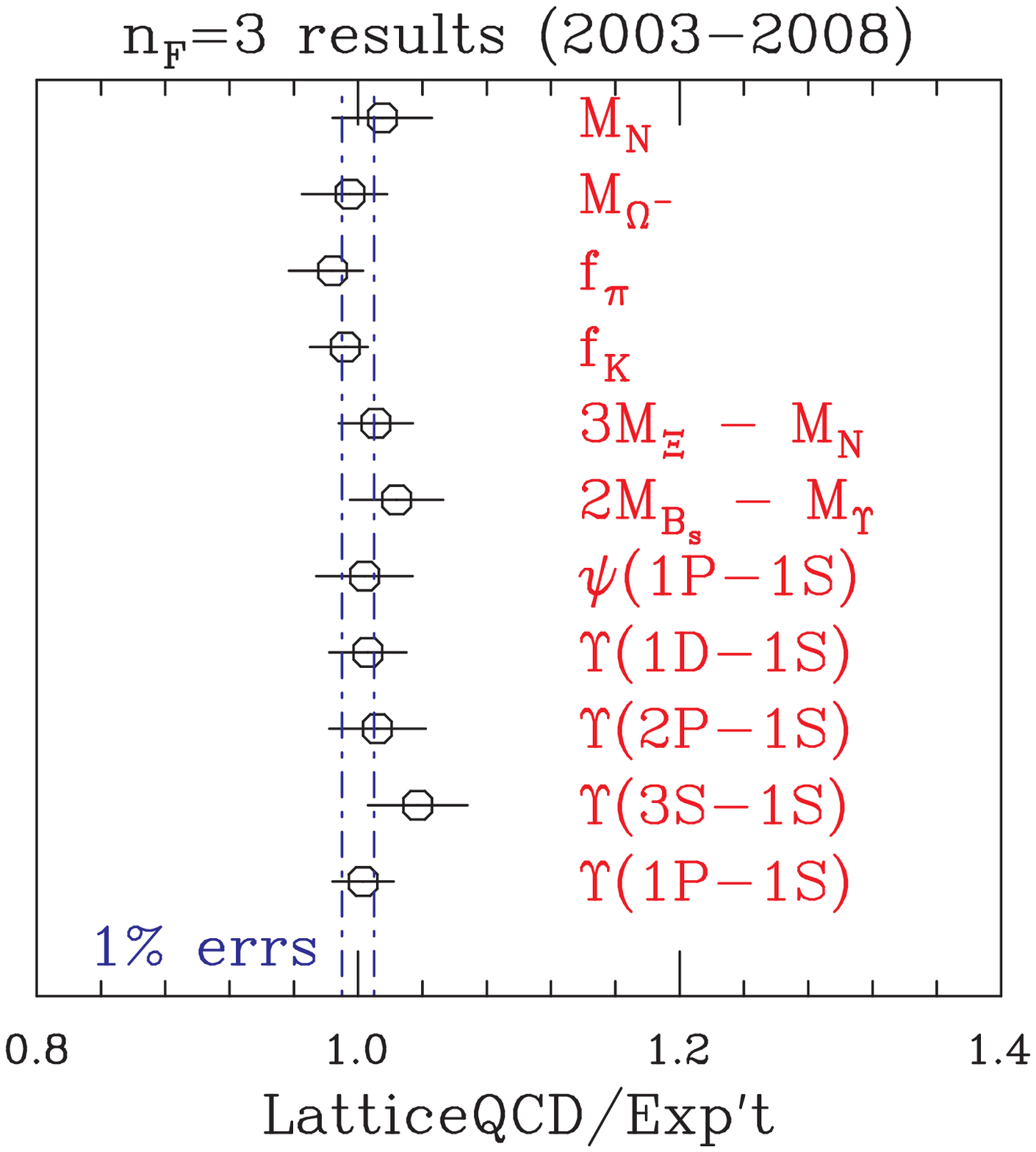}
\end{tabular}
\end{center}
\caption{Comparison of the ratio of lattice QCD and experimental values
for several observables, where the lattice QCD calculations are done
in the quenched approximation (left) and with $2+1$ flavors of asqtad
sea quarks (right). This is an updated version of a figure from
\textcite{Davies:2003ik}.
}
\label{fig:ratio-plots}
\end{figure}

The asqtad improved staggered fermion approach has enjoyed
considerable success. Its comparatively high degree of improvement and
its relatively low computational cost enabled a broad set of
full QCD  phenomenological calculations earlier 
than was possible with
other fermion methods. In Fig.~\ref{fig:ratio-plots} we illustrate the
dramatic effects of including sea quarks in a variety of physical
quantities \cite{Davies:2003ik}.
Computations with asqtad sea quarks are able 
to account for a wide variety of known decay constants, some hadronic
masses, and several quarkonium mass splittings to a precision of a few
percent \cite{Davies:2003ik}.  Their predictions for a few heavy-light
leptonic \cite{Aubin:2005ar} and semileptonic decays
\cite{Aubin:2004ej} have been experimentally confirmed.  They provide
values for the strong fine structure constant $\alpha_s$
\cite{Davies:2008sw}, the charm quark mass \cite{Davies:2008nq},
the CKM matrix elements $|V_{us}|$
\cite{Bernard:2007ps}, $|V_{cb}|$ \cite{Bernard:2008dn}, and $|V_{ub}|$
\cite{Bailey:2008wp}, and the $D^+$ and $D_s$ leptonic decay constants
\cite{Follana:2007uv} that are competitive with the most accurate
determinations to date.

In Sec.~\ref{sec:Asqtad-formalism}, we begin with a brief review of
lattice gauge theory, discussing gauge field and fermion field
formulations and numerical simulation methods.  We end
Sec.~\ref{sec:Asqtad-formalism} with an overview of the asqtad and the
more recent HISQ fermion formulations.

Section~\ref{sec:schpt-and-root} first discusses the inclusion of
staggered discretization errors in chiral perturbation theory,
resulting in ``staggered chiral perturbation theory'' (\schpt).
The application to the light pseudoscalar meson sector is described in
detail; the applications to heavy-light mesons and to a mixed-action
theory (with chiral valence quarks and staggered sea quarks) are treated
more briefly. We then turn attention to the
procedure we use to deal with the extra species that occur
for staggered
fermions.  Each staggered field (each flavor of quark)
normally gives rise to four species in the continuum limit.  
The additional
degree of freedom is called ``taste.'' To obtain the correct counting
of sea quarks it is necessary  to take the fourth-root of the fermion
determinant. This rooting procedure has been
shown to produce a theory that is nonlocal on the lattice, leading to the
legitimate question of whether the nonlocality persists as the lattice
spacing goes to zero. Such nonlocality would spoil the continuum limit,
giving a theory inequivalent to QCD. In recent years, however, there
has been a considerable amount of work on this issue, and there is now
a substantial body of theoretical and computational evidence that the
fourth-root methodology is indeed correct. We discuss some of that work
in detail in Sec.~\ref{sec:schpt-and-root}, and also explain how to take
rooting into account properly in the chiral effective theory.

In Sec.~\ref{sec:ensembles_1}, we list the ensembles of publicly
available asqtad gauge configurations, and describe tests
of their intended properties, including the determination of the lattice
scale and the topological susceptibility.
In the following sections, we review physics results obtained with 
them.
In Sec.~\ref{sec:spec_other}, we review
the spectroscopy of light hadrons other than the pseudoscalar mesons,
including vector and scalar mesons and baryons.  Section~\ref{sec:fpi}
is devoted to properties of the pseudoscalar mesons, including masses,
decay constants and Gasser-Leutwyler low energy constants.
We turn in Secs.~\ref{sec:h_l_mass} and \ref{sec:semilept} to the
masses and decays of mesons containing one heavy (charm or bottom)
quark and one light antiquark.  Section~\ref{sec:h_l_mass} treats
masses and leptonic decays; Sec.~\ref{sec:semilept}, semileptonic
decays.

In Sec.~\ref{sec:other_comps}, we review a variety of other calculations,
including the determination of the strong coupling $\alpha_s$, quarkonium
spectroscopy, the spectroscopy of baryons containing one or two heavy
quarks, $K_0-\bar K_0$
and $B_0-\bar B_0$ mixing, the muon anomalous
magnetic moment, and quark and gluon propagators.

Finally, in Sec.~\ref{sec:future}, we discuss further improvements under
way or under consideration, including the incorporation of electromagnetic
effects and the implementation of the HISQ action, and briefly
comment on future prospects for the field.

We do not review applications of the asqtad formulation to
QCD thermodynamics.
A recent article by DeTar and Heller \cite{DeTar:2009ef}
contains a review of high temperature
and nonzero density results, including those obtained using the asqtad
fermion action.

%% file: RMP_sec2.tex
% File for section 2 for RMP article
%
%\section{Section 2}
\section{Fermions on the lattice: Improved staggered formalism}
\label{sec:Asqtad-formalism}

\subsection{Brief introduction to lattice gauge theory}
\label{sec:LGT_intro}

\subsubsection{Basic setup}
\label{sec:Basics}

Euclidean, {\it i.e.}, imaginary time, field theories
can be regulated by formulating them on a
space-time lattice, with the lattice points, called sites, separated
by the lattice spacing $a$. This introduces an ultraviolet cutoff
$\pi/a$ on any momentum component. Matter fields then reside only
on the lattice sites, while the gauge fields are associated with the
links joining neighboring sites. 
The gauge fields are represented by
gauge group elements $U_\mu(x)$ on the links, which represent parallel
transporters from site $x$ to the neighboring site $x + a \hat{\mu}$,
where $\hat{\mu}$ is the unit vector in the direction $\mu$, with $\mu=1,
\dots, d$ for a $d$-dimensional lattice:
\begin{eqnarray}
U_\mu(x) &=& \mathcal{P} \exp\left\{ ig \int_{x}^{x+a \hat\mu}
 dy_\nu \, A_\nu(y) \right\}
= \exp\left\{ iga \left[A_\mu(x+a \hat{\mu}/2) + \frac{a^2}{24}
 \partial_\mu^2 A_\mu(x+a \hat{\mu}/2) + \dots \right] \right\} \nonumber \\
 &=& 1 +iag A_\mu(x+a \hat{\mu}/2) + \dots ~.
\label{eq:Umu}
\end{eqnarray}

Under gauge transformations $V(x)$, restricted to the sites of the
lattice, the gauge links transform as
\begin{equation}
U_\mu(x) \rightarrow V(x) U_\mu(x) V^\dagger(x + a \hat{\mu}) ~.
\label{eq:GT_U}
\end{equation}
The traces of products of gauge links around closed loops on the lattice,
so-called Wilson loops, are then gauge invariant. The gauge action can be
built from the sum over the lattice of combinations of small Wilson loops
with coefficients adjusted such that in the continuum limit, $a \to 0$,
it reduces to $\int d^dx \frac{1}{2} \mathrm{Tr} F^2_{\mu\nu}$ up to
terms of $\mathcal{O}(a^2)$. The simplest gauge action, the original
action introduced by \textcite{Wilson:1974sk}, consists of a sum
over plaquettes ($1 \times 1$ Wilson loops)
\begin{equation}
S_G = \frac{\beta}{N} \sum_{pl} \mathrm{Re Tr} ( 1 - U_{pl} ) ~,
\label{eq:S_G_Wils}
\end{equation}
where $\beta = 2N/g^2$, for gauge group SU($N$), with $g^2$ the bare
coupling constant.

Fermions, in Euclidean space, are represented by Grassmann fields
$\psi_x$ and $\bar\psi_x$, which in the lattice formulation reside
on the sites of the lattice. A generic fermion action 
can be written as
\begin{equation}
S_F = \sum_{x,y} \bar\psi_x M_{F;x,y} \psi_y ~,
\label{eq:S_F}
\end{equation}
where the fermion matrix $M_{F;x,y}$ is some lattice discretization of
the continuum Dirac operator $D + m$. Details of lattice fermion actions are described below.

The lattice gauge theory partition function is then given by
\begin{equation}
Z(\beta) = \int \prod_{x,\mu} dU_\mu(x) \prod_x [d\bar\psi_x d\psi_x]
\exp\{ - S_G - a^4 S_F \} ~,
\label{eq:Z_orig}
\end{equation}
where $dU_\mu(x)$ is the invariant SU($N$) Haar measure and
$d\bar\psi_x d\psi_x$ indicate integration over the Grassmann fields.

Since $S_F$ is quadratic in the fermion fields, the integration over
the Grassmann fields can be carried out, leading to (up to a trivial overall factor)
\begin{equation}
Z(\beta) = \int \prod_{x,\mu} dU_\mu(x) \det M_F \exp\{ - S_G \}
 = \int \prod_{x,\mu} dU_\mu(x) \exp\{ - S_{eff} \} ~,
\label{eq:Z}
\end{equation}
with $S_{eff} = S_G - \mathrm{Tr} \log M_F$.

The expectation value of some observable $O$ is given by
\begin{eqnarray}
\langle O \rangle &=& \frac{1}{Z(\beta)} \int \prod_{x,\mu} dU_\mu(x)
\prod_x [d\bar\psi_x d\psi_x] O \exp\{ - S_G - a^4 S_F \} \nonumber \\
&=& \frac{1}{Z(\beta)} \int \prod_{x,\mu} dU_\mu(x) O
\det M_F \exp\{ - S_G \}
 = \frac{1}{Z(\beta)} \int \prod_{x,\mu} dU_\mu(x) O \exp\{ - S_{eff} \} ~.
\label{eq:ev_O}
\end{eqnarray}
If the observable $O$ involves fermion fields $\psi_x$ and
$\bar\psi_y$ then, in the second line of Eq.~(\ref{eq:ev_O}) each
pair is replaced by $M^{-1}_{F;x,y}$ in all possible combinations
with the appropriate minus signs for Wick contractions of fermion
fields.

\subsubsection{Improved action}
\label{sec:Imp_acts}

As mentioned before Eq.~(\ref{eq:S_G_Wils}), the typical gauge action
on the lattices reduces to the continuum action up to terms of
$\mathcal{O}(a^2)$. These terms lead to $\mathcal{O}(a^2)$ deviations from
the continuum result of physical observables computed at finite lattice
spacing. These $\mathcal{O}(a^2)$ effects can be reduced by using an
improved gauge action (together with improved operators, where necessary)
in an improvement program initiated by
\textcite{Symanzik:1980UH, Symanzik:1983dc}.

\begin{figure}[h]
\centering
\includegraphics[scale=0.4]{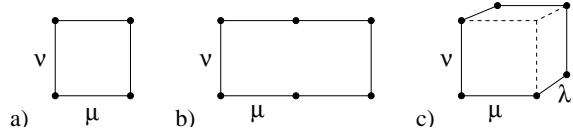}
\caption{\small L\"uscher-Weisz action Wilson loops: a) standard
plaquette, b) $2\times1$ rectangle and c) $1\times1\times1$ parallelogram}
\label{fig:lwactloops}
\end{figure}

For the gauge action, the improvement can be achieved by adding $2\times1$
(planar) rectangle (labeled ``$rt$'') and generalized 3-d all $1\times1\times1$
parallelogram (labeled ``$pg$'') Wilson loop terms (see
Fig.~\ref{fig:lwactloops}) to the Wilson action, Eq.~(\ref{eq:S_G_Wils}),
with coefficients computed, at one-loop order in perturbation theory,
by \textcite*{Luscher:1985zq, Luscher:1984xn},
\begin{equation}
S_{LW} = \frac{\beta}{N} \left\{
 \sum_{pl} c_{pl} \mathrm{Re Tr} ( 1 - U_{pl} ) +
 \sum_{rt} c_{rt} \mathrm{Re Tr} ( 1 - U_{rt} ) +
 \sum_{pg} c_{pg} \mathrm{Re Tr} ( 1 - U_{pg} ) \right\} ~.
\label{eq:S_LW_1}
\end{equation}
The coefficients, $c_i=c_i^{(0)} + 4\pi\alpha_0 c_i^{(1)}$ at one loop,
can be found in Table 1 of \textcite{Luscher:1985zq}.

Bare lattice perturbation theory results generally converge slowly
but can be improved by using tadpole-improved perturbation
theory \cite{Lepage:1992xa}. This starts with using a more
continuum-like gauge link $U_\mu \rightarrow \tilde U_\mu =
u_0^{-1} U_\mu$. The so-called tadpole factor $u_0$ is determined
in numerical simulations either as the expectation value of $U_\mu$
in Landau gauge or, more commonly, from the expectation value of the
average plaquette
\begin{equation}
u_0= \langle \frac{1}{N}\mathrm{Re Tr}U_{pl} \rangle^{1/4}.
\label{eq:u0_plaq}
\end{equation}
The L\"uscher-Weisz action can now be tadpole improved by explicitly
pulling a $u_0^{-1}$ factor out of each link and replacing $\alpha_0$
in the one-loop perturbative coefficients $c_i$ with a nonperturbatively
renormalized coupling $\alpha_s$ defined,
for gauge group SU(3), in terms of the
measured lattice value of $u_0$ by
\begin{equation}
\alpha_s \equiv - 1.303615 \log u_0 ~,
\label{eq:alpha_u0}
\end{equation}
where the proportionality factor is determined by
the one-loop expression for $\log u_0$.
Defining $\beta_{LW} \equiv u_0^{-4} \beta c_{pl}$, since $U_{pl}$
involves the product of four links, the improved action can be written as
\cite{Alford:1995hw}
\begin{equation}
S_{LW} = \frac{\beta_{LW}}{3} \left\{
 \sum_{pl} \mathrm{Re Tr} ( 1 - U_{pl} ) -
 \sum_{rt} \frac{[1+0.4805 \alpha_s]}{20u_0^2} \mathrm{Re Tr} ( 1 - U_{rt} ) -
 \sum_{pg} \frac{0.03325 \alpha_s}{u_0^2} \mathrm{Re Tr} ( 1 - U_{pg} )
 \right\} ~.
\label{eq:S_LW_2}
\end{equation}
Since higher perturbative orders in the coefficients are neglected,
the one-loop improved L\"uscher-Weisz action, Eq.~(\ref{eq:S_LW_2}),
leads to remaining lattice artifacts of $\mathcal{O}(\alpha_s^2 a^2)$.
Sometimes, only a tree-level improved action without the terms
proportional to $\alpha_s$ in Eq.~(\ref{eq:S_LW_2}) is used, leading to
lattice artifacts of $\mathcal{O}(\alpha_s a^2)$. Since the
parallelogram terms are then absent such simulations are somewhat
faster.
It should be noted that Eq.~(\ref{eq:S_LW_2}) does not include
the one-loop contributions from dynamical fermions, which were unknown
at the time the MILC collaboration started the $2+1$-flavor simulations
reviewed in this article. Therefore, for those simulations, the
leading lattice artifacts in the gauge sector are
$\mathcal{O}(\alpha_s a^2)$ as in the fermion sector, described later.
The one-loop fermion contribution has recently been computed by
\textcite*{Hao:2007iz}.

\subsection{Fermions on the lattice}
\label{sec:Latt_ferm}

\subsubsection{The doubling problem}
\label{sec:Doubling}

Putting fermions on a lattice, one replaces the covariant derivative
in the continuum fermion action with a covariant (central) difference
\begin{equation}
S_{naive} = \sum_x \bar\psi(x) \left\{ \sum_\mu \gamma_\mu \nabla_\mu \psi(x)
 +  m\, \psi(x) \right\} ~,
\label{eq:S_naive}
\end{equation}
where
\begin{equation}
\nabla_\mu \psi(x) = \frac{1}{2a} \left( U_\mu(x) \psi(x + a \hat{\mu})
 - U^\dagger_\mu(x - a \hat{\mu})  \psi(x - a \hat{\mu}) \right) ~.
\label{eq:D_naive}
\end{equation}
The inverse propagator in momentum space derived from the action
Eq.~(\ref{eq:S_naive}) in the free case, with all link fields $U_\mu = 1$,
is
\begin{equation}
a S^{-1}(ap) = i \sum_\mu \gamma_\mu \sin(ap_\mu) + a m ~.
\label{eq:S_inv_naive}
\end{equation}
In the massless case, this inverse propagator not only vanishes when
$p = 0$, but also when $p_\mu = 0$ or $p_\mu = \pi/a$ for each
$\mu = 1, \dots, 4$, {\it i.e.,} on all 16 corners of the Brillouin zone
in $d=4$ dimensions.  Thus, when we try to put one fermion on the lattice
we actually get 16 in the continuum limit. This is the
infamous doubling problem of lattice fermions.

\subsubsection{Wilson fermions}
\label{sec:Wilson_ferm}

This doubling problem was recognized by Wilson when he first formulated
lattice gauge theories. He also proposed a solution: adding an irrelevant
term --- a term that vanishes in the continuum limit, $a \to 0$
\cite{Wilson:1975hf}
\begin{equation}
S_W = S_{naive} - \frac{a r}{2} \sum_x \bar\psi(x) \sum_\mu \Delta_\mu \psi(x)
 = \bar\psi D_W(m) \psi~,
\label{eq:S_W}
\end{equation}
where $r$ is a free parameter, usually set to $r=1$, and the Laplacian is
\begin{equation}
\Delta_\mu \psi(x) = \frac{1}{a^2} \left( U_\mu(x) \psi(x + a \hat{\mu})
+ U^\dagger_\mu(x - a \hat{\mu})  \psi(x - a \hat{\mu}) - 2 \psi(x)
\right) ~.
\label{eq:Lap}
\end{equation}
The free inverse propagator now is
\begin{equation}
a S^{-1}(ap) = i \sum_\mu \gamma_\mu \sin(ap_\mu) + a m
 - r \sum_\mu \left( \cos(ap_\mu) - 1 \right) ~.
\label{eq:S_inv_W}
\end{equation}
The doublers, with $n$ momentum components $p_\mu = \pi/a$, now attain
masses $m + 2nr/a$, and only one fermion, with $p \approx 0$, remains
light.

We note that the Wilson Dirac operator is $\gamma_5$-Hermitian,
\begin{equation}
D_W^\dagger(m) = \gamma_5 D_W(m) \gamma_5 ~.
\label{eq:g5_h}
\end{equation}
Thus $\det D_W^\dagger(m) = \det D_W(m)$, implying that two flavors --- and
by extension any even number of flavors of Wilson fermions  --- lead to a
manifestly positive (semi-) definite fermion determinant,
$\det [D_W^\dagger(m) D_W(m)]$.

The price for eliminating the doubling problem in this Wilson fermion
approach is that the action Eq.~(\ref{eq:S_W}) violates the chiral
symmetry $\delta \psi = i \alpha \gamma_5 \psi$, $\delta \bar\psi
= i \alpha \bar\psi \gamma_5$ of massless fermions (with $\alpha$
an infinitesimal parameter). As a consequence,
the massless limit of fermions is no longer protected --- the mass gets
an additive renormalization; to get massless quarks requires a fine
tuning of the bare mass parameter.

According to the usual, renormalization group based universality
arguments, the chiral symmetry, broken at finite lattice spacing only by
an irrelevant, dimension-five operator, will be recovered in the continuum
limit after fine tuning of the bare mass parameter. But
the explicit violation
of chiral symmetry allows the generation of
other contributions to dimension-five operators
which are suppressed by only one power of the lattice spacing $a$.
The lattice artifacts for Wilson fermions are therefore of
$\mathcal{O}(a)$, rather than $\mathcal{O}(a^2)$ as in the pure
gauge sector.

Besides $\bar\psi(x) \Delta \psi(x)$, with $\Delta = \sum_\mu \Delta_\mu$,
there is a second dimension-five (chiral symmetry breaking) operator
\begin{equation}
S_{SW} = \frac{iag}{4} c_{SW} \sum_x \bar\psi(x) 
\sigma_{\mu\nu} \mathcal{F}_{\mu\nu}(x) \psi(x) ~,
\label{eq:S_SW}
\end{equation}
where $\mathcal{F}_{\mu\nu}(x)$ is a lattice representation of the
field strength tensor $F_{\mu\nu}(x)$, and $\sigma_{\mu\nu} =
\frac{i}{2}[ \gamma_\mu, \gamma_\nu]$. Inclusion of Eq.~(\ref{eq:S_SW})
into the fermion action, with properly adjusted coefficient $c_{SW}$, was
proposed by \textcite*{Sheikholeslami:1985ij}
to eliminate the $\mathcal{O}(a)$ effects of the Wilson fermion action.
Since $\mathcal{F}_{\mu\nu}(x)$ on the lattice is usually represented
by a ``clover leaf'' pattern of open plaquettes, the action including
the term Eq.~(\ref{eq:S_SW}) is commonly referred to as the clover
action.

The appropriate coefficient $c_{SW}$ of the clover term, Eq.~(\ref{eq:S_SW}),
can be computed in perturbation theory \cite{Wohlert:1987rf, Luscher:1996vw},
or even better, nonperturbatively \cite{Luscher:1996sc, Luscher:1996ug}
-- truly reducing the remaining lattice effects from $\mathcal{O}(a)$
to $\mathcal{O}(a^2)$.

Another problem with Wilson fermions is that, because of the additive
mass renormalization, the fermion determinant $\det D_W(m)$ is not
positive definite even for putative positive quark mass. Configurations
with $\det D_W(m) \approx 0$ can occur, called exceptional configurations,
which can slow down numerical simulations considerably. A formulation
that removes such exceptional configurations, introduced by Frezzotti
{\it et al.}~\cite{Frezzotti:1999vv, Frezzotti:2000nk, Frezzotti:2003ni} is called
``twisted-mass QCD''. For two flavors one considers the Dirac operator
\begin{equation}
D_{twist} = D + m + i \mu \gamma_5 \tau_3 ~,
\label{eq:S_twist}
\end{equation}
where the isospin generator $\tau_3$ acts in flavor space. In the
continuum, the twisted-mass Dirac operator is equivalent to a usual
Dirac operator with mass $\sqrt{m^2 + \mu^2}$. On the lattice,
however, with $D$ replaced by the (massless) Wilson Dirac operator
$D_W(0)$ of Eq.~(\ref{eq:S_W}), the twisted-mass term ensures a
positive-definite two-flavor determinant, 
$\det[D_W^\dagger(m) D_W(m) + \mu^2] > 0$.
An added benefit of the twisted-mass (Wilson) fermion formulation
is, that at maximal twist $\tan \alpha = \mu/m$, the twisted-mass
Wilson Dirac operator is automatically $\mathcal{O}(a^2)$ improved
\cite{Frezzotti:2003ni}. Unfortunately, the real part of the mass $m$
still receives an additive renormalization so that achieving maximal
twist requires a fine tuning. Furthermore, at finite lattice spacing,
isospin symmetry is broken, making the $\pi^0$ mass different from
the mass of the $\pi^\pm$.

\subsubsection{Staggered fermions}
\label{sec:Staggered_ferm}

Another way of dealing with the doubling problem, alleviating though
not eliminating it, is the staggered fermion formalism
\cite{Kogut:1974ag, Banks:1975gq, Banks:1976ia, Susskind:1976jm}.
One introduces a new fermion field by
\begin{equation}
\psi(x) = \Gamma_x \chi(x) ~~~,~
 \bar\psi(x) = \bar\chi(x) \Gamma^\dagger_x ~,
\label{eq:chi}
\end{equation}
with
\begin{equation}
\Gamma_x = \gamma_1^{(x_1/a)}\; \gamma_2^{(x_2/a)}\; \gamma_3^{(x_3/a)}\; \gamma_4^{(x_4/a)} ~.
\label{eq:Omeg}
\end{equation}
Using $\Gamma^\dagger_x \Gamma_x = 1$ and
\begin{equation}
\Gamma^\dagger_x \gamma_\mu \Gamma_{x+a\mu}  =
 (-1)^{(x_1 + \dots + x_{\mu-1}) / a}  \equiv \eta_\mu(x) ~,
\label{eq:eta_mu}
\end{equation}
the naive fermion action, Eq.~(\ref{eq:S_naive}), can be written as
\begin{equation}
S_{KS} = \sum_x \bar\chi(x) \left\{ \sum_\mu \eta_\mu(x)\; \nabla_\mu\; \chi(x)
 +  m \chi(x) \right\} \equiv \bar \chi\; (D_{KS}+m)\;\chi ~,
\label{eq:S_KS}
\end{equation}
where matrix multiplication is implied in the final expression.
Here, the four Dirac components decouple from each other, and the
fermion field $\chi(x)$ can be restricted to a single component,
thereby reducing the doubling by a factor of four, from sixteen
to four. 
It is, in principle, possible to interpret these four
remaining degrees of freedom as physical flavor  ($u$, $d$, $s$, $c$),
but, in order to give different masses to the flavors,
one must introduce general mass terms coupling nearby sites
\cite{Golterman:1984cy,Gockeler:1984rq}.
That approach then leads to a variety of practical problems including complex
determinants
and the necessity of fine tuning.

Instead, we follow modern usage and refer to the quantum number labeling
the four remaining fermion species as ``taste,''
which, unlike flavor, is an unwanted degree of freedom that must be removed.
We describe how this removal is accomplished by the so-called
``fourth-root procedure'' at the end of this section, and discuss it in more detail
in Sec.~\ref{sec:Rooting}.
If more than one physical
flavor is required, as is, of course, the case for simulations of QCD, one
then needs to introduce a separate staggered field for each flavor. 
For example, for QCD with three light flavors, one employs 
three staggered fields, $\chi_u$, $\chi_d$, and $\chi_s$.%
\footnote{In practice, since one usually takes $m_u=m_d\not=m_s$, the $u$ and $d$
fields can be simulated together, and one can use 
only two staggered fields.  For clarity, we ignore 
this technical detail in our exposition.}
However, for simplicity, we consider only a single staggered field (one flavor)
in the remainder of this section.

 The one-component fermions
with action Eq.~(\ref{eq:S_KS}) are referred to as (standard) staggered
or Kogut-Susskind fermions. The ``standard'' distinguishes them from
improved versions, described later on.

An important discrete symmetry of the staggered fermion action, Eq.~(\ref{eq:S_KS}),
is shift symmetry \cite{vandenDoel:1983mf, Golterman:1984cy}
\begin{eqnarray}
\chi(x) &\to& \rho_\mu(x)\; \chi (x +a\hat \mu) \nonumber \\
\bar \chi(x) &\to& \rho_\mu(x)\; \bar \chi (x +a\hat \mu) \nonumber \\
U_\nu(x ) &\to& U_\nu(x +a\hat \mu) \ , 
\eqn{shifts}
\end{eqnarray}
with the phase $\rho_\mu(x)$ defined by
$\rho_\mu(x) =  (-1)^{(x_{\mu+1}+ \dots + x_4) / a}$.
Additional
discrete symmetries of the staggered action are $90^\circ$ rotations, axis
inversions, and charge conjugation. 
In the continuum limit, these symmetries are expected to enlarge to a direct product
of the Euclidean Poincar\'e group and a vector SU(4)$_V$ among the 
tastes (plus parity and charge conjugation) \cite{Golterman:1984cy}.

For massless quarks, $m=0$, the staggered fermion action 
also has a continuous even/odd U(1)$_e\times$U(1)$_o$ chiral symmetry \cite{Kawamoto:1981hw,
KlubergStern:1981wz, KlubergStern:1982bs}, a remnant
of the usual chiral symmetry for massless fermions in the continuum.
The U(1)$_e\times$U(1)$_o$ symmetry is
\begin{eqnarray}
\chi(x) \rightarrow \exp\{ i \alpha_e \} \chi(x) ~,~~~
 \bar\chi(x) \rightarrow \bar\chi(x) \exp\{ -i \alpha_o \} ~~~~ &~&
 \mathrm{for} ~ x = \mathrm{even} ~, \nonumber \\
\chi(x) \rightarrow \exp\{ i \alpha_o \} \chi(x) ~,~~~
 \bar\chi(x) \rightarrow \bar\chi(x) \exp\{ -i \alpha_e \} ~~~~ &~&
 \mathrm{for} ~ x = \mathrm{odd} ~,
\label{eq:U1_x_U1}
\end{eqnarray}
where $\alpha_e$ and $\alpha_o$ are the symmetry parameters, and a site $x$ is 
called even or odd if $\sum_\mu (x_\mu/a)$ is even or odd.
The ``axial part'' of this symmetry, $\alpha_e = -\alpha_o \equiv \alpha_\epsilon$,
is known as U(1)$_\epsilon$ symmetry \cite{Kawamoto:1981hw} and takes the form
\begin{equation}
\chi(x) \rightarrow \exp\{ i \alpha_\epsilon \epsilon(x) \} \chi(x) ~,~~~
 \bar\chi(x) \rightarrow \bar\chi(x) \exp\{i \alpha_\epsilon \epsilon(x) \}\qquad
 \mathrm{with}\  \epsilon(x)\equiv (-1)^{\sum_\mu (x_\mu/a) }\ . 
\eqn{U1eps_chi}
\end{equation}
The chiral symmetry, Eq.~(\ref{eq:U1_x_U1}) or \eq{U1eps_chi},
protects the mass term in \eq{S_KS} from
additive renormalization, while the discrete symmetries (especially shift symmetry)
are also needed to prevent other mass terms (coupling $\chi$ and $\bar \chi$
at nearby sites) from arising \cite{Golterman:1984cy}. In particular, an alternative
version of staggered quarks called  the ``Dirac-K\"ahler action'' \cite{Becher:1982ud} 
does not have shift symmetry and therefore generates a mass term at one loop even
when $m=0$ \cite{Mitra:1983bi}.

The even/odd symmetry
is spontaneously broken to the diagonal vector U(1)$_V$ (quark number)
symmetry, $\alpha_e = \alpha_o$, with an ensuing Goldstone boson. In addition, the
mass term breaks the U(1)$_e\times$U(1)$_o$ symmetry explicitly, giving mass
to the Goldstone boson, $m^2_G \propto m$.

The staggered Dirac operator $D_{KS}$ in \eq{S_KS} obeys \cite{Smit:1986fn}
\begin{equation}
D^\dagger_{KS} =  -D_{KS} = \epsilon\; D_{KS} \;\epsilon  \ ,
\eqn{DKS-eps}
\end{equation}
where $\epsilon$ is a diagonal matrix in position space with $\epsilon(x)$ along the diagonal,
and the second equality follows 
from the fact that $D_{KS}$ connects only
even and odd sites.
The fact that $D_{KS}$ is antihermitian implies that its eigenvalues are purely imaginary;
the $\epsilon$ relation in \eq{DKS-eps}
then tells us that 
the nonzero eigenvalues come in complex-conjugate pairs.
For $m>0$, which is the case of interest here,
this ensures that the staggered determinant $\det(D_{KS}+m)$ is 
strictly positive.%
\footnote{We do not expect any exact zero modes on generic configurations, even
those with net topological charge. Such configurations will in general
have only some near-zero ($\cO(a)$ or smaller) eigenvalues. So, in fact, the determinant
should be positive even for $m<0$. This is different
from the case of chiral fermions discussed in \secref{Chiral_ferm}.} Note that
the continuum Euclidean Dirac operator $D_{cont}$ is also
antihermitian and obeys a corresponding equation
\begin{equation}
D^\dagger_{cont} =  -D_{cont} = \gamma_5 \; D_{cont} \;\gamma_5   \ ,
\eqn{Dcont-gamma5}
\end{equation}
which similarly (but now only formally) results in a positive determinant for positive
quark mass.

The one-component staggered fermion fields $\chi(x)$ can be assembled
into Dirac fields $q(y)$, living on $2^4$ hypercubes of the original
lattice, labeled by $y$, with corners $x = 2y + aA$, where $A_\mu = 0, 1$
\cite{Gliozzi:1982ib, Duncan:1982xe, KlubergStern:1983dg}.  One has
\begin{equation}
q(y)_{\alpha i} = \frac{1}{8} \sum_A (\Gamma_A)_{\alpha i}\; U_A(y) \;\chi(2y + aA) ~,
\qquad \bar q(y)_{i \alpha} = \frac{1}{8} \sum_A \bar \chi(2y + aA) \;U^\dagger_A(y) \;(\Gamma_A)^\dagger_{i\alpha} ~,
\label{eq:q_Kluberg}
\end{equation}
where $\alpha,\; i$ label the Dirac and taste indices, respectively, and
$U_A(y)$ is a product of the gauge links over some fixed path from
$2y$ to $2y+aA$.
Bilinear quark operators, with spin structure $\gamma_s = \Gamma_s$
and taste structure $\xi_t = \Gamma^*_t$ are defined by \cite{Sharpe:1993ur}
\begin{equation}
\mathcal{O}_{st} = \bar q(y) (\gamma_s \otimes \xi_t) q(y)
 = \frac{1}{16} \sum_{A,B} \bar\chi(2y+aA)\; U^\dagger_A(y) \;U_B(y) \; \chi(2y+aB) \;
 \frac{1}{4} \mathrm{tr} \left( \Gamma^\dagger_A \Gamma_s\Gamma_B
 \Gamma^\dagger_t \right ) ~.
\label{eq:O_ST}
\end{equation}

In the free case (all $U_\mu(x)=1$),
the quark action in \eq{S_KS} can be expressed 
in terms of the fields $q(y)$ as \cite{KlubergStern:1983dg}
\begin{equation}
S_{KS} = 16  \sum_y \bar q(y) \left\{  m  (I\otimes I)  + \sum_\mu \left[ 
\left(\gamma_\mu \otimes I\right )\nabla_\mu  
+ a\left(\gamma_5 \otimes \xi_\mu\xi_5\right)\Delta_\mu\right]
\right\} q(y) \ ,
\eqn{action_Kluberg}
\end{equation}
where $I$ is the identity matrix, the factor of $16$ arises from the fact that
there are $1/16$ as many $y$ points as $x$ points,
and $\nabla_\mu$ and $\Delta_\mu$ are
the free-field versions of \eqs{D_naive}{Lap}, but acting on the doubled ($y$)
lattice:
\begin{eqnarray}
\nabla_\mu f(y) &=&  \frac{1}{4a}\left[f(y + 2a\hat\mu) - f(y - 2a\hat\mu)\right] \ ,\nonumber \\
\Delta_\mu f(y) &=&  \frac{1}{4a^2}\left[f(y + 2a\hat\mu) -2f(y) + f(y - 2a\hat\mu)\right]  \ .
\eqn{derivatives}
\end{eqnarray}
These derivatives go to $\partial_\mu f(y)$ and  $\partial^2_\mu f(y)$, respectively,
in the continuum limit.  
In the interacting case there is another dimension-five, $\cO(a)$, term, involving the
field-strength tensor $\cF_{\mu\nu}$, in addition to the
$\Delta_\mu$ term in \eq{action_Kluberg}. There are also higher
contributions of $\cO(a^2)$ starting at dimension six \cite{KlubergStern:1983dg}.

In the $\nabla_\mu$ (first derivative) kinetic energy term of \eq{action_Kluberg},
the even/odd U(1)$_e\times$U(1)$_o$ symmetry is enlarged to a full continuous chiral symmetry,
U(4)$_L \times$U(4)$_R$, acting on the taste indices of the right and left
fields,
$q_R(y) = \frac{1}{2}(1+\gamma_5) q(y)$ and
$q_L(y) = \frac{1}{2}(1-\gamma_5) q(y)$.  
The mass term breaks this down to
an SU(4)$_V$ vector taste symmetry (plus the U(1)$_V$ of quark number).  
On the other hand, because of the explicit taste matrices,
the second derivative term in \eq{action_Kluberg} breaks the 
full chiral symmetry to the U(1)$_e\times$U(1)$_o$ symmetry (plus the 
discrete staggered symmetries).  Because these are all symmetries of the original
staggered action, they remain symmetries in the taste basis, even when the additional
terms that appear in  \eq{action_Kluberg} in the interacting case are taken into account.

The key point is that, in the interacting theory, 
one can split the staggered Dirac operator in the taste basis
as:
\begin{equation}
D_{KS}  = D \otimes I  + a\Delta \ ,
\eqn{D_KS}
\end{equation}
where $I$ is here the ($4\times4$) identity matrix in taste space, and $\Delta$
is the taste-violating (traceless) part,
with minimum dimension five.
One expects the SU(4)$_V$ vector taste symmetry to be restored in the continuum limit
because $\Delta$ should be
irrelevant in the renormalization-group sense.

In the free case, 
the shift symmetry, \eq{shifts}, takes the form for the Dirac fields $q(y)$ \cite{Luo:1996vt}:
\begin{eqnarray}
q(y)& \to & \frac{1}{2}\left((I\otimes\xi_\mu+\g_5\g_\mu\otimes\xi_5)q(y)
+(I\otimes\xi_\mu-\g_5\g_\mu\otimes\xi_5)q(y+2a\hat\mu)\right)\ , \\
\bar q(y)& \to & \frac{1}{2}\left(\bar q(y) (I\otimes\xi_\mu-\g_5\g_\mu\otimes\xi_5)
+\bar q(y+2a\hat\mu) (I\otimes\xi_\mu+\g_5\g_\mu\otimes\xi_5))\right)\ .
\eqn{tasteshifts}
\end{eqnarray}
As the continuum limit is approached, shifts become simply multiplication by
the taste matrix $\xi_\mu$, plus higher-dimension terms involving derivatives.
Thus shifts are basically discrete vector taste transformations, coupled with
translations.  

In the taste basis, 
the axial U(1)$_\epsilon$ symmetry is 
\begin{equation}
q(y) \rightarrow \exp\left\{ i \alpha_\epsilon
\left(\gamma_5\otimes\xi_5\right) \right\} q(y) ~,\qquad
 \bar q(y) \rightarrow \bar q(y) \exp\left\{ i \alpha_\epsilon  
       \left(\gamma_5\otimes\xi_5\right)  \right\}  \ .
\eqn{U1eps_q}
\end{equation}
Because of the $\xi_5$, this is clearly a taste nonsinglet axial symmetry, and hence is 
nonanomalous. 
The anomalous axial symmetry U(1)$_A$ must be a taste-singlet: 
\begin{equation}
q(y) \rightarrow \exp\left\{ i \alpha_A
\left(\gamma_5\otimes I\right) \right\} q(y) ~,\qquad
 \bar q(y) \rightarrow \bar q(y) \exp\left\{ i \alpha_A
       \left(\gamma_5\otimes I\right)  \right\}  \ .
\eqn{U1_A}
\end{equation}
Indeed, this symmetry is not an invariance of the staggered lattice action 
in the massless limit, and the
symmetry violations generate, through the triangle graph, the correct axial anomaly in
the continuum limit \cite{Sharatchandra:1981si}.

The bilinear quark operators in \eq{O_ST} can create (or annihilate) mesons.
Therefore, for staggered quarks, each meson kind with given spin
(Dirac) structure $\Gamma_s$ ({\it e.g.}, $\Gamma_s = \gamma_5$ for
the pion, $\Gamma_s = \gamma_k$ for the rho, {\it etc.}) comes in sixteen
varieties, labeled by the taste index $t$. In the continuum limit all
nonsinglet mesons of a given spin are degenerate%
\footnote{Mesons that are singlets under taste and any additional
flavor symmetries need not be degenerate with the nonsinglet
mesons, since they can have physically
distinct disconnected contributions to their propagators. The most
important example is the $\eta'$, which will get a contribution
from the anomaly and have a
mass in the continuum limit different from that of all other pseudoscalars.}
  -- SU(4)$_V$ taste symmetry
connects them.
But at nonzero lattice spacing, there is only the staggered symmetry
group, the group of the discrete symmetries of the staggered
action (shifts, $90^\circ$ rotations, axis inversions, charge conjugation) plus
the U(1)$_V$  of quark number,  which are remnants of the 
continuum  Poincar\'e, taste SU(4)$_V$, quark number, and discrete
symmetries.
Meson states may be classified under  the subgroup of the staggered symmetry group,
the ``staggered rest frame symmetry
group,'' which is the symmetry group of the
transfer matrix \cite{Golterman:1985dz, Golterman:1986jf}.
The sixteen tastes
of a meson with given spin structure are not degenerate at finite
lattice spacing, but are split according to irreducible representations
of the rest frame group. In particular, only the pion with
pseudoscalar taste structure $\xi_t = \gamma^*_5$ is a Goldstone boson,
denoted by $\pi_P$ ($P$ stands for pseudoscalar taste), 
whose mass vanishes for massless quarks, $m = 0$.
To leading order in the chiral expansion (see \secref{SChPT})
the other tastes have masses
\begin{equation}
m^2_{\pi_t} = m^2_{\pi_P} + a^2 \delta_t  = 2Bm + a^2 \delta_t ~,
\label{eq:taste-split}
\end{equation}
with $B$ a low energy constant and $\delta_t$ a taste-dependent splitting
that is independent of $a$ (up to logarithms) for small $a$.
The non-Goldstone pions become degenerate with the
Goldstone pion only in the continuum limit. The taste violations in the pion
system are found to be larger than those for other hadrons \cite{Ishizuka:1993mt}.

Since staggered fermions have only one (spin) component per lattice site,
and since they have a remnant chiral symmetry that insures positivity
of the fermion determinant at positive quark mass, they are one of the cheapest
fermion formulations to simulate numerically. 
The main drawback is the need to 
eliminate the unwanted extra tastes, using 
the so-called ``fourth-root procedure.'' Each continuum fermion species
gives a factor of $\det M_F$ in the partition function, \eq{Z}.
Therefore, to reduce the contribution from four tastes to a single one,
we take the fourth root of the determinant,
$\left( \det M_{KS} \right)^{1/4}$, where $M_{KS}= D_{KS}+m\otimes I$, with
$D_{KS}$ given in \eq{D_KS}.
The procedure was first introduced
in the two dimensional version of staggered fermions (where it is a
``square-root procedure'' because there are only two tastes) by
\textcite*{Marinari:1981qf}. 
The point is that the Dirac operator $D_{KS}$ (and hence $M_{KS}$) 
should become block diagonal in taste space in the continuum limit because
$\Delta$ is an irrelevant operator.
The fourth-root procedure then becomes equivalent to
replacing the $D_{KS}$ by its restriction to a single taste.
Conversely, the nontriviality of the prescription arises because taste
symmetry is broken at nonzero lattice spacing.
In Sec.~\ref{sec:Rooting},
we discuss the status of this procedure and the evidence that it
accomplishes the goal of producing, in the continuum limit,
a single quark species with a local action.

\subsubsection{Chirally invariant fermions}
\label{sec:Chiral_ferm}

None of the ways of dealing with the fermion doubling problem outlined so far
are entirely satisfactory. Wilson-type fermions explicitly break
chiral symmetry, and staggered fermions have a remaining doubling
problem, requiring the fourth-root procedure, that continues to be somewhat
controversial because of the broken taste symmetry at finite lattice
spacing.

Indeed, the chiral anomaly implies that no lattice action can have
an exact flavor-singlet chiral symmetry \cite{Karsten:1980wd}.
There is even a no-go theorem
\cite{Nielsen:1981hk} that states that the doubling can not be avoided
with
an ultralocal\footnote{We denote by ``ultralocal'' an action that
couples only sites a finite number of lattice spacings apart. A ``local''
action is either ultralocal, or the coupling falls off exponentially
with distance with a range of the order of a few lattice spacings,
so that the action becomes local in the continuum limit. Such ``local''
actions are believed not to change the universality class in the
renormalization group sense. Any other action is called ``nonlocal.''}
and unitary fermion action.
However, actions with a modified form of chiral symmetry on the lattice can
avoid doubling while retaining most of the desirable features of chiral symmetry.
Such actions couple arbitrarily distant
points on the lattice but with exponentially suppressed couplings,
$\exp\{ -r/r_d \}$, where $r_d$ should be of the order the lattice
spacing to ensure a local action in the continuum limit. There are three
known ways of achieving this.

The first goes under the name of ``domain-wall fermions'' and was
developed by \textcite*{Kaplan:1992bt}, \textcite*{Shamir:1993zy}, and 
\textcite*{Furman:1994ky}. The construction of Furman and Shamir is usually
used nowadays. One introduces an additional, fifth dimension of length
$L_s$ and considers 5-d Wilson fermions with no gauge links in the
fifth direction, and the 4-d gauge links independent of the fifth
coordinate $s$,
\begin{equation}
S_{DW} = \sum_{s=0}^{L_s-1} \sum_x \bar\psi(x,s) \left\{ 
 \sum_\mu \left( \gamma_\mu \nabla_\mu - \frac{1}{2} \Delta_\mu \right)
 \psi(x,s) - M \psi(x,s) - P_- \psi(x,s+1) - P_+ \psi(x,s-1) \right\} ~,
\label{eq:S_DW}
\end{equation}
where $P_\pm = \frac{1}{2} (1 \pm \gamma_5)$ are chiral projectors and
we have set $r = a = 1$. $M$, introduced here with a sign opposite that
of the mass term for Wilson fermions (\ref{eq:S_W}), is often referred
to as the domain-wall height and needs to be chosen such that $0 < M < 2$. For free
fermions, $M=1$ is the optimal choice, while in the interacting case
$M$ should be somewhat larger. The fermion fields satisfy the boundary
condition in the fifth direction,
\begin{equation}
P_- \psi(x,L_s) = - m_f P_- \psi(x,0) ~,~~~
 P_+ \psi(x,-1) = 
 - m_f P_+ \psi(x,L_s-1) ~,
\label{eq:DW_bc}
\end{equation}
where $m_f$ is a bare quark mass.

For $m_f=0$, the domain-wall action, Eq.~(\ref{eq:S_DW}), has 4-d chiral
modes bound exponentially to the boundaries at $s=0$ and $s=L_s-1$,
which are identified with the chiral modes of 4-d fermions as
\begin{equation}
q^R(x) = P_+ \psi(x,L_s-1) ~,~~~ q^L(x) = P_- \psi(x,0) ~,~~~
 \bar q^R(x) = \bar\psi(x,L_s-1) P_- ~,~~~ \bar q^L(x) = \bar\psi(x,0)
 P_+ ~.
\label{eq:DW_4d_q} \end{equation}
When $L_s \to \infty$ the chiral modes become exact zero modes, the left and
right handed modes $q^L$ and $q^R$ do not interact for $m_f=0$, and
the domain-wall action has a chiral symmetry. At finite $L_s$ the chiral
symmetry is slightly broken. Often $L_s = \mathcal{O}(10-20)$ is large
enough to keep the chiral symmetry breaking negligibly small. The
computational cost of domain-wall fermions is roughly a factor of $L_s$
larger than that for Wilson-type fermions.

Related to these domain-wall fermions are the so-called overlap
fermions developed by 
\textcite*{Narayanan:1994gw, Neuberger:1997fp}. The overlap Dirac operator
for massless fermions can be written as \cite{Neuberger:1997fp},
\begin{equation}
a D_{ov} = M \left[ 1 + \gamma_5 \Theta \left( \gamma_5 D_W(-M) \right)
\right] ~,
\label{eq:D_ov}
\end{equation}
where $D_W(-M)$ is the usual Wilson Dirac operator with negative mass
$m = -M$, and again $0 < M < 2$ should be used. 
$\Theta(X)$ is the
matrix sign function, for a Hermitian matrix $X$, that can be defined as
\begin{equation}
\Theta(X) = \frac{X}{\sqrt{X^2}} ~.
\label{eq:eps_X}
\end{equation}

Using the fact that $\Theta^2(X) = 1$, it is easy to see that the
Neuberger Dirac operator satisfies the so-called Ginsparg-Wilson
relation \cite{Ginsparg:1981bj},
\begin{equation}
\left\{ \gamma_5, D_{ov} \right\} = a D_{ov} \gamma_5 R D_{ov} ~,
\label{eq:G_W}
\end{equation}
with $R = 1/M$, or equivalently, when the inverse of $D_{ov}$ is
well defined,
\begin{equation}
\left\{ \gamma_5, D^{-1}_{ov} \right\} = a \gamma_5 R ~.
\label{eq:G_W_inv}
\end{equation}
In the continuum, chiral symmetry implies that the massless fermion
propagator anticommutes with $\gamma_5$. The massless overlap
propagator violates this only by a local term that vanishes in the
continuum limit. Ginsparg and Wilson argued that this is the mildest
violation of the continuum chiral symmetry on the lattice possible.
In fact, any Dirac operator satisfying the Ginsparg-Wilson relation
(\ref{eq:G_W}) has a modified chiral symmetry at finite lattice
spacing \cite{Luscher:1998pqa},
\begin{equation}
\delta \psi = i \alpha \gamma_5 \left( 1 - \frac{a}{2M} D \right) \psi ~,~~~
 \delta \bar\psi = i \alpha \bar\psi \left( 1 - \frac{a}{2M} D \right)
 \gamma_5 ~.
\label{eq:chi_sym}
\end{equation}
or
\begin{equation}
\delta \psi = i \alpha \gamma_5 \left( 1 - \frac{a}{M} D \right) \psi
 = i \alpha \hat\gamma_5 \psi ~,~~~
 \delta \bar\psi = i \alpha \bar\psi \gamma_5 ~,
\label{eq:chi_sym2}
\end{equation}
with $\hat\gamma_5 = \gamma_5 \left( 1 - \frac{a}{M} D \right)$
satisfying $\hat\gamma_5^\dagger = \hat\gamma_5$ and, using the
G-W relation, Eq.~(\ref{eq:G_W}), $\hat\gamma_5^2 = 1$.

The close connection between domain-wall and overlap fermions can be
made more explicit by integrating out the ``bulk fermions'', which have
masses of the order of the cutoff $1/a$, from the domain-wall action,
Eq.~(\ref{eq:S_DW}), see
\textcite{Neuberger:1997bg,Borici:1999da,Edwards:2000qv,Kikukawa:1999sy}.
In the limit $L_s \rightarrow \infty$, one ends up with the overlap
Dirac operator, but with the Hermitian Wilson kernel $H_W = \gamma_5 D_W$
in Eq.~(\ref{eq:D_ov}) replaced by a more complicated Hermitian kernel,
\begin{equation}
H_T = \frac{1}{1 + 2 a_5 H_w \gamma_5} H_W =
 H_W \frac{1}{1 + 2 a_5 H_w \gamma_5} ~.
\label{eq:H_T}
\end{equation}
Here we 
denote the lattice spacing in the fifth direction by $a_5$.
It is usually chosen to be the same as the 4-d lattice spacing, $a_5 = a$,
which, in turn, is usually set to 1. From Eq.~(\ref{eq:H_T}) we see
that domain-wall fermions in the limit $L_s \rightarrow \infty$,
followed by the limit $a_5 \rightarrow 0$ become identical to overlap
fermions with the standard Neuberger Dirac operator.

The difficulty with numerical simulations using overlap fermions is the
evaluation of the sign function $\Theta(H_W)$ of the Hermitian
Wilson Dirac operator $H_W = \gamma_5 D_W$ in Eq.~(\ref{eq:D_ov}).
This can be done with a Lanczos-type algorithm \cite{Borici:1998mr}.
Alternatively, $\Theta(H_W)$ can be represented as a polynomial,
or, more efficiently, as a rational function that can be rewritten as
a sum over poles \cite{Neuberger:1998my, Edwards:1998yw}, with the
optimal approximation, using a theorem of Zolotarev, first given in
\textcite{vandenEshof:2002ms},
\begin{equation}
\Theta(H_W) = H_W \frac{\sum_j a_j H^{2j}_W}{\sum_j b_j H^{2j}_W}
 = H_W \left[ c_0 + \sum_{k=1}^n \frac{c_k}{H^2_W + d_k} \right] ~.
\label{eq:eps_poles}
\end{equation}
All $d_k$'s are positive, and the necessary inversions with the sparse
matrix $H^2_W$ are done using a multishift conjugate gradient inverter
\cite{Frommer:1995ik, Jegerlehner:1996pm, Jegerlehner:1997rn}.

Finally, two versions of fermions that satisfy the Ginsparg-Wilson
relation approximately have been considered. One, the so-called fixed
point action \cite{Hasenfratz:1997ft}, approximates the fixed point
of a renormalization group transformation by truncating to a small
range. \textcite{Hasenfratz:1998ri} have shown that (untruncated) fixed
point fermion actions satisfy the Ginsparg-Wilson relation. The second
version \cite{Gattringer:2000js}, directly minimizes deviations from
the Ginsparg-Wilson relation by adjusting the parameters in an arbitrary
Dirac operator with a finite (small) number of terms.

\subsection{Numerical simulations}
\label{sec:Num_simul}

After having chosen a gauge and fermion action one computes expectation
values of interesting observables, Eq.~(\ref{eq:ev_O}), by numerical
Monte Carlo simulations. For this one creates a sequence of gauge
field configurations $\{ U_\mu^{(i)}(x) \}$, $i = 1, \dots, N$, distributed
with probability distribution
\begin{equation}
P( \{ U_\mu^{(i)}(x) \} ) = \frac{1}{Z(\beta)} (\det M_F(U))^\delta
 \exp\{ - S_G(U) \} = \frac{1}{Z(\beta)} \exp\{ - S_{eff}(U) \} ~.
\label{eq:prob_U}
\end{equation}
Here, $\delta = n_f$, the number of flavors, for Wilson and chirally
invariant fermions, and $\delta = n_f/4$ for (rooted) staggered
fermions,\footnote{The sketch here is somewhat schematic: each fermion
with a different mass would get its own determinant factor. Furthermore,
$M_F$ should be Hermitian and positive semi-definite. For Wilson fermions
one therefore takes $M_F = D^\dagger_W D_W$ and uses $\delta = n_f/2$,
while for staggered fermions one takes $M_F = [D^\dagger_{KS} D_{KS}]_{ee}$
where the subscript ``ee'' refers to the matrix restricted to the even
sublattice. This is possible, since $D^\dagger_{KS} D_{KS}$
block-diagonalizes to even and odd sublattices. Restricting to only one
sublattice removes the doubling introduced by the ``squaring.''}
and now
\begin{equation}
S_{eff}(U) = S_G(U) - \delta \mathrm{Tr} \log M_F(U) ~.
\label{eq:S_eff}
\end{equation}
Expectation values $\langle O \rangle$ are then computed as an average
over the ensemble of gauge field configurations,
\begin{equation}
\langle O \rangle = \frac{1}{N} \sum_{i=1}^N O^{(i)} ~,
\label{eq:av_O}
\end{equation}
where $O^{(i)} = O(U_\mu^{(i)})$ is the observable evaluated on the
gauge field configuration $i$.

For pure gauge simulations, when no fermions are present, or in the
quenched approximation, where the fermion determinant is set to one
($\det M_F = 1$), the action is local (in the gauge fields) and the
sequence of configurations can be generated with a local updating
algorithm, such as the Metropolis algorithm \cite{Metropolis:1953am}
or a heatbath algorithm \cite{Creutz:1980zw, Kennedy:1985nu}.

With the fermion determinant present, all gauge fields are coupled
and the local updating algorithms become impractical. Molecular
dynamics based algorithms \cite{Callaway:1982eb, Callaway:1983ee}
have become the standards for simulations with dynamical fermions.
For a scalar lattice field theory with action $S(\phi_x)$ one
introduces a fictitious momentum $p_x$ on each lattice site,
and considers the Hamiltonian
\begin{equation}
H(p,\phi) = \sum_x \frac{p^2_x}{2} + S(\phi) ~.
\label{eq:H_p_phi}
\end{equation}
This Hamiltonian defines a classical evolution in a fictitious time
$\tau$ by,
\begin{equation}
\dot\phi_x = p_x ~,~~ \dot p_x = - \frac{\partial S}{\partial \phi_x} ~,
\label{eq:eom_p_phi}
\end{equation}
where the dot denotes the derivative with respect to $\tau$. Given some
initial values $(p_x(0), \phi_x(0))$ these equations of motion define
a trajectory $(p_x(\tau), \phi_x(\tau))$ through phase space. The
classical partition function corresponding to the set of all such
trajectories is
\begin{equation}
Z = \int \prod_x [dp_x d\phi_x] \exp\{ - H(p,\phi) \} =
 \mathcal{N} \int \prod_x d\phi_x \exp\{ - S(\phi) \} ~,
\label{eq:Z_phi}
\end{equation}
where in the second step the quadratic integration over the $p_x$ has
been carried out, and $\mathcal{N}$ is an unimportant normalization
factor.
The integration of Hamilton's equations, Eq.~(\ref{eq:eom_p_phi}), conserves
the Hamiltonian, Eq.~(\ref{eq:H_p_phi}), up to numerical errors.  To get
the correct distribution corresponding to the canonical
partition function (\ref{eq:Z_phi}), the fictitious momenta are
``refreshed'' periodically by replacement with new
Gaussian random numbers \cite{Duane:1985hz, Duane:1986iw}.
This algorithm goes under the name of Hybrid Molecular Dynamics (HMD).

Relying on the ergodicity
hypothesis, the expectation value of observables can then be computed
by averaging over many MD trajectories
\begin{equation}
\langle O \rangle = \frac{1}{T} \int_{\tau_0}^{T+\tau_0} d\tau
 O(\phi(\tau)) ~.
\label{eq:ev_O_tau}
\end{equation}

Integration of the equations of motion, Eq.~(\ref{eq:eom_p_phi}), is done
numerically by introducing a finite step size $\Delta \tau$ and using
a volume-preserving integration algorithm, such as leapfrog. Due to the
finite step size, the Hamiltonian is not exactly conserved during
the MD evolution, leading to finite step size errors in observables,
including the Hamiltonian itself, of $\mathcal{O}((\Delta \tau)^2)$
for the leapfrog integration algorithm. These step size errors can
be eliminated --- the algorithm made exact --- by combining the
refreshed MD evolution with a Metropolis accept/reject step at the
end of each trajectory \cite{Duane:1987de}, resulting in the so-called
Hybrid Monte Carlo (HMC) algorithm.

For a lattice gauge theory the equations of motion have to be set up
such that the gauge fields remain group elements. This is ensured by
writing
\begin{equation}
\dot U_\mu(x) = i H_\mu(x) U_\mu(x) ~,
\label{eq:emo_U}
\end{equation}
with $H_\mu(x) = \sum_a t^a h^a_\mu(x)$ a traceless Hermitian matrix
and $t^a$ the SU($N$) generators, see, {\it e.g.,} \cite{Gottlieb:1987mq}.
The MD Hamiltonian is given by
\begin{equation}
H(H_\mu(x), U_\mu(x)) = \sum_{x, \mu} \frac{1}{2} \mathrm{Tr} H^2_\mu(x)
 + S_{eff}(U_\mu(x)) ~.
\label{eq:H_h_U}
\end{equation}
The equation of motion for $H_\mu(x)$ is then, somewhat schematically,
\begin{equation}
\dot H_\mu(x) = \left. i U_\mu(x)
 \frac{\partial S_{eff}(U)}{\partial U_\mu(x)} \right|_{TH} ~,
\label{eq:emo_H}
\end{equation}
where ``TH'' denotes the traceless Hermitian part. The term on the
right-hand side of (\ref{eq:emo_H}) is usually referred to as the
force term. With $S_{eff}$ of Eq.~(\ref{eq:S_eff}) we have
\begin{equation}
\frac{\partial S_{eff}(U)}{\partial U_\mu(x)} =
 \frac{\partial S_G(U)}{\partial U_\mu(x)} 
 - \delta \mathrm{Tr} \left[ \frac{\partial M_F(U)}{\partial U_\mu(x)}
   M^{-1}_F(U) \right] ~.
\label{eq:Tr_M_inv_dM}
\end{equation}
To evaluate (\ref{eq:Tr_M_inv_dM}) we need to know all matrix elements
of $M^{-1}_F(U)$, a dense matrix, even though the fermion matrix $M_F(U)$
is sparse. This would be prohibitively expensive. Instead, one estimates
the inverse stochastically. Let $R$ be a Gaussian random field such that
\begin{equation}
\overline{R^*_A(x) R_B(y)} = \delta_{AB} \delta_{xy} ~,
\label{eq:Gaus_av}
\end{equation}
where $A,B$ denote color indices, and for Wilson-type fermions also
Dirac indices. Then,
\begin{equation}
\mathrm{Tr} \left[ \frac{\partial M_F(U)}{\partial U_\mu(x)}
 M^{-1}_F(U) \right] =
 \overline{ R^\dagger \frac{\partial M_F(U)}{\partial U_\mu(x)}
 M^{-1}_F(U) R} ~,
\label{eq:R_M_inv_dM_R}
\end{equation}
and for each random vector $R$ only a single inversion, $M^{-1}_F(U) R$
is needed. Typically, for each time step in the MD evolution one uses
just one Gaussian random vector, and hence one inversion. This
algorithm goes under the name of ``HMD R-algorithm'' \cite{Gottlieb:1987mq}.

Instead of doing molecular dynamics starting with $S_{eff}$ of
Eq.~(\ref{eq:S_eff}) one can first represent the fermion determinant
by an integral over bosonic fields, called pseudofermions
\begin{equation}
\det M_F(U) = \int \prod_x [d\Phi^\dagger(x) d\Phi(x)]
 \exp\{ - \Phi^\dagger M^{-1}_F(U) \Phi \} ~.
\label{eq:det_M_Phi}
\end{equation}
HMD using (\ref{eq:det_M_Phi}), referred to as the $\Phi$-algorithm
\cite{Gottlieb:1987mq}, consists in creating, together with the
momenta refreshments, a $\Phi$-field distributed according to
Eq.~(\ref{eq:det_M_Phi})\footnote{For $M_F = D^\dagger D$ this can
be achieved by creating random Gaussian variables $R$ and then
setting $\Phi = D^\dagger R$.} and then integrating the molecular
dynamics equations for the effective action
\begin{equation}
S_{eff}(U, \Phi) = S_G(U) + \Phi^\dagger M^{-1}_F(U) \Phi ~,
\label{eq:S_eff_Phi}
\end{equation}
with the $\Phi$-field fixed.
Now the force term becomes
\begin{equation}
\frac{\partial S_{eff}(U, \Phi)}{\partial U_\mu(x)} =
 \frac{\partial S_G(U)}{\partial U_\mu(x)} 
 - \Phi^\dagger M^{-1}_F(U) \frac{\partial M_F(U)}{\partial U_\mu(x)}
   M^{-1}_F(U) \Phi ~.
\label{eq:Phi_M_inv_dM}
\end{equation}
This again requires one inversion, $M^{-1}_F(U) \Phi$, in each step of
the MD evolution. One major benefit of the $\Phi$-algorithm formulation
is that an accept/reject Metropolis step is easily implemented at the
end of each trajectory resulting in an exact HMC algorithm.

The representation of the fermion determinant by an integral over
pseudofermion fields, Eq.~(\ref{eq:det_M_Phi}), can formally be extended
to fractional powers $\delta = n_f/4$, as needed for rooted staggered
fermions, and $\delta = n_f/2$, as needed for odd number of flavors for
Wilson fermions,
\begin{equation}
(\det M_F(U))^\delta = \int \prod_x [d\Phi^\dagger(x) d\Phi(x)]
 \exp\{ - \Phi^\dagger M^{-\delta}_F(U) \Phi \} ~.
\label{eq:det_M_Phi_frac}
\end{equation}
The problem then is, how to deal with $M^{-\delta}_F$.
In the HMD R-algorithm this is handled by weighting the fermionic
contribution to the force by a factor of $\delta$ and evaluating
$M^{-1}R$ at a point in the integration time chosen so that
the errors in observables remain order $\epsilon^2$, where
$\epsilon$ is the step size in the molecular dynamics integration
\cite{Gottlieb:1987mq}.
Clark and Kennedy recently proposed using a rational function
approximation rewritten as a sum over poles \cite{Clark:2003na,Clark:2004cq},
\begin{equation}
M^{-\delta}_F(U) \approx r(M_F(U)) =a_0 + \sum_{k=1}^n
 \frac{a_k}{M_F(U) + b_k} ~,
\label{eq:rat_M}
\end{equation}
with suitable constants $a_k$ and $b_k$.
A $\Phi$-algorithm
can then easily be constructed, resulting
in the so-called rational hybrid molecular dynamics (RHMD) algorithm,
or, with inclusion of the Metropolis accept/reject
step to eliminate errors from nonzero $\epsilon$,
the rational hybrid Monte Carlo (RHMC) algorithm.
Elimination of the noisy estimator yields smaller errors than in the
HMD R-algorithm at a given integration step size.

Several improvements of the HMD-type algorithms over the last several
years have made them substantially more efficient. These improvements
include ``multiple time step integration schemes'' \cite{Sexton:1992nu},
preconditioning of the fermion determinant by multiple pseudofermion
fields \cite{Hasenbusch:2001ne,Hasenbusch:2002ai}, and replacing the
leapfrog integration scheme with more sophisticated ``Omelyan integrators''
\cite{Sexton:1992nu,Omelyan:2002E1,Omelyan:2002E2,Omelyan:2003CC,Takaishi:2005E1}

\subsection{Asqtad improved staggered fermions}
\label{sec:Asqtad_ferm}

Staggered fermions, with only one component per lattice site, and the
massless limit protected by a remnant even/odd U(1)$_e\times$U(1)$_o$ chiral
symmetry, are numerically very fast to simulate. One of the major
drawbacks is the violation of taste symmetry. At a lattice
spacing $a$ of order $0.1$ fm, which until recently was typical of numerical
simulations, the smallest pion taste splitting Eq.~(\ref{eq:taste-split})
for standard staggered fermions
is of order $\Delta (m^2_{P}) = a^2 \delta_{P} \sim (300\; \MeV)^2$,
{\it i.e.}, more than twice the physical pion mass.
Even when the lattice spacing is reduced to about $0.05$ fm this  
smallest splitting is still the size of the physical pion mass.
It is therefore important to reduce taste violations. Since the
different taste components live on neighboring lattice sites and
in momentum space have momentum components that differ by $\pi/a$,
emission or absorption of gluons with (transverse) momentum components
close to $\pi/a$ can change the taste of a quark. Exchange of such
ultraviolet gluons thus leads to taste violations.

Suppressing the coupling to such UV gluons thus should reduce the
taste violations
\cite{Blum:1996uf, Lepage:1997id, Orginos:1998ue, Lagae:1998pe, Orginos:1999cr}.
This can be achieved by replacing the link field $U_\mu$ in the
covariant difference operator $\nabla_\mu$
(see Eq.~(\ref{eq:D_naive}))
by a smeared link built from 3-link staples (``fat3'')
\begin{equation}
U_\mu(x) \rightarrow U^{f3}_\mu(x)\equiv {\cal F}^{f3}U_\mu(x)
  = U_\mu(x) +
 \omega a^2 \sum_{\nu \ne \mu} \Delta^\ell_\nu U_\mu(x) ~,
\label{eq:fat3}
\end{equation}
where the superscript $\ell$ indicates that the Laplacian acts on a
link field,
\begin{equation}
\Delta^\ell_\nu U_\mu(x) = \frac{1}{a^2} \left(
 U_\nu(x) U_\mu(x + a \hat{\nu}) U^\dagger_\nu(x + a \hat{\mu})
 + U^\dagger_\nu(x - a \hat{\nu}) U_\mu(x - a \hat{\nu})
   U_\nu(x - a \hat{\nu} + a \hat{\mu}) - 2 U_\mu(x) \right) ~.
\label{eq:Lap_l}
\end{equation}
In momentum space, expanding to first order in $g$, Eq.~(\ref{eq:fat3})
leads to
\begin{equation}
A_\mu(p) \rightarrow A_\mu(p) + \omega \sum_{\nu \ne \mu} \left\{
 2 A_\mu(p) \left[ \cos(ap_\nu) - 1 \right] +
 4 \sin(ap_\mu/2) \sin(ap_\nu/2) A_\nu(p) \right\} ~.
\label{eq:fat3_p}
\end{equation}
Choosing $\omega=1/4$ eliminates the coupling to gluons $A_\mu(p)$
with a single momentum component $p_\nu = \pi/a$.
Adding a 5-link staple (``fat5'')
\begin{equation}
U_\mu(x) \rightarrow U^{f5}_\mu(x) \equiv {\cal F}^{f5}U_\mu(x)
 = U^{f3}_\mu(x) +
 \frac{a^4}{32} \sum_{\rho \ne \nu \ne \mu} \Delta^\ell_\rho
 \Delta^\ell_\nu U_\mu(x) ~,
\label{eq:fat5}
\end{equation}
eliminates the coupling to gluons with two momentum components
$p_\nu = \pi/a$ and adding a 7-link staple (``fat7'')
\begin{equation}
U_\mu(x) \rightarrow U^{f7}_\mu(x) \equiv {\cal F}^{f7}U_\mu(x)
 = U^{f5}_\mu(x) +
 \frac{a^6}{384} \sum_{\sigma \ne \rho \ne \nu \ne \mu} \Delta^\ell_\sigma
 \Delta^\ell_\rho \Delta^\ell_\nu U_\mu(x) ~,
\label{eq:fat7}
\end{equation}
eliminates the coupling to gluons with all three transverse momentum
components $p_\nu = \pi/a$.

For smooth gauge fields, with $p \approx 0$, the Laplacian,
Eq.~(\ref{eq:Lap_l}), becomes
\begin{equation}
\Delta^\ell_\nu U_\mu(x) = a D_\nu F_{\nu\mu} + \cdots ~,
\label{eq:Lap_l_a}
\end{equation}
where $\cdots$ represent higher order terms in $a$.  The change in
\eq{fat3} thus produces a change $\sim a^2D_\nu F_{\nu\mu} $ to the
gauge field $A_\mu$.  This is a new $\mathcal{O}(a^2)$ lattice artifact,
and will occur when using  fat3, fat5 or fat7 links. It, in turn, can
be canceled by a ``straight 5-link staple'' \cite{Lepage:1998vj}
\begin{eqnarray}
\Delta^{2\ell}_\nu U_\mu(x) &=& \frac{1}{4a^2} \left(
   U_\nu(x) U_\nu(x + a \hat{\nu}) U_\mu(x + 2a \hat{\nu})
   U^\dagger_\nu(x + a \hat{\nu} + a \hat{\mu}) U^\dagger_\nu(x + a \hat{\mu})
 \right. \nonumber \\
 && + \left. U^\dagger_\nu(x - a \hat{\nu}) U^\dagger_\nu(x - 2a \hat{\nu})
   U_\mu(x - 2a \hat{\nu}) U_\nu(x - 2a \hat{\nu} + a \hat{\mu})
   U_\nu(x - a \hat{\nu} + a \hat{\mu}) - 2 U_\mu(x) \right) \nonumber \\
 &=& a D_\nu F_{\nu\mu} + \cdots ~,
\label{eq:Lap_2l}
\end{eqnarray}
referred to as the ``Lepage-term.''
In momentum space, expanding to first order in $g$, this becomes
\begin{equation}
\frac{1}{2a} \left\{ A_\mu(p) \left[ \cos(2ap_\nu) - 1 \right] +
 2 \sin(ap_\mu/2) \left[ \sin(ap_\nu/2) + \sin(3ap_\nu/2) \right]
 A_\nu(p) \right\} ~,
\label{eq:Lap_2l_p}
\end{equation}
and thus does not affect the coupling to gluons with momentum components
at the corners of the Brillouin zone. Therefore, replacing
\begin{equation}
U_\mu(x) \rightarrow U^{f7L}_\mu(x) \equiv {\cal F}^{f7L}U_\mu(x)
 = U^{f7}_\mu(x) -
 \frac{a^2}{4} \sum_{\nu \ne \mu} \Delta^{2\ell}_\nu U_\mu(x) ~,
\label{eq:fat7L}
\end{equation}
eliminates, at tree level, the coupling to gluons with any of the
transverse momentum components $p_\nu = \pi/a$ without introducing new
lattice artifacts.

Finally, for a complete $\mathcal{O}(a^2)$ improvement we include a
so-called ``Naik-term'' \cite{Naik:1986bn} to improve the free
propagator, and hence the free dispersion relation. To keep the
structure of the couplings to the different tastes unchanged, this
involves adding a 3-hop term,
\begin{eqnarray}
\nabla_\mu \chi(x) &\rightarrow& \nabla_\mu \chi(x) -
 \frac{a^2}{6} (\nabla_\mu)^3 \chi(x) \\
 &=& \left( 1 + \frac{1}{8} \right) \nabla_\mu \chi(x)
  - \frac{1}{48a} \left( U_\mu(x) U_\mu(x + a \hat{\mu})
    U_\mu(x + 2a \hat{\mu}) \chi(x + 3a \hat{\mu}) \right. \nonumber \\
  && \left. \qquad - U^\dagger_\mu(x - a \hat{\mu})
    U^\dagger_\mu(x - 2a \hat{\mu}) U^\dagger_\mu(x - 3a \hat{\mu})
    \chi(x - 2a \hat{\mu}) \right)
 ~. \nonumber
\label{eq:Naik}
\end{eqnarray}
In the free inverse propagator this changes
\begin{equation}
\frac{1}{a} \sin(ap_\mu) \rightarrow \frac{1}{a} \sin(ap_\mu)
 \left[ 1 + \frac{1}{6} \sin^2(ap_\mu) \right] = p_\mu + \mathcal{O}(a^4) ~.
\label{eq:Naik_p}
\end{equation}
The fermion action with only the improvement in Eq.~(\ref{eq:Naik}) is
referred to as the ``Naik action''. This is also the free (noninteracting)
limit of the asq and asqtad fermion actions, defined next.

\begin{figure}[t]
\begin{center}
\includegraphics[width=2.5in]{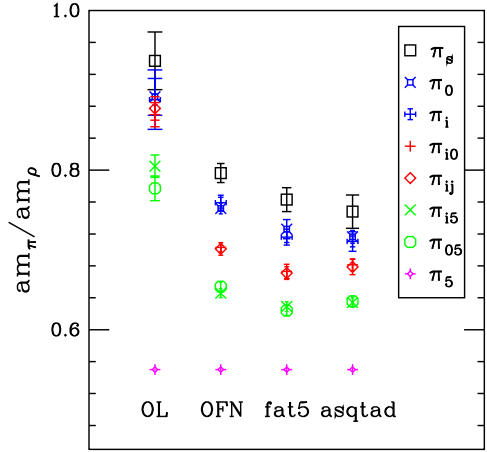}
\end{center}
\caption{Illustration of taste violations for staggered fermion actions
with various link fattenings. The valence quark masses were adjusted to
give the same $m_{\pi_5}/m_\rho = 0.55$ for all fermion actions. The
results are for quenched gauge field configurations with a Symanzik improved
gauge action using $\beta= 7.30$. The staggered fermion actions are
standard, or one-link (OL), ``fat3+Naik'' (OFN) , ``fat5'' and ``asqtad''.
The pions are labeled by the taste structure, with the taste singlet the
heaviest, and the taste pseudoscalar ($\pi_5$), the pseudo-Goldstone
boson, the lightest.
For more comparisons see \cite{Orginos:1999kg}.
}
\label{fig:taste-improv}
\end{figure}

We now have all the ingredients for an improved staggered fermion action,
called the ``asq'' action ($\mathcal{O}(a^2)$ improved action): use the
covariant derivative with the Naik term, Eq.~(\ref{eq:Naik}), and in the
one-link term replace the gauge links $U_\mu$ by the fat7 links with
Lepage term $U^{f7L}_\mu$ of Eq.~(\ref{eq:fat7L}).
Replacing the various coefficients in the asq action by tadpole
improved coefficients finally gives the ``asqtad'' fermion action.
The reduction of taste violations for pions with increasing amount of
link fattening is illustrated in Fig.~\ref{fig:taste-improv}.

\begin{figure}
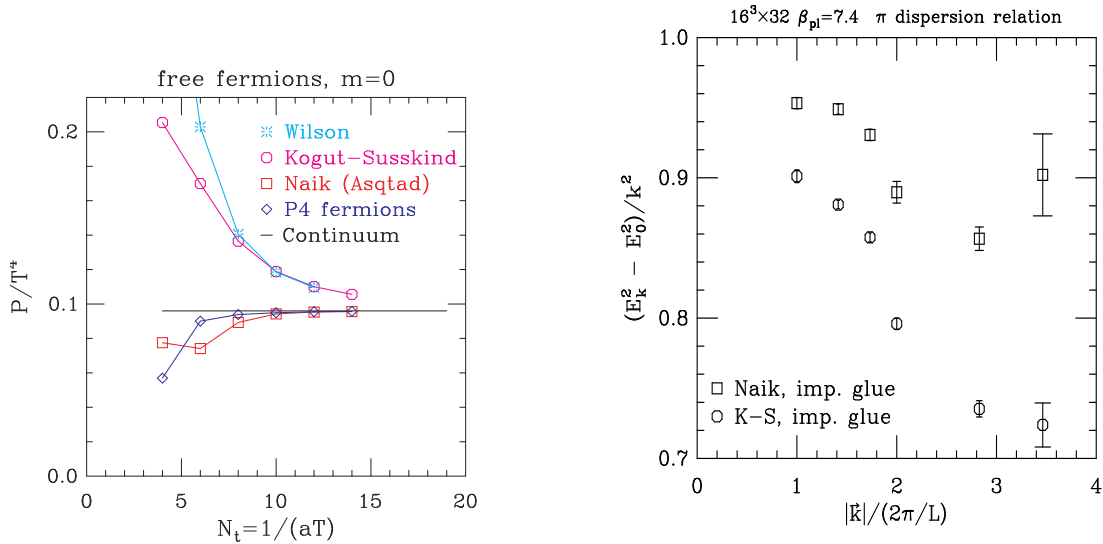

\begin{center}
\vspace{-0.5in}
\begin{tabular}{c c}
\epsfxsize=2.5in
\epsfysize=2.5in
\epsfbox[0 0 4096 4096]{figs/f_pressure.ps}
&
\epsfxsize=3.15in
\epsfysize=3.15in
\epsfbox[0 400 4411 4347]{figs/csqr_pi.ps}
\end{tabular}
\end{center}
\caption{
\label{fig:press-csq}
The pressure (left) per fermion degree of freedom for free Kogut-Susskind,
Naik, Wilson and ``p4'' \protect{\cite{Heller:1999xz}} fermions as a function of
$N_T = 1/(aT)$. The continuum value is shown as the horizontal solid line.
Figure from \textcite{Bernard:2004je}; an earlier version appeared in
\textcite{Bernard:1997mz}.
The `speed of light squared', (right), calculated
from the pion dispersion relation, for Naik and K-S pions.
Figure from \textcite{Bernard:1997mz}.
}
\end{figure}

The Naik term, Eq.~(\ref{eq:Naik}), reduces the lattice artifacts in
the pressure for free fermions, and thus in the very high temperature
limit of QCD as illustrated in Fig.~\ref{fig:press-csq}, left panel,
and in the `speed of light' determined from the pion dispersion relation,
right panel, from \textcite{Bernard:1997mz}.
In Fig.~\ref{fig:press-csq}, left panel, ``p4'' fermions are another
variant of improved staggered fermions \cite{Heller:1999xz} designed to
improve the dispersion relation and high temperature behavior.
The speed of light, shown in the right panel, is determined from
pion energies $E_\pi(\vec p)$ for various momenta as
\begin{equation}
c^2 = \frac{E^2_\pi(\vec p) - E^2_\pi(\vec 0)}{\vec p^2} ~.
\label{eq:c_sq}
\end{equation}

\begin{figure}
\begin{center}
\begin{tabular}{c c}
\includegraphics[width=2.5in]{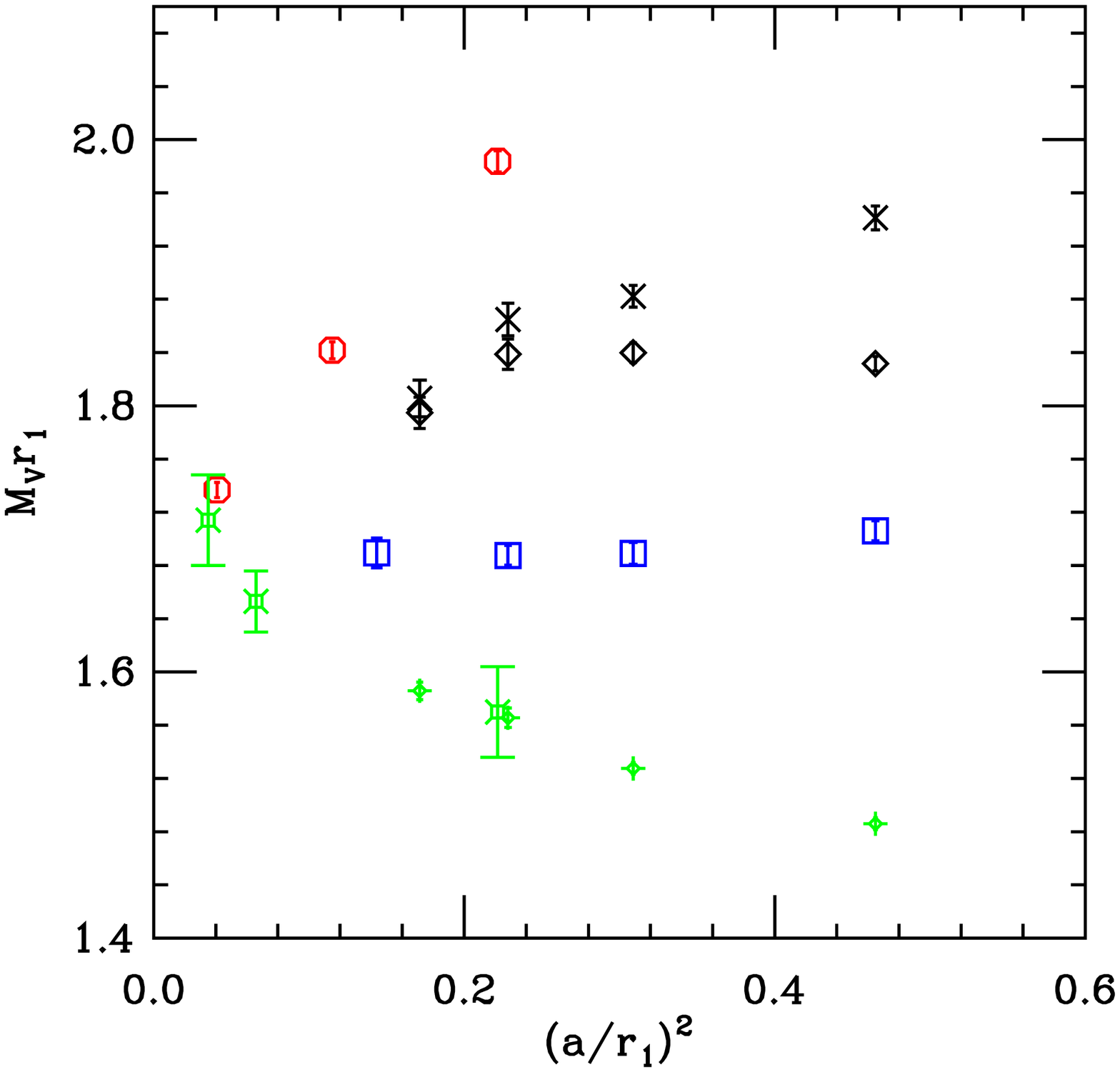}
&
\includegraphics[width=2.5in]{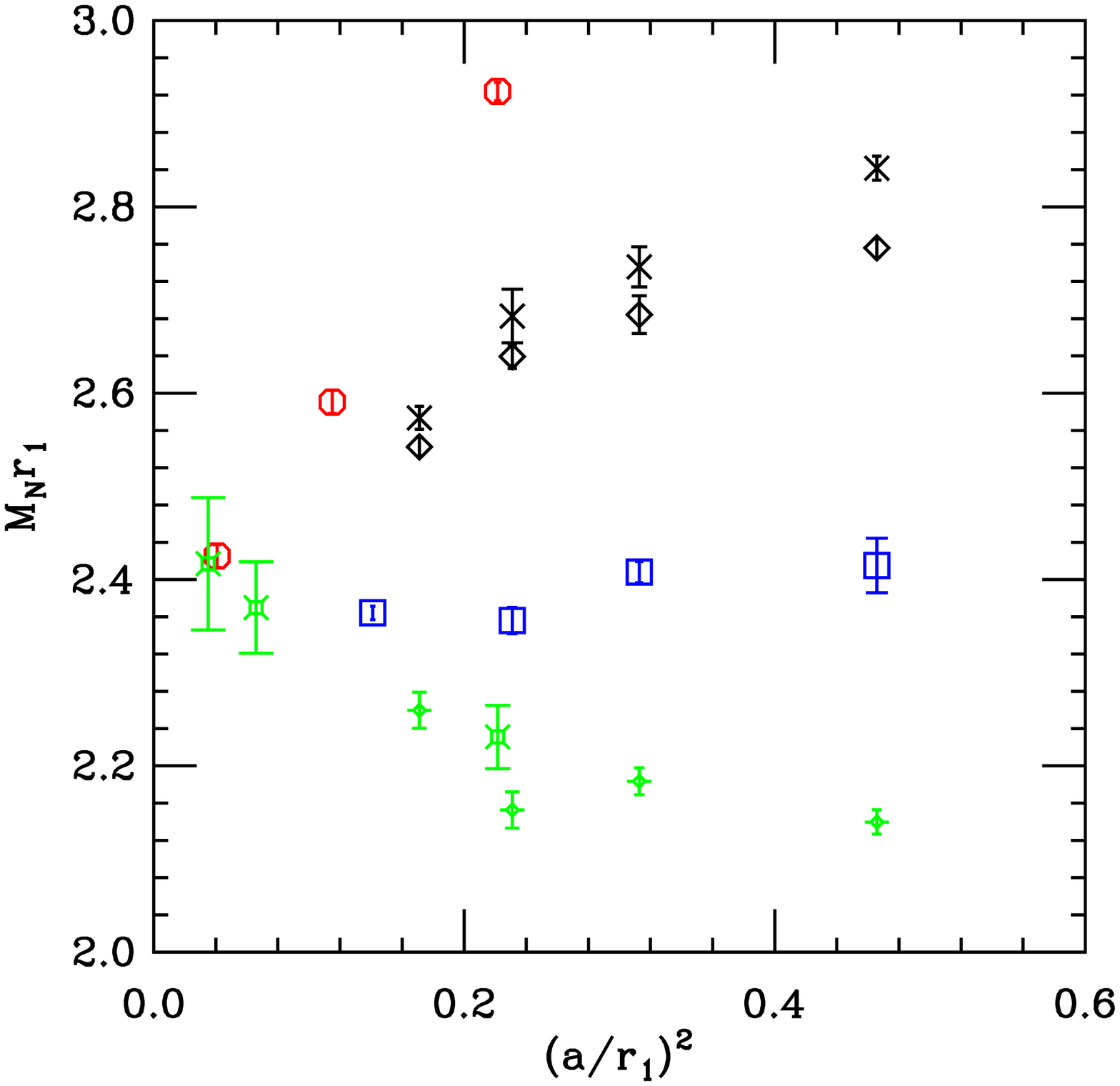}
\end{tabular}
\end{center}
\caption{
\label{fig:rho-nuc-scal}
Rho masses (left) and nucleon masses (right) in units of $r_1 \approx 0.32$
fm, in a slight update from \textcite{Bernard:1999xx}.
Octagons are unimproved staggered fermions with Wilson gauge action,
diamonds are unimproved staggered fermions with Symanzik improved gauge
action, crosses are Naik fermions and squares are asqtad fermions, both
with Symanzik improved gauge action.
For comparison we also show tadpole clover improved Wilson fermions with
Wilson gauge action \cite{Bowler:1999ae} (fancy squares)
and with Symanzik improved gauge action \cite{Collins:1996zc} (fancy diamonds).
}
\end{figure}

The $\mathcal{O}(a^2)$ improvement of the asqtad action gives
a staggered fermion formulation with good scaling properties, as
shown in Fig.~\ref{fig:rho-nuc-scal} for a quenched study
\cite{Bernard:1999xx}.

\subsection{Highly improved staggered fermions}
\label{sec:HISQ_ferm}

The largest contribution to the $O(a^2)$ error in the asqtad action
originates from the taste-exchange interactions. This error
can be completely eliminated at one-loop level by adding four-quark
interactions (which are hard to implement in dynamical simulations)
or greatly reduced by additional smearings.
Multiple smearings,
for instance
\begin{equation}
U_\mu(x) \rightarrow X_\mu(x) = {\cal F}^{f7L}
 {\cal F}^{f7L} U_\mu(x)
\label{eq:fat7Lfat7L}
\end{equation}
are found to further reduce mass splittings between pions of different taste.
However, they increase the number
of products of links in the sum for $X_\mu(x)$ links and
effectively enhance the contribution of two-gluon vertices
on quark lines (see \textcite{Follana:2006rc} for a more detailed
discussion). Thus, an operation that bounds smeared links
needs to be introduced:
\begin{equation}
U_\mu(x) \rightarrow X_\mu(x) = {\cal F}^{f7L}
 {\cal U}{\cal F}^{f7L} U_\mu(x) ~,
\label{eq:fat7LUfat7L}
\end{equation}
where ${\cal U}$ is an operation that projects smeared links onto the
U(3) or SU(3) group. Cancellation of the $O(a^2)$ artifacts introduced
by fat7 smearing with the Lepage term can be achieved on the outermost
level of smearing, and \eq{fat7LUfat7L} can be simplified:
\begin{equation}
U_\mu(x) \rightarrow X_\mu(x) = {\cal F}^{f7L}
 {\cal U}{\cal F}^{f7} U_\mu(x)\equiv
 {\cal F}^{HISQ}U_\mu(x) ~.
\label{eq:FHISQ}
\end{equation}
Introducing smeared and reunitarized links that arise after each operation
in \eq{FHISQ}
\begin{eqnarray}
V_\mu(x) &=& {\cal F}^{f7} U_\mu(x) ~,\label{eq:Vlinks}\\
W_\mu(x) &=& {\cal U} V_\mu(x)={\cal U}{\cal F}^{f7} 
U_\mu(x) ~,\label{eq:Wlinks}\\
X_\mu(x) &=& {\cal F}^{f7L} W_\mu(x) = {\cal F}^{HISQ} U_\mu(x) ~,
\label{eq:Xlinks}
\end{eqnarray}
we can write the covariant derivative that replaces the naive one:
\begin{equation}
\nabla_\mu [U] \chi(x) \rightarrow \nabla_\mu(x)[X] \chi(x) -
 \frac{a^2}{6}(1+\epsilon)(\nabla_\mu)^3[W] \chi(x)~.
\label{eq:derHISQ}
\end{equation}
Equation (\ref{eq:derHISQ}) is a recently proposed ``Highly improved
staggered quark", or ``HISQ", discretization scheme \cite{Follana:2006rc}.
In square brackets we indicate which links are used as gauge
transporters in the derivatives. The second term is the Naik term
evaluated using the reunitarized links $W_\mu(x)$. Its coefficient
includes a correction $\epsilon$ introduced to compensate for the order
$(am)^4$ and $\alpha_s(am)^2$ errors. This correction is negligible
for light quarks, but may be relevant for charm physics if a level of
accuracy better than 5--10\% is desired. The correction $\epsilon$ can
be either tuned nonperturbatively or calculated in perturbation theory
\cite{Follana:2006rc}.

The HISQ action suppresses the taste-exchange interactions by a factor of
about 2.5 to 3 compared to the asqtad action, which makes it a very good
candidate for the next generation of simulations with 2+1 or
2+1+1 flavors of dynamical quarks, where in the latter case the last quark
is the charm quark. We discuss preliminary studies
of the HISQ action in more detail in \secref{future}.

%% file: RMP_sec3.tex
% File for section 3 for RMP article
%

\section{Staggered chiral perturbation theory and ``rooting''}
\label{sec:schpt-and-root}
\subsection{Chiral effective theory for staggered quarks}
\label{sec:SChPT}

Because simulation costs increase with decreasing quark mass, most
QCD simulations are done with the masses of the two lightest quarks
(up and down) larger than their physical values.
The results, therefore, have to
be extrapolated to the physical light quark masses. This is done using
chiral perturbation theory, the effective field theory that describes
the light quark limit of QCD \cite{Weinberg:1978kz,Gasser:1983yg,Gasser:1984gg}.

Even with the asqtad improvement of staggered fermions, taste-symmetry violations
are not negligible in current simulations. It is therefore important to include
the effects of discretization errors
in the chiral perturbation theory forms one uses to extrapolate
lattice data to physical light quark masses and to infinite volume; in other
words, one needs to use ``staggered chiral perturbation theory'' (\schpt).  
Indeed, it is not possible to fit the mass dependence of the staggered
data to continuum chiral forms \cite{Aubin:2004fs}.  Once the
discretization effects are included explicitly by making \schpt\ fits, one can 
gain good control of the errors from the continuum extrapolation.  Furthermore,
the effects of taking the fourth root of the staggered determinant can be included
in \schpt. The resulting ``rooted staggered chiral perturbation theory'' (\rschpt)
allows us to understand the nonlocal and nonunitary consequences 
of rooting on the lattice and
to test that these sicknesses go to zero as $a\to 0$.

\textcite{Lee:1999zxa} first developed \schpt\ for a single
staggered flavor (a single staggered field) at $\cO(a^2)$; this was generalized to an arbitrary number of flavors by
\textcite{Aubin:2003mg,Aubin:2003uc}. 
Here, we outline the theory with $N_f$ flavors to this
order; for the next order we refer the reader to the literature  \cite{Sharpe:2004is}.  

To derive \schpt, one starts by determining, to the desired order in
$a^2$, the Symanzik effective theory (SET) \cite{Symanzik:1983dc} for
staggered quarks.  The SET is an effective theory for physical momenta $p$
small compared with the cutoff ($p\ll 1/a)$; it parametrizes discretization
effects by adding higher-dimensional operators to  continuum QCD.
In particular, taste violations appear to $\cO(a^2)$ in the SET as
four-quark (dimension six) operators.  These operators arise from the
exchange of gluons with net momenta $\sim\!\pi/a$ between two quark lines.
Such gluons can change taste, spin, and color, but not flavor.  Therefore,
the operators generated have the form
\begin{equation}
O_{ss'tt'} = \bar q_i (\gamma_s \otimes \xi_t) 
q_i\; \bar q_j (\gamma_{s'} \otimes \xi_{t'}) q_j \ ,
\eqn{dim6op}
\end{equation}
where $i$ and $j$ are flavor indices, spin and taste matrices have the notation
of \eq{O_ST}, and color indices are omitted because they play no role in what follows.
The SU($N_f$) vector flavor symmetry guarantees that
$O_{ss'tt'}$ is a flavor singlet, which means that $i,j$ are (implicitly) summed
over their $N_f$ values in \eq{dim6op}.  

The possible choices of the spin and taste matrices in \eq{dim6op} are constrained by
the staggered symmetries.  
First of all, we can use the separate
U(1)$_\epsilon$ 
for each flavor.
This forces each of the bilinears making
up $O_{ss'tt'}$, for example
$\bar q_i (\gamma_s \otimes \xi_t) q_i $, to be U(1)$_\epsilon$ invariant by itself for each
$i$. 
From \eq{U1eps_q},  we then have that $\{\g_5\otimes \xi_5, \gamma_s \otimes \xi_t\}=0$,
which gives twelve choices for $\g_s$ and $\xi_t$:
One of them must be a scalar, tensor, or pseudoscalar ($S$, $T$
or $P$) and the other must be a vector or axial vector ($V$ or $A$).  For example, we might
have $A\otimes T$, that is, $\gamma_s \otimes \xi_t= \g_{\mu5} \otimes \xi_{\nu\lambda}$,
with the notation $\g_{\mu5} \equiv \g_\mu \g_5$ (and similarly for tastes), and
$\xi_{\nu\lambda} = \frac{1}{2}[\xi_\nu,\xi_\lambda]$ (and similarly for spins).
Such operators are called ``odd'' because, in the original one-component form of
\eq{S_KS}, the fields $\chi$ and $\bar \chi$ are separated by an odd number of links
(1 or 3) within an elementary hypercube.  This is easily seen from the equivalence given
in \eq{O_ST}.

Shift symmetry gives the next constraint.
As mentioned following \eq{tasteshifts}, shift symmetries are a combination of
discrete taste symmetries and translations.  In the SET, however, where external
momenta are always small compared with the cutoff, it is possible to redefine the
fields $q(y)$
to make the action invariant under arbitrary translations, like in any continuum theory
\cite{Bernard:2007ma}.
The shifts then have the form:
\begin{equation}
q(y)\to  (I\otimes\xi_\mu) q(y)\ ;\qquad\quad \bar q(y)\to  \bar q(y) (I\otimes\xi_\mu) \ .
\eqn{SETshifts}
\end{equation}
Thus,
for each of the
sixteen possibilities for $\xi_t$,  the 
bilinear $\bar q_i (\gamma_s \otimes \xi_t) q_i $ undergoes
a unique set of sign changes under shifts in the four directions $\hat \mu$. 
Since the only bilinears that are invariant under all shifts are those with
$\xi_t=I$, this immediately shows why taste symmetry cannot be broken by
bilinear operators.  
Moreover, it forces
$\xi_t=\xi_{t'}$
in the four-quark operators of the SET, \eq{dim6op}.

We now consider the implications of rotations and parity.  
Rotational symmetry requires that Lorentz (Euclidean) indices be repeated and summed
over, but since the lattice action is invariant only under $90^\circ$  rotations,
an index can be repeated any even number of times before summing, not just twice.
Further, with staggered quarks, the lattice rotational symmetry transforms the
taste indices together with the space-time (and spin) 
indices \cite{vandenDoel:1983mf,Golterman:1984cy}.  Since, 
$\xi_t=\xi_{t'}$,
the spin indices on $\g_{s'}$ must be the same as those on $\g_s$.  
Parity then forces $\g_s$ and $\g_{s'}$ to be identical;
combinations such as $\g_s=\g_\nu$,
$\g_{s'}=\g_{\nu5}$ are forbidden.  There 
are now only two choices:  either the spin indices and taste indices are separately
summed over, or there are some indices that are common to both the spin and taste matrices.
\textcite{Lee:1999zxa} called the former class of operators ``type A,'' and the
latter, ``type B.''

Because there are twelve choices for an odd bilinear, there are a total of twelve
type-A operators. 
An example is
\begin{eqnarray}
O_{[V\times P]}  &=&  a^2 \bar q_i (\g_\mu \otimes \xi_5) q_i\; 
\bar q_j (\g_\mu \otimes \xi_5) q_j\ , %
\eqn{typeA}
\end{eqnarray}
with the repeated index $\mu$ summed over.
The fields here have standard continuum dimensions, 
so we write explicit factors
of $a$ to give the operator
dimension four.
Note that type-A operators are 
invariant over the full Euclidean space-time 
rotation group, SO(4), as well as a corresponding SO(4) of taste, 
a subset of the complete SU(4)$_V$ of taste that appears in the continuum limit.

In order to have a sufficient number of indices to construct a type-B operator, 
either $\g_s=\g_{s'}$ or $\xi_t=\xi_{t'}$ must be  a tensor ($T$); the other
set is then either $V$ or $A$.  Thus there are four type-B operators.
An example is
\begin{equation}
O_{[V_\mu\times T_\mu]}  =  a^2 \left[ \bar q_i (\g_\mu\otimes \xi_{\mu\nu}) q_i\; 
\bar q_j (\g_\mu \otimes \xi_{\nu\mu}) q_j -
\bar q_i (\g_\mu\otimes \xi_{\mu\nu}\xi_5) q_i\; 
\bar q_j (\g_\mu \otimes \xi_5\xi_{\nu\mu}) q_j\right] \ ,
\eqn{typeB}
\end{equation}
where the second term ensures
that the operator has
no separate spin- or taste-singlet piece.
Since the index
$\mu$ is repeated four times, one
sees explicitly from \eq{typeB} that type-B operators
are invariant only under joint  
$90^\circ$ rotations of spin and taste.

The SET to $\cO(a^2)$ for $N_f$ flavors of (unrooted) staggered fermions
is then simply the continuum QCD Lagrangian for $4N_f$ species together with
the above type-A and type-B operators.%
\footnote{There are additional $\cO(a^2)$ terms in the SET, for example
from the gluon sector, that we ignore here for simplicity.  Such terms
are taste invariant, and at leading order only produce ``generic''
effects in the chiral Lagrangian: $\cO(a^2)$ changes in the physical
low energy constants.}
Given this SET, the $\cO(a^2)$ 
chiral Lagrangian is constructed by finding --- with a ``spurion'' analysis, outlined below ---
the chiral operators that break
the full SU($4N_f$)$_L\times$SU($4N_f$)$_R$ symmetry in the same way as the
four-quark operators in the SET do.  
However, the
symmetry
is also broken by the quark mass terms in the SET.
In order to arrive at a consistent expansion scheme (a consistent power counting)
for the chiral theory, 
we must first decide how the breaking by $a^2$ terms compares with the breaking
by mass terms.  

The standard power counting, which we follow here,
takes $a^2 \sim m$, where $m$ is a generic quark mass.  More precisely, 
we assume that  (see \eq{taste-split})
\begin{equation}
a^2 \delta \sim m^2_{\pi_P} = 2Bm \ ,
\eqn{power-counting}
\end{equation}
where $a^2\delta$ is a typical pion taste-splitting. 
 The taste splittings and squared Goldstone
pion masses are indeed comparable in current MILC 
ensembles.  
Goldstone pion masses range from about $240\;\MeV$ to $600\;\MeV$;
while, on the ``coarse'' ($a\approx 0.12\;$fm) ensembles,
 the average taste splitting is about $(320\;\MeV)^2$.  This splitting drops
to about $(210\;\MeV)^2$ on the ``fine'' ($a\approx 0.09\;$fm) ensembles
and to about $(125\;\MeV)^2$ on the ``superfine'' ($a\approx 0.06\;$fm) ensembles.
It is clear that \eq{power-counting} is appropriate in the range of lattice
spacings and masses we are working on. However, for future analysis of data
that include still finer lattices and omit the coarse and possibly the fine ensembles,
it might be reasonable to use a power counting where $a^2$ is taken to be smaller than $m$.

To derive the leading order (LO) chiral Lagrangian, 
we
start with the Lagrangian in the continuum limit, \ie in the absence of taste-breaking
operators. In Euclidean space, we have
\begin{equation}
        \cL^{\rm cont}  =  \frac{f^2}{8} \Tr(\partial_{\mu}\Sigma 
        \partial_{\mu}\Sigma^{\dagger}) - 
        \frac{1}{4}B f^2 \Tr(\cM\Sigma+\cM\Sigma^{\dagger})
        + \frac{m_0^2}{24}(Tr(\Phi))^2  \ ,
\eqn{Lcont}
\end{equation}
where the meson field $\Phi$, $\Sigma=\exp(i\Phi / f)$, and the quark
mass matrix $\cM$ are $4N_f \times 4N_f$
matrices,  
and $f$ is the pion decay constant at LO.
The field $\Sigma$ transforms
under SU($4N_f$)$_L\times$SU($4N_f$)$_R$ as $\Sigma \rightarrow L\Sigma
R^{\dagger}$.
The field $\Phi$ is given by:
\begin{eqnarray}\label{eq:Phi}
        \Phi = \left( \begin{array}{cccc}
                 U  & \pi^+ & K^+ & \cdots \\*
                 \pi^- & D & K^0  & \cdots \\*
                 K^-  & \bar{K^0}  & S  & \cdots \\*
                \vdots & \vdots & \vdots & \ddots \end{array} \right),
\end{eqnarray}
where each entry is a $4\times 4$ matrix in taste space, with, for example,
$\pi^+\equiv\sum_{a=1}^{16} \pi^+_a T_a$.
The 16 Hermitian taste generators $T_a$ are
$T_a = \{ \xi_5, i\xi_{\mu5}, i\xi_{\mu\nu} (\mu>\nu), \xi_{\mu}, I\}$.
Since the normal staggered mass term is taste invariant
(see \eq{action_Kluberg}), the mass matrix has the form
$\cM = {\rm diag}( m_u I,  m_d I, m_s I, \cdots)$.

The quantity  $m_0$ in \eq{Lcont} is  the anomaly
contribution to the mass of the taste- and flavor-singlet meson,
the $\eta'\propto \Tr(\Phi)$.  As usual, the
$\eta'$ decouples in the limit $m_0\to\infty$. However, one may
postpone taking the limit and keep the
$\eta'$
 as a dynamical field \cite{Sharpe:2001fh} in order to avoid putting conditions on the
diagonal elements of $\Phi$.  
These diagonal fields, $U,D,\dots$, are then simply the $u\bar u$, $d\bar d$ bound states,
which makes it easy 
to perform a ``quark flow'' analysis \cite{Sharpe:1989bb,Sharpe:1992ft}
by following the
flow of flavor indices through diagrams.

Since a typical pion four-momentum $p$ obeys $p^2\sim m_\pi^2 \sim 2Bm$, both the kinetic
energy term and the mass term in \eq{Lcont} are $\cO(m)$.  By
our power counting scheme, \eq{power-counting},
we need to add $\cO(a^2)$ chiral operators to
complete the LO Lagrangian.  These are induced by the $\cO(a^2)$ operators in the SET.
We start with the type-A operator
$O_{[V\times P]}$ of \eq{typeA}. Using $q_i = q_i^R + q_i^L$, with $q^{R,L}_i =
 [(1\pm\g_5)/2] q_i$, and similarly for $\bar q_i$ with $\bar q^{R,L}_i =
 \bar q_i [(1\mp\g_5)/2] $, we have
\begin{equation}
O_{[V\times P]}  = a^2 \left[\bar q^R_i \left(\g_\mu\otimes \xi_{5}\right) q^R_i +
\bar q^L_i \left(\g_\mu\otimes \xi_{5}\right) q^L_i\right]^2 
 \equiv  \left[\bar q^R \left(\g_\mu\otimes F_R\right) q^R +
\bar q^L \left(\g_\mu\otimes F_L\right) q^L\right]^2  \ ,
\eqn{OVP}
\end{equation}
where flavor indices are implicit in the 
last expression.
The spurions $F_R$ and $F_L$ will eventually take the values%
\begin{equation}
F_R = a\; \xi^{(N_f)}_5\equiv a\; \xi_5 \otimes I_{\rm flavor}\ ,\qquad
F_L = a\; \xi^{(N_f)}_5\equiv a\; \xi_5 \otimes I_{\rm flavor}\ ,
\eqn{spurions}
\end{equation}
but for the moment are given spurious SU($4N_f$)$_L\times$SU($4N_f$)$_R$ transformation 
properties $F_R \to RF_R R^\dagger$ and   $F_L \to LF_L L^\dagger$ in 
order to make $O_{[V\times P]}$ ``invariant.''   

The corresponding $\cO(a^2)$ operators in the chiral Lagrangian are then
invariants constructed only from $\Sigma$, $\Sigma^\dagger$, and  quadratic factors
in $F_R$ and/or $F_L$. We cannot use derivatives or factors of the mass
matrix $\cM$ because such terms would be higher order.  It turns out that there
is only one such operator:  
\begin{equation}
        C_1 \Tr(F_L\Sigma F_R \Sigma^{\dagger})
 = C_1 a^2 \Tr(\xi^{(N_f)}_5\Sigma\xi^{(N_f)}_5\Sigma^{\dagger}) \ ,
\eqn{O1}
\end{equation}
where $C_1$ is
a constant that can be determined 
in principle by fits to staggered lattice data.

The eleven other type-A operators can be treated in the same way, though
of course different operators will have different spurions with
different transformation properties.  Some
of the type-A operators give more than one chiral operator, but, because of repeats,
a total of only eight chiral operators are generated.  

The type-B operators couple space-time and taste indices, and are invariant
only under $90^\circ$ rotations.  
Their chiral representatives must therefore have derivatives to carry the space-time
indices; an example is $\Tr(\Sigma\partial_\mu\Sigma^\dagger \xi^{(N_f)}_\mu
\Sigma^\dagger \partial_\mu\Sigma  \xi^{(N_f)}_\mu)$ \cite{Sharpe:2004is}.
Because of the derivatives, however, these operators are higher order and do
not appear in the LO chiral Lagrangian.
This was an important insight of \textcite{Lee:1999zxa}. It means that at LO the 
physics has the 
``accidental'' SO(4) taste symmetry of the type-A operators.

We can now write down the complete LO chiral Lagrangian:
\begin{equation}
        \cL  =  \frac{f^2}{8} \Tr(\partial_{\mu}\Sigma 
        \partial_{\mu}\Sigma^{\dagger}) - 
        \frac{1}{4}B f^2 \Tr(\cM\Sigma+\cM\Sigma^{\dagger})
        + \frac{m_0^2}{24}(\Tr(\Phi))^2 +a^2\cV \ ,
\eqn{LSChPT}
\end{equation}
where the taste-violating potential $\cV$ is given by
\begin{eqnarray}
        -\cV  & = & C_1
         \Tr(\xi^{(N_f)}_5\Sigma\xi^{(N_f)}_5\Sigma^{\dagger}) 
+\frac{C_3}{2} [ \Tr(\xi^{(N_f)}_{\nu}\Sigma
        \xi^{(N_f)}_{\nu}\Sigma) + h.c.] \nonumber \\*
        & & +\frac{C_4}{2} [ \Tr(\xi^{(N_f)}_{\nu 5}\Sigma
        \xi^{(N_f)}_{5\nu}\Sigma) + h.c.] 
+\frac{C_6}{2}\ \Tr(\xi^{(N_f)}_{\mu\nu}\Sigma
        \xi^{(N_f)}_{\nu\mu}\Sigma^{\dagger}) \nonumber \\*
        & & + \frac{C_{2V}}{4} 
                [ \Tr(\xi^{(N_f)}_{\nu}\Sigma)
        \Tr(\xi^{(N_f)}_{\nu}\Sigma)  + h.c.] 
        +\frac{C_{2A}}{4} [ \Tr(\xi^{(N_f)}_{\nu
         5}\Sigma)\Tr(\xi^{(N_f)}_{5\nu}\Sigma)  + h.c.] \nonumber \\*
        & & +\frac{C_{5V}}{2} [ \Tr(\xi^{(N_f)}_{\nu}\Sigma)
        \Tr(\xi^{(N_f)}_{\nu}\Sigma^{\dagger})]
         +\frac{C_{5A}}{2} [ \Tr(\xi^{(N_f)}_{\nu5}\Sigma)
        \Tr(\xi^{(N_f)}_{5\nu}\Sigma^{\dagger}) ],
\eqn{V}
\end{eqnarray}
with implicit sums over repeated indices. 

Expanding \eq{LSChPT} to quadratic
order in the meson field $\Phi$, we find, as expected, that pions with nonsinglet
flavor fall into SO(4) taste multiplets, labeled by $P$, $A$, $T$, $V$, $S$.
We show numerical evidence for this
in \secref{Rooting}.  
The splittings $\delta_{t}$ of \eq{taste-split},
with $t=P$, $A$, $T$, $V$, $S$, are given in terms of $C_1$, $C_3$, $C_4$ and $C_6$.   
The presence of two traces in the terms multiplied by $C_{2V}$, $C_{2A}$,
$C_{5V}$, and $C_{5A}$ means that they cannot contribute at this order 
to the masses of (flavor-)charged mesons.  
\textcite{Aubin:2003mg} showed that
such terms do
generate ``taste hairpins,''  which mix the flavor-neutral mesons
however,
of taste $V$ (and separately, taste $A$).  In other words, there are terms 
of form  $ \frac{a^2\delta'_V}{2} ( U_{\mu}+D_{\mu}+S_{\mu}+\cdots)^2$
and
$\frac{a^2 \delta'_A}{2} ( U_{\mu5}+D_{\mu5}+S_{\mu5}+\cdots)^2$
in the expansion of \eq{LSChPT}, where $\delta'_V$ and $\delta'_A$ are functions
of $C_{2V}$, $C_{2A}$, $C_{5V}$, and $C_{5A}$.  These terms have been indirectly observed
\cite{Aubin:2004fs}
in fits to charged pion masses and decay constants to one-loop
expressions derived from \eq{LSChPT}.  Because of the practical difficulties in
simulating disconnected diagrams, taste-hairpins have not yet been studied directly
in two-point functions of neutral mesons.

So far, the entire discussion of \schpt\ has been in the context of unrooted
staggered quarks.  
\textcite{Bernard:2001yj} and \textcite{Aubin:2003mg}
 proposed that
rooting could be taken into account by using quark flow
to determine
the presence of closed sea-quark loops in an \schpt\ 
diagram, and then multiplying the diagram by a factor of $1/4$ for each such loop.
This is a natural assumption,  because it is exactly what happens in
weak coupling
perturbation theory \cite{Bernard:1993sv}. 
In the chiral theory, however, the validity of
the prescription is not obvious. 

To study in more detail how rooting should be handled in \schpt, it is convenient
to replace the quark-flow picture with a more systematic way to find and
adjust the sea-quark loops.  This is provided by a ``replica rule,''
introduced for this problem by \textcite{Aubin:2003rg}.  Since rooting
is defined as an operation on sea quarks,
it is useful first to separate off the valence quarks by
replacing the original theory with a partially-quenched one: 
introduce
new (valence) quarks along with ghost (bosonic) quarks to cancel the
valence determinant.  The adjustment to the \schpt\ theory, \eq{LSChPT}, is
the standard one
for a partially-quenched theory \cite{Bernard:1993sv}: just add some additional quark flavors 
and corresponding bosonic flavors.  
The masses of the valence quarks may be equal to or different from the
sea masses. The latter case is clearly unphysical, but is useful for
getting more information out of a given set of sea-quark configurations.

We 
now
replicate each sea-quark flavor $n_r$
times, where $n_r$ is a positive integer, so that there are total of $n_rN_F$
flavors.  We then calculate as usual with the replicated (and partially-quenched)
version of \eq{LSChPT}, going to some given order in chiral perturbation theory.
The result will be a polynomial in $n_r$,
where factors of $n_r$
arise from summing over the indices in chiral loops.
Finally, we put $n_r=1/4$ in the polynomial. We thus take into account the rooting
by effectively counting each sea-quark flavor
as $1/4$ of a flavor, which
cancels the factor of $4$ that arises from the taste degree of freedom.  The
chiral theory obtained by applying this replica rule to \schpt\ is called
``rooted staggered chiral perturbation theory'' (\rschpt).  

Note that
we have done nothing to the valence quarks.
Since
the number
of tastes of the sea quarks has been reduced
by a factor of 4,  it is clear that there is a mismatch, even when the valence
masses are taken equal to the sea masses.
This is
still
true 
in the continuum limit, where 
the issue is particularly transparent.  When taste symmetry is exact,
rooting removes three of the
four tastes from the quark sea for each physical flavor, 
but leaves the valence quarks unaffected.
It is therefore possible
to construct Greens functions, either at the quark or the chiral level, which
are unphysical, in the sense that the external particles have no counterpart
in the intermediate states. 
Sharpe has called this the ``valence-rooting problem'' \cite{Sharpe:2006re}.
The solution is however straightforward \cite{Sharpe:2006re,Bernard:2006vv,Bernard:2007eh}:
the physical subspace can be obtained simply by choosing all external particles
to have a single value of taste (taste 1, say).  Using flavor and taste symmetries,
other Greens functions may also be constructed that happen to equal these physical
correlators in the continuum limit \cite{Bernard:2006vv}. Nevertheless, most 
Greens functions will 
be unphysical.  This is not a cause for concern
as long as there is a physical subspace.  In fact such a situation has nothing,
{\it per se}, to do with rooting:  it will happen in continuum QCD, or in any
lattice version thereof, if we introduce arbitrary numbers of valence quarks.

We emphasize   
that the replica rule tells us to take into account only the explicit factors of $n_r$
from chiral loops. Putting $n_r=1/4$ in the polynomial resulting from the \schpt\ 
calculation is 
thus a
 well-defined procedure.
We are not concerned with
the fact that, if replication is done in the fundamental, QCD-level theory, 
the low energy constants (LECs) such as $f$ and $B$ will be (implicit) functions
of $n_r$.  Such dependence is in general unknown and nonperturbative, and not
amenable to analytic continuation in $n_r$.  
Instead, as is always the case in chiral perturbation theory, we treat the LECs
as free parameters. After setting explicit factors of $n_r$ to $1/4$ in our calculations,
the LECs can be determined   
by fitting the lattice 
data to the resulting chiral forms.
The unknown dependence of
the LECs on $n_r$ is however an obstacle in trying to show, directly
from the fundamental theory, that \rschpt\ is  
the correct chiral theory.  
This is discussed further in  \secref{Rooting}.

\subsection{Extensions of staggered chiral perturbation theory}
\label{sec:HQSChPT}

The purely staggered theory discussed thus far is often insufficient
for calculations of many physical quantities.
It would be very difficult, for example, to simulate heavy quarks
with the asqtad action at currently-available lattice 
spacings because of the large discretization errors 
that appear when $am\sim 1$.
Thus, the determination of phenomenologically important properties
of heavy-light mesons and baryons has usually been carried out by adding
a heavy valence quark with
the Fermilab \cite{ElKhadra:1996mp} or NRQCD  \cite{Thacker:1990bm} action
to asqtad simulations of the sea quarks and light valence quarks.
Alternatively, HISQ valence quarks have been used on the asqtad sea configurations
to get precise results for charmed mesons \cite{Follana:2007uv}. 
To the accuracy strived for in current calculations, the effects of heavy 
sea quarks can be neglected; that is, these quarks can be treated in the 
quenched approximation.

For several other quantities, the complicated effects of taste-symmetry
violation make staggered quarks difficult to use.  Since these effects
often present the greatest obstacle in the valence sector,
a very successful compromise, first introduced in \textcite{Renner:2004ck}, has been to 
add domain-wall valence quarks on top of the
MILC sea-quark ensembles.  Such ``mixed-action'' simulations are being used
to study scalar mesons \cite{Aubin:2008wk},
$B_K$ and related quantities \cite{Aubin:2008wk,Aubin:2008ie,Aubin:2007pt},
nucleon properties \cite{Bratt:2008uf,Renner:2007pb,Hagler:2007xi,Edwards:2005ym},
hadron spectroscopy \cite{WalkerLoud:2008bp,Edwards:2006zza},
meson scattering \cite{Beane:2007xs,Beane:2007uh},
and nuclear-physics topics \cite{Beane:2007es,Detmold:2008yn,Detmold:2008fn,Beane:2006gf}.

To take full advantage of simulations with heavy valence quarks or
mixed actions, it is useful to have  chiral effective theories that
properly include the discretization effects.  We briefly discuss such
theories, starting with the mixed-action case of domain-wall valence
quarks on a staggered sea. The basic ideas of mixed-action chiral
perturbation theory were developed in \textcite{Bar:2002nr,Bar:2003mh}
and \textcite{Golterman:2005xa} for the case of 
Wilson fermions in the sea and chiral fermions in the valence sector. By
chiral fermions we mean overlap or domain wall quarks, where we assume for
domain wall quarks that $L_s$ is large enough
that the residual mass is negligible.
The extension to
chiral valence
fermions on staggered sea quarks \cite{Bar:2005tu} is then
fairly straightforward.
Features of mixed-action chiral theory that are universal, in
the sense that they are independent of the sea-quark action, have been
discussed in \textcite{Chen:2006wf,Chen:2007ug}.

Because the valence and sea quarks have different actions,
a mixed-action theory lacks the symmetries that would normally rotate valence into
sea quarks (or {\it vice versa}) in a standard theory.  Since we assume that both the
valence and sea sectors approach the expected continuum theories as $a\to0$,
these symmetries should be restored in the continuum limit.
At the level of the Symanzik effective action, the violation of these symmetries
first appears at $\cO(a^2)$ in the existence of
independent ``mixed'' four-quark
operators:  in our case, the product of a domain-wall (valence) bilinear and a
staggered (sea) bilinear.  We know, following the development in \secref{SChPT},
that each bilinear must be separately chirally invariant, and that any
staggered bilinear must be taste invariant. It is then simple to see that only
two mixed four-quark operators are possible:
\begin{equation}
O_{V}  =  a^2 \bar \psi_a \g_\mu \psi_a\; 
\bar q_i (\g_\mu \otimes I ) q_i\ , \qquad\qquad
O_{A}  =  a^2 \bar \psi_a \g_\mu\g_5 \psi_a\; 
\bar q_i (\g_5\g_\mu \otimes I ) q_i\ ,
\eqn{mixed-ops}
\end{equation}
where $\psi_a$ is a domain-wall quark 
or ghost
of valence flavor $a$, and $q_i$ is a staggered
quark of sea flavor $i$, and $a$ and $i$ are summed over.  
As in the pure staggered case, the color indices in these operators 
are irrelevant.

In addition to the operators in  \eq{mixed-ops}, there are the full complement of
standard, purely staggered four-quark operators in the sea sector, and standard, purely domain-wall
four-quark operators 
involving valence quarks and valence ghosts.
In a  
normal
theory, the relative coefficients of corresponding sea-sea,
valence-valence, and valence-sea operators 
would be fixed by the symmetries.
But in the mixed case, 
all such operators are independent
and must be treated separately.

In the corresponding chiral effective theory, the
purely sea-quark
sector is the same as the sea-quark sector of a standard staggered theory.  Similarly, the
purely valence-quark sector  is the same as the valence-quark sector
of a standard domain-wall theory.
Mixed valence-sea mesons are affected by various operators, including the operator corresponding to \eq{mixed-ops}:
\begin{equation}
-a^2 C_{\rm Mix} \Tr(\tau_3 \Sigma \tau_3 \Sigma^\dagger) \ ,
\eqn{chiral-mix}
\end{equation}
where $\Sigma$ is the complete chiral field involving both sea and valence (and ghost-valence)
quarks, and $\tau_3$ is a diagonal matrix that takes the value $+1$ in the sea sector
and $-1$ in the valence sector.  
At LO one finds \cite{Bar:2005tu,Chen:2009su}
\begin{eqnarray}
m^2_{\pi,ab} &=&  B(m_a+m_b) \nonumber \\
m^2_{\pi,ij,t} &=&  B(m_i+m_j) + a^2 \delta_t 
\label{eq:mixed-split} \\
m^2_{\pi,ia} &=&  B(m_i+m_a) + a^2 \delta_{\rm Mix}\ ,  \nonumber 
\end{eqnarray}
where $a,b$ are domain-wall (valence) flavors,
$i,j$ are staggered (sea) flavors, 
$t$ is the taste of a sea-sea meson, as in \eq{taste-split}, and
$\delta_{\rm Mix}$ is a function of $C_{\rm Mix}$ and other low energy constants.
\textcite{Orginos:2007tw} and \textcite{Aubin:2008wk}
have determined  $\delta_{\rm Mix}$ numerically by measuring the masses of
mixed mesons.  

The mixed-action chiral Lagrangian thus developed can 
be used to calculate
one-loop effects in pseudoscalar masses and decay constants \cite{Bar:2005tu},
in $B_K$ \cite{Aubin:2006hg} and $I=2$ $\pi-\pi$ scattering \cite{Chen:2005ab}.

Next, we consider the case of heavy-meson staggered chiral perturbation
theory (HM\schpt), the relevant chiral theory for a heavy meson made out
of a heavy valence quark 
and a
light staggered valence quark, on the background of staggered sea quarks.
HM\schpt\ is designed to parameterize the light quark chiral extrapolation
and the light quark discretization effects. Discretization errors due
to the heavy quark are not included; it is assumed that they can be
estimated independently by using heavy-quark effective theory (HQET)
\cite{Isgur:HQET,Neubert:1993mb} to
describe the lattice heavy quark \cite{Kronfeld:2000ck,Kronfeld:2003sd}.

At the level of the   
SET,
the first nontrivial effect of combining
the heavy quark with the staggered theory is again the generation of mixed four-quark
operators (a heavy-quark bilinear times a light-quark one).  As before, such
operators do not break taste symmetry.  Furthermore, unlike the mixed-action case,
symmetry between heavy and light quarks is already strongly broken by the
heavy-quark mass.  So the mixed 
operators have no important effects in this
case.

The power counting 
for heavy-light mesons in \chpt\ 
makes the
HM\schpt\ at LO rather simple \cite{Aubin:2005aq}.
In the continuum, the chiral Lagrangian for heavy-light mesons
\cite{Manohar:2000dt} starts at $\cO(k)$, with $k$ the residual momentum
of the heavy quark.  The light meson momentum $p$ should also be $\cO(k)$.
In our 
staggered power counting,
\eq{power-counting}, we take
$ p^2\sim m_\pi^2 \sim a^2$.
This means that the
LO heavy-light meson terms are lower order than
the 
$\cO(a^2)$ discretization errors
in the light quark action.
The
LO heavy-light meson propagator and vertices are thus the same as in the
continuum, as are the heavy-light currents that enter, \eg
in leptonic and semileptonic decays. The  
light-quark discretization errors in heavy-light meson quantities
first appear at one loop (NLO),
through the taste violations in the light meson propagators in the loop.
These 
corrections have been calculated for heavy-light leptonic
decay constants \cite{Aubin:2005aq},  for   
semileptonic heavy-to-light
decays, \eg $B\to\pi$, \cite{Aubin:2007mc}, and for semileptonic heavy-to-heavy
decays, \eg $B\to D$ and $B\to D^*$ \cite{Laiho:2005ue}.
There are also
analytic NLO  corrections to physical processes,
coming
both from light-quark mass corrections (as in
the continuum) and from taste-violating corrections to the LO Lagrangian and
currents.
In practice,
it is usually
easy to guess these analytic NLO corrections
from symmetry arguments, so it is not necessary
to use the  
complicated
NLO heavy-light
Lagrangian \cite{Aubin:2005aq}.

\subsection{The issue of rooting}
\label{sec:Rooting}

The extra tastes are eliminated in staggered dynamical simulations
by taking
the fourth root of the fermion determinant --- the fourth-root
procedure. 
In the past few years there has been progress in understanding and validating this procedure,
and we give a 
brief overview of this progress here.  For more detailed
discussion, and full lists of references, see recent reviews by \textcite{Sharpe:2006re}, \textcite{Kronfeld:2007ek} and \textcite{Golterman:2008gt}.

The fourth-root
procedure
would be unproblematic if
the action had full SU(4)$_V$ taste symmetry, which would give a
Dirac operator that was block-diagonal in taste space.  Indeed, this is
the ``cartoon version'' of
what we expect 
in the continuum limit.  Assuming taste symmetry
is restored, the positive fourth root
of the positive staggered determinant 
would
then become equal to the determinant
of a single continuum species.  

\begin{figure}
\begin{center}
\begin{tabular}{c c}
\includegraphics[width=2.5in]{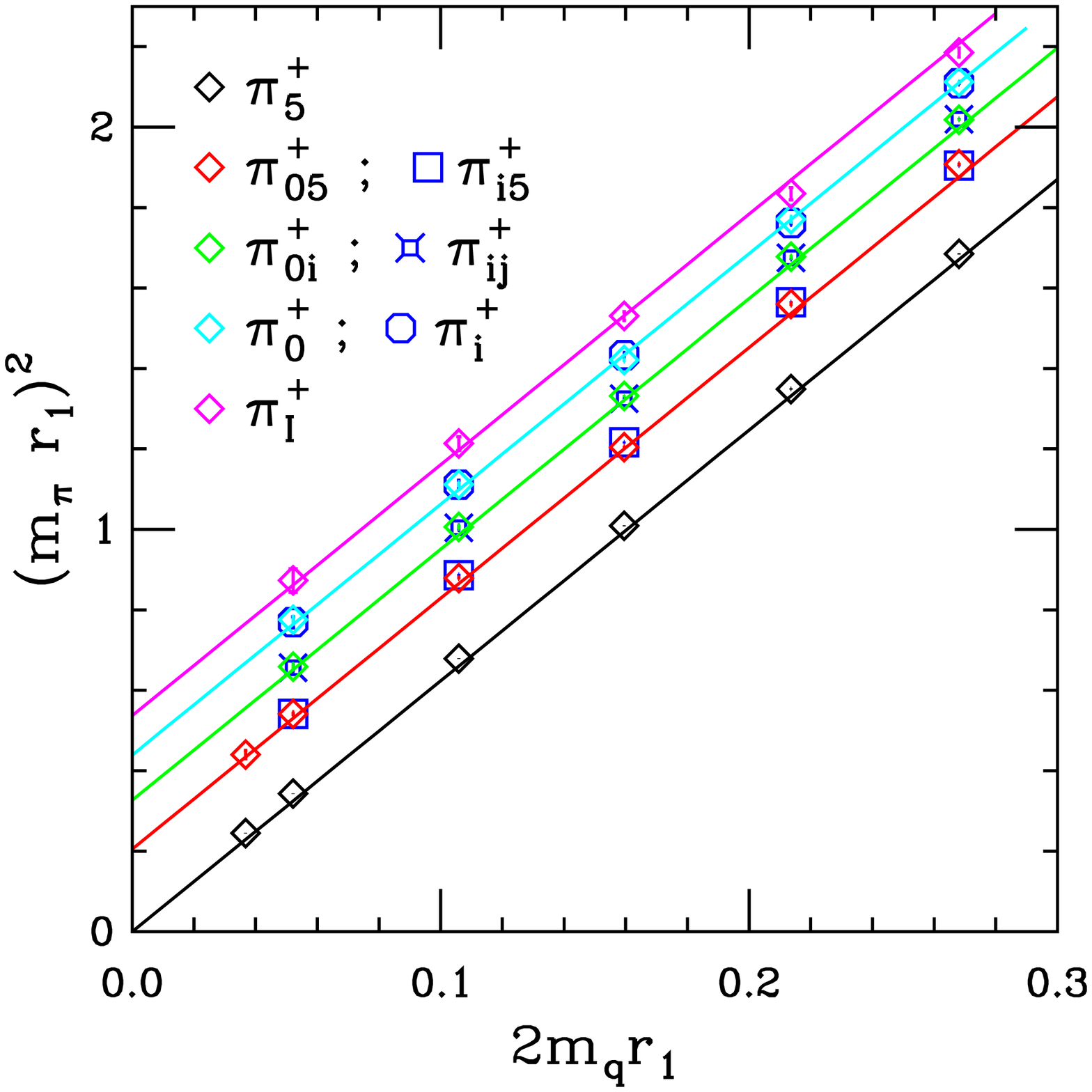}
&
\includegraphics[width=2.66in]{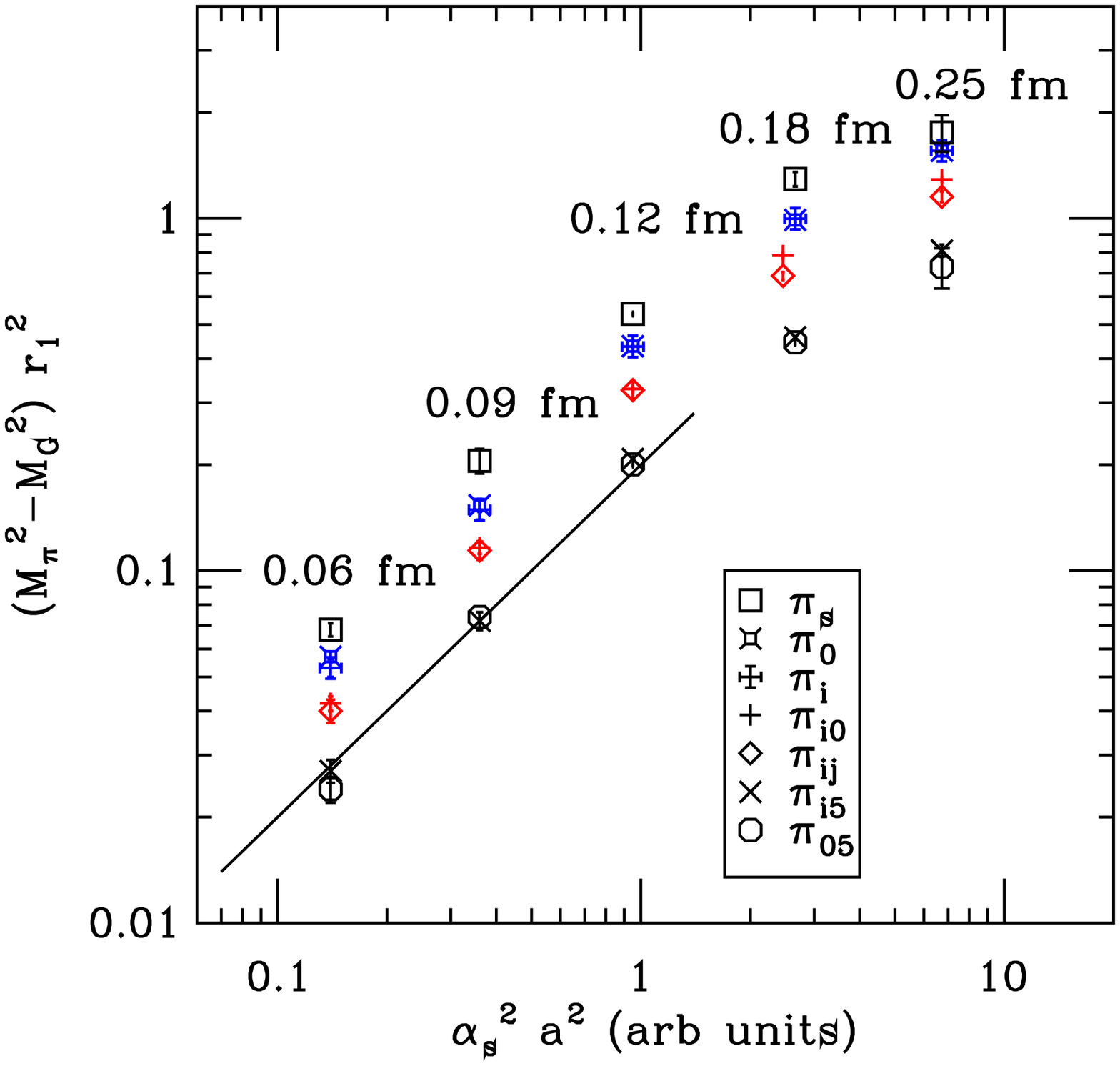}
\end{tabular}
\end{center}
\caption{
Squared charged pion masses, in units of $r_1$, as function of quark
mass (left).  Figure from \textcite{Bernard:2006wx,Bernard:2007ez}. A previous version
appeared in \textcite{Bernard:2001av}.  The splittings appear to be
independent of the quark mass.
The taste splittings as function of $\alpha^2 a^2$ (right) in a log-log
plot, showing the expected behavior, indicated by the diagonal straight
line. A slightly different version of this figure appeared in
\textcite{Bernard:2006nj}.} 
\label{fig:taste-split}
\end{figure}

However, at nonzero lattice spacing $a$,
taste symmetry is broken and the Dirac operator is not block-diagonal
(see \eq{action_Kluberg}). From \eq{D_KS}, one has
\begin{equation}
\ln\, \det(D_{KS}+ m \otimes I) = 4\, \ln\,\det(D+m) + \ln\, \det\{ I +
 [(D+m)^{-1}\otimes I]a\Delta\} \ .
\end{equation}
Since $ (D+m)^{-1}$ is nonlocal, we should not expect the rooted theory to be
local for $a\not=0$.
In fact it is possible to prove \cite{Bernard:2006ee}
that the fourth root of the determinant is not equivalent
to the determinant
of any local lattice Dirac operator.%
\footnote{``Equivalent'' here means equal up to a factor
of the exponential of some local effective action of the gauge field.
This is enough to guarantee that the two theories 
have the same physics at distances much larger than the lattice spacing
\cite{Adams:2004mf,Shamir:2004zc}.
}
The idea of the proof is 
simple:
If there were such a local operator, then one could
construct a theory with four degenerate quarks, each one with that 
local action. Calling this introduced degree of freedom ``taste,'' one now
has a local theory with exact SU(4)$_V$ taste symmetry by construction, and whose determinant
is equivalent to that of the original staggered theory.  This is a contradiction,
because the taste symmetry of the constructed theory guarantees that it has fifteen
pseudo-Goldstone bosons (pions), whereas the staggered pions are known to split
up into nondegenerate irreducible representations \cite{Golterman:1985dz,Lee:1999zxa}.
Indeed,
\figref{taste-split} shows our lattice measurements of the pion splittings as a function
of quark mass (left) and lattice spacing (right).  The left plot clearly shows 
the characteristic splitting of the charged pion ($\pi^+$) 
multiplet into the five nondegenerate submultiplets
with tastes $P$, $A$, $T$, $V$, $S$. This is as predicted at
$\cO(a^2)$ in the chiral expansion, as discussed in \secref{SChPT}. Further splitting at higher order into
a total of eight submultiplets is allowed by the lattice symmetries \cite{Golterman:1985dz},
but we see little evidence of that at the current level of statistics.

The same features of the rooted theory that imply nonlocality
also imply nonunitarity on the 
lattice \cite{Prelovsek:2005rf,Bernard:2006vv,Bernard:2006zw,Bernard:2007qf}.  
The issue is particularly sharp
in the rooted one-flavor theory.  The physical one-flavor theory should have
no light pseudoscalar mesons (pions) but only a heavy $\eta'$.  
In a rooted theory with
exact taste symmetry (\eg with four copies of rooted overlap quarks), this 
works automatically: the fourth power of the fourth root of a (positive) determinant
is equal to the determinant itself.
Alternatively, one can check directly
in the rooted four-taste theory that, in physical correlators, the pion intermediate
states cancel and only the $\eta'$ remains \cite{Bernard:2006vv}. 
On the other hand, in the rooted one-flavor staggered theory, the
pions have different masses at nonzero lattice spacing and cannot
cancel, leaving light intermediate states with both positive
and negative weights. 
This is a clear violation of unitarity.

In the continuum limit, we expect that all the pions become  degenerate.
For the tree-level improved asqtad fermions,
generic lattice artifacts are of order $\mathcal{O}(\alpha_s a^2)$.
Taste violations, however, require exchange of at least two UV gluons, since
the coupling of a quark to a single gluon with any momentum components equal to $\pi/a$
vanishes.
Therefore taste violations with the asqtad action
should vanish as $\alpha_s^2 a^2$ as 
$a\to0$.
The lattice-spacing dependence of the pion splittings, 
shown in the right-hand plot of \figref{taste-split}, 
agrees very well with this expectation. Note that since
we are looking here at flavor-nonsinglet pions, 
the taste-singlet $\pi^+_I$ also becomes degenerate with the other fifteen pions
as the continuum limit is approached. 

Thus, the rooted staggered theory is inherently nonlocal and nonunitary
at nonzero lattice spacing, but
should become local and unitarity in the continuum limit if taste symmetry is restored.
This is because, 
in the limit
of exact taste symmetry, 
rooting of the sea quarks is equivalent to restriction to
a single taste, which is a local operation.
Clearly, the numerical evidence for taste-symmetry restoration in the continuum is strong,
and accords with the theoretical expectation coming from the fact that taste violation
is due to an operator with dimension five.
How, then, could rooting go wrong?  The main problem is that the theoretical
expectation is based on
standard lore of the renormalization group (RG) that operators with dimension 
greater than four are irrelevant in the
continuum limit.  This standard lore for the scaling of operators 
assumes a local lattice action, which does not
apply here.
The numerical results indicate that the lore is not leading us
astray, but of course numerical evidence does not constitute a proof. 

There is a further problem in the formal argument we have made so far
that rooting is equivalent in the continuum limit to restriction to a single taste.
The argument seems to require
that taste symmetry is restored for the Dirac operator $D_{KS}$, \eq{D_KS}, itself.
In \figref{taste-split}, however, we are only testing the restoration of
taste symmetry at physical scales, those much larger than the lattice spacing.
At the scale of the cutoff, there is actually no reason to expect that taste symmetry
is restored.  Indeed, direct studies of the eigenvalues 
of $D_{KS}$ on the lattice \cite{Follana:2004sz,Durr:2004as} find only approximate
quartets of eigenvalues (indicating approximate taste symmetry) for {\it low-lying}
eigenvalues, those corresponding to long (physical) distance scales.

\textcite{Shamir:2004zc, Shamir:2006nj} has set
up an RG framework for both  unrooted and rooted staggered theories, and used it to address
the potential problems of rooting.
The renormalization group is clearly the natural framework to study
the scaling of operators, and it also makes possible a more precise treatment
of the continuum limit.
As one blocks $D_{KS}$ to longer distance
scales, the eigenvalues at the scale of the cutoff are removed, and one 
may then expect that taste symmetry is truly restored.

Shamir's RG scheme starts with unrooted staggered quarks, and blocks them on the
hypercubic lattice by a factor of 2 at each step, integrating
out the 
finer 
quark fields.  The gauge fields are also blocked, but the integration
over them is postponed until the end, so that the quark action stays quadratic
at every step.  
The starting  ``fine'' lattice spacing $a_f$ is blocked $n$ times to a final
``coarse'' lattice spacing $a_c$.
As $n$ is increased, the coarse spacing is held fixed but small,
with $a_c \ll 1/\Lambda_{QCD} $.  The fine lattice spacing thus obeys
$a_f = 2^{-n} a_c$, and the continuum limit is $n\to\infty$, which sends $a_f$ to zero.
In this unrooted theory, the scaling of $\Delta$ like $a_f$ is guaranteed by the
standard lore, since the action is local.

The rooted theory cannot be blocked in the same way because
rooted quarks are not defined by a standard
Lagrangian, but by a rule to replace the fermion determinant by its fourth root
in the path integral. We can, however, apply the rule at every stage of the (unrooted) blocking,
obtaining, at the $n^{\rm th}$ step, the theory given by
\begin{equation}
        Z_n^{\rm KSroot} = \int d\cA \det{}^{\frac{1}{4}}(D_{KS,n}+m_n
	\otimes I ) \ ,
\eqn{ZrootKS}
\end{equation}
where $D_{KS,n}$ is the blocked staggered Dirac operator,
$m_n$ is the (renormalized) mass on the blocked lattice,  and
$d\cA$ is the full gauge measure (which includes integrals over gauge fields
at each level of blocking, as well as Jacobian terms
coming from integrating out the fermions on the coarse lattices).
This defines a RG for the rooted theory. However, it is difficult to make
progress directly from \eq{ZrootKS}, because of the problem of nonlocality.

Shamir's key insight is that one may define, at each stage of blocking,
an intermediate,  ``reweighted theory,'' which becomes closer
and closer to the rooted staggered theory but retains locality. 
Define $D_n$ to be the taste-singlet part of $D_{KS,n}$, and $a_f \Delta_n$
to be the remainder: 
\begin{eqnarray}
D_{n}&=&\frac{1}{4}\;\tr_{ts}\left(D_{KS,n}\right)\ ,\nonumber \\
D_{KS,n}&=&D_{n}\otimes I +a_f \Delta_n\ ,
\eqn{split}
\end{eqnarray}
where $\tr_{ts}$ is the trace over taste, and $I$ is the identity in taste space.
This parallels \eq{D_KS}.
We will see below the explicit $a_f$ in the second term  of \eq{split}
does not mislead us about the scaling of $a_f\Delta$.
The operator $D_n$ is local because $D_{KS}$ is.
 Further,
 $\det(D_n+m_n)=\det^{1/4}((D_n+m_n)\otimes I)$.
The (rooted) reweighted theory is then defined by
\begin{equation}
	Z_n^{\rm reweighted} = \int d\cA \det(D_n + m_n) \ ,
\eqn{Zreweigh}
\end{equation}

Now, since the reweighted theory is QCD-like, {\it albeit}\/ with a more complicated 
gauge integration than usual, we expect it to be renormalizable and asymptotically free.
The running of the  operator $a_f \Delta_n$ from $a_f$ to $a_c$
can then be calculated perturbatively because in this range the lattice
spacings are all much less than $1/\Lambda_{QCD}$. Because the theory
is local, 
the standard lore tells us that the perturbative running will be a reliable guide
to the complete, nonperturbative behavior. Thus we expect that the operator norm
of $a_f \Delta_n$ will obey, in an ensemble-average sense,
\begin{equation}
||a_f\Delta_n|| \ltwid \frac{a_f}{a^2_c} = \frac{2^{-n}}{a_c} \ ,
\eqn{Delta-norm}
\end{equation}
where the $\ltwid$ sign implies that the scaling is true up to logs. For the same reasons,
the mass $m_n$ should run logarithmically, just as in QCD.
From this and \eq{split}, we have
\begin{eqnarray}                                                                
\det{}^{\frac{1}{4}}(D_{KS,n} + m_n\otimes I ) &=& \det(D_n + m_n )\exp         
\left({\scriptstyle \frac{1}{4}}\                                               
\tr\ln\left[I + ((D_n+m_n)^{-1}\otimes I)a_f\Delta_n\right]\right)              
\nonumber \\                                                                    
&=& \det(D_n+m_n)\left( 1 + \cO\left(\frac{a_f}{a_c^2 m_n}\right)               
\right) \ ,                                                                     
\eqn{KS-to-reweigh}                                                             
\end{eqnarray}                                                                  
where the quark mass provides a lower bound to the absolute value of the eigenvalues of 
$D_n+m_n$.
Thus, 
\begin{equation}
\lim_{n\to\infty}   Z_n^{\rm KSroot} =  \lim_{n\to\infty} Z_n^{\rm reweighted}\ .
\eqn{root-limit}
\end{equation}
In other words, the nonlocal rooted staggered theory coincides with a local, one-taste, theory
in the continuum limit, as desired.

Note that \eq{KS-to-reweigh} makes it clear that one must take 
the continuum ($a_f\to0$) limit before the chiral ($m\to0$) limit
for rooting to work.  This is not surprising, since it is already 
well known \cite{Smit:1986fn,Durr:2004ta,
Bernard:2004ab,Bernard:2006vv} that the two limits do not
commute for all physical quantities, and that taking the chiral limit first can give
incorrect answers.  This is true even for the unrooted staggered theory.
As a trivial example, consider the low energy constant $B$ (see \eq{taste-split})
defined at a given lattice spacing $a$ by
$ B(a) \equiv m^2_{\pi_t}/(2m)$
for some taste $t$.  Unless $t=P$, giving the Goldstone pion, one has
$\lim_{a\to0}\; \lim_{m\to0}\; B(a) = \infty$; while the desired
result is $\lim_{m\to0}\; \lim_{a\to0}\; B(a) = B$.

The reader may worry that the argument thus far 
presumes too much about how perturbation
theory works in the reweighted theory.  After all, the perturbation 
theory involves multiple levels of
gauge integrations, making it quite complicated.  Indeed, no such perturbative calculations
have 
been performed to date.  \textcite{Shamir:2006nj}
has pointed out, however, that we may avoid the details of perturbation theory
in the reweighted theory by leaning a bit more on the standard lore and on
perturbation theory in the {\it unrooted}\/ staggered theory, which is fairly
well understood --- see
\textcite{Sharpe:2006re} and the references therein.
One starts by considering the unrooted staggered
theory replicated $n_r$ times, where $n_r$ is an integer.  In this theory
the $\beta$ function and the logarithmic anomalous dimension of $a_f\Delta_n$ will
be the standard functions of the total number of fermion species, and $a_f\Delta_n$
will scale as expected as long as $n_r$ is not so large that asymptotic freedom is lost. 

Now, $a_f\Delta_n$ is just the difference between the (replicated) unrooted 
staggered theory and a
(replicated) {\it unrooted}\/ reweighted theory defined by the Dirac operator
$(D_n + m_n)\otimes I$.  Since $a_f\Delta_n$ gets small as $n\to\infty$ in one theory,
it must get small in the other theory.  Both theories are local, so the standard
lore says that $a_f\Delta_n$ scales as expected in perturbation theory in
the unrooted reweighted theory --- however complicated such calculations
would actually  be in practice.    The results
of perturbation theory to any fixed order are polynomial in $n_r$, with the power
of $n_r$ just counting the number of closed quark loops.  So in this
perturbation theory, we may put $n_r=1/4$ to obtain the perturbation theory
for the rooted reweighted theory, \eq{Zreweigh}.   Thus we do not have to calculate
explicitly in either the unrooted or rooted reweighted theories; we know
that $a_f \Delta_n$ will scale to zero as expected in perturbation theory.
Now the standard lore takes over, as above, for the local, rooted reweighted theory,
and says  $a_f \Delta_n$ will scale to zero as $n\to\infty$ even nonperturbatively.

A numerical test of the scaling of $a_f\Delta_n$ 
was attempted in \textcite{Bernard:2005gf}. The results were encouraging
but far from conclusive, due to quite large statistical errors. 

Of course, although the above arguments make it plausible that rooting
works, they do not constitute a rigorous proof.  As always in lattice QCD,
one relies heavily on the standard lore about RG running of irrelevant
operators, which is what ``guarantees'' universality.   
Further, we are unable to do justice here
to all the arguments
and assumptions involved in the perturbative analysis.  We have also  
ignored the nontrivial issues involving the Jacobian obtained by
integrating out the fermions at each level of blocking.   
The
Jacobian can
be written as the exponential of an effective action for the gauge fields.
The claim is that this effective action is local, basically because it
comes from short-distance fluctuations of the fermions.
The reader is urged to see \textcite{Shamir:2006nj} and the 
reviews  by \textcite{Sharpe:2006re}, \textcite{Kronfeld:2007ek} and
\textcite{Golterman:2008gt} for details and discussion.

We now consider the question of whether \rschpt\ is the correct chiral theory
for rooted staggered QCD.  This is important
first of all because \rschpt\ allows us to fit lattice data and take the limits $a\to0$
and $m\to0$ in the correct order and with controlled errors. In addition, 
the validity of \rschpt, coupled with the strong numerical evidence for the
restoration of taste symmetry for $a\to0$ (see \figref{taste-split}), guarantees that 
rooted staggered QCD produces
the desired results for the pseudoscalar meson sector in the continuum limit.
This is because \rschpt\ becomes continuum \chpt\ when taste symmetry is restored.

\begin{figure}
\begin{center}
\includegraphics[width=1.1in]{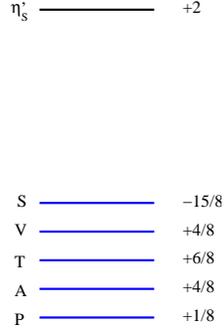}
\end{center}
\caption{Relative weights (shown at the right of each line) of two-particle intermediate states in the scalar, taste-singlet correlator
in the one-flavor case. 
The two-$\eta_S'$ state ($S$ indicates taste singlet) is shown at top; while
the various two-pion states below are labeled by the pion taste ($S$, $V$, $T$, $A$, $P$).
The height of each line represents, qualitatively, the relative mass of the state.}
\label{fig:weights}
\end{figure}

Before discussing the arguments supporting \rschpt, we note that \rschpt\ has 
the main features desired for a chiral effective theory of the rooted theory.
In particular \rschpt\ reproduces the nonunitarity and nonlocality of rooted staggered
QCD at nonzero lattice spacing.  This comes about because \rschpt,
like the rooted staggered theory itself, is not an
ordinary Lagrangian
theory, but a Lagrangian theory {\it with a rule}.  For \rschpt\ the rule is:  
calculate in the
replicated theory for integer $n_r$ number of replicas, and then set $n_r=1/4$.
Setting $n_r=1/4$ gives ``funny'' relative weights for different diagrams,
which can result ultimately in negative weights for some intermediate states
in an ostensibly positive correlator.
For example, \figref{weights} shows the weights of various two-meson intermediate states
coming from a \rschpt\ calculation \cite{Bernard:2007qf,Bernard:2006zw}
of the scalar, taste-singlet correlator in
a one-flavor rooted staggered theory.  The physical theory should only have
a two-$\eta'$ intermediate state, but here we have various light pion states, with
the taste-singlet pions%
\footnote{The taste-singlet pion is distinct from the $\eta'$ here because
it is a flavor nonsinglet arising at the arbitrary, integral $n_r$ values at which
the calculation is done.}
having a negative weight.  In the continuum limit, however, all the pions become
degenerate, and they decouple since their weights add to zero.

The first argument for the validity of \rschpt\  
was given by
\textcite{Bernard:2006zw}.
The starting point is the observation that the case of four degenerate flavors of 
rooted staggered quarks is particularly simple because it is the same as the case of
one flavor of unrooted staggered quarks.  Thus we know the chiral theory: it is exactly
that obtained by \textcite{Lee:1999zxa} for one unrooted flavor. 
This chiral theory is equivalent to that of \rschpt\ for four degenerate 
flavors.  The equivalence is manifest order 
by order in the chiral theory: Since the
result for any physical quantity is polynomial in the number of degenerate flavors, taking
$4n_R$ degenerate flavors and then putting $n_R=1/4$ gives the same chiral expansion
as a one-flavor theory.

The case of four nondegenerate flavors may
then be treated by expanding around the degenerate limit.  The expansion is however
somewhat subtle.  Once we move away from the degenerate limit, nontrivial weighting
factors of various diagrams, caused by the 
fourth root of the determinant of the sea quarks, come into play.  
This means that it is impossible to write all needed
derivatives with respect to the quark masses as derivatives in the one-flavor unrooted
theory of Lee and Sharpe.  The solution is to keep the sea quarks degenerate, but
to introduce arbitrary numbers of valence quarks.  Bernard then shows that it is possible
to rewrite all derivatives with respect to sea quark masses as sums of various combinations
of derivatives with respect to the valence quark masses. This approach allows us to remain
in the degenerate sea-quark limit, where the chiral theory is known.  
It is however necessary to assume that partially-quenched
chiral perturbation theory (PQ\chpt) \cite{Bernard:1993sv} is valid in the unrooted case.
Since the unrooted case is local, this is very plausible. Further,
there is a significant amount of numerical work that supports
the validity of PQ\chpt\
for local theories, using
other fermion discretizations, not just staggered quarks.   
But it should be pointed
out that partially-quenched chiral theories rest on shakier ground than the standard chiral theory
for QCD, as emphasized recently by \textcite{Sharpe:2006pu}.  
For example, the argument by \textcite{Weinberg:1978kz} for QCD
invokes
unitarity, which partially-quenched theories do not have.
On the other hand, the argument by \textcite{Leutwyler:1993iq} emphasizes
cluster decomposition instead of unitarity and may be possible to apply
to a partially-quenched Euclidean theory.  Work on putting PQ\chpt\ on a firmer
foundation is in progress \cite{BGinprogress}.

An additional, technical assumption for this approach is that the mass
expansion does not encounter any singularities.  This is reasonable
because the expansion is about
a {\it massive} theory, and one therefore does not expect infrared problems. 

To reach the 
more interesting case of three light flavors,
Bernard raises the mass of one of the four quarks (call it the charm quark,
with mass $m_c$) to the cutoff, decoupling it from the theory.  This requires
an additional technical assumption, arising from the fact that there is a
range of masses, which begins roughly at $m_c\sim 2m_s$ (with $m_s$ the strange 
quark mass),  where the charm quark has decoupled from the chiral theory, but
not yet from the QCD-level theory. 
While the resulting three-flavor chiral theory
has the same form as that of QCD when $a\to0$, the assumption does
leave open the possible ``loophole'' that the LECs have different numerical values from
those of QCD.

The above argument takes place entirely within the
framework of the chiral theory.  It has the nice feature that the recovery 
of the correct QCD chiral expressions, and the vanishing of nonlocal and
nonunitary effects, only requires taste violations to vanish in the continuum
limit in the unrooted, and hence local, theories with integral $n_r$.
The vanishing of these taste violations in the rooted chiral theory
then follows.  On the other hand, because the argument does not connect
\rschpt\ to the QCD-level rooted staggered theory, the replica rule ends
up emerging rather mysteriously.
The chain of reasoning also depends on several technical assumptions.

An argument for the validity of \rschpt\ directly from the fundamental
rooted staggered theory is therefore desirable.  It has been developed
by \textcite{Bernard:2007ma} by starting from the RG framework of Shamir.
The basic idea is to generalize the fundamental (lattice-level) theory to one in which
the dependence on the number of replicas $n_r$ is polynomial to any given
order in the fine lattice spacing $a_f$.  Then we can find the chiral theory
for each integer $n_r$ in a standard way (because the theories are local), and
apply the replica rule to get the rooted staggered theory at the fundamental level
and \rschpt\ at the chiral level.  

For simplicity we focus on a target theory with $n_s$ degenerate
quarks in the continuum limit.  Unlike the previous argument, the extension here
to quarks with nondegenerate masses is straightforward. Consider
\eq{ZrootKS}, the rooted staggered theory at the $n^{\rm th}$ step
of blocking, but with $n_s$ degenerate staggered flavors:
\begin{equation}
        Z_n^{\rm KSroot}(n_s) = \int d\cA\;  \det{}^{\frac{n_s}{4}}\;
	(D_{KS,n}+m_n\otimes I) \ ,
\eqn{ZrootKS-ns}
\end{equation}
Now generalize this, using the definitions of \eq{split}, to
\begin{equation}
	Z_n^{\rm gen}(n_s,n_r) = \int d\cA\; \det{}^{n_s}(D_n+m_n)\;\frac{ \det{}^{n_r}[(D_n+m_n)\otimes I +ta_f\Delta_n]}
{\det{}^{n_r}[(D_n+m_n)\otimes I] }\ ,
\eqn{Zgen}
\end{equation}
where $t$ is a convenient interpolating parameter.
When $t=1$ and $n_r=n_s/4$, this reduces to \eq{ZrootKS-ns} because the determinants 
of the reweighted fields (those involving $D_n +m$ or $(D_n+m)\otimes I$ only) cancel,
and the remaining determinant is just that of the 
rooted staggered theory.
When $t=0$, on the other hand, \eq{Zgen} gives a local theory of $n_s$ reweighted one-taste quarks.

\Equation{Zgen} has an
important advantage over \eq{ZrootKS-ns}.  While the dependence on $n_s$ is unknown
and nonperturbative in both cases, the dependence on $n_r$ of $Z_n^{\rm gen}(n_s,n_r)$
is well controlled because it vanishes when the taste violations vanish
($a_f\Delta_n=0$ or $t=0$). This makes it possible to apply a replica
rule on $n_r$ at
the fundamental QCD level.  To see this,
we first write
\begin{equation}
\frac{ \det{}^{n_r}[(D_n+m_n)\otimes I +ta_f\Delta_n]}
{\det{}^{n_r}[(D_n+m_n)\otimes I] } =
\exp\left( n_r\; \tr\, \ln \left [ 1 + \left((D_n+m_n)^{-1}\otimes I\right)ta_f\Delta_n\right]\right) \ .
\eqn{nr-expand}
\end{equation}
We 
then
expand in powers of the fine lattice spacing $a_f$.  These can come from the
explicit factor $a_f$ in the taste-violating term 
or from the implicit dependence on $a_f$ of 
the gluon action and the lattice operators $D_n$ and $\Delta_n$.  The parameter $t$ serves
to keep track of the explicit dependence; the power of $t$ must be less than or equal to
the power of $a_f$ to which we expand. From \eq{nr-expand}, the power of $n_r$ must in turn
be less than or equal to the power of $t$.  Thus, to any fixed order in $a_f$,
the dependence of the theory on $n_r$ must be polynomial.
This means that $n_r$ is a valid replica parameter of the fundamental theory
(again to any fixed order in $a_f$).  We can in principle find the polynomial
dependence  of
any correlation function by calculations for integer values of $n_r$ only, 
and then
determine the correlation function in the rooted staggered 
theory by simply setting $n_r=n_s/4$ (and $t=1$).

We now discuss the effective theories, the SET and the chiral theory.
For convenience, we can  
work at $t=1$.
For $n_r$ and $n_s$ (positive) integers,
$Z_n^{\rm gen}(n_s,n_r)$ is a local, but partially-quenched,
theory that can be written directly as a path integral.  It is
partially quenched because the determinant in the denominator
needs to be generated as an integral over ghost (bosonic) quarks.
Finding
the SET and the chiral effective theory for such local theories is standard,
although the {\it caveats}\/ about the foundations of
PQ\chpt\ apply. 
All
that we really need to know is that the effective theories exist for
any integer $n_r$ and $n_s$, and that their dependence on $n_r$ is polynomial (because the
dependence in the underlying theory is polynomial).  In the chiral theory
we can then set $n_r=n_s/4$. At the QCD level this just
gives the rooted staggered theory for $n_s$ flavors.
At the chiral level, 
the reweighted parts of the theory again cancel order by order at $n_r=n_s/4$, 
because we have $n_s$ flavors of one-taste quarks and $n_r$ flavors of
four-taste ghost quarks, with exact taste symmetry.  We are then left
with exactly the result we would have gotten from \rschpt.

This argument avoids the  ``loophole'' and technical assumptions of the argument 
of
\textcite{Bernard:2006zw}.  It also makes clear how the replica rule arises from 
the fundamental theory.  On the other hand, it inherits the assumptions of
\textcite{Shamir:2006nj}, since it is based on that framework. 
Both arguments rely on the standard PQ\chpt\ for local theories. 
This is not surprising since rooted staggered QCD inherently shares some features
of a partially-quenched theory:  
Since rooting is done only on the sea quarks,  
there is an excess of valence quarks. 
As noted earlier, however, this ``valence-rooting'' issue is not
a fundamental problem because there is a physical subspace.

A nice feature of the current argument is that, by coupling \rschpt\ directly to
the RG framework, it makes numerical tests of \rschpt\ into tests of the RG framework, and hence
of the validity of rooting at the fundamental level. We discuss such tests in
\secref{fpi}.

\bigskip

We now turn to the objections raised to rooted staggered quarks by
\textcite{Creutz:2006ys,Creutz:2006wv,Creutz:2007yg,Creutz:2007pr,Creutz:2007rk,Creutz:2008kb,Creutz:2008nk}.
Since these objections have been refuted, \cite{Bernard:2006vv,Bernard:2007eh,Bernard:2008gr,Adams:2008db,Golterman:2008gt} 
--- see also the reviews by \textcite{Sharpe:2006re} and \textcite{Kronfeld:2007ek} --- we give only a very brief
discussion here.  The main point is that most of Creutz's claims apply equally well
to the proposed continuum limit theory of rooted staggered  quarks: 
 a rooted four-taste theory with exact taste symmetry, which
is called a ``rooted continuum theory'' (RCT)
by 
\textcite{Bernard:2007eh}. 
Such a theory provides a tractable framework in which to examine Creutz's claims.
Because, as emphasized before, $\det^{1/4}((D+m)\otimes I) = \det(D+m)$, the RCT is clearly
equivalent to a well-behaved, one-taste theory, and gives a counterexample to most of
Creutz's objections.  Alternatively, \textcite{Adams:2008db} has found counterexamples
to Creutz's claims in a simple lattice context, namely a version of twisted Wilson quarks.

While the RCT is 
equivalent to a one-taste theory, it is not exactly the same in the following sense:
In the RCT, with its four tastes, one can couple sources to various tastes and generate
Green functions that have no analogue in the one-taste theory.
Such unphysical Greens functions are at the basis of many of the ``paradoxes'' Creutz finds.
For example, one can find 't Hooft vertices that are singular in the limit $m\to 0$.
Nevertheless these unpleasant effects exist purely in the unphysical sector of the
RCT;
in the physical sector all  't Hooft vertices are well behaved.  

Finally, Creutz has noticed that there is a subtlety involving rooted staggered quarks
for negative quark mass, and this is in fact true.  Independent of the sign of the quark
mass, the staggered determinant is positive, as discussed following \eq{DKS-eps}. 
The fourth root of the determinant generated by the dynamical 
algorithms, \secref{Num_simul}, is then automatically positive for any sign 
of $m$. In other words, the rooted staggered theory is actually a function of $|m|$, not $m$.
This means that rooted staggered fermions cannot be used straightforwardly to investigate
the effects that are expected \cite{Dashen:1970et,Witten:1980sp} to occur 
for an odd number of negative quark masses.%
\footnote{In principle, the negative mass region can be simulated by adding a $\theta$ term
to the action.  Because of the sign problem, this would be extremely challenging in four dimensions.
However, it has been shown to work well in the Schwinger model \cite{Durr:2006ze}.}
A 
related problem occurs when one adds a chemical potential to the theory ---
the determinant becomes complex, and the fourth root, ambiguous 
\cite{Golterman:2006rw}.  Nevertheless, these problems have no relevance to the
validity of the rooted staggered theory in the usual case of positive quark mass and
zero
chemical potential.  For more details, see \textcite{Bernard:2006vv}.

%% file: RMP_sec4.tex
% File for section 4 for RMP article
%
%\section{Section 4}

\section{Overview of the MILC lattice ensembles}
\label{sec:ensembles_1}

In this program of QCD simulations, ensembles of lattices were
generated at several lattice spacings and several
light quark masses. This allows extrapolations to zero
lattice spacing (the ``continuum extrapolation'') and to the
physical light quark mass (the ``chiral extrapolation'').
In all ensembles the masses of the up and down quarks are
set equal, which has a negligible effect ($< 1\%$)
on isospin-averaged quantities.
The original goals of the program were to simulate with three dynamical
quark flavors, with a large enough physical volume to make finite size
effects small, and to vary the quark masses to study the effects of
``turning on'' the dynamical quarks.   It later became clear that
more lattice spacings were needed to understand the continuum limit.
Fortunately, computer power was increasing rapidly, which made the
simulations with smaller $a$ practical.

Currently, the lattice spacings of the ensembles fall into six
sets, with lattice spacings approximately 0.18 fm, 0.15 fm, 0.12 fm,
0.09 fm, 0.06 fm and 0.045 fm. 
In many places these are called ``extracoarse,'' ``medium coarse,''
``coarse,'' ``fine,''
``superfine,'' and ``ultrafine,'' respectively.
The 0.12 fm lattices were the first to be generated. Over time, as
computer power permitted, the lattice spacing was reduced progressively
by $\approx 1/\sqrt{2}$ so that in each reduction the estimated leading
finite lattice spacing artifacts were a factor of two smaller than in the
previous set. The coarser lattices were added to support thermodynamics
studies and to provide further leverage for continuum extrapolations.
The medium coarse ensemble was added after coarse and fine and has a
better tuned strange quark mass based on analysis of the other ensembles.

For comparison,
at $a \approx 0.12$ fm, $a \approx 0.09$ fm and $a \approx 0.06$ fm,
quenched ensembles  with the same gauge action were also generated.
For each of these lattice spacings, the gauge coupling $\beta=10/g^2$ was
adjusted as the light quark mass was changed to keep the lattice
spacing approximately fixed.  However, the lattice spacing could
only be determined accurately after the large ensembles were
generated, so it is necessary to take into account the
small differences in lattice spacing among the ensembles
in the same set. In Sec.~\ref{sec:determine_a}, we then describe measurement
of the lattice spacing on each ensemble, and a parameterized fit to smooth
out statistical fluctuations.

The strange quark mass in lattice units, $a m_s$, was estimated before simulations
began, and was held fixed as the light quark mass and gauge coupling were varied.
Later analysis, described in  Sec.~\ref{sec:fpi}, determined the correct strange quark mass
much more accurately, and in fact the initial estimates turned out
to be wrong by as much as 25\%.

In the $a \approx 0.12$ fm set, several ensembles have a large dynamical
quark mass --- as large as 
eleven times the physical strange quark mass.   This was done
to investigate the physics of continuously ``turning on''
the quarks by lowering their masses from infinity.
There are also a number of ensembles with a lighter-than-physical
strange quark mass.
These were generated to explicitly
study dependence on the sea strange quark mass, and,
since the lighter
strange quark implies less sensitivity to higher orders in
SU(3) chiral perturbation theory,
enable improved determinations of the parameters in the
chiral expansion, particularly of the low-energy constants
(see Sec.~\ref{sec:fpi}).

The fields satisfy periodic boundary conditions in the space directions,
while the boundary condition in the Euclidean time direction is periodic
for the gauge fields and antiperiodic for the quark fields.

\begin{figure}
\begin{center}
\epsfxsize=4.0in
\epsfbox[0 0 4096 4096]{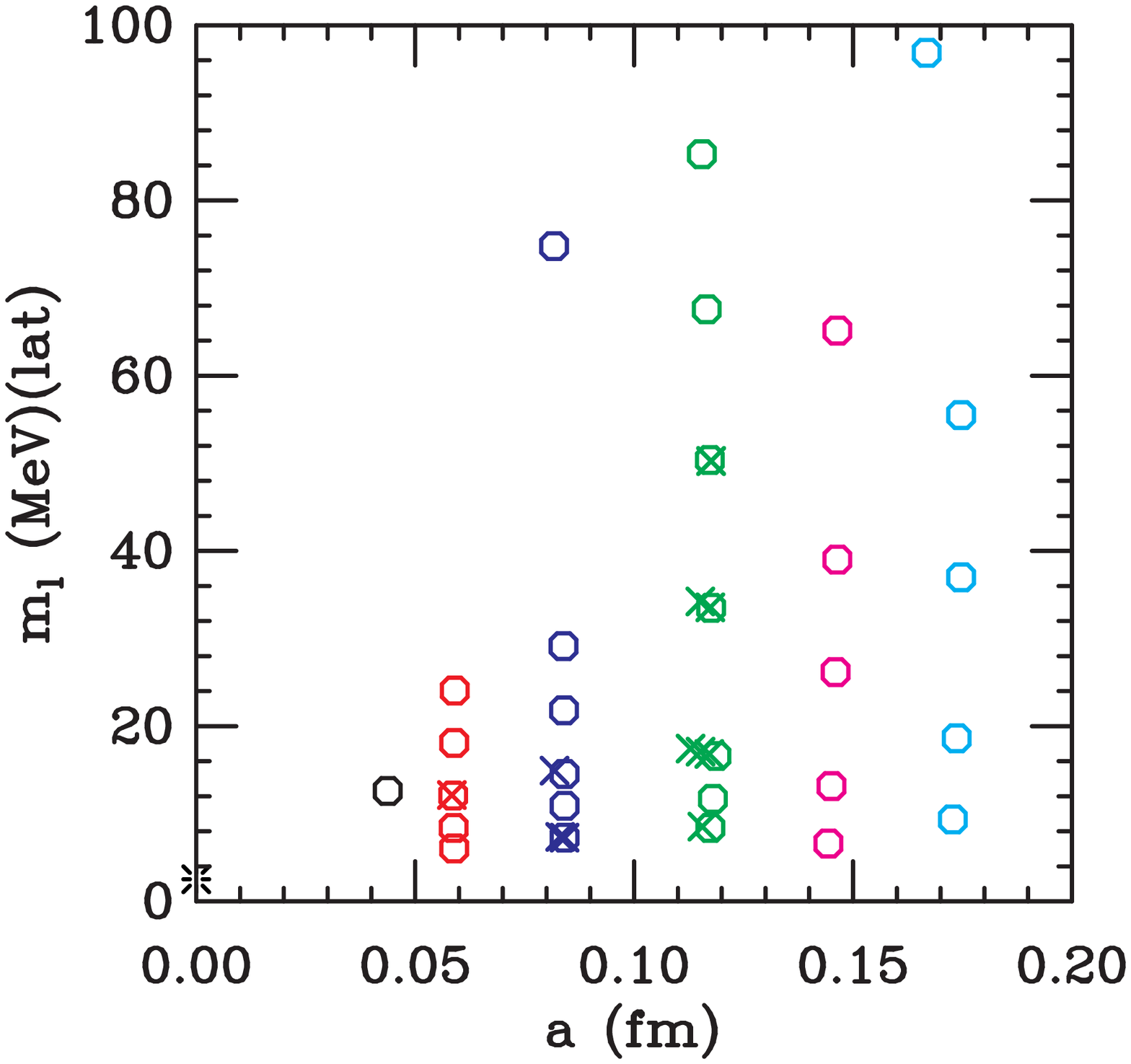}
\end{center}
\caption{
Lattice spacings and quark masses used.  The octagons
indicate ensembles with the strange quark near its physical
value, 
while  the crosses indicate those with an unphysically light
strange quark.  The burst at lower left shows the physical
light quark mass.   Here the quark masses are in units of
MeV, but using the asqtad action lattice regularization.
}
\label{fig:runs1}
\end{figure}

Table \ref{table:runtable1} shows the parameters
of the asqtad ensembles (a few short ``tuning'' ensembles are not
included).
Here $am_l$ is the dynamical light quark mass in lattice units and $am_s$
is the strange
quark mass.
Figure~\ref{fig:runs1} plots the quark masses
and lattice spacings of these ensembles.
\\

\input runtable4.tex

\subsection{Algorithms and algorithm tests}
\label{sec:algorithms}
The earlier ensembles were generated using the ``R'' algorithm~\cite{Gottlieb:1987mq}
described in Sec.~\ref{sec:Num_simul}.
The molecular dynamics step size was generally
set at about two thirds of the light quark mass in lattice units.
More recent lattice generation has used rational function approximations
for the fractional powers
described in Sec.~\ref{sec:Num_simul}.
In those simulations, we have used the Omelyan second order integration
algorithm \cite{Sexton:1992nu,Omelyan:2002E1,Omelyan:2002E2,Omelyan:2003CC,Takaishi:2005E1}.
We used different step sizes for the fermion and gauge forces ~\cite{Sexton:1992nu},
with the step
size for the fermion force three times that of the gauge force.   We used
four sets of pseudofermion fields and corresponding rational functions
\cite{Hasenbusch:2001ne, Hasenbusch:2002ai}.
The first set implements the ratio of the roots of the determinants for the physical light
and strange quarks to the determinant for three heavy ``regulator'' quarks with
mass $am_r=0.2$.
That is, it corresponds to the weight $\det \LP M(m_l) \RP^{1/2} \det \LP M(m_s) \RP^{1/4} \det \LP M(m_r) \RP^{-3/4} $.
The remaining three pseudofermion fields each
implement the force from one flavor of the regulator quark, or the fourth root of
the corresponding determinant.  These choices are known to be
reasonably good, but could be optimized further.

For all but the largest lattices generated with rational function
methods, we included the Metropolis accept/reject decision
to eliminate step size errors, or the RHMC algorithm.
Because the integration error is extensive, use of the RHMC algorithm for 
the largest lattices would have forced us to use very small step sizes and 
double precision in many parts of the integration. For these lattices
it was much more efficient to run at a small enough step size that the 
integration error was less than other expected errors in the calculation
(the RHMD algorithm).

\begin{figure}
\begin{center}
\epsfxsize=4.0in
\epsfbox[0 0 4096 4096]{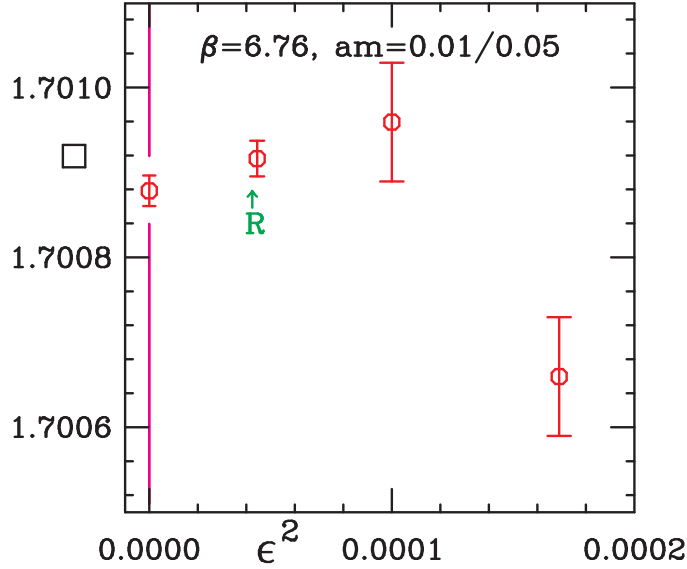}
\end{center}
\caption{
The plaquette as a function of integration step size squared for
$20^3\times 64$ lattices with $\beta=6.76$ and $am_q=0.01/0.05$.
The point at $\epsilon^2=0$ is from the RHMC algorithm, and
the point indicated by $R$ is the value used in the R algorithm
production runs.  The remaining two points are from short test
runs described in \protect\textcite{Aubin:2004wf}.
}
\label{fig:plq_vs_step}
\end{figure}

Errors from the integration step size in the R algorithm were originally
estimated from short runs with different step sizes, as described
in \textcite{Bernard:2001av} and \textcite{Aubin:2004wf}.
In several cases, ensembles originally generated with the R algorithm were later extended
with the RHMC algorithm.   This allows an {\it ex post facto} test of the step size
errors in the R algorithm, with much higher statistics than possible
for a tuning run.
Figure~\ref{fig:plq_vs_step} shows the average plaquette for one
$a \approx 0.12$ fm run as a function of step size squared, combining
the early tuning runs with the R and RHMC algorithm production runs.
Table~\ref{table:r_vs_rhmc1} compares the expectation values of
the plaquette and the light quark condensate
and, in some cases, the lattice spacing and pion mass,
for the ensembles where both algorithms were used.
The differences are small and in most cases are comparable to the statistical errors.

\begin{table}
\begin{center}
\begin{tabular}{|l|l|l|l|lll|lll|}
\hline
$\beta$ & $am_l$ & $am_s$ & $\epsilon$ & $\Plaq(R)$ & $\Plaq(RHMC)$ &
   difference & $\pbp(R)$ & $\pbp(RHMC)$ & difference \\
\hline
6.79	& 0.020	& 0.050	& 0.01333	& 1.709160(26) & 1.708805(16) & -0.000355(30) &
	0.052553(61) & 0.052306(28) & 0.000251(67) \\
6.76	& 0.010	& 0.050 & 0.00667	& 1.700917(21)	& 1.700879(18) & -0.000038(28)	&
	0.036875(43)	& 0.037174(36)	& 0.000300(56)	\\
6.76	& 0.007	& 0.050	& 0.00500	& 1.701183(22)	& 1.701177(18)	& -0.000006(29)	&
	0.031388(54)	& 0.031306(38)	& -0.000082(66)	\\
6.76	& 0.005	& 0.050	& 0.00300	& 1.701181(17)	& 1.701211(11)	& 0.000030(20)	&
	0.027551(50)	& 0.027597(25)	& 0.00045(56)	\\
\hline
7.11	& 0.0124	& 0.031	& 0.00800	& 1.789213(19)	& 1.789075(7)	& -0.000138(20)	&
	0.024584(22)	& 0.024620(10)	& 0.000036(24)	\\
7.09	& 0.0062	& 0.031	& 0.00400	& 1.784552(9)	& 1.784541(6)	& -0.000011(11)	&
	0.015622(17)	& 0.015608(14)	& -0.00015(22)	\\
7.08	& 0.0031	& 0.031	& 0.00200	& 1.782300(8)	& 1.782254(11)	& -0.000046(11)	&
	0.010664(18)	& 0.010860(19)	& 0.000196(26)	\\
\hline
\hline
$\beta$ & $am_l$ & $am_s$ & $\epsilon$ & $\frac{r_1}{a}(R)$ &
   $\frac{r_1}{a}(RHMC)$ & difference & $a m_\pi(R)$ & $a m_\pi(RHMC)$ &
   difference \\
\hline
7.11    & 0.0124        & 0.031 & 0.00800       & 3.708(13)     & 3.684(17)     & -0.024(21)    &
        0.20640(20)     & 0.20648(20)    & 0.00008(28)   \\
7.09    & 0.0062        & 0.031 & 0.00400       & 3.684(12)     & 3.681(8)      & -0.003(14)    &
        0.14797(20)     & 0.14767(13)   & -0.00030(24)  \\
7.08    & 0.0031        & 0.031 & 0.00200       & 3.702(8)      & 3.682(7)      & -0.020(11)    &
        0.10528(9)      & 0.10545(9)    & 0.00017(13)   \\
\hline
\end{tabular}
\caption{Comparison of plaquette and light quark 
condensate for
ensembles run partly with the R algorithm and partly with the RHMC
algorithm.  For the $a \approx 0.09$ fm ensembles, we also show $r_1/a$
and the pion mass.
}
\label{table:r_vs_rhmc1}
\end{center}
\end{table}

In one case, $a\approx 0.12$ fm and $am_q=0.01/0.05$, an ensemble with
a larger
spatial size ($28^3$), was generated to check for effects of the spatial
size.  In general, these effects were found to be small as expected, although
the effects on $f_\pi$ and $f_K$ differ significantly
from one-loop chiral perturbation theory estimates, as discussed in Sec.~\ref{sec:fpi}.

\subsection{The static potential and determining the lattice spacing}
\label{sec:determine_a}

\begin{figure}
\begin{center}
%\epsfxsize=5.7in
%\epsfxsize=4.0in
%\epsfbox[0 0 4096 4096]{figs/pot.b709m0062m031.ps}
\includegraphics[scale=0.5]{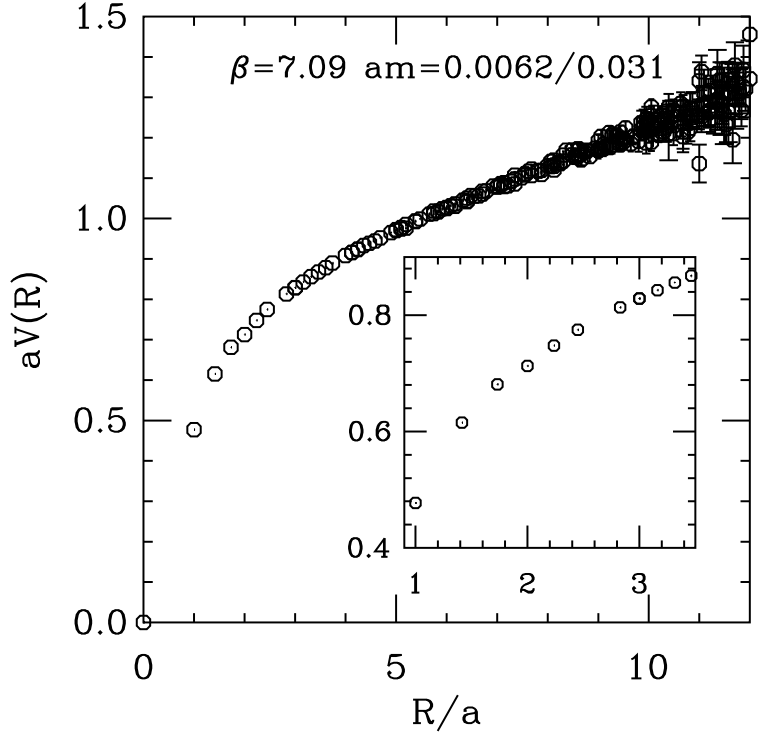}
\end{center}
\caption{
The static quark potential for the ensemble with $a \approx 0.09$ fm
and $m_l \approx 0.2 m_s$.   This was obtained from time range
five to six.  The inset magnifies the short distance part, showing
a lattice artifact which is discussed in the text.
}
\label{fig:pot.b709m0062m031}
\end{figure}

\begin{figure}
\begin{center}
%\epsfxsize=5.7in
%\epsfxsize=4.0in
%\epsfbox[0 0 4096 4096]{figs/spot.b746m0018m018.ps}
\includegraphics[scale=0.5]{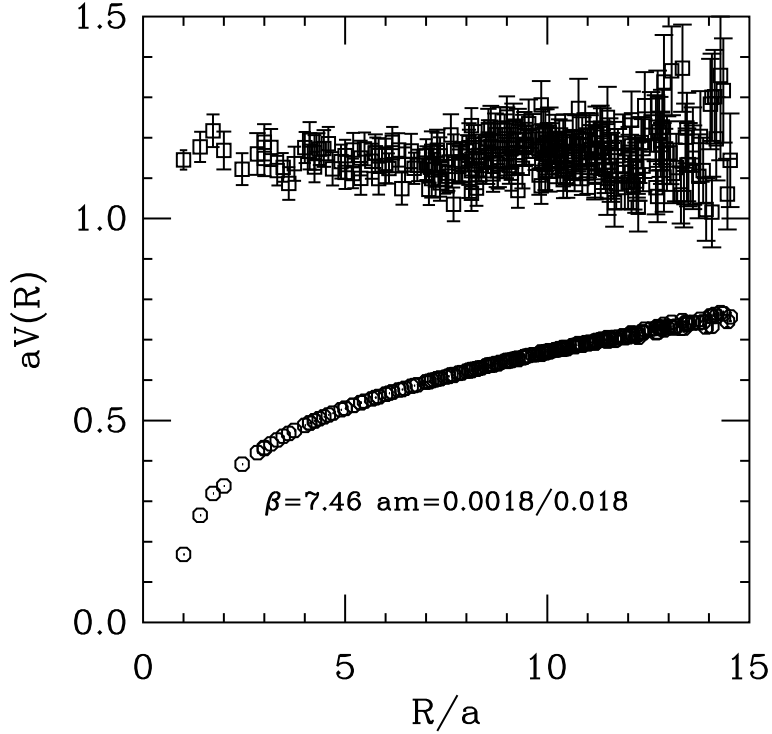}
\end{center}
\caption{
The static quark potential  and first excited state potential
for the ensemble with $a \approx 0.06$ fm
and $m_l \approx 0.1 m_s$.   This was obtained from time range
three to twenty, using the APE smeared time links discussed in
the text.
}
\label{fig:spot.b746m0018m018}
\end{figure}

\begin{figure}
\begin{center}
%\rule{0.0in}{0.1in}\vspace{-1.5in}\\

%\epsfxsize=9.0in
%\epsfxsize=4.0in
%\epsfbox[0 0 4096 4096]{figs/potmatchphys.sb781sb747b709b676b6572.ps}
\includegraphics[scale=0.5]{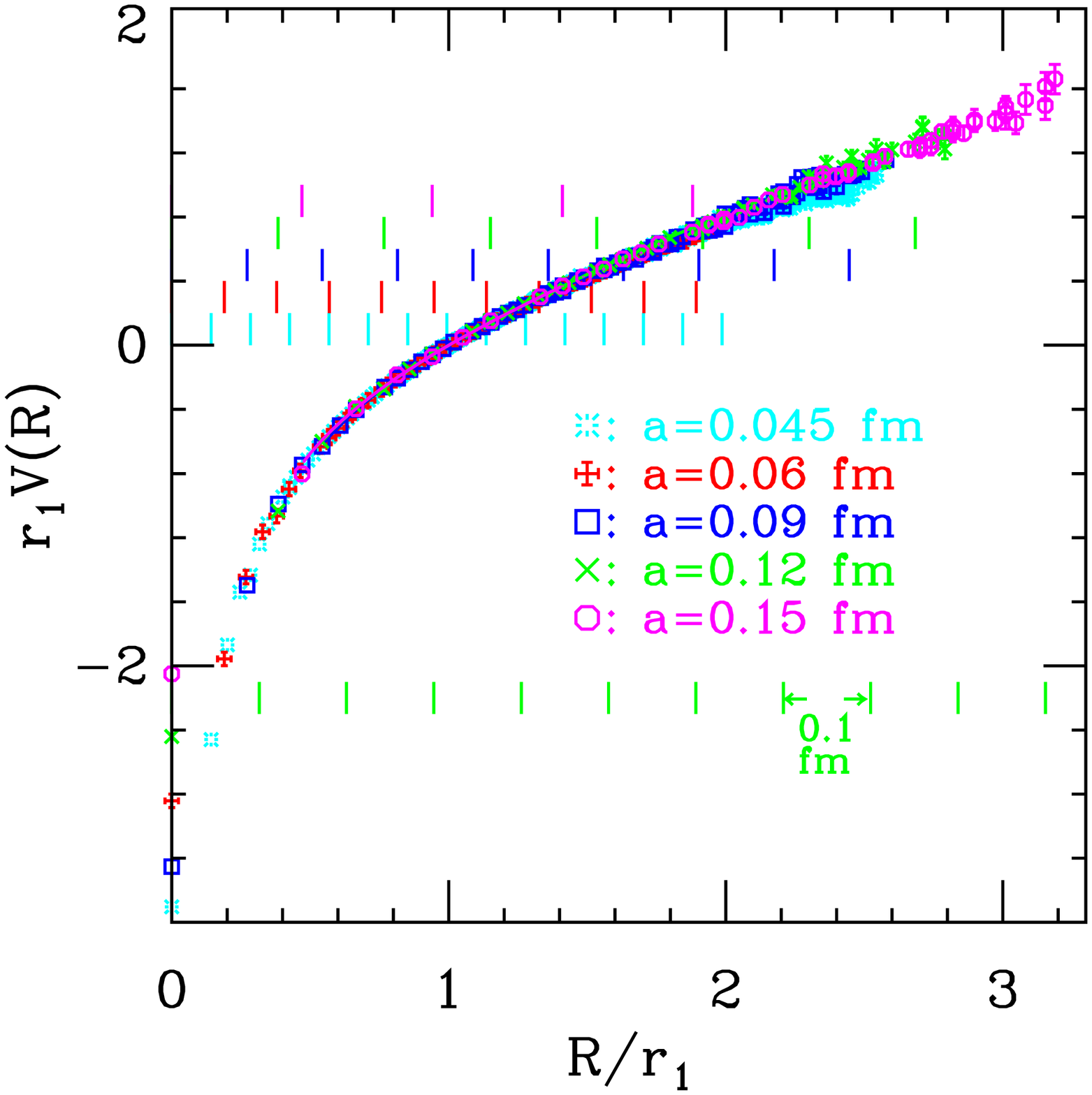}
\end{center}
%\rule{0.0in}{0.1in}\vspace{-0.5in}\\
\caption{
The static quark potential in units of $r_1$ for five different
lattice spacings.  In all cases, these are for light quark mass
of two tenths the simulation strange quark mass.
For each lattice spacing, a constant has been subtracted to
set $V(r_1)=0$.
The ruler near the bottom of the plot shows distance in units
of fm, using $r_1=0.318$ fm.
The multiple rulers in the upper half of the plot show distance
in units of the lattice spacings for the different ensembles.
}
\label{fig:potmatch_phys.sb781sb747b709b676b6572}
\end{figure}

Since results of lattice QCD simulations are initially in units
of the lattice spacing, knowing the lattice spacing is crucial
to calculating any dimensionful quantity.
However, since ratios of dimensionful quantities (mass ratios)
calculated on the lattice will only have their physical
values at the physical quark masses and in the continuum limit,
there is arbitrariness in the determination of the lattice
spacing except in the physical limit.
Some dimensionful quantity must be taken
to be equal to its physical value or to some {\it a priori}
model.

Following the practice of most current lattice simulation programs,
we use a Sommer scale~\cite{Sommer:1993ce} as the quantity kept fixed,
and determine this scale from some well controlled measurement.

A Sommer scale is defined as the length where the force
between a static (infinitely heavy) quark and antiquark
satisfies 
$r^2 F(r) = C$,
where $C$ is a constant.
Intuitively, this is a length where this static
potential changes behavior from the short distance
Coulomb form to the long distance linear form.
In particular, the most common choice is $r_0$, defined
by $C=1.65$.
We have chosen to use $r_1$, defined by $C=1$.  This choice
was made based on early simulations at $a\approx 0.12$ fm
where it was found that $r_1$ had smaller statistical
errors than $r_0$ \cite{Bernard:2000gd}.

The calculation of the static potential on the earlier ensembles
is described in \textcite{Bernard:2000gd}.
We begin by fixing the lattice
to Coulomb gauge.  In this gauge, we can evaluate
the potential from correlators of (nonperiodic) Wilson
lines,  where the line at ($\vec x,t$) with length $T$ is
$W_T(\vec x,t) = \prod_{i=0}^{T-1} U_4(\vec x, t+i)$.
The Coulomb gauge fixing, which makes the spatial links as 
smooth as possible, is an implicit way of averaging
over all spatial paths closing the loop at the top and
bottom.   Because we do not explicitly construct the spatial parts,
it is easy to average over all lattice points $(\vec x,t)$ and to
get the potential at all spatial separations $\vec R$.

The first step in determining $r_1$ is to extract $V(\vec R)$ from
the expectation value of the correlators of Wilson lines.  We expect
\BE L(\vec R,T) \ = \ \LL W_T^\dagger(\vec x,t) W_T(\vec x + \vec R,t) \RR \ =\ 
 A e^{-V(\vec R)T} + A^\prime e^{-V^\prime(\vec R)T} + \ldots ~,
\EE
where $V^\prime$, etc. are potentials for excited states.
For $a \ge 0.09$ fm, the excited states are negligible for
fairly small $T$, and we simply take $V(\vec R) = \log( L(\vec R,T)/L(\vec R,T+1) )$.
Specifically, we use $T=3$ for $a \approx 0.15$ fm, $T=4$ for $a\approx 0.12$ fm
and $T=5$ for $a\approx 0.09$ fm.
Figure~\ref{fig:pot.b709m0062m031} shows the resulting potential
for the run at $a \approx 0.09$ fm and $m_l = 0.2 m_s$.
The inset in this figure shows the short distance part of the
potential.  In this inset, there is a visible lattice artifact
where the point at $R=2$, or separation $(2,0,0)$ is slightly
below a smooth curve through the nearby points
with off-axis distances $\vec R$.
However, at $R=3$ the lattice artifacts are quite small.  In fact,
what appears to be a single point at $R=3$ is actually two points,
one for $\vec R=(3,0,0)$ and another for $\vec R = (2,2,1)$.
The small objects in the center of the plot symbols are the
statistical error bars on $V(R)$.

For $a \approx 0.06$ fm, the above procedure for finding
$V(R)$ gave large statistical errors.
This is primarily because a large constant term in the potential causes
a rapid falloff of $L(\vec R,T)$ with $T$.
This constant can be considered a self energy of
the static quark, diverging as $1/a$.
To fix this, the timelike links were smeared by adding
a multiple of the three link ``staples'' \cite{Albanese:1987ds},
namely ``fat3 links''  defined in Eq.~(\ref{eq:fat3}) with $\omega = 0.1$.
The Wilson line correlators $L(\vec R,T)$ were computed from
the smeared time direction links as described above.
As expected, this reduces the constant term in $V(R)$, and
comparison with the potential from unsmeared links suggests that
any systematic effects on $r_1/a$ are less than $0.005$ at $a \approx 0.06$
fm, smaller than the statistical errors.
With the smeared time links, the correlators $L(\vec R,T)$ are
statistically significant out to $T$ as large as twenty (for small
$R$).  It is then advantageous to do a two state fit to
$L(\vec R,T)$.  For the $a\approx 0.06$ fm ensembles, we generally
fit these two states over a time range $3 \le T \le 20$.
An example of the potential from this procedure is shown in Fig~\ref{fig:spot.b746m0018m018}.
The first excited state potential is also shown, but we caution the reader
that in addition to having large statistical errors this excited state
potential has not been carefully checked for stability under varying fit ranges, or under
addition of a third state to the fit.

Once $V(R)$ is determined, we find $r_1$ by fitting $V(R)$ to a
range of $R$ approximately centered at $r_1$.  
We use a fit form
\BE\label{eqn:potform} V(R) = C + \frac{B}{R} + \sigma R +
 \lambda \LP \left. \frac{1}{R} \right|_{lat} - \frac{1}{R} \RP \EE
Here $C$ is part of the quarks' self energy, $\sigma$ is the string tension
and $B$ is $\frac{-3}{4}\alpha_s$ for a potential definition of $\alpha_s$.
The last term, $\left. \frac{1}{R} \right|_{lat} - \frac{1}{R}$, is the difference between
the lattice Coulomb potential, 
$\left. \frac{1}{R} \right|_{lat} = 4\pi \int \frac{d^3p}{(2\pi)^3}
D^{(0)}_{00}(p) e^{ipR}$ with $D^{(0)}_{00}(p)$ the free lattice gluon
propagator calculated with the Symanzik improved
gauge action, and the continuum Coulomb potential $1/R$.
Use of this correction term was introduced by the UKQCD collaboration~\cite{Booth:1992bm}.
This correction was used for $R<3$.
The scale $r_1$ (or $r_0$) was then found from solving $r^2 F(r) = C$
with $\lambda$ set to zero,
$r_1 = \sqrt{\frac{1+B}{\sigma}}$.
Since we often want lattice spacing estimates from only
a few lattices, and there are a large number of distances
to be fit, these fits were generally done without including
correlations among the different $\vec R$.
Errors on $r_1$ are estimated by the jackknife method, where the
size of the blocks eliminated ranges from 30 to
100 simulation time units.
Spot checks comparing fits including the correlations confirmed
that the jackknife errors are consistent with derivative errors
in the correlated fits, and that the fit function does fit the
data well over the chosen range.

For the $a \approx 0.18$ fm ensembles, we used the spatial range
from $1.4$ or $1.5$ to $6.0$;
for the $a\approx 0.15$ fm ensembles, $\sqrt{2} \le R \le 5$;
for the $a \approx 0.12$ fm ensembles,
$\sqrt{2} \le R \le 6$; and for the $a \approx 0.09$ fm ensembles,
$2 < R \le 7$.
For the $a\approx 0.06$ fm ensembles, where the two state fits
with smeared links were used, the spatial range was $4 < R \le 7$,
and for the $a \approx 0.045$ fm run, it was $5 < R \le 10$.

The static quark potentials for different lattice spacings can
be overlaid after rescaling to check for lattice effects and to
plot the potential over a large range.
Figure~\ref{fig:potmatch_phys.sb781sb747b709b676b6572} shows such a
plot in units of $r_1$ for five different lattice
spacings, using the ensembles with $m_l = 0.2\, m_s$ at each lattice
spacing.
In \textcite{Bernard:2000gd}, it was found that including the
dynamical quarks modifies the static potential in the expected
way.  This can be seen by plotting dimensionless quantities such
as $r_0/r_1$ or $r_1\sqrt{\sigma}$.  When this is done in a region
where the potential is approximated by Eq.~(\ref{eqn:potform})
and $r_1$ is found from $r_1 = \sqrt{\frac{1+B}{\sigma}}$, this amounts to
plotting the coefficient of $1/R$ in the fit.

Once $r_1$ is estimated for each ensemble, the estimate
can be improved by fitting all values of $r_1/a$ to a smooth function of
the gauge coupling and quark masses.  We have used two different
forms for this smoothing.   In the first form, we fit $\log(r_1/a)$
to a polynomial in $\beta$ and $2am_l+am_s$.  The second form is
a function based on work of \textcite{Allton:1996kr}:
\BE\label{eq:smoothform}
\frac{a}{r_1} = \frac{C_0 f + C_2 g^2 f^3 + C_4 g^4 f^3}{1 + D_2 g^2 f^2}
\EE
where
\begin{eqnarray}
f  =   (b_0 g^2) ^ {(-b_1/(2 b_0^2 ))}  \exp(-1/(2 b_0 g^2) ) ~, &&
 b_0 = (11 - 2 n_f/3)/(4 \pi)^2 ~, \nonumber \\
b_1 = (102 - 38 n_f/3)/(4 \pi)^4 ~, &&
 am_{tot} = 2 am_l/f + am_s/f ~, \\
C_0 = C_{00} + C_{01l} am_l/f + C_{01s} am_s/f + C_{02} (am_{tot})^2 ~, &&
 C_2 = C_{20} + C_{21} am_{tot} ~. \nonumber
\label{eq:r1_fit_params}
\end{eqnarray}
Here $C_{00}$, $C_{01l}$, $C_{01s}$, $C_{02}$,
$C_{20}$, $C_{21}$, $C_4$, and $D_2$ are parameters.
The second form is a slightly better fit, and we have used it
for the $r_1/a$ values in Table~\ref{table:runtable1}.
Errors on the smoothed $r_1/a$ are estimated by a bootstrap for which
artificial data sets were generated. In these data sets the value
of $r_1/a$ for each ensemble was chosen from a Gaussian distribution
centered at the value for the ensemble given by the fit, and the standard 
deviation was given by the statistical error in $r_1/a$ for the ensemble.

To find $r_1$ in physical units, we use a quantity
that is both well known experimentally and
accurately determined in a lattice calculation.
One such quantity, and the one used in most of
our work, is the splitting between two energy levels of the
$b \bar b$ mesons.  These splittings have been calculated
on several of the asqtad ensembles by the HPQCD/UKQCD
collaboration~\cite{Gray:2002vk,Wingate:2003gm,Gray:2005ur}.
{}From fitting the 2S-1S splittings on the $a\approx 0.12$ fm ensembles
with quark masses $am_l/am_s=0.01/0.05$, $0.02/0.05$, $0.03/0.05$ and $0.05/0.05$,
and the $a\approx 0.09$ fm ensembles with light masses
$am_l/am_s = 0.0062/0.031$ and $0.0124/0.031$,
to the form $r_1(a,am_l,am_s) = r_1^{\mathrm{phys}}+c_1 a^2+c_2 am_l/am_s$,
we find $r_1^{\mathrm{phys}}=0.318$ fm with
an error of $0.007$ fm.
(\textcite{Gray:2005ur}
used a different fitting procedure to estimate $r_1^{\mathrm{phys}}=0.321(5)$ fm.)

More recently, analysis of the light pseudoscalar meson masses
and decay constants gave an accurate value of $f_\pi$.
The fitting procedure to arrive at this is complicated --- see
Sec.~\ref{sec:fpi}.  Requiring that $f_\pi$
in the continuum and chiral limits
match its experimental value gives $r_1 =0.3108(15)({}^{+26}_{-79})\; {\rm fm}$,
where the errors are statistical and systematic, respectively.

To summarize, we set the scale for each ensemble by
$a\equiv (a/r_1) \times r^{\rm phys}_1$, where $(a/r_1)$ is the output of the smoothing
function, Eq.~(\ref{eq:smoothform}),  at the ensemble values of $am_l$, $am_s$, and $g^2$, and
$r^{\rm phys}_1$ is the physical value of $r_1$, obtained either from $b\bar b$ mesons
splittings or $f_\pi$.  The scheme is useful for generic chiral extrapolations, and
tends to result in fairly small dependence of physical quantities on the sea-quark masses.
However, chiral perturbation theory assumes a mass-independent scale setting scheme,
because all dependence on quark masses is supposed to be explicit.  So detailed fits to
chiral perturbation theory forms require a mass-independent scale procedure, especially if
one hopes to extract low energy constants that govern mass dependence.
Once the $r_1$ smoothing form is known, though, it is easy to
modify the procedure to make it mass independent:  instead of using the
ensembles' values of $am_l$ and $am_s$ in the smoothing function, Eq.~(\ref{eq:smoothform}), use
the physical values.  This mass-independent scheme is used for the analysis of
light pseudoscalars described in \secref{fpi}.

\subsection{Tuning the strange quark mass}
\label{sec:strangemass}

In most of these ensembles, the original intent was to fix the
strange quark mass at its correct value, and to set the light
quark mass to a fixed fraction of the strange quark mass.
The correct strange quark
mass, however, is actually not known
until the lattices are analyzed.
In practice, the best that can be done is to estimate the
correct strange quark mass from short tuning runs or
by scaling arguments from results of earlier runs.
As described in \secref{fpi}, the physical strange and up/down
quark masses are determined by demanding that the
light pseudoscalar meson masses take their physical values.
For the strange mass, we find
$a m_s = 0.0439 (18)$ at $a \approx 0.15$ fm,
$a m_s = 0.0350 (7)$ at $a \approx 0.12$ fm,
$a m_s = 0.0261 (5)$ at $a \approx 0.09$ fm
and
$a m_s = 0.0186 (4)$ at $a \approx 0.06$ fm.
For the up/down mass, we find
$a m_l = 0.00158 (7)$ at $a \approx 0.15$ fm,
$a m_l = 0.00126 (2)$ at $a \approx 0.12$ fm,
$a m_l = 0.000955 (8)$ at $a \approx 0.09$ fm
and
$a m_l = 0.000684 (8)$ at $a \approx 0.06$ fm.
The errors include statistical and systematic effects, but they are
dominated by the systematic effects.

\subsection{The topological susceptibility}
\label{sec:topological_susceptibility}

The topological structure of the QCD vacuum is an important 
characteristic of the theory.
A stringent test for lattice simulations consists in correctly
capturing the dependence of the topological susceptibility on the
number of quarks and their masses, since this susceptibility reveals
the effect of the quarks on the nonperturbative vacuum structure.
Chiral perturbation theory predicts $\Chi_{\rm
topo}(n_f, m_i)$ in the chiral limit \cite{Leutwyler:1992yt}. 
Lattice calculations, however, have struggled to reproduce this
dependence satisfactorily because the topological charge is not uniquely
defined and the fermion action typically breaks chiral symmetry.
The asqtad action combined with \rschpt\ gives
us good control over the taste and chiral symmetry breaking effects;
thus we expect that a careful treatment of the topological charge
will lead to an accurate computation of the topological
susceptibility. 
This has been explored in
\textcite{Bernard:2003gq}, \textcite{Billeter:2004wx} and \textcite{Bernard:2007ez}.

As explained in \textcite{Aubin:2003mg} and \textcite{Billeter:2004wx},
the chiral anomaly couples to the {\it taste-singlet} meson, not the
Goldstone pion, which is the usual focus of hadron spectroscopy
calculations. (Of course, in the continuum limit these mesons are
degenerate.)  To leading order in \rschpt, the topological
susceptibility depends on this mass as
\begin{equation}
\Chi_{\rm topo} = \frac{f_\pi^2 m_{\pi,I}^2/8}
{1 + m^2_{\pi,I}/(2m^2_{ss,I}) + 3m^2_{\pi,I}/(2m^2_0)} ~,
\label{eq:toposusc}
\end{equation}
where $m_{\pi,I}$ is the taste-singlet pion mass, and $m_0$ comes from
the term representing the coupling of the anomaly to the $\eta^\prime$ 
in the chiral
Lagrangian, \eq{Lcont}. The strange flavor-singlet, taste-singlet meson mass
is denoted $m_{ss,I}$.

Equation (\ref{eq:toposusc}) interpolates smoothly between the
infinite sea-quark-mass (quenched) prediction 
\cite{Witten:1979vv,Veneziano:1979ec},
$\Chi = f_\pi^2 m_0^2/12$,
which we can use to set $m_0$, and the chiral limit, $m_{l} \rightarrow 0$,
which is dominated by the pion,
$\Chi = f_\pi^2 m_\pi^2/8$.
Hence, to this order, we simply replace the Goldstone pion mass
with the mass of the taste-singlet (non-Goldstone) pion in the
Leutwyler-Smilga formula. Note that this means that, at nonzero
lattice spacing, the topological susceptibility does not vanish
as $m_l \rightarrow 0$, a reminder that the continuum limit must
be taken before the $m_l\rightarrow 0$ extrapolation.

\begin{figure}[t]
\centering
\includegraphics[scale=0.5]{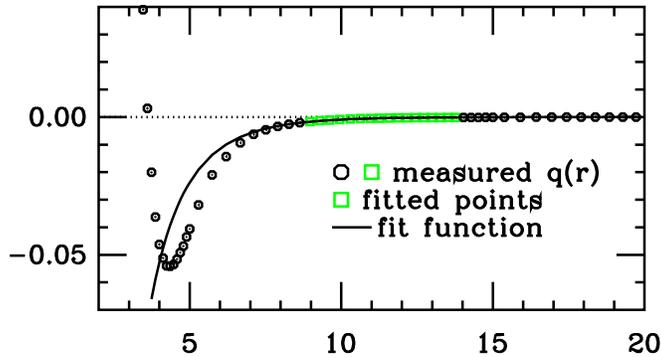}
\caption{Points used to compute $\LL q(r)q(0)\RR$. Measured points
(open symbols)
are used for $r\leq r_c\sim 9a$. For $r > r_c$ the fit function
(solid curve)
is used in \eq{topodef}. From \textcite{Bernard:2007ez}.}
\label{fig:qqfit}
\end{figure}

In order to compute the topological charge density $q(x)$ on our lattice
ensembles, we use three iterations of the Boulder HYP smoothing
method \cite{DeGrand:1997gu, Hasenfratz:2001hp},
which we have found \cite{Bernard:2003jq, Bernard:2003gq} 
compares well with the
improved cooling method of \textcite{deForcrand:1997sq}. We define the topological
susceptibility from the correlator of $q(x)$ via
\begin{equation}
\Chi_{\rm topo} = \LL Q^2 \RR/V = \int d^4r \LL q(r)q(0) \RR ~.
\eqn{topodef}
\end{equation}
On our lattices, the short-distance part of the density correlator
has a strong signal, but the correlator at large separation is noisy.
To reduce the resulting variance, we define a cutoff distance $r_c$. In the
integral above, for $r\leq r_c$ where the signal is strong, we use the
measured values of the correlator $\LL q(r)q(0) \RR$. For $r > r_c$ we
integrate a function obtained by fitting the measured correlator to a
Euclidean scalar propagator 
\begin{equation}
  \left\langle q(r)q(0) \right\rangle \sim
    A_\eta K_1(m_\eta r)/r + A_{\eta^\prime} K_1(m_{\eta^\prime} r)/r ~,
\end{equation}
where we use priors for the masses of the $\eta$ and $\eta^\prime$,
and $K_1$ is a Bessel function.
This significantly
reduces the variance in $Q^2$. An example of the measured values of 
$q(r)$, the fit function, and the fitting range are shown in 
\figref{qqfit}.

\begin{figure}[t]
\centering
\includegraphics[scale=0.5]{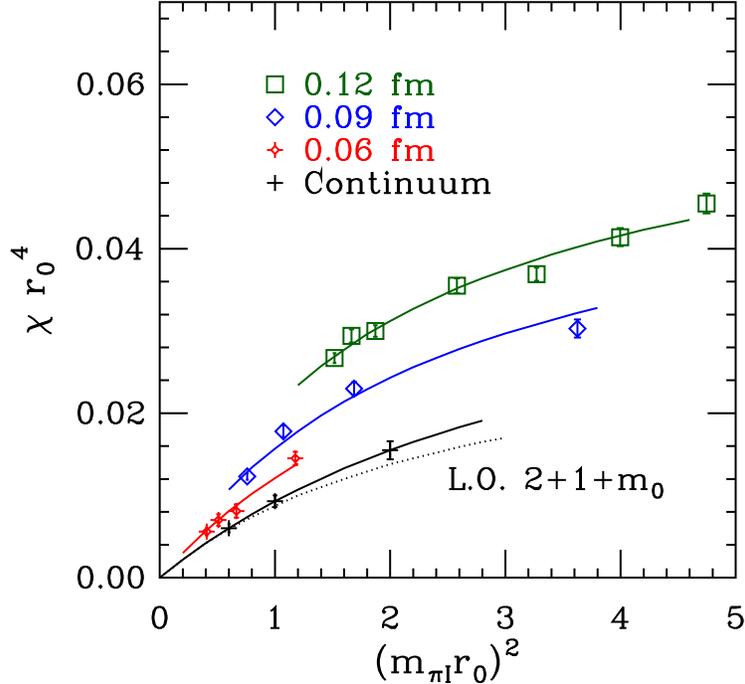}
\caption{Topological susceptibility data, and its continuum 
extrapolation, compared with the prediction of Eq.~(\protect\ref{eq:toposusc}).
Update of figure from \textcite{Bernard:2007ez}.}
\label{fig:toposuscfig}
\end{figure}

Figure~\ref{fig:toposuscfig} shows this definition of $\Chi_{\rm topo}$ computed on our
coarse
($a \approx 0.12$ fm), fine ($a \approx 0.09$ fm), and superfine
($a \approx 0.06$ fm) lattices.
The continuum limit is taken first by fitting the susceptibility data to 
\begin{equation}
\frac{1}{\Chi_{\rm topo}r_0^4}(m^2_{\pi,I}, a)
            = A_0 + A_1 a^2 + (A_2 + A_3 a^2 + A_4 a^4)/m_{\pi,I}^2 ~.
\end{equation}
The solid black line in \figref{toposuscfig} shows the $a\rightarrow 0$ 
form of this function. Some
representative points along this line are shown with error bars
reflecting the errors of the continuum extrapolation. Finally,
the chiral perturbation theory prediction of Eq.~(\ref{eq:toposusc}),
shown as a dotted line,
is based on the value for
$m_0$ set by the quenched data.

With the addition of the new $a \approx 0.06$ fm data, we see that the
topological susceptibility is behaving as expected in the
$m^2_{\pi,I}\rightarrow 0$ limit of rooted staggered chiral perturbation
theory.

These results lend further credibility to the use of the fourth root
procedure to simulate single flavors, since aberrant results from this
procedure would be expected to arise first in anomalous behavior of
topological quantities and correlations, as these are rather sensitive to
the number of flavors.

%% file: runtable4.tex
%
% see ~/Asq/3flav_fitting/Masstable, ~/Asq/3flav_fitting/pion_masses.nov7_06
%
\begin{table}
\begin{tabular}{|c|l|l|c|r|l|l|l|}
\hline
\multicolumn{1}{|c|}{$\beta=10/g^2$} & \multicolumn{1}{c|}{$am_l$} &  \multicolumn{1}{c|}{$am_s$} &
      \multicolumn{1}{c|}{$(L/a)^3 \times (T/a)$} & \multicolumn{1}{c|}{Lats.} &  \multicolumn{1}{c|}{$u_0$} &
      \multicolumn{1}{c|}{$r_1/a$} &  \multicolumn{1}{c|}{$m_\pi L$} \\
\hline
\multicolumn {8}{|c|}{$a \approx 0.18$ fm} \\
\hline
6.503 & 0.0492 & 0.0820 & $16^3\times 48$ & 250 & 0.85636 & 1.778(8) & 9.07 \\
6.485 & 0.0328 & 0.0820 & $16^3\times 48$ & 334 & 0.85585 & 1.785(7) & 7.47 \\
6.467 & 0.0164 & 0.0820 & $16^3\times 48$ & 416 & 0.85492 & 1.801(8) & 5.36 \\
6.458 & 0.0082 & 0.0820 & $16^3\times 48$ & 484 & 0.85489 & 1.813(8) & 3.84 \\
\hline
\multicolumn {8}{|c|}{$a \approx 0.15$ fm} \\
\hline
6.628 & 0.0484 & 0.0484 & $16^3\times 48$ & 621 & 0.8623 & 2.124(6) & 8.48 \\
6.600 & 0.0290 & 0.0484 & $16^3\times 48$ & 596 & 0.8614 & 2.129(5) & 6.63 \\
6.586 & 0.0194 & 0.0484 & $16^3\times 48$ & 640 & 0.8609 & 2.138(4) & 5.46 \\
6.572 & 0.0097 & 0.0484 & $16^3\times 48$ & 631 & 0.8604 & 2.152(5) & 3.93 \\
6.566 & 0.00484 & 0.0484 & $20^3\times 48$ & 603 & 0.8602 & 2.162(5) & 3.50 \\
\hline
\multicolumn {8}{|c|}{$a \approx 0.12$ fm} \\
\hline
8.000 & $\infty\ $ & $\infty\ $ & $20^3\times 64$ & 408 & 0.8879 & 2.663(6)$\null^*$ & na \\
7.350 & 0.4000 & 0.4000 & $20^3\times 64$ & 332 & 0.8822 & 2.661(7)$\null^*$ & 29.4 \\
7.150 & 0.2000 & 0.2000 & $20^3\times 64$ & 341 & 0.8787 & 2.703(7)$\null^*$ & 19.6 \\
6.960 & 0.1000 & 0.1000 & $20^3\times 64$ & 340 & 0.8739 & 2.687(0)$\null^*$ & 13.7 \\
6.850 & 0.0500 & 0.0500 & $20^3\times 64$ & 425 & 0.8707 & 2.686(8) & 9.70 \\
6.830 & 0.0400 & 0.0500 & $20^3\times 64$ & 351 & 0.8702 & 2.664(5) & 8.70 \\
6.810 & 0.0300 & 0.0500 & $20^3\times 64$ & 564 & 0.8696 & 2.650(4) & 7.56 \\
6.790 & 0.0200 & 0.0500 & $20^3\times 64$ & 1758 &  0.8688 & 2.644(3) & 6.22 \\
6.760 & 0.0100 & 0.0500 & $20^3\times 64$ & 2023 & 0.8677 & 2.618(3) & 4.48 \\
6.760 & 0.0100 & 0.0500 & $28^3\times 64$ & 275 & 0.8677 & 2.618(3) & 6.27 \\
6.760 & 0.0070 & 0.0500 & $20^3\times 64$ & 1852 & 0.8678 & 2.635(3) & 3.78 \\
6.760 & 0.0050 & 0.0500 & $24^3\times 64$ & 1802 & 0.8678 & 2.647(3) & 3.84 \\
6.790 & 0.0300 & 0.0300 & $20^3\times 64$ & 367 & 0.8689 & 2.650(7) & 7.56 \\
6.750 & 0.0100 & 0.0300 & $20^3\times 64$ & 357 & 0.8675 & 2.658(3) & 4.48 \\
6.715 & 0.0050 & 0.0050 & $32^3\times 64$ & 701 & 0.8671 & 2.697(5) & 5.15 \\
\hline
\end{tabular}
\caption{
Table of asqtad ensembles. $u_0$ is the input tadpole factor,
Eq.~(\ref{eq:u0_plaq}), rather than the value determined from the
ensemble average of the plaquette.
Lattice spacings are from the smoothed
fit described in the text, except where indicated by a
``$*$''.  For these ensembles, $r_1/a$ 
is from this ensemble alone, rather than the smoothed fit.
To convert to physical units,
use $r_1\approx 0.31$ fm.
A $\dagger$ indicates that the run is
in progress.
This list of ensembles and counts of archived lattices are as of December 2008.
}
\label{table:runtable1}
\end{table}

\begin{table}
\begin{tabular}{|c|l|l|c|r|l|l|l|}
\hline
\multicolumn{1}{|c|}{$\beta=10/g^2$} & \multicolumn{1}{c|}{$am_l$} &  \multicolumn{1}{c|}{$am_s$} &
      \multicolumn{1}{c|}{$(L/a)^3 \times (T/a)$} & \multicolumn{1}{c|}{Lats.} &  \multicolumn{1}{c|}{$u_0$} &
      \multicolumn{1}{c|}{$r_1/a$} &  \multicolumn{1}{c|}{$m_\pi L$} \\
\hline
\multicolumn {8}{|c|}{$a \approx 0.09$ fm} \\
\hline
8.400 & $\infty\ $ & $\infty\ $ & $28^3\times 96$ & 396 & 0.89741 & 3.730(7)$\null^*$ & na \\
7.180 & 0.0310 & 0.0310 & $28^3\times 96$ & 500 & 0.8808 & 3.822(10) & 8.96 \\
7.110 & 0.0124 & 0.0310 & $28^3\times 96$ & 1996 & 0.8788 & 3.712(4) & 5.78 \\
7.100 & 0.0093 & 0.0310 & $28^3\times 96$ & 1138 & 0.8785 & 3.705(3) & 5.04 \\
7.090 & 0.0062 & 0.0310 & $28^3\times 96$ & 1946 & 0.8782 & 3.699(3) & 4.14 \\
7.085 & 0.00465 & 0.0310 & $32^3\times 96$ & 540$\null^\dagger$ & 0.8781 & 3.697(3) & 4.11 \\
7.080 & 0.0031 & 0.0310 & $40^3\times 96$ & 1012 & 0.8779 & 3.695(4) & 4.21 \\
7.075 & 0.00155 & 0.0310 & $64^3\times 96$ & 530$\null^\dagger$ & 0.877805 & 3.691(4) & 4.80 \\ % max=4518
7.100 & 0.0062 & 0.0186 & $28^3\times 96$ & 985 & 0.8785 & 3.801(4) & 4.09 \\
7.060 & 0.0031 & 0.0186 & $40^3\times 96$ & 642 & 0.8774 & 3.697(4) & 4.22 \\
7.045 & 0.0031 & 0.0031 & $40^3\times 96$ & 440$\null^\dagger$ & 0.8770 & 3.742(8) & 4.20 \\
\hline
\multicolumn {8}{|c|}{$a \approx 0.06$ fm} \\
\hline
7.480 & 0.0072 & 0.0180 & $48^3\times 144$ & 625 & 0.8881 & 5.283(8) & 6.33 \\
7.475 & 0.0054 & 0.0180 & $48^3\times 144$ & 617 & 0.88800 & 5.289(7) & 5.48 \\
7.470 & 0.0036 & 0.0180 & $48^3\times 144$ & 771 & 0.88788 & 5.296(7) & 4.49 \\
7.465 & 0.0025 & 0.0180 & $56^3\times 144$ & 800 & 0.88776 & 5.292(7) & 4.39 \\
7.460 & 0.0018 & 0.0180 & $64^3\times 144$ & 826 & 0.88764 & 5.281(8) & 4.27 \\
7.460 & 0.0036 & 0.0108 & $64^3\times 144$ & 483 & 0.88765 & 5.321(9) & 5.96 \\
\hline
\multicolumn {8}{|c|}{$a \approx 0.045$ fm} \\
\hline
7.810 & 0.0028 & 0.0140 & $64^3\times 192$ & 861 & 0.89511 & 7.115(20) & 4.56 \\
\hline
%\vfill
\end{tabular}
\caption{
Table \ref{table:runtable1} continued.
}
\end{table}

%% file: RMP_sec5.tex
% File for section 5 for RMP article
%
%\section{Section 5}

\section{Spectroscopy of light hadrons}
\label{sec:spec_other}

Computing the masses of the light hadrons is a classic problem for
lattice QCD, since the masses and structures of these particles are
highly nonperturbative.  By this point, hadron mass computations,
including the effects of light and strange dynamical quarks, have been
done for several different lattice actions, including staggered quarks,
Wilson quarks \cite{Ukita:2008mq,Ukita:2007cu,Durr:2008rw,Durr:2008zz}
and domain-wall quarks \cite{Ukita:2007cu,Allton:2008pn}.  It has long
been apparent from these and other studies that lattice QCD reproduces
the experimental masses within the accuracy of the computations.
For most of the light hadrons, however, this accuracy is not as good
as for many of the other quantities discussed in this review.  The
reasons for this are that these masses have a complicated dependence on
the light quark mass, making the chiral extrapolation (to the physical
light quark mass) difficult, and that all but a few of these hadrons
decay strongly.   Most of the lattice simulations are at heavy enough
quark masses or small enough volumes that these decays cannot happen,
so the chiral extrapolation crosses thresholds.   With staggered
quarks there is the additional technical complication that for all but
the pseudoscalar particles  with equal mass quarks the lattice correlators
contain states with both parities, with one of the parities contributing
a correlator that oscillates in time.

\begin{figure}[tbph]
\begin{center}
\epsfxsize=5.0in
\epsfbox[0 0 4096 4096]{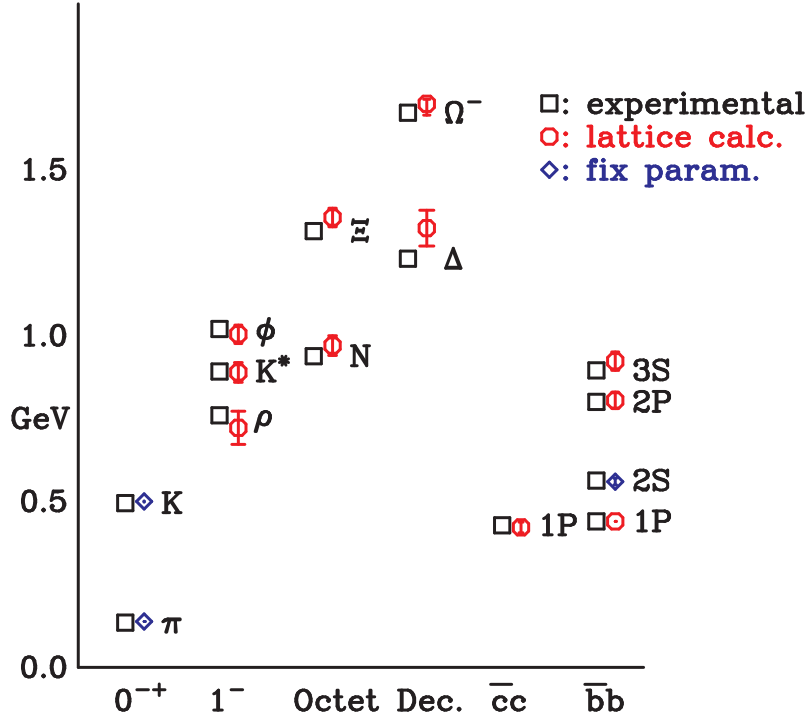}
\end{center}
\caption{
The ``big picture'' --- comparison of masses calculated on the
asqtad ensembles with experimental values.
For the light quark hadrons we plot the hadron mass, and for
the $\bar c c$ and $\bar b b$ masses the difference from
the ground state (1S) mass.
The continuum and chiral extrapolations of the pion and kaon masses
are described in Sec.~\ref{sec:fpi}, and most other
meson masses were extrapolated to the continuum and physical
light quark masses using simple polynomials. Masses of
hadrons containing strange quarks were adjusted for the
difference in the strange quark mass used in generating the
ensembles from the correct value.  The nucleon mass extrapolation,
described in \protect\textcite{Bernard:2007ux},
used a one-loop chiral perturbation theory form.
The charmonium mass splitting is from \protect\textcite{Follana:2007uv},
and the $\bar b b$ splittings from \protect\textcite{Gray:2002vk},
\protect\textcite{Wingate:2003gm} and \protect\textcite{Gray:2005ur}.
Experimental values are from \protect\textcite{Amsler:2008zzb}.
The $\Upsilon$ 2S-1S splitting and the $\pi$ and $K$ masses are
shown with a different symbol since these quantities were used
to fix $r_1$ in physical units and the light and strange quark masses.
Earlier versions of the plot appeared in \textcite{Aubin:2004wf}
and the PDG ``Review of Particle Physics'' \cite{Amsler:2008zzb}.
}
\label{fig:Bigpic}
\end{figure}

Masses of the lowest-lying light-quark hadrons have been computed
on almost all of the MILC asqtad ensembles.   Hadron masses from the
$a\approx 0.12$ fm ensembles were reported in \textcite{Bernard:2001av},
masses from the $a\approx 0.09$ fm ensembles were added in
\textcite{Aubin:2004wf}, and nucleon and $\Omega^-$ masses from
the $a \approx 0.06$ fm ensembles in \textcite{Bernard:2007ux}.
Simple extrapolations of these masses to the continuum limit and
physical quark mass, including results from several of the $a \approx
0.06$ fm ensembles, are compared to experiment in Fig.~\ref{fig:Bigpic}.
In addition, this figure shows charm and bottom meson mass splittings
\cite{Gray:2002vk,Wingate:2003gm,Gray:2005ur} compared with experimental
values \cite{Amsler:2008zzb}.

\begin{figure}[t]
\begin{center}
\begin{tabular}{c c}
\epsfxsize=3.5in
\epsfysize=3.5in
\epsfbox[0 0 4096 4096]{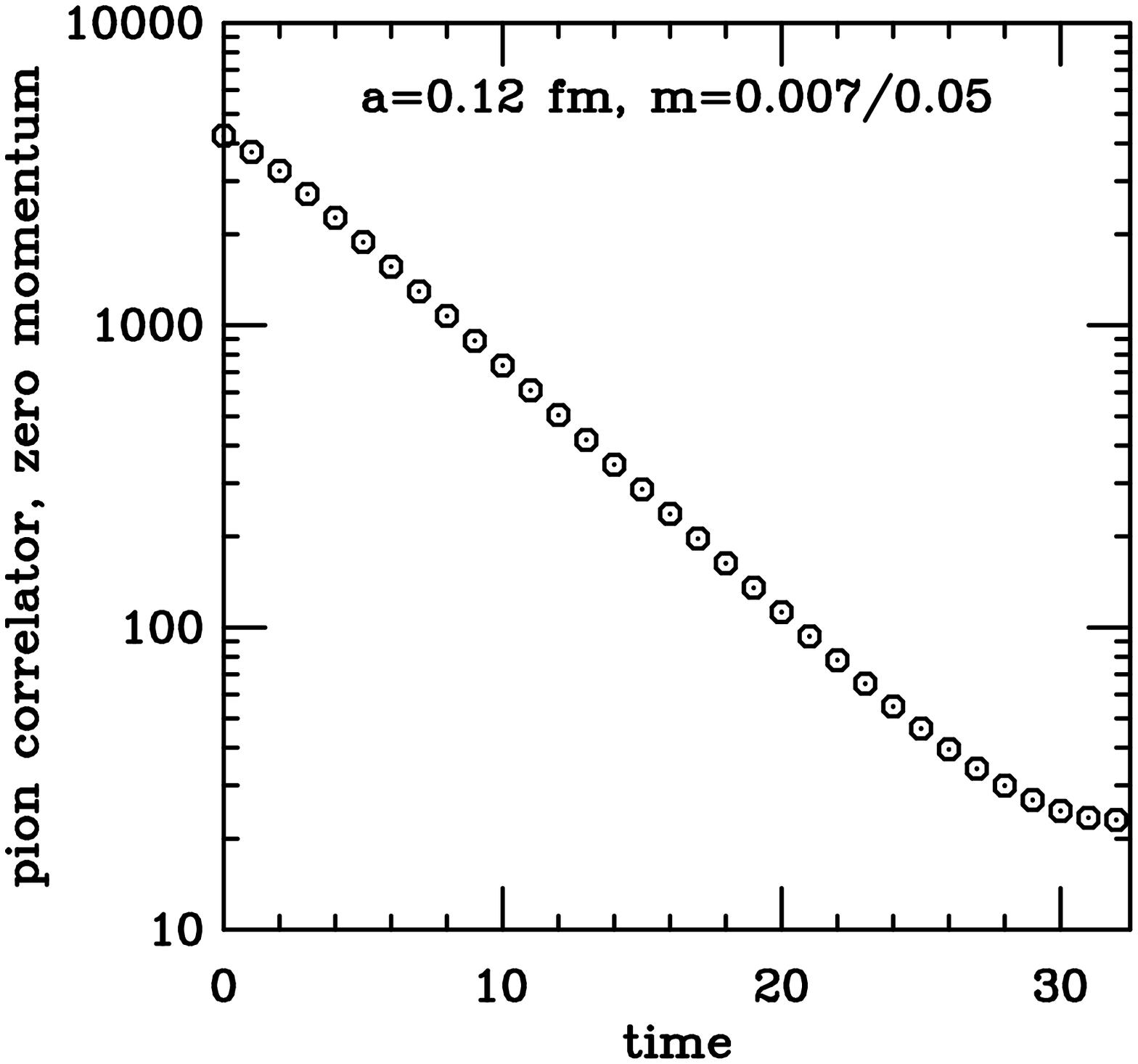}
&
\epsfxsize=3.5in
\epsfysize=3.5in
\epsfbox[0 0 4096 4096]{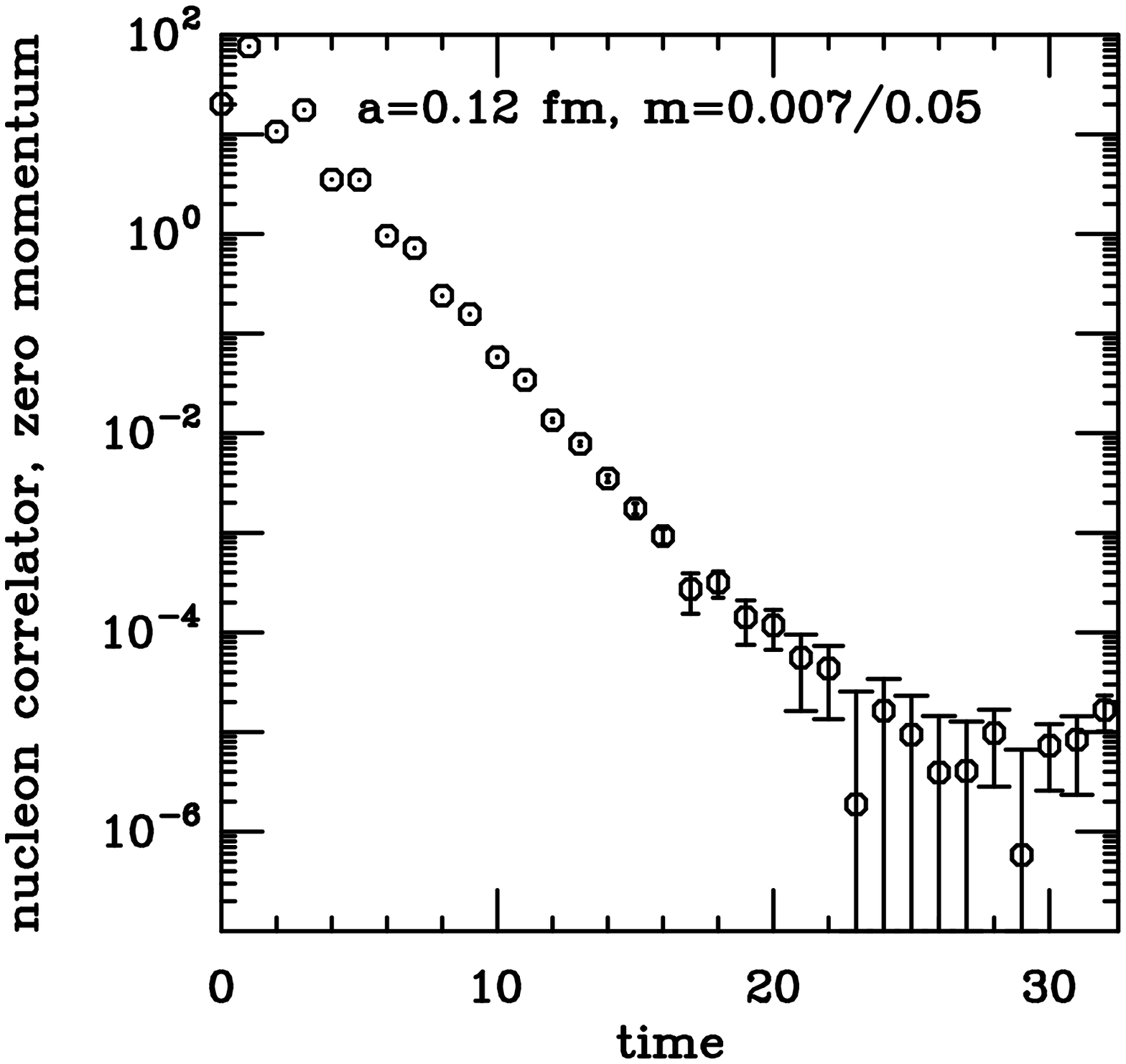}
\end{tabular}
\end{center}
\caption{Pion and nucleon correlators plotted {\it vs}. the distance
from the source.  These correlators are from the $\beta=6.76$,
$am_l/am_s=0.007/0.05$ ensemble. The small symbols in the center of the
octagons in the pion correlator are error bars.  Note the increasing
fractional errors with distance in the nucleon correlator, and the
constant fractional errors in the pion correlator.}
\label{fig:pion_nuc_propagator}
\end{figure}

\subsection{Hadron mass computations}
\label{sec:hadron_mass_theory}

The theory behind hadron mass computations with staggered quarks was
developed in \textcite{KlubergStern:1983dg}, \textcite{Golterman:1985dz}
and \textcite{Golterman:1984dn} (see also \textcite{Kilcup:1986dg}).
Early implementations, in which technical aspects were addressed,
include \textcite{Marinari:1981nu}, \textcite{Bowler:1986fw},
\textcite{Gupta:1990mr}, and \textcite{Fukugita:1992hr}.

The calculation begins with a
Euclidean-time correlation function for any operator that can produce the
desired state from the vacuum.  For instance, if an operator $\cal O$ can
annihilate a particle $p$ and the adjoint $\cO^\dagger$ can create $p$,
then we study the zero-momentum correlation function, or ``correlator''
$C_{\cO^\dagger \cO}$ given by
\begin{equation}
C_{\cO^\dagger \cO}(t) = 
\sum_x \langle \cO(x,t) \cO^\dagger (0,0)  \rangle \ .
\label{eq:corr_generic}
\end{equation}
By putting in a complete set of states between the two operators, we find
\begin{equation}
C_{\cO^\dagger \cO}(t) =
\sum_n \langle  0|\cO|n\rangle \langle n|\cO^\dagger|0 \rangle \exp( -M_n t) \ .
\label{eq:corr_complete}
\end{equation}
If the particle $p$ is the lowest-energy state $n$, then for
large Euclidean time, the dominant contribution will be $|\langle
0|\cO|p\rangle |^2 \exp( -M_p t)$.  Generally, there will be additional
contributions from higher mass states, and with staggered quarks there
are usually contributions from opposite parity states of the form $(-1)^t
\exp( -M^\prime t)$.  In addition, because of the antiperiodic boundary
conditions in time for the quarks, there will be additional terms of the
form $\exp( -M_n (T-t))$, where $T$ is the time extent of the lattice.
Thus, with staggered quarks a meson correlator generically has the form
\BEA C_{\cO^\dagger \cO}(t) &=& 
A_0 \LP e^{-M_0 t} + e^{-M_0 (T-t)} \RP
+ A_1  \LP e^{-M_1 t}  + e^{-M_1 (T-t)} \RP + \ldots \EL
&+& (-1)^t A_0^\prime \LP e^{-M_0^\prime t}  + e^{-M_0^\prime (T-t)} \RP + \ldots
\EEA
Here the primed masses and amplitudes with the factor of $(-1)^t$
correspond to particles with parity opposite that of the unprimed.
For baryons the form is similar, except that the backwards propagating
terms ($e^{-M(T-t)}$) have an additional factor of $(-1)^{t+1}$.
Here the overall minus sign in the backwards propagating part is due to
the antiperiodic boundary conditions for the quarks in the Euclidean time
direction.  Figure~\ref{fig:pion_nuc_propagator} shows correlators for
the pion and nucleon in a sample asqtad ensemble.  Statistical errors on
the pion correlator are the tiny symbols in the center of the octagons.
The effect of periodic (for a meson correlator) boundary conditions in
time is clearly visible.  For short times, there are contributions from
heavier particles.

For hadrons other than glueballs, evaluating this correlator requires
computing $M^{-1}_{x,y}$ where $M$ is the matrix defining the quark
action.  This can be done by making a ``source'' vector $b$ which is
nonzero only at lattice point $y$, or in some
small region, and solving the sparse matrix equation
$Ma=b$, usually using the conjugate gradient algorithm.  (Here $a$ and
$b$ are vectors with one component for each color at each lattice site
in the system -- {\it i.e.,} $3V$ complex components.  With Wilson-type
quarks there would also be four spin components per lattice site.)

The simplest possibility for $\cO$ is an operator built from quarks and
antiquarks located in the same $2^4$ hypercube, often even on the same
lattice site.   This is usually called a point source.  Because the point
operator ${\cal O}_P$ tends to have a large overlap with excited states,
it is usually advantageous to take a  ``smeared'' source operator ${\cal
O}^\dagger$, where the quarks in the hadron may be created at different
lattice sites.   One common approach
is to choose a smeared operator that creates quarks
and antiquarks with a distribution similar to that of
the expected quark model wave function of the
desired hadron.  A cruder and simpler approach used in most of the MILC
light hadron mass calculations is to take a ``Coulomb wall'' source,
where the lattice is first gauge transformed to the lattice Coulomb
gauge, making the spatial links as smooth as possible.   Then a source
is constructed which covers an entire time slice, for example, with a $1$
in some corner of each $2^3$ cube in the time slice.  This works because
with Coulomb gauge fixing contributions from source components
within a typical hadronic correlation length interfere coherently, while
contributions average to zero if the quarks created by
${\cal O}^\dagger$ are
widely separated (although they do contribute to the statistical noise).
In other words,
$\LL M^{-1}_{\vec x_1,t_i;\vec y,t_f} M^{-1}_{\vec y,t_f;\vec x_2,t_i} \RR$
is significant only when $|\vec x_1 - \vec x_2|$ is less than a typical
hadronic size.
For example, a Coulomb wall operator appropriate for a Goldstone pion is
\begin{equation} 
{\cal O}_W (t) = \sum_{\vec x, \vec y} \bar\chi(\vec x,t)(-1)^{\vec x + t} \chi(\vec y,t)\ .
\label{eq:coulomb-wall-operator1}
\end{equation}

In a mass calculation, we want the state with zero spatial
momentum, which is isolated by summing the sink position over all spatial
points on a time slice.   In many matrix element studies, we need hadron
states with nonzero momenta, and they are isolated by summing over the
spatial slice with the appropriate phase factors.

Statistics are usually further enhanced by averaging correlators from
wall sources, or other types of sources, from several time slices in
the lattice.   In general, each different source requires a new
set of sparse matrix inversions.

For most hadrons, statistical error is the limiting factor in the mass
computations.   At long Euclidean time $t$, a correlator with hadron
$H$ as its lowest mass constituent is proportional to $e^{-M_H t}$.
The variance of this correlator can itself be thought of as the correlator
of the square of the operator
\BE\LL {\cal O}_H(x) {\cal O}_H^\dagger(x) \ 
        {\cal O}^\dagger_H(y) {\cal O}_H(y) \RR \ \ , \EE
where in this correlator for flavor-nonsinglet hadrons it is understood
that quark lines all run from the operators at $x$ to those at $y$
\cite{Lepage:1989hd}.  The behavior of the variance at long distances
is dominated by the lowest mass set of particles created by ${\cal
O}_H(x) {\cal O}_H^\dagger(x)$.  Thus for mesons ${\cal O}_H(x)
{\cal O}_H^\dagger(x)$ creates two quarks and two antiquarks  which
can propagate as two pseudoscalar mesons. Then the variance decreases
approximately as $e^{-2M_{PS}t}$, where $M_{PS}$ is the mass of the
pseudoscalar meson made from the quarks in ${\cal O}^\dagger_H {\cal
O}_H$.  For baryons there are three quarks and three antiquarks, and
the variance decreases approximately as $e^{-3M_{PS}t}$.  This behavior
can be seen in Fig.~\ref{fig:pion_nuc_propagator}, where the fractional
error on the pion correlator does not increase with distance, while the
fractional error on the nucleon correlator grows quickly.

As discussed in Sec.~\ref{sec:Staggered_ferm}, hadrons with staggered
quarks come with different ``tastes,'' all of which are degenerate in the
continuum limit.  For pseudoscalar mesons, the mass differences between
different tastes are large, but they are well understood as discussed in
Sec.~\ref{sec:SChPT}.  For the other hadrons, for which chiral symmetry
is not the most important factor in determining the mass, taste symmetry
violations are much smaller.  In particular, we have computed masses
for four different tastes of the $\rho$ meson on many of our ensembles,
and have failed to find any statistically significant taste splittings.
(See also \textcite{Ishizuka:1993mt}.)

\subsection{Correlated fits}
\label{correlated_fits}

There are several kinds of correlations in the numerical results of lattice
gauge theory simulations.  The Markov chain that produces the
configurations produces correlated configurations.  Thus,
there are correlations in ``simulation time.''  The correlations vary
with the algorithm, and one can reduce them by increasing the simulation
time gap between the configurations that are analyzed. Generation of
configurations is computationally expensive, however, and
the autocorrelation length is unknown until the run and some analysis is completed,
so one usually saves configurations with some degree of correlation.
A simple way to deal with these correlations is to block successive
configurations together and then to estimate errors from the variance
of blocks.  However, if the number of blocks is not many times larger
than the number of degrees of freedom, the finiteness of the sample size
must be considered when estimating goodness-of-fit or statistical errors
on the parameters in a fit \cite{Michael:1993yj,Toussaint:2008ke}.
In cases where blocking is not practical, notably the pseudoscalar
meson analysis in Sec.~\ref{sec:fpi}, we have estimated elements of the
covariance matrix by using the measured autocorrelations in the data to
rescale a covariance matrix based on unblocked data.

Even if successive configurations are not correlated, different
physical quantities are correlated with each other.  For example, if
the pion correlator is larger than average at a separation $t$ from
the source on a particular configuration, it is likely to be larger
at $t+1$.
Thus, when extracting hadron masses,
or other fit parameters, we must use the full correlation matrix in the
fit model, not just the variance in each particular element fit.   To be
specific, let the values of the independent parameters be denoted
$x_i$ and corresponding lattice ``measured'' values be $y_i$.  The fitting
procedure requires varying the model parameters $\{\lambda\}$ that define
the model function $y_M(x_i,\{\lambda\})$ in order to minimize $\chi^2$.
For uncorrelated data,
\begin{equation}
\chi^2 = \sum_i { \LP y_M \LP x_i,\{\lambda\} \RP -y_i \RP ^2 / \sigma_i^2} \ ,
\label{eq:chisqnocorr}
\end{equation}
where $\sigma_i$ is the standard deviation of $y_i$.
When the data is correlated, let $C_{ij} = \mathrm{Cov}(y_i,y_j) $ and then
\begin{equation}
\chi^2 = \sum_{i,j} {  \LP y_M \LP x_i,\{\lambda\} \RP -y_i \RP C_{ij}^{-1} \LP y_M \LP x_j,\{\lambda\} \RP -y_j \RP }
\label{eq:chisqcorr}
\end{equation}
(In practice $C_{ij}$ is almost always estimated from the same data as
the $y_i$, and in this case $\chi^2$ is more properly called $T^2$.)
Uncorrelated data reduces to $C_{ij} = \delta_{ij} \sigma_i^2 $.
If $C_{ij}$ has positive off-diagonal entries, then the data will
look smoother than it would if uncorrelated.

\begin{figure}[t]
\begin{center}
\begin{tabular}{c c}
\epsfxsize=3.0in
\epsfysize=3.0in
\epsfbox[0 0 4096 4096]{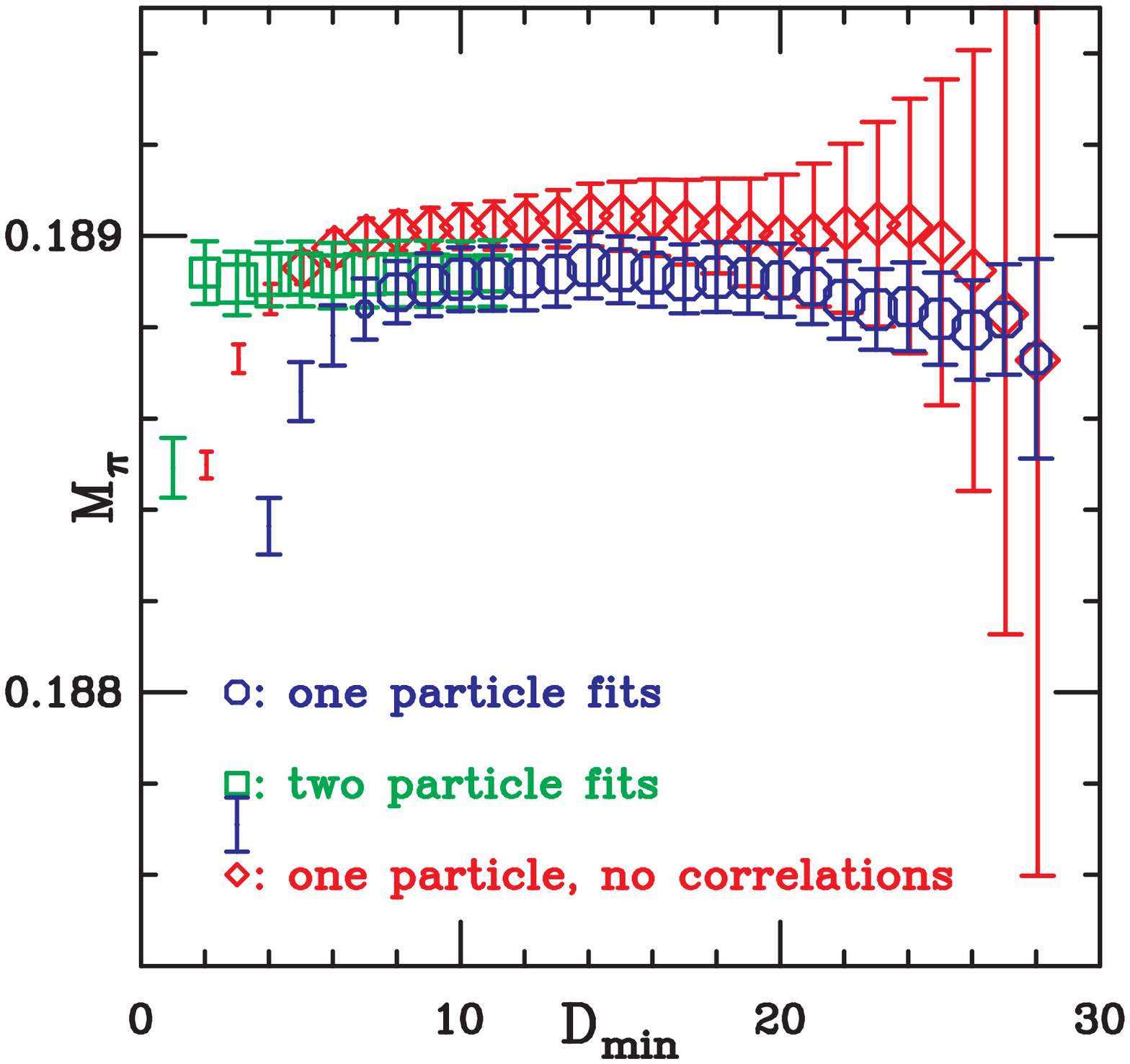}
&
\epsfxsize=3.0in
\epsfysize=3.0in
\epsfbox[0 0 4096 4096]{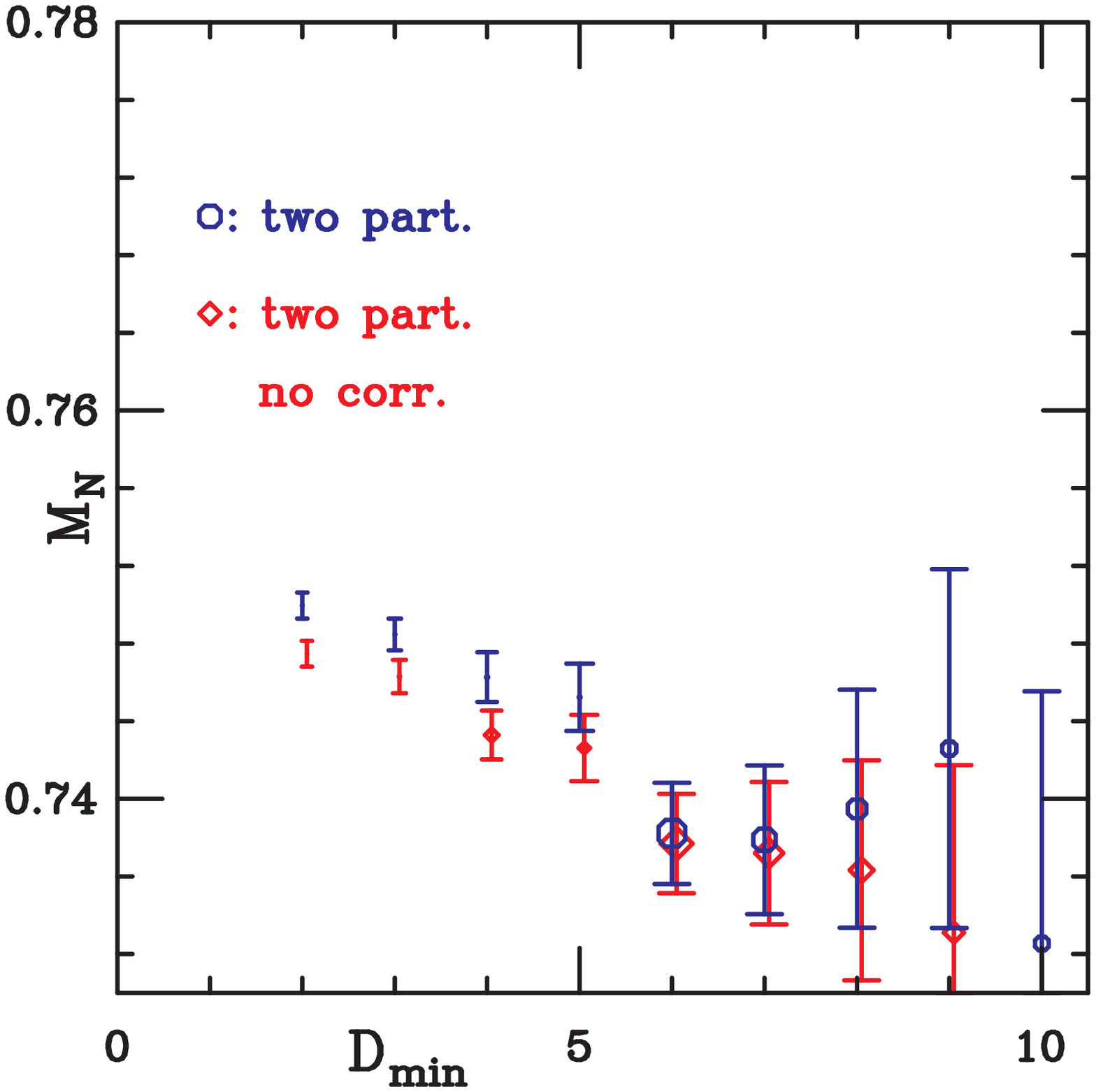}
\end{tabular}
\end{center}
\caption{
Result of fitting the correlators in Fig.~\protect\ref{fig:pion_nuc_propagator}
from a minimum distance to the center of the lattice (for the pion)
or distance at which the correlator loses statistical significance
(for the nucleon).  For the pion correlator (left panel), octagons
correspond to single-particle fits and squares to two-particle fits.
The diamonds are from single-particle fits ignoring correlations among
the data points.  For the nucleon fits (right panel), all the fits use
two particles, one of each parity.  Octagons are correlated fits, and
diamonds are fits ignoring the correlations.  The sizes of the symbols are
proportional to the confidence level of the fits, with the symbol size
in the legends corresponding to 50\% confidence.
}
\label{fig:masses_vs_dmin}
\end{figure}

In Fig.~\ref{fig:masses_vs_dmin}, we show how the fitted pion and nucleon
masses vary with the minimum distance from the source that is included
in the fit.  The octagons and squares are correlated fits, minimizing
$\chi^2$ in Eq.~(\ref{eq:chisqcorr}).   For the pion, the octagons
correspond to a single-particle (two-parameter) fit, and the squares
correspond to a two-particle (four-parameter) fit.  For the nucleon,
the octagons are fits including one particle of each parity.  We need
to decide which fit is best, and we do that based on the confidence
levels of the fits, which are roughly indicated by the symbol size.
Figure~\ref{fig:masses_vs_dmin} also contains fits ignoring correlations
while
minimizing the $\chi^2$ in Eq.~(\ref{eq:chisqnocorr}).
It can be seen that the error bars on these points are in general incorrect ---
they are neither a correct estimate of how much the parameters would
likely vary if the calculation were repeated, nor
of how much the parameters are likely to differ from the true value.
We also see that the confidence levels are generally too large for the
uncorrelated fits.  In particular, based on its confidence level, one
might accept the uncorrelated pion fit with minimum distance five.  But in
fact it can be seen that it differs significantly from the asymptotic
value.  The effects on the confidence level from ignoring correlations
can be quite extreme.  For example, in the single-particle pion fits
with $D_{min}=5$, the correlated fit has $\chi^2=180$ for 25 degrees
of freedom, for a confidence of $10^{-24}$, while the uncorrelated fit
has $\chi^2=14$ for 25 degrees of freedom, or an (erroneous) confidence
of $0.96$.

Jackknife or bootstrap methods are often used with correlated data.
These methods give estimates of the errors in fit parameters, but they
do not provide information about goodness of fit.

Once the hadron propagators are fit, we still need to perform chiral or
continuum extrapolations.  In these cases, it is also imperative to deal
with the correlations among the fitted quantities that come from the
same ensemble.   With partial quenching these covariance matrices can
become quite large, so it is essential to have enough configurations in
each ensemble to be able to get a good estimate of the covariance matrix.

\subsection{Results for some light hadrons}
\label{sec:hadron_mass_results}

\begin{figure}[t]
\begin{center}
\epsfxsize=4.5in
\epsfbox[0 0 4096 4096]{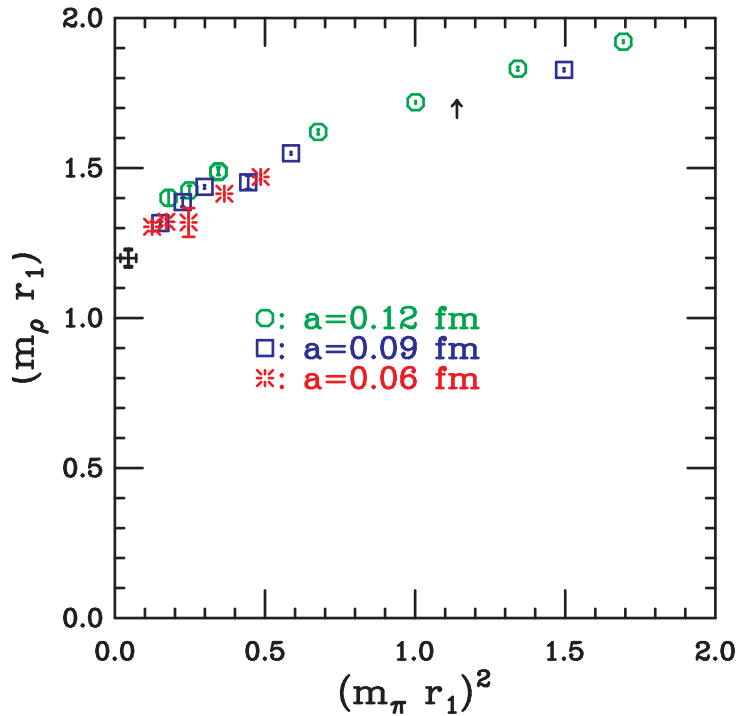}
\end{center}
\caption{
The $\rho$ mass in units of $r_1$, plotted versus the squared pion mass.
Since $m_\pi^2 \propto m_q$, this is effectively a plot versus light
quark mass.  The octagons are from ensembles with $a \approx 0.12$ fm,
the squares from ensembles with $a \approx 0.09$ fm, and the bursts from
ensembles with $a \approx 0.06$ fm.  The decorated plus at the left is
the physical $\rho$ mass, with the error on this point coming from the
error in $r_1$.  For reference, the upward arrow indicates approximately
where the quark mass equals the strange quark mass.
}
\label{fig:mrho}
\end{figure}

The pseudoscalar mesons are special for several reasons.  First, very
accurate mass computations are possible. This is because the statistical
error in the correlator (square root of the variance) decreases with the
same exponential as the correlator itself -- the fractional error is
nearly independent of $t$, and accurate correlators can be computed
out to the full extent of the lattice.  Second, for equal mass quarks
the pseudoscalar correlator does not have oscillating
contributions from opposite parity particles, and the oscillating
contributions are negligible for the kaon.  Third, because of the pions'
role as the approximate Goldstone bosons for broken chiral symmetry,
the breaking of taste symmetry leads to large mass splittings among the
different taste combinations.  Finally, because it is related to the decay
constant of the meson, the amplitude of the pseudoscalar correlator is
as interesting as the mass.  Because of the exact U(1) chiral symmetry
of the staggered quark action, the axial-vector current corresponding to
the Goldstone (taste pseudoscalar) pion needs no renormalization, so the
decay constants can also be calculated to high precision.  For these
reasons, discussion of the light pseudoscalar mesons is deferred to
Sec.~\ref{sec:fpi}.

For the vector mesons, the fractional statistical error in the correlator
increases as $e^{(M_V-M_{PS})t}$.  Also, the vector mesons decay strongly.
On the lattice, conservation of momentum and angular momentum forbids
the mixing of a zero-momentum vector meson with two zero momentum
pseudoscalars, so the vector meson is ``stable on the lattice'' for pion
masses large enough that $2\sqrt{M_{PS}^2 + (2\pi/L)^2} > M_V$. (Taste
breaking adds some additional complications to this.)  For all of
the asqtad ensembles except those with the smallest quark masses, this
condition is satisfied, and the vector meson masses can be easily, if not
accurately, found.   However, the problem of extrapolation through the
decay threshold to the physical quark mass has not been fully addressed.
Figure~\ref{fig:mrho} shows the $\rho$ meson mass as a function of light
quark mass for three different lattice spacings.  Results for the $K^*$
and $\phi$ are similar, except that there is an added complication in
that the mass needs to be adjusted to compensate for the fact that the
strange quark mass used in the correlator computations differs
from the physical $m_s$.  While the values in \textcite{Bernard:2001av}
and \textcite{Aubin:2004wf} use the same valence and sea strange quark
masses, the masses in Fig.~\ref{fig:Bigpic} have been interpolated to
the correct valence strange quark mass.

\begin{figure}[t]
\begin{center}
\epsfxsize=4.5in
\epsfbox[0 0 4096 4096]{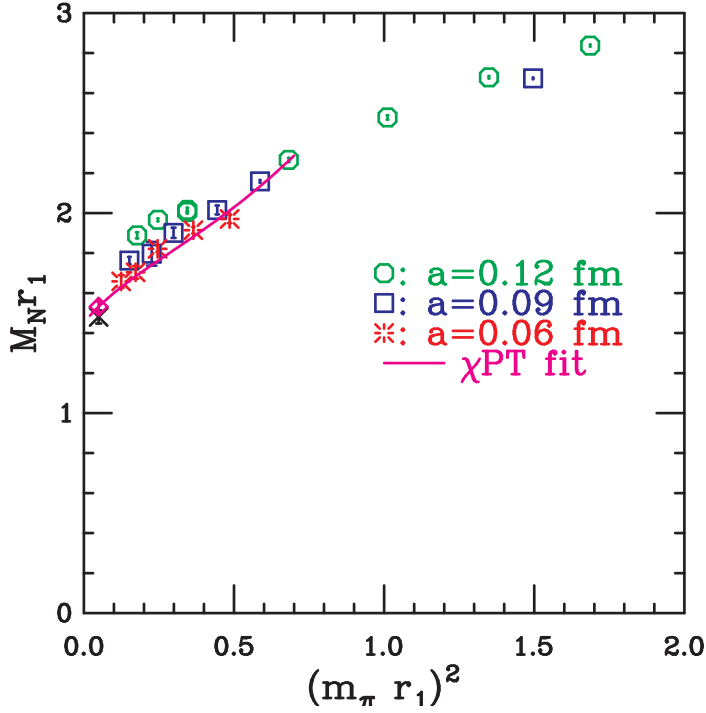}
\end{center}
\caption{
The nucleon and a chiral fit.  Nucleon masses are shown for
different light quark masses at three lattice spacings.  The cross
at the left is the experimental value.  The slightly curved line
and the diamond at the physical quark mass are a continuum and
chiral extrapolation.   Lattice spacing errors are assumed to be
linear in $a^2 \alpha_s$.  The particular chiral form used here is a
one-loop calculation with $\pi-N$ and $\pi-\Delta$ intermediate states
\protect\cite{Jenkins:1991ts,Bernard:1993nj}.  This plot is an updated
version of one in \protect\textcite{Bernard:2007ux}.
}
\label{fig:mnuc}
\end{figure}

\begin{figure}[t]
\begin{center}
\epsfxsize=4.5in
\epsfbox[0 0 4096 4096]{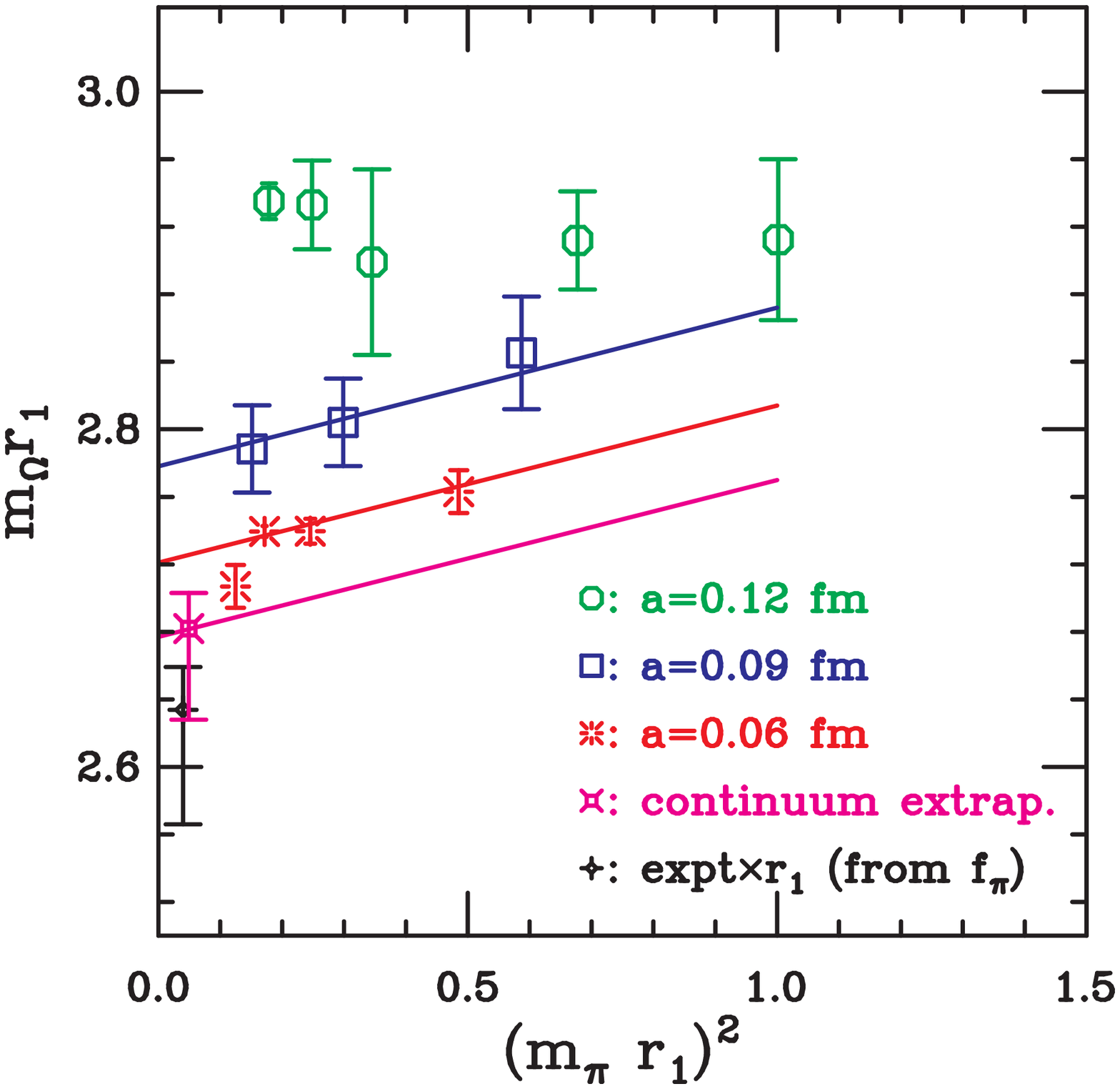}
\end{center}
\caption{
The $\Omega^-$ mass.  Results are shown for three different lattice
spacings.  The points with $a \approx 0.09$ fm and $a \approx 0.06$
fm were fit to the form $M_\Omega r_1 = A + Ba^2\alpha_s + C (m_\pi
r_1)^2$. The sloping lines show this fit form evaluated at the values
of $a^2\alpha_s$ for these lattice spacings, and at $a=0$.  Finally,
the fancy cross with error bars is the fit form evaluated at the
physical pion mass, and the small diamond is the experimental value.
Note that in this case the vertical axis does not begin at zero.
Earlier versions of the plot appeared in \textcite{Toussaint:2004cj}
and in \textcite{Bernard:2007ux}.
}
\label{fig:Omega}
\end{figure}

The nucleon is stable and chiral perturbation theory is available to
guide the extrapolation in quark mass.   However, computation of reliable
masses is difficult because the fractional error in the nucleon propagator
increases as $e^{(M_N-\frac{3}{2}M_{PS})t}$.  Also, there are excited
states  with masses not too far above the nucleon mass that contribute
to the correlator.  In fact, with staggered quarks the simplest baryon
source operators couple to the $\Delta$ as well as the nucleon, so the
lowest positive-parity excited state in the correlator is the $\Delta$
\cite{Golterman:1984dn}.  Figure \ref{fig:mnuc} shows nucleon masses for
three lattice spacings versus quark mass, together with a continuum and
chiral extrapolation.

Another hadron of particular interest is the $\Omega^-$
\cite{Toussaint:2004cj}.  This particle is stable against strong decays.
Also, in one-loop chiral perturbation theory there are no pion-baryon
loops, so at this order there are no logarithms of $m_\pi$ in the chiral
extrapolation of the mass.  Therefore, we expect that a simple polynomial
extrapolation in light quark mass should be good.  Unfortunately, the
$\Omega^-$ is a difficult mass computation with staggered quarks, first
because it is a heavy particle and second because a baryon operator
that has the $\Omega^-$ as its lowest energy state has its three
quarks at different lattice sites \cite{Golterman:1984dn,Gupta:1990mr}.
The $\Omega^-$ mass is strongly dependent on the strange quark mass,
and in principle provides an independent way to determine the correct
lattice strange quark mass.
Figure \ref{fig:Omega} contains $\Omega^-$ mass estimates, using strange
valence quark masses at each lattice spacing that were independently
determined from the pseudoscalar meson analysis in Sec.~\ref{sec:fpi}.
To do this, $\Omega^-$ correlators were generated using two different
strange quark masses near the desired one, and the $\Omega^-$mass was
obtained by linearly interpolating to the strange quark mass determined
separately.  This plot also shows a continuum and chiral extrapolation
using the simple form $M_\Omega r_1 = A + Ba^2\alpha_s + C (m_\pi r_1)^2$.

Masses of other particles, such as the $a_1$ and $b_1$
and particles including strange quarks were calculated in
\textcite{Bernard:2001av,Bernard:2007ux}, and the excited
state of the pion was identified in \textcite{Bernard:2007ux}.
Light hybrid mesons with exotic quantum numbers were studied in
\textcite{Bernard:2003jd,Bernard:2002rz}, and exotic hybrid mesons
with nonrelativistic heavy quarks in \textcite{Burch:2003zf}, and
\textcite{Burch:2001nk,Burch:2001tr}.

%%%%%%%%%%%%%%%%%%%%%%%%%%%%%%%%%%%%%%%%%%%%%%%%%%%%%%%%%%%%%%%%%%%%%%
\subsection{Flavor singlet spectroscopy}
\label{sec:Flav_sing}

Determining the masses of flavor-singlet mesons is, perhaps, the most
challenging endeavor in lattice QCD light hadron spectroscopy. The difficulty
in doing so
has three main sources:

(i) Flavor-singlet correlators have two different contributions:
quark-line connected and quark-line disconnected.
The quark-line disconnected piece requires
so-called ``all-to-all'' correlators.  To avoid the $\mathcal{O}(V)$
inversions to compute these all-to-all propagators, stochastic methods
are used. \textcite{Kuramashi:1994aj} used a unit source at each site
and let gauge invariance do the averaging. More common now is the use
of random sources \cite{Dong:1993pk,Venkataraman:1997xi} similar to
Eqs.~(\ref{eq:Gaus_av}), (\ref{eq:R_M_inv_dM_R}), with various noise
reduction techniques
\cite{Wilcox:1999ab,McNeile:2000xx,Struckmann:2000bt,Mathur:2002sf,Foley:2005ac},
including low-eigenmode preconditioning
\cite{Venkataraman:1997yj,DeGrand:2002gm}.

(ii) While the stochastic noise of the quark-line connected correlators
falls off exponentially (albeit with a smaller exponent than the signal),
the noise in the quark-line disconnected part is constant. So the signal
to noise ratio falls off much faster for the disconnected part.

(iii) The quark-line connected correlator is the same as for a
flavor-nonsinglet meson -- in particular the pion for the pseudoscalar
channel.  Therefore, the very noisy disconnected correlator first has
to cancel the connected correlator before giving the desired singlet
correlator whose falloff gives the flavor-singlet mass.

Since much larger statistics are needed for the computation of the
flavor-singlet correlators, the UKQCD collaboration has extended
a couple of the MILC lattice ensembles to around 30000 trajectories
\cite{Gregory:2008mn,Gregory:2007ce,Gregory:2007ev}.  Their simulations
are still on-going. So far, the only result given is for the $0^{++}$
glueball, whose correlator can be constructed from gauge field operators
and requires no noisy estimators and Dirac operator inversions. For two
different lattice spacings, $a \approx 0.12$ and $0.09$ fm, the UKQCD
collaboration finds $m_{0^{++}} = 1629(32) \, \MeV$ and $1600(71) \,
\MeV$ \cite{Gregory:2008mn}, respectively.

It is important to continue this investigation. In particular, obtaining
the correct $\eta^\prime$ mass would further support the correctness of
the rooting procedure to eliminate the unwanted tastes for staggered
fermions.

\subsection{Scalar mesons $f_0$ and $a_0$}
\label{sec:Scalar_mesons}

In this subsection, we describe briefly the analysis of correlators
for two light, unstable scalar mesons, namely, the isosinglet $f_0$
and the isovector $a_0$.

With the first good measurements of the $a_0$ channel in the staggered
fermion formulation a peculiarity was encountered: it was found that
on coarse lattices the $a_0$ correlator appeared to have a spectral
contribution with an anomalously low mass, lighter than any physical
decay channel \cite{Aubin:2004wf,Gregory:2005yr}.

For sufficiently light $u$ and $d$ quark masses, the $f_0$ decays to
two pions.  Likewise, the isovector scalar meson $a_0$ decays to a pion
and an $\eta$.  On the lattice, the open decay channels complicate the
analysis of the scalar meson correlators.  They are dominated by the
spectral contributions of the significantly lighter decay channels.  As a
flavor singlet, the $f_0$ also suffers from the quark-line disconnected
contributions described in the previous subsection.  Finally, with
staggered fermions at nonzero lattice spacing, the splitting of the
pseudoscalar meson taste multiplets in the decay channel deals a seeming
{\it coup de gr\^ace}.

Fortunately, one can make progress using
\rschpt\ described in Sec.~\ref{sec:SChPT}
\cite{Prelovsek:2005qc,Prelovsek:2005rf,Bernard:2006gj}.  The essential
idea is to match definitions of the desired correlator of local
interpolating operators in the lattice QCD formulation and in \rschpt.
The lattice definition is the basis for the numerical simulation of
the correlator, and the \rschpt\ definition provides a model
for fitting the result of the simulation, including all taste-breaking
effects in the decay channels.  If we take the taste-multiplet masses from
separate calculations, then, despite the rather complicated
set of two-meson channels, that portion of the fit model depends on
only three low energy constants.  In principle, even these constants
can be determined from independent measurements, leaving no free
parameters. So this fit
provides a further test of the viability of \rschpt\ as a low
energy effective theory for the staggered action.

The hadron propagator from lattice site $0$ to $y$ is defined in the same
way from the generating functionals for both QCD and the chiral theory:
\begin{equation}
  \frac{\partial^2 \log Z}
       {\partial m_{f,f^\prime}(y) 
         \partial m_{e^\prime,e}(0)}~.
\label{eq:matching}
\end{equation}
In QCD, the source $m_{f,f^\prime}(y)$ generalizes the usual quark
mass term and includes off-diagonal flavor mixing $f,f^\prime$.
The same correlator is defined in \rschpt,  where the local
source $m_{f,f^\prime}(y)$ appears in the generalized meson mass matrix.
This establishes a correspondence between the correlator defined
in terms of the quark fields $\bar q(y) q(y)$ in QCD and in terms of
the local meson fields $B \Phi^2(y)$.

\begin{figure}[t]
\centering
\includegraphics[scale=0.4]{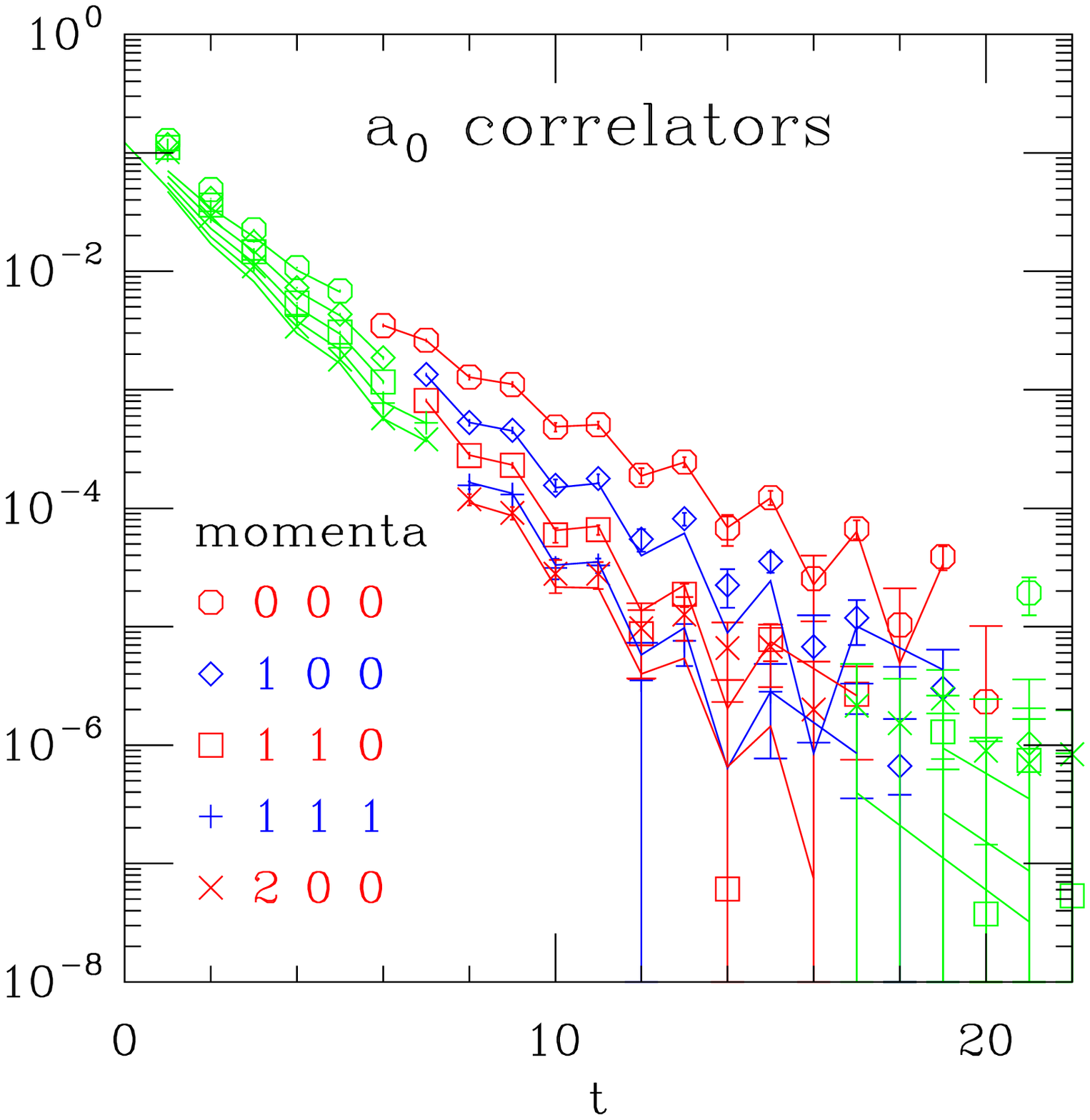}
\includegraphics[scale=0.4]{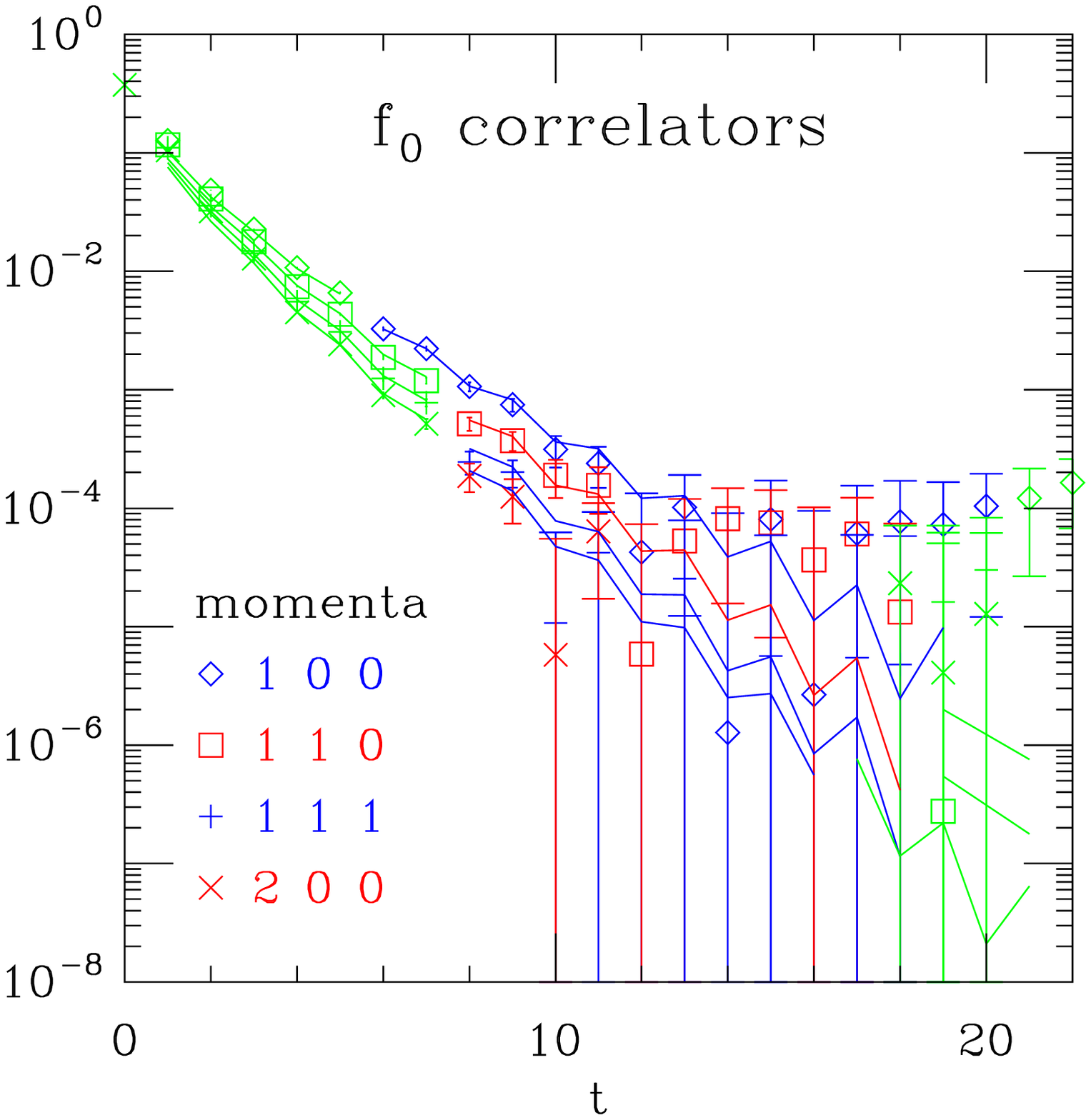}
\caption{Best fit to the $a_0$ correlator (left panel) for five
  momenta and the $f_0$ correlator (right panel) for four momenta. The
  fitting range is indicated by points and fitted lines in red and blue
  (darker points and lines). Occasional points with negative central
  values are not plotted. Data are determined from the $a \approx 0.12$
  fm (coarse) ensemble with $am_l = 0.005$ and $am_s = 0.05$. Figures
  from \textcite{Bernard:2007qf}.}
\label{fig:a0f0fits}
\end{figure}

To lowest order in \rschpt, the meson correlator is described by a bubble
diagram, which gives the contributions of the two-pseudoscalar-meson
intermediate states, including all taste multiplets and hairpins.  These
contributions are determined from the multiplet masses and the \rschpt\
low energy constants $B$, $\delta^\prime_A$, and $\delta^\prime_V$
described in Sec.~\ref{sec:SChPT}.  In addition to the bubble diagram,
one adds an explicit quark-antiquark $a_0$ or $f_0$ state to complete
the fit model.
Results are shown in Fig.~\ref{fig:a0f0fits},
and results for the low energy constants are listed in Table~\ref{tab:LEC}.
\begin{table}
  \begin{center}
  \begin{tabular}{|l|rr|}
  \hline
          & \quad $f_0$ and $a_0$ correlators & \quad Meson masses and decays \\
  \hline
    $r_1 m_\pi^2/(2 m_{u,d})$ & 7.3(1.6) & 6.7 \\
    $\delta_V              $ & (prior)  & $-0.016(23)$ \\
    $\delta_A              $ & $-0.056(10)$ & $-0.040(6)$ \\
  \hline
  \end{tabular}
  \end{center}
  \caption{Comparison of our fit parameters for the \rschpt\ low energy
    constants with results from \protect\textcite{Aubin:2004fs} \label{tab:LEC}}
\end{table}

It is particularly instructive to examine the variety of
two-pseudoscalar-meson taste channels contributing to the scalar
meson correlators.  To be physical states, the external scalar mesons
$a_0$ and $f_0$ must be taste singlets.  Taste selection rules then
require that they couple only to pairs of pseudoscalar mesons of the
same taste.  Thus, for example, for the $a_0$, each flavor channel,
such as $\pi-\eta$, comes with a multiplicity of sixteen taste pairs,
although lattice symmetries reduce the number of distinct thresholds
to six.  There is also a set of $\pi-\eta^\prime$ channels.  To get
the energies of the thresholds, we look at the taste splitting of the
component hadrons.  We have already seen how the pion taste multiplet
splits into the Goldstone state and a variety of higher-lying states,
all of which become degenerate in the continuum limit.  The $\eta$ and
$\eta^\prime$, on the other hand, have unusual splitting because they
mix with the chiral anomaly.  Since the anomaly is a taste singlet, only
the taste-singlet $\eta$ and $\eta^\prime$ mix with it in the usual way.
Thus, in the continuum limit only the taste singlet states
are expected to have the correct masses.  They are the only physical
states. The fifteen taste nonsinglet $\eta$'s and $\eta^\prime$'s remain
light.
The pseudoscalar-taste eta pairs with the pseudoscalar-taste pion.
The unphysical pseudoscalar-taste $\pi-\eta$ channel
gives an anomalously light spectral contribution to the $a_0$ correlator
\cite{Prelovsek:2005qc,Prelovsek:2005rf}.  A similar complication occurs
in the $f_0$ correlator, but it is masked by the expected physical
two-pion intermediate state.

\begin{figure}[t]
\begin{center}
\includegraphics[width=3.5in]{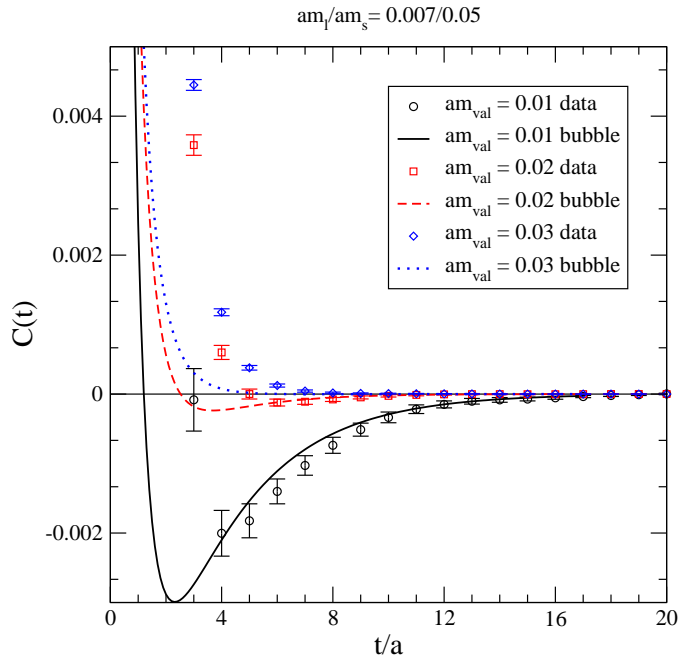}
\caption{
The isovector scalar ($a_0$) correlator on the MILC coarse $am_l /
am_s = 0.007/0.05$ ensemble with three different domain-wall valence
masses. Overlaid on the data are the predicted bubble contributions,
which should dominate over the exponentially-decaying contributions at
sufficiently large times. Figure from \textcite{Aubin:2008wk}.}
\label{fig:MA_a0}
\end{center}
\end{figure}

The unphysical taste contributions provide a concrete illustration of
the breakdown of unitarity at nonzero lattice spacing as a result of the
fourth-root.  The theory heals the scalar meson
correlators in the continuum limit by a mechanism that parallels exactly
the one described for the one-flavor model in Sec.~\ref{sec:Rooting}.
The pseudoscalar meson bubble diagram contains a
negative-norm channel.  This unphysical ghost channel has the
weight needed to cancel the contributions of all the unphysical taste
components in the continuum limit.  Thus in the continuum limit only
the physical intermediate two-meson states survive.

The behavior of the isovector scalar correlator has also been analyzed
for the case of domain-wall valence quarks on the MILC staggered
ensembles~\cite{Aubin:2008wk}.  In the mixed-action case, the $a_0$
correlator receives contributions from two-particle intermediate states
with mesons composed of two domain-wall quarks, mixed mesons composed
of one domain-wall and one staggered quark, and mesons composed of two
staggered quarks.  Because the symmetry of the external valence quarks
restricts the sea-sea mesons to be taste singlets, the correlator does not
receive contributions from all of the taste channels.  As in the purely
staggered case, the one-loop bubble contribution is
determined by three low-energy constants~\cite{Prelovsek:2005rf},
which are known from tree-level \chpt\ fits to meson masses.
For domain-wall quarks on the coarse and fine MILC lattices, the
contribution from the bubble term is predicted to be large and negative
for several time slices.  Thus a comparison of the mixed-action \chpt\
prediction for the behavior of the $a_0$ correlator with 
lattice data provides a strong consistency check.

\textcite{Aubin:2008wk} compare the mixed-action \chpt\ prediction for
the bubble contribution with the lattice $a_0$
correlator for several domain-wall valence masses on the coarse and fine
MILC lattices.  They find that, in all cases the size of the bubble
contribution is quantitatively consistent with the data, and that
the behavior of the data cannot be explained if mixed-action lattice
artifacts are neglected.  For fixed light sea quark mass, the size of the
bubble term decreases as the valence quark mass increases (see
Fig.~\ref{fig:MA_a0}).  The bubble contribution also decreases as $a
\to 0$.  These results of \textcite{Aubin:2008wk} support the
claim that mixed-action \chpt\ is indeed the low-energy effective theory
of the domain-wall valence, staggered sea lattice theory.  Furthermore,
mixed-action \chpt\ describes the dominant unitarity-violating
effects in the mixed-action theory even when such effects are larger
than the continuum full QCD contributions that one wishes to extract.
Thus mixed-action \chpt\ fits can be used to remove taste-breaking and
unitarity-violating artifacts and recover physical quantities.

\subsection{Summary}
\label{sec:spectrum_summary}

In general these and other lattice spectrum calculations confirm
that QCD does predict the hadron spectrum.   However, although we
can see the effects of decay thresholds as the quark mass is varied
({\it e.g.,} Sec.~\ref{sec:Scalar_mesons}), and though some scattering
lengths can be indirectly determined through chiral perturbation theory
\cite{Leutwyler:2006qq}, most hadronic decay rates and cross sections
remain to be calculated in the future.

%% file: RMP_sec6.tex
% File for section 6 for RMP article
%
%\section{Section 6}
\section{Results for the light pseudoscalar mesons}
\label{sec:fpi}

\subsection{Motivation}
\label{sec:motivation-light-pseudoscalars}

Precise
computations are possible for
light
pseudoscalar mesons
(see Sec.~\ref{sec:hadron_mass_results}),
and they lead to
interesting
physics. 
If lattice
calculations of light pseudoscalar mesons and decay constants can
approach the chiral and continuum limits, 
we can determine the up, down and strange
quark masses and many of the low energy constants (LECs) of the chiral
Lagrangian, including several combinations of the NLO Gasser-Leutwyler
constants $L_i$ \cite{Gasser:1983yg}.  From the ratio
$f_K / f_\pi$,
we can extract $|V_{us}|$ from the kaon leptonic branching fraction, 
providing a test of CKM matrix unitarity 
for the first row of the matrix.

\subsection{From correlators to lattice masses and decay constants}

Study of the light pseudoscalar mesons 
on MILC lattices began in 2004
\cite{Aubin:2004fs} and has included several updates at the annual Lattice
conferences \cite{Bernard:2005ei,Bernard:2006wx,Bernard:2007ps}.  
We first review the methodology 
of \textcite{Aubin:2004fs}.
In the Goldstone (taste
pseudoscalar) case,
we can use the PCAC relation to relate the decay constant
$f_{PS}$ to matrix elements of the spin- and taste-pseudoscalar operator
${\cal O}_P (t) =\bar \psi (\gamma_5\otimes\xi_5) \psi$ between the vacuum and the meson.
In terms of the one-component staggered quark formalism, 
\begin{equation} {\cal O}_P (t) = \bar\chi^a(\vec x,t) (-1)^{\vec x + t} \chi^a(\vec x,t) \ ,
\label{eq:pion_point_operator}
\end{equation}
where $a$ is the (summed) color index.
As in
Eqs.~(\ref{eq:corr_generic},\ref{eq:corr_complete}), we define a
correlator by
\begin{equation} C_{PP}(t) = \frac{1}{V_s}\sum_{\vec y} \langle 
{\cal O}_P(\vec y,t) {\cal O}^\dagger_P(\vec x,0) \rangle
= c_{PP} e^{-m_{PS} t} + \ldots \ \ ,
\label{eq:pion_point_correlator}
\end{equation}
where $m_{PS}$ is the mass of the (lightest) pseudoscalar and $V_s$
is the spatial volume.  After fitting the correlator to this form,
we can find the decay constant from
\begin{equation}\label{eq:fpi_eq}
 f_{PS} = (m_x+m_y) \sqrt{\frac{V_s c_{PP}}{4 m_{PS}^3} } \ ,
\end{equation}
where $m_x$ and $m_y$ are the two valence quark masses.

Although the decay constant is found from the overlap of the point-source
operator with the meson state, 
most directly obtained
from the point-point correlator Eq.~(\ref{eq:pion_point_correlator}),
it is useful to
use the Coulomb wall source
Eq.(\ref{eq:coulomb-wall-operator1}) and point sink to calculate the
correlator
\begin{equation}  
C_{WP}=\langle {\cal O}_P(\vec x,t) {\cal O}^\dagger_W(0) \rangle  = c_{WP}
e^{-m_{PS} t}  + \ldots \ \ .
\label{eq:coulomb-wall-correlator}
\end{equation}
The advantage of
this correlator
is that it
has less contamination from excited states than does $C_{PP}$, and
helps in fixing the pseudoscalar mass.

A random-wall source can also be used instead
of a point source to calculate $C_{PP}$, giving smaller
statistical errors.  The source for the quark on each site of a time slice
is a three component
complex unit vector
with a random direction in color space.
Thus, contributions where the quark and antiquark in a meson
originate on different spatial sites average to zero.
After dividing by the spatial lattice volume, this source is used instead
of ${\cal O}^\dagger_P$ in $C_{PP}$.  The preferred method is then to fit
$C_{WP}$ and
the random-wall point-sink
$C_{PP}$ 
with three free parameters $A_{PP}$, $A_{WP}$ and $m_{PS}$: 
\begin{eqnarray} \label{eq:FITFORM_EQ}
     C_{PP} &=& m_{PS}^3 \, A_{PP} e^{-m_{PS} t} \, , \nonumber\\
     C_{WP} &=& m_{PS}^3 \, A_{WP} e^{-m_{PS} t} \, , \end{eqnarray}
so that $A_{PP}$ is the desired combination $c_{PP}/m_{PS}^3$ that
appears in Eq.~(\ref{eq:fpi_eq}).  An appropriate range of Euclidean
time must be selected to get a good confidence level of the fit.
If the minimum distance from 
the source point is too small, there will be excited state contamination.
It is essential to
use the full correlation matrix 
of the data
to get a meaningful confidence level and 
avoid contamination.

For chiral fits
used
to extract LECs that govern
the mass-dependence of physical quantities, it is important to fix
the scale in a mass-independent manner.   This is because all mass
dependence should be explicit in \chpt, and none should be hidden in
the scale-fixing scheme.  As described in Sec.~\ref{sec:determine_a}, 
a mass independent method is used to determine $a$ in which
$r_1/a$ is extrapolated to the physical, rather than simulated, quark
masses on the given ensemble.

Partial quenching 
is very useful 
in order to obtain
enough data to perform the required chiral fits.
For the valence masses on 
a typical ensemble,
nine different masses 
from $0.1m'_s$ to $m'_s$ ($m'_s$ is the simulated strange sea mass)
may be used.
This yields 45 distinct
pairs of valence masses, and 
hence 90 values (meson masses and decay constants) for the chiral fit.
 Without partial
quenching, we would have only 
four
values.  Of course, 
the correlations among the 90 values must be taken 
into account.

Finite volume corrections are included in the one-loop
\rschpt\ forms used to fit the lattice data.
Since the
spatial box sizes are at least 2.4 fm,
and for the smallest light sea-quark masses they are increased to
about 2.9 fm or larger,
these corrections are always less than 1.5\%.
Smaller, additional corrections representing ``residual'' effects from
higher-loop contributions are applied at the end of the calculation, as 
described below.  
The results cannot be fit without the
one-loop finite volume corrections, nor can they be fit with continuum
\chpt.  In \textcite{Aubin:2004fs}, five coarse and two fine ensembles
were fit with continuum \chpt; however, the confidence level of the fit
was $10^{-250}$!

In the remainder of this section, we present methods and results from
\textcite{Bernard:2007ps}.  A final version of the analysis, using added ensembles
and two-loop chiral logarithms \cite{Bijnens:2004hk,Bijnens:2005ae,Bijnens:2006jv},
is in progress.

The fitting is done in two stages.  In the first stage, the leading
order (LO) and next-to-leading order (NLO) low energy constants (LECs)
are determined by fitting a restricted set of data that is closer to
the 
chiral and continuum
limits than the additional points included 
later.
Specifically, the largest
lattice spacing ($a\approx 0.15$ fm) is omitted and the valence quark
masses 
are required to obey
$am_x +am_y \ltwid 0.39 \ am_s $ (for $a\approx 0.12$ fm),
$am_x +am_y \ltwid 0.51 \ am_s $ (for $a\approx 0.09$ fm), and
$am_x +am_y \ltwid 0.56 \ am_s $ (for $a\approx 0.06$ fm).
Further, for $a\approx 0.12$ fm three 
higher-mass combinations of sea-quark masses
are omitted.  Despite the restrictions,
it is found 
that due to the high precision of
the data it is 
necessary to add NNLO analytic terms in order to get good fits.
In the second stage,
the range of valence and
sea-quark masses is extended to include the region around the 
strange quark mass.
The LO and NLO low energy constants 
are constrained to be within the range
determined by the first stage of fitting.  In this stage, 
NNNLO analytic terms 
are needed
to get good fits.

\begin{figure}[t]
\begin{center}
\includegraphics[width=0.5\textwidth]{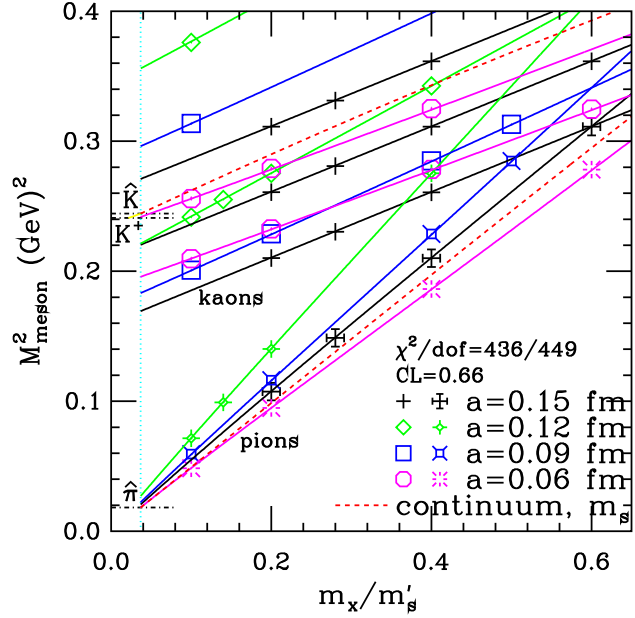}
\end{center}
\caption{NNNLO fit to partially-quenched squared meson masses.
Only the lightest sea-quark ensemble for each
lattice spacing is shown.  The data fit includes the results
for decay constants and is reflected in the number of
degrees of freedom. Figure from \textcite{Bernard:2007ps}.
\label{fig:piandkmasses}}
\end{figure}

In \figref{piandkmasses}, we show the squared meson masses in units of
$(\GeV)^2$.  
For the ``pions'' $m_x=m_y$.  For the ``kaons''
a few
fixed 
values of $m_y$ are picked for illustration,
and
$m_x$ is varied.
The horizontal axis is $m_x/m_s'$.
Only a small fraction of the points used in the fit
are shown.  For each lattice spacing,
the plot contains only the
lightest sea-quark mass ensemble, and no decay constant data is plotted.
For this fit,  $\chi^2=436$
with 449 degrees of freedom, corresponding to a confidence
level of 0.66.  The dashed red line shows the continuum prediction after
all lattice spacing dependence in the fit parameters is extrapolated
away, the strange sea-quark mass is fixed to its physical value and
the light valence and sea masses are set equal.  The physical values of
$m_s$ and $\hat m = (m_u+m_d)/2$ are required to simultaneously yield
the kaon and pion masses denoted $\hat K$ and $\hat\pi$ in the figure.
These masses correspond to what the kaon and pion masses would be with
isospin and electromagnetic effects removed.  Some phenomenological
input is needed to account for the electromagnetic effects.  This is
explained in detail in \textcite{Aubin:2004fs}.  The vertical dotted
line is drawn at $\hat m/m_s$.

The ``residual'' finite volume corrections are then applied.
\textcite*{Colangelo:2005gd} have
shown that higher 
than one-loop
\chpt\ corrections can be significant in the
current range of quark masses and volumes.  For $a\approx0.12$ fm with
sea masses $am_l / am_s' = 0.01/0.05 $,
there is a direct test of finite
volume effects on $20^3$ and $28^3$ volumes that correspond to 2.4 and
3.4 fm box sides.
\textcite{Bernard:2007ps}
detail the 
comparison between these calculations and the one-loop result.
On this basis,
a small correction 
is applied to the continuum prediction.
This amounts to 0.25\% for $f_\pi$, 0.05\% for $f_K$, $-$0.15\% for
$m_\pi^2$, and $-$0.10\% for $m_K^2$.  These values are also added to
the systematic error.

By extending the kaon extrapolation line in \figref{piandkmasses}, 
one finds the value of $m_u$ that corresponds to the $K^+$ mass (see
\textcite{Aubin:2004fs}).  
Two important
mass ratios are determined:
\begin{equation}
m_s/\hat m = 27.2(1)(3)(0) \;,\qquad m_u/m_d = 0.42(0)(1)(4)\;.
\label{eq:quark-mass-ratios}
\end{equation}
The errors are statistical, lattice-systematic, and electromagnetic (from
continuum estimates).  
Note that the $m_u=0$ solution to the strong CP problem
is ruled out at the 10 $\sigma$ level.

\begin{figure}[t]
\begin{center}
\begin{tabular}{c c}
\includegraphics[width=0.5\textwidth]{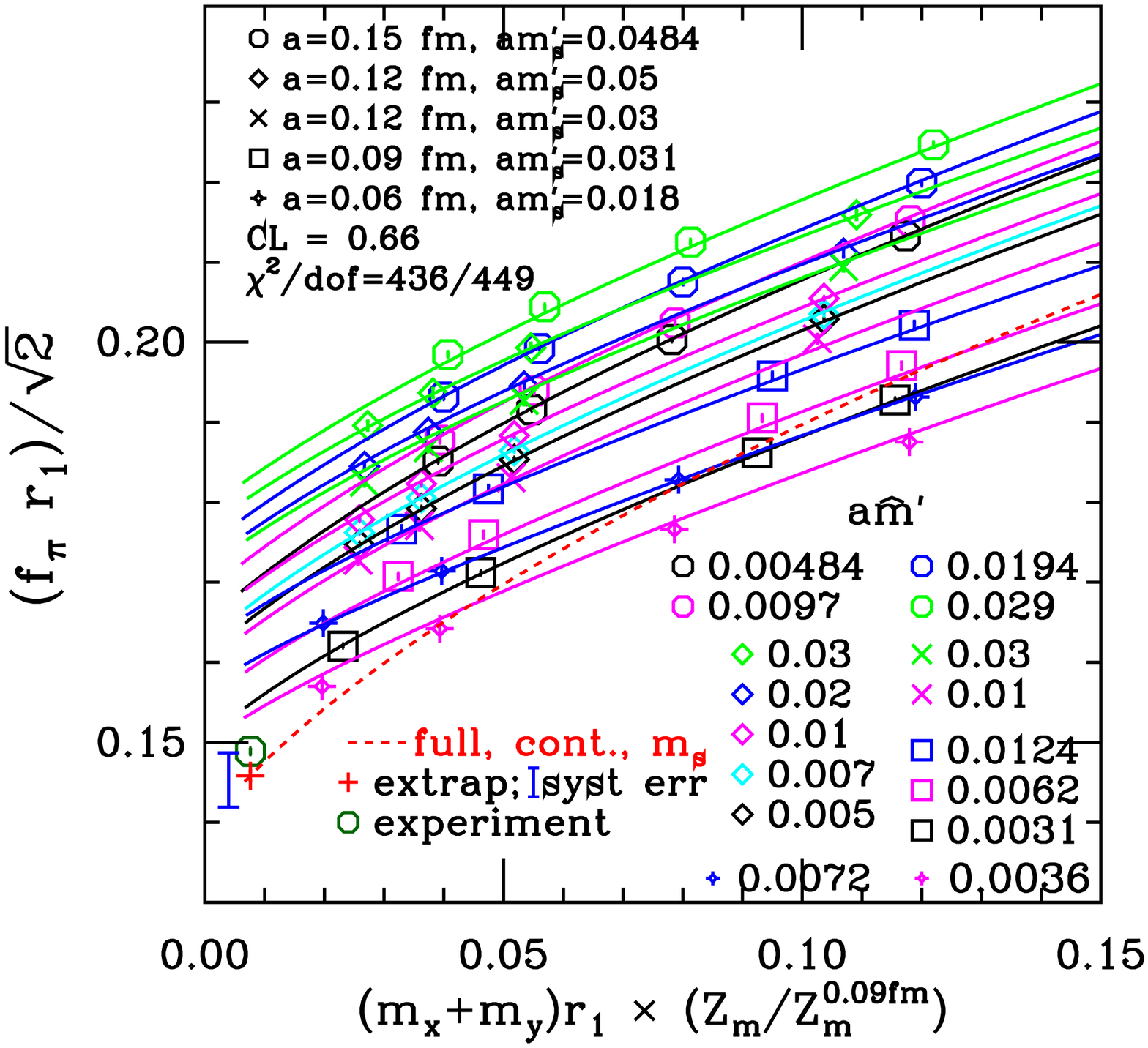}
&
\includegraphics[width=0.5\textwidth]{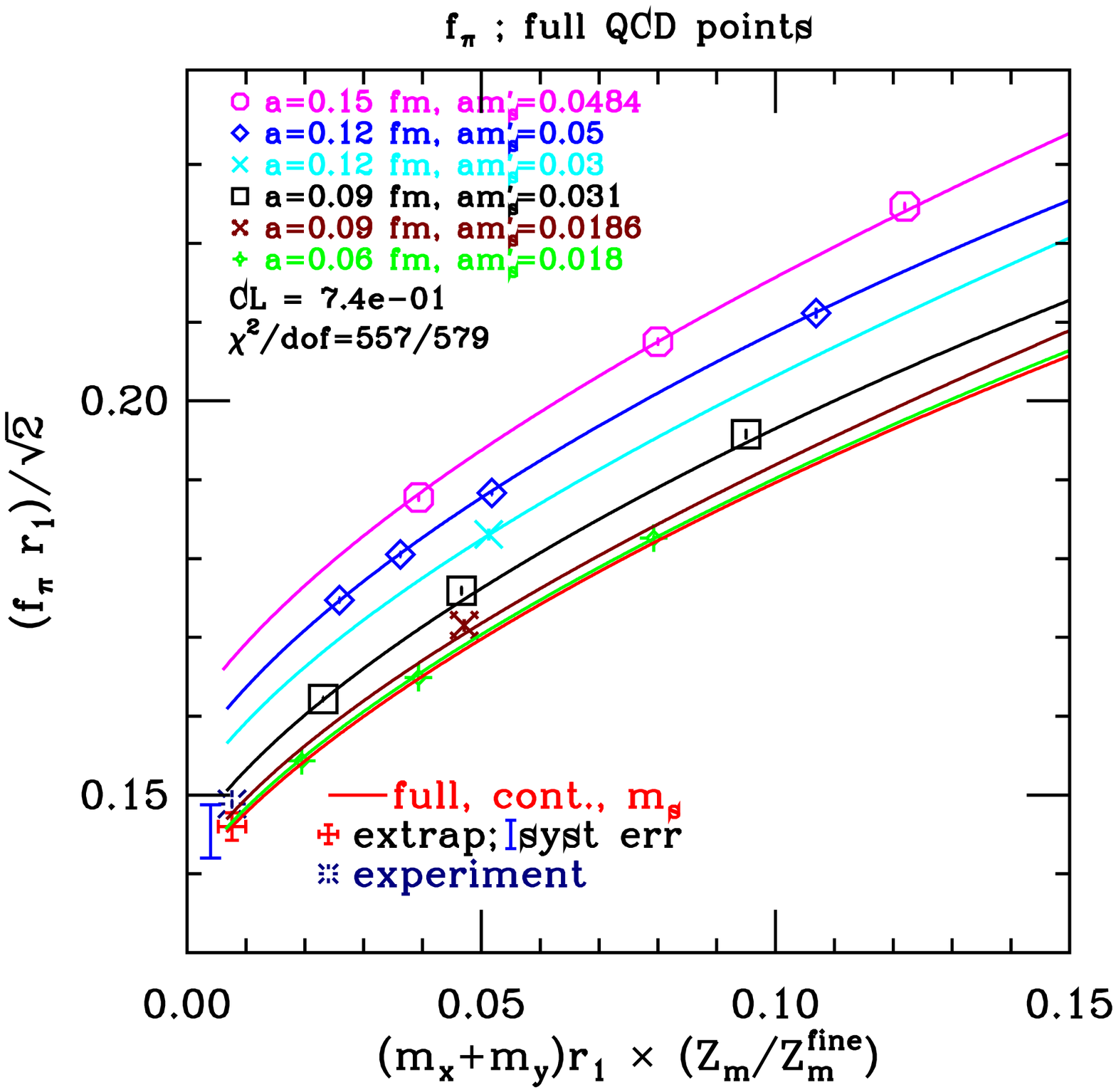}
\end{tabular}
\end{center}
\caption{The meson decay constants are plotted along with the NNNLO fit
that was shown 
for the masses in \figref{piandkmasses}.
The left plot shows partially quenched data from more ensembles than
in \figref{piandkmasses}, but still only a fraction of the data fit. 
On the right, still more ensembles are included, but only full QCD data points are plotted.
Both figures are from \textcite{Bernard:2007ps}.
\label{fig:decay-constants}}
\end{figure}

Having 
found
the continuum fit parameters and the 
physical
quark masses, 
the decay constants are predicted.
Figure~\ref{fig:decay-constants} (left) shows (some of) the decay
constant data 
and the fit through the displayed data.
For 
the continuum prediction 
(dashed red line),
the strange sea-quark mass is set to its physical value and the light valence
and sea masses are set equal.  The left end of the curve corresponds
to $m_x=m_y=\hat m$.  The vertical error bar to the left of the $+$
shows the systematic error.  The experimental result is shown as
an octagon.  It comes from the decay $\pi^+ \rightarrow \mu^+\nu_\mu$
with the assumption that $|V_{ud}|=0.97377(27)$ \cite{Amsler:2008zzb}.
Figure~\ref{fig:decay-constants} (right) shows 
the full QCD points from a slightly different
fit with data from additional ensembles.  
Note that the data points at $a \approx 0.06$
fm are quite close to the full QCD continuum extrapolated curve.

Up to this point,
the lattice spacing 
is set by calculation of the
heavy-quark potential parameter $r_1$, which
yields relative lattice
spacings between ensembles, and the continuum extrapolation of $\Upsilon$
splittings determined by the HPQCD collaboration \cite{Gray:2005ur},
which gives 
an absolute scale.  These results yield a value
$r_1=0.318(7)\;$fm.  On this basis,
\begin{eqnarray}\label{eq:decay-constant-results}
f_\pi & = &  128.3 \pm 0.5\; {}^{+2.4}_{-3.5} \; \MeV \, , \nonumber \\
f_K & = &  154.3 \pm 0.4\; {}^{+2.1}_{-3.4}  \; \MeV \, , \nonumber \\
f_K/f_\pi  & = &  1.202(3)({}^{+\phantom{1}8}_{-14})\ ,
\end{eqnarray}
where the errors are from statistics and lattice systematics.
This value for $f_\pi$ is consistent with the experimental result,
$f_\pi^{\rm expt} = 130.7\pm 0.1\pm 0.36 \; \MeV$ \cite{Amsler:2008zzb}.

An alternative approach is to set the scale from  $f_\pi$ itself.  In 
this case, there are small changes in the quark masses and
\begin{equation}
\label{eq:r1-hpqcd}
 r_1 =0.3108(15)({}^{+26}_{-79})\; {\rm fm}\ ,
\end{equation}
which is 1-$\sigma$ lower (and with somewhat smaller errors) than the
value from the $\Upsilon$ system.  For the decay constants,
\begin{eqnarray}
\label{eq:fK-with-scale-from-fpi}
f_K  =    156.5 \pm 0.4\; {}^{+1.0}_{-2.7}  \; \MeV \, , \nonumber \\
f_K/f_\pi   = 1.197(3)({}^{+\phantom{1}6}_{-13}) \;, 
\end{eqnarray}
where the errors are statistical and systematic.

\textcite{Marciano:2004uf}
noted that the lattice value 
of $f_K/f_\pi $ can be combined with
measurements of the
kaon branching fraction \cite{Ambrosino:2005ec,Ambrosino:2005fw} 
to obtain $|V_{us}|$.
{}From Eq.~(\ref{eq:fK-with-scale-from-fpi}),
\begin{equation}
\label{eq:Vus-from-fpi}
|V_{us}|=0.2246({}^{+25}_{-13}) \;,
\end{equation}
which is consistent with (and competitive with) the world-average value
$|V_{us}|=0.2255(19)$ \cite{Amsler:2008zzb} coming from semileptonic
$K$-decay coupled with non-lattice theory.

Using the two-loop perturbative calculation of 
the mass renormalization constant $Z_m$ \cite{Mason:2005bj}%
\footnote{With this two-loop $Z_m$-factor a tadpole improved definition
of the bare quark mass should be used, in which what we have denoted
by $am_q$ throughout this review should be replaced by $u_0 am_q$.
The tadpole factors for the various MILC ensembles are listed in
Table~\ref{table:runtable1}},
absolute quark masses can be found:
\begin{eqnarray}
\label{eq:light-quark-masses}
m_s = 88(0)(3)(4)(0)\;\MeV \;, &\quad\qquad&  \hat m = 3.2(0)(1)(2)(0)\;\MeV
 \;, \nonumber \\
m_u = 1.9(0)(1)(1)(1)\;\MeV \;, &\quad\qquad&  m_d = 4.6(0)(2)(2)(1)\;\MeV \;.
\end{eqnarray}
The errors are statistical, lattice-systematic, perturbative, and
electromagnetic (from continuum estimates).
Nonperturbative computations of $Z_m$ are in progress.

The chiral fits also determine various Gasser-Leutwyler low energy
constants and chiral condensates:
\begin{eqnarray}
2L_6 - L_4 = 0.4(1)({}^{+2}_{-3}) \;, &\quad\qquad& 2L_8 - L_5 = -0.1(1)(1)
 \;, \nonumber \\
L_4 = 0.4(3)({}^{+3}_{-1}) \;, &\quad\qquad& L_5 =2.2(2)({}^{+2}_{-1})
 \;, \nonumber \\
L_6 = 0.4(2)({}^{+2}_{-1}) \;, &\quad\qquad& L_8 =1.0(1)(1) \;,  \nonumber \\
   f_\pi/f_{2} =  1.052(2)({}^{+6}_{-3}) \;, &\quad\qquad&
  \langle\bar uu\rangle_{2} =   -(\, 278(1)({}^{+2}_{-3})(5)\;\MeV\,)^3
 \;, \nonumber \\
 f_\pi/f_3 = 1.21(5)({}^{+13}_{-\phantom{1}3}) \;, &\quad\qquad&
  \langle\bar uu\rangle_{3}  =  -(\, 242(9)({}^{+\phantom{2}5}_{-17})(4)\;\MeV\,)^3
 \;, \nonumber \\
f_{2}/f_{3}  =  1.15(5)({}^{+13}_{-\phantom{1}3})  \;, &\quad\qquad& \langle\bar uu\rangle_{2}/\langle\bar uu\rangle_{3}  =  1.52(17)({}^{+38}_{-15}) \;.
\label{eq:low-energy-constants}
\end{eqnarray}
The errors are statistical, lattice-systematic and perturbative for the
condensates.
$f_2$ ($f_3$)
denotes 
the three-flavor decay constant
in the two (three) flavor chiral limit, and $\langle\bar uu\rangle_{2}$
($\langle\bar uu\rangle_{3}$) is the corresponding condensate.  The low
energy constants $L_i$ are in units of $10^{-3}$ and are evaluated at
chiral scale $m_\eta$; the condensates and masses are in the $\msbar$
scheme at scale $2\,\GeV$.  
The indications are that the $L_i$ will
change significantly when the two-loop logarithms are included, just as they
do in phenomenological estimates \cite{Bijnens:2007yd}.  Other results are very stable,
however.

The \rschpt\ formalism relies on the replica
proceedure, and taking the fourth
root corresponds to $n_r=1/4$ where $n_r$ is the number of replicas.
The fact that
there are
good fits with the \rschpt\ formulae, but not with
continuum \chpt, is a 
test of staggered chiral perturbation theory.
A further test of \rschpt\ is to allow $n_r$ to be a free parameter in
the fits.
For the low mass data,
$n_r= 0.28(2)(3)$
where the first error is statistical and the second systematic coming from
varying the details of the chiral fits.  We are encouraged by this strong
constraint on $n_r$, and the success of \rschpt\ in describing 
the MILC data.

\subsection{Other computations of $f_\pi$ and $f_K$}
\label{sec:other_fpi}

Since the MILC collaboration's initial calculation of the light
pseudoscalar meson masses, decay constants, and quark masses using the
$a\approx 0.12$
and
$0.09$ fm lattices~\cite{Aubin:2004fs},
several other groups have also computed $f_\pi$ and $f_K$ on the
MILC ensembles using different valence quark formulations.   All of
the results are consistent with those of the MILC collaboration,
Eq.~(\ref{eq:decay-constant-results}), and with each other.

The HPQCD collaboration uses HISQ staggered valence quarks and the
MILC asqtad staggered sea quark ensembles with lattice spacings
$a \approx 0.15$, 0.12 and 0.09 fm
fm~\cite{Follana:2007uv}.  
They generate one ``pion" point and one ``kaon" point per ensemble,
matching the masses of the Goldstone HISQ pion to the asqtad one, and the
mass of the HISQ $s\bar s$ meson to 696 MeV, the \chpt\ value.
Although \textcite{Follana:2007uv} are performing a mixed action lattice
simulation, they extrapolate to the physical light quark masses and
the continuum using continuum NLO \chpt\ augmented by analytic terms
constrained with Bayesian priors.  Terms proportional to $\alpha_s
a^2$ and $a^4$ are included to test for conventional discretization
errors, while those proportional to $\alpha_s^3 a^2$, $\alpha_s^3 a^2
\textrm{log}(m_q)$, and $\alpha_s^3 a^2 m_q$ are intended to test for
residual taste-changing interactions with the HISQ valence quarks.
HPQCD obtains the following results for $f_\pi$, $f_K$, and the ratio:
\begin{equation}
	f_\pi = 132(2) \textrm{MeV}, \quad f_K = 157(2)  \textrm{MeV},
	\quad f_K/f_\pi = 1.189(7) ,
\end{equation}
where the largest source of error is the uncertainty in the scale $r_1$
(1.4\% for $f_\pi$ and 1.1\% for $f_K$).

The NPLQCD collaboration uses domain-wall valence quarks and
four $a \approx 0.12$ fm ensembles with $m_l / m_s' =$ 0.14 --
0.6~\cite{Beane:2006kx}.  
They tune to match the valence pion and kaon to the corresponding
asqtad particles. Due to the mixed
action, there are still unitarity-violating artifacts 
that vanish only in the limit $a \to 0$. They
compute only the ratio $f_K/f_\pi$, which has a milder dependence upon the
quark mass than the individual decay constants,
and
extrapolate to the
physical light quark masses using the NLO continuum \chpt\ expression,
which depends only on one free parameter, $L_5$.  
The result is
\begin{eqnarray}
	f_K/f_\pi = 1.218 \pm 0.002^{+0.011}_{-0.024} \;, \nonumber \\
	L_5(m_\eta) = 2.22 \pm 0.02^{+0.18}_{-0.54} \times 10^{-3} \;,
\end{eqnarray}
where the first error is statistical and the second error is the sum of
systematic errors added in quadrature.  The dominant source of uncertainty
is from the truncation of the \chpt\ expression ($^{+0.011}_ {-0.022}$
for the ratio), which they estimate by varying the fit function through
the addition of NNLO analytic terms and double logarithms.  Although they
do not include an error due to their use of only a single lattice spacing,
this is likely a small effect in the ratio $f_K/f_\pi$.

\textcite{Aubin:2008ie} also use domain-wall valence quarks.  In contrast
with NPLQCD, however, they compute many partially quenched points on the
$a\approx 0.12$
and 0.09 fm ensembles, and use NLO mixed
action \chpt\ with higher-order analytic terms to extrapolate to physical
quark masses and the continuum~\cite{Bar:2005tu}.  Their preliminary
results for the light pseudoscalar meson decay constants are
\begin{equation}
	f_\pi = 129.1(1.9)(4.0) \textrm{MeV}, \quad f_K = 153.9(1.7)(4.4)
	\textrm{MeV}, \quad  f_K/ f_\pi = 1.191(16)(17),
\end{equation}
where the first error is statistical and the second is the sum of
systematic errors added in quadrature.  The dominant source of error
is from the chiral extrapolation procedure (2.2\% for $f_\pi$ and 2.3\%
for $f_K$), and is estimated by varying the analytic terms included in
the fit function.

\begin{figure}
\begin{center}
\includegraphics[width=3.5in]{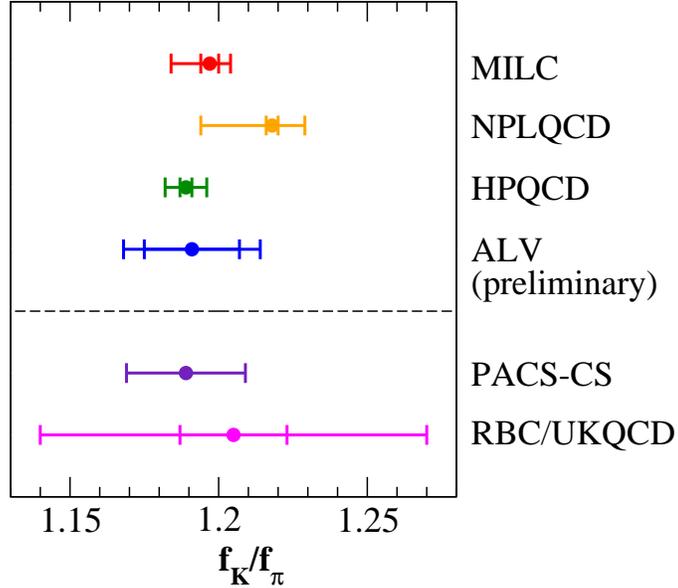}
\end{center}
\caption{The ratio of light decay constants $f_K/f_\pi$ from six calculations.
The top four use MILC asqtad configurations and the lower two use other types of
quarks.  Details and references can be found in the text.}
\label{fig:fK_fpi}
\end{figure}

In Fig.~\ref{fig:fK_fpi}, we compare results for $f_K/f_\pi$ from a
variety of 2+1 flavor calculations.  The top four results all use MILC
aqstad configurations and were discussed above.  The two lower results
from the PACS-CS Collaboration \cite{Aoki:2008sm} and the RBC/UKQCD
Collaboration \cite{Allton:2008pn} use clover quarks with the Iwasaki
gauge action and domain-wall quarks, respectively.

%% file: RMP_sec7.tex
% File for section 7 for RMP article
%
%\section{Section 7}
\section{Heavy-light mesons: masses and decay constants}
\label{sec:h_l_mass}

Calculations of $B$- and $D$-meson masses and decay constants using
the 2+1 flavor MILC configurations have been performed
in joint work by the Fermilab Lattice and MILC collaborations, and by
the HPQCD collaboration.
Of numerical quantities involving heavy $b$ and $c$ quarks, meson masses
and decay constants are among the simplest quantities to compute
numerically and are often well measured experimentally. Thus
they provide valuable cross-checks
of lattice QCD methods.  In particular, once the treatment of the light
sea and valence quarks has been validated within the light pseudoscalar
sector, calculations of heavy-light meson masses and decay constants
allow tests of the various lattice QCD formalisms used for heavy quarks.
In this section, we describe the 2+1 flavor calculations by Fermilab/MILC
and HPQCD of  heavy-light meson masses and decay constants, and show that,
with one exception, they are consistent with experiment.  These results
give confidence in other lattice QCD calculations involving $b$ and
$c$ quarks, such as those of semileptonic form factors described in
Sec.~\ref{sec:semilept}.

\subsection{Heavy quarks on the lattice}
\label{sec:heavy_q}

Heavy quarks, {\it i.e.}, those for which the quark mass in lattice
units $am$ is large, present special challenges.  As long as $am\ll 1$,
heavy quarks on the lattice can be treated with
light-quark formalisms such as staggered fermions.  At the lattice spacings
in common use, we have $am_c\sim$ 0.5--1.0 and $am_b\sim$ 2--3.
For charm quarks, light-quark methods can only be used if they are highly
improved to remove discretization errors.  Bottom quarks 
still require
special heavy-quark methods.

\subsubsection{Nonrelativistic QCD}

A straightforward way of formulating heavy quarks on the lattice is to
rewrite the Dirac-like light-quark action as a sum in a nonrelativistic
operator expansion, as is done in HQET \cite{Isgur:HQET,Neubert:1993mb}
and in nonrelativistic expansions in QED \cite{Caswell:1985ui,Lepage:1992tx}:
\begin{equation}
S_{NRQCD} = \sum_{x}\, \phi^\dagger(x)\,\left(-\nabla_0^{(+)} 
+ \frac{1}{2m}\sum_i\Delta_i +\frac{1}{2m}\sigma\cdot B(x) 
+  \frac{1}{8m^3}(\sum_i\Delta_i)^2 +\ldots \right)\,\phi(x),
\label{eq:NRQCD}
\end{equation}
where 
\begin{equation}
\nabla_\mu^{(+)}\psi(x)\equiv 
\frac{1}{a}\left(U_\mu(x)\psi(x+a\hat{\mu})-\psi(x)\right),
\end{equation}
and where the $\phi$ are two-component fermions representing the quarks.
An analogous term in the action governs the antiquarks.  The leading
heavy-quark mass dependence is absorbed into the fermion field and
vanishes from explicit calculations.  For $b$ quarks in particles
with a single heavy quark, the first term in this action yields the
static approximation \cite{Eichten:1989zv}.  In heavy-light systems,
the importance of operators in this expansion is ordered according to
HQET power counting ($\lambda\sim\Lambda/m_Q$).
In quarkonium systems, operators are ordered by heavy-quark velocity.

\subsubsection{Wilson fermions with the Fermilab interpretation}

In NRQCD, the kinetic energy operator of the Dirac action, 
$\overline{\psi}(x)\sum_i \gamma_i \nabla_i \psi(x) $
is replaced by the leading kinetic energy operator
$
\phi^\dagger(x)\,\frac{1}{2m}\sum_i\Delta_i\,\phi(x)
$
plus a series of higher dimension operators.  The action for Wilson
fermions contains the leading kinetic energy operators of both the Dirac
and the nonrelativistic actions, as in Eq.~(\ref{eq:S_W}):
\begin{equation}
S_W  = \sum_{x}\, \overline{\psi}(x)\,\left(\sum_\mu\gamma_\mu\nabla_\mu-
\frac{ar}{2}\sum_\mu\Delta_\mu + m\right)\,\psi(x).
\end{equation}
The effects of the Laplacian term, which eliminates the doubler states,
vanish in the limit $am\rightarrow 0$.  As $am$ becomes larger,
the importance of the Laplacian term grows.  When $am \gg 1$, the
Laplacian term dominates the Dirac-like kinetic energy term, and the
theory behaves like a type of nonrelativistic theory in which the rest
mass $m_1\equiv E(p^2=0)$ does not equal the kinetic mass $m_2\equiv
1/(2 \partial E/\partial p^2)$.  (Note that we use lower-case $m$ to
refer to quarks and capital $M$ to refer to mesons in this section.)
As $am\rightarrow 0$, the two masses converge to the bare quark mass $m$.
For heavy quarks the kinetic mass controls the physics, and the rest mass
may be absorbed into a field redefinition.  This means that the Wilson
action and related actions can be used as actions for heavy quarks as
long as $m_2$, with contributions from both terms in the kinetic energy,
is adjusted to equal the desired physical mass \cite{ElKhadra:1996mp}.
It is possible to set $m_1=m_2$ by breaking time-space axis-interchange
symmetry in the Lagrangian.  If this is not done, $m_1$ and $m_2$ have
the tree-level form
\begin{equation}
am_1=\log(1+am_0)
\end{equation}
and
\begin{equation}
\frac{1}{am_2}=\frac{2}{am_0(2+am_0)} + \frac{1}{1+am_0} ~.
\end{equation}

The action of the nonrelativistic  expansion can be viewed as arising from
a field transformation of the Dirac field, the Foldy-Wouthuysen-Tani (FWT)
transformation.  The Wilson action, with both types of kinetic energy
operators, can be viewed as arising from  a partial FWT transformation.
Like the action of NRQCD, it produces the same physics as the Dirac
action as long as a series of correction operators is added to sufficient
precision \cite{Oktay:2008ex}.  The leading dimension-five correction
operator has the same form for heavy Wilson fermions as for light
clover/Wilson fermions [Eq.~(\ref{eq:S_SW})], $S_{SW} = \frac{iag}{4}
c_{SW} \sum_x \bar\psi(x) \sigma_{\mu\nu} \mathcal{F}_{\mu\nu}(x)
\psi(x)$. All simulations to date using this approach to heavy quarks 
have therefore used clover/Wilson fermions.  A systematic improvement
program is possible as outlined in Sec.~\ref{sec:imp_FNAL}.

\subsubsection{The HISQ action}

Because $0.5 \ltwid am_c \ltwid 1$ at currently accessible lattice spacings,
it is possible to use ordinary light-quark actions to treat the charm
quark.  However, to obtain high precision it is necessary to correct
the action to a high order in $am$.  This approach is followed with
``highly improved staggered quarks'' \cite{Follana:2006rc}, as explained
in Sec.~\ref{sec:HISQ_ferm}.

\subsection{Lattice calculations of masses and decay constants}

As in the light pseudoscalar meson case, the heavy-light decay constant
is proportional to the matrix element of the axial current:
	\begin{equation}
	\langle 0 | A_\mu | H_q(p) \rangle = i f_{H_q} p_\mu ,
	\end{equation}
where $A_\mu  = \bar{q} \gamma_\mu \gamma_5 Q$.
Because of the heavy-quark normalization in HQET, it is often useful to
consider the combination decay amplitude
	\begin{equation}
        	\phi_{H_q} = f_{H_q} \sqrt{M_{H_q}} ,
	\end{equation}
which is computed from the correlators
	\begin{equation}
	C_0(t) = \langle O_{H_q}(t) O^\dagger_{H_q} (0) \rangle , \quad C_{A_4}(t) = \langle A_4(t) O^\dagger_{H_q}(0) \rangle .
 	\end{equation}
For the case of Fermilab heavy quarks or NRQCD $b$ quarks, the
heavy-light meson mass is obtained from the kinetic mass ($M_2$) in the
dispersion relation, whereas for HISQ charm quarks $M_1=M_2$, so both
are simultaneously set to the $D$- or $D_s$-meson mass.

The Fermilab Lattice and MILC collaborations' calculation of heavy-light
meson decay constants~\cite{Aubin:2005ar,Mackenzie:Lat08} employs
the Fermilab action for the heavy $b$ and $c$ quarks and the asqtad
staggered action for the light $u$, $d$, and $s$ quarks.  They construct
the heavy-light meson interpolating operator and axial vector current
$A_\mu$ using the method for combining four-component Wilson quarks with
1-component staggered quarks described in \textcite{Wingate:2003nn}.
Their most recent determination from Lattice 2008~\cite{Mackenzie:Lat08}
uses data on the medium-coarse, coarse, and fine lattices, with 8--12
partially quenched valence masses per ensemble.  The clover coefficient
$c_{SW}$ and hopping parameter $\kappa$ in the Fermilab action are
tuned to remove errors of $\cO(1/m_Q)$ in the heavy-quark action.
In particular, they set $c_{SW} = u_0^{-3}$, the value given by tree-level
tadpole-improved perturbation theory~\cite{Lepage:1992xa}.  They choose
the charm quark hopping parameter $\kappa_c$ so that the spin-averaged
(kinetic) $D_s$-meson mass is equal to its physical value, and choose the
bottom quark hopping parameter $\kappa_b$ to reproduce the $B_s$-meson
mass in an analogous manner; this implicitly fixes the $b$ and $c$ quark
masses.  They also remove errors of $\cO(1/m_Q)$ from the heavy-light
axial vector current $A_\mu$ by rotating the heavy-quark field in the
two-point correlation function:
\begin{equation}
	\psi_b \to \Psi_b =  \left( 1 + a d_1  \vec{\gamma} \cdot
	\vec{D}_\textrm{lat} \psi_b \right) ,
\end{equation}	
where $\vec{D}_\textrm{lat}$ is the symmetric, nearest-neighbor, covariant
difference operator, and the tadpole-improved tree-level value for $d_1$
is given by~\cite{ElKhadra:1996mp}:
\begin{equation}	 
	d_1 = \frac{1}{u_0} \left( \frac{1}{2 + am_0} +
	\frac{1}{2(1 + am_0)} \right) .
\end{equation}	
They obtain the renormalization factor needed to match the lattice
heavy-light current onto the continuum using the method of
\textcite{Hashimoto:1999yp}:
\begin{equation}
	Z^{Qq}_{A_4} = \rho^{Qq}_{A_4} \sqrt{Z^{QQ}_{V_4} Z^{qq}_{V_4}} ,
\end{equation}	
where the flavor-conserving factors $Z^{QQ}_{V_4}$ and
$Z^{qq}_{V_4}$ are determined nonperturbatively and the
remaining factor is determined to 1-loop in lattice perturbation
theory~\cite{Lepage:1992xa,ElKhadra:2007qe}.

The Fermilab/MILC collaboration fits its decay constant data as a
function of light-quark sea and valence masses to the one-loop form
given by HM\schpt\ (see \secref{HQSChPT}), supplemented by analytic
NNLO terms, which are quadratic in the light valence and/or sea masses.
This is very similar to the approach taken in the light pseudoscalar
sector, as described in \secref{fpi}.  While pure NLO fits are adequate
to describe the data for very light valence mass, once this mass gets
to be roughly half the strange quark mass or higher, at least some NNLO
terms are necessary to obtain acceptable fits.

\begin{figure}
\begin{center}
\includegraphics[width=4.5in]{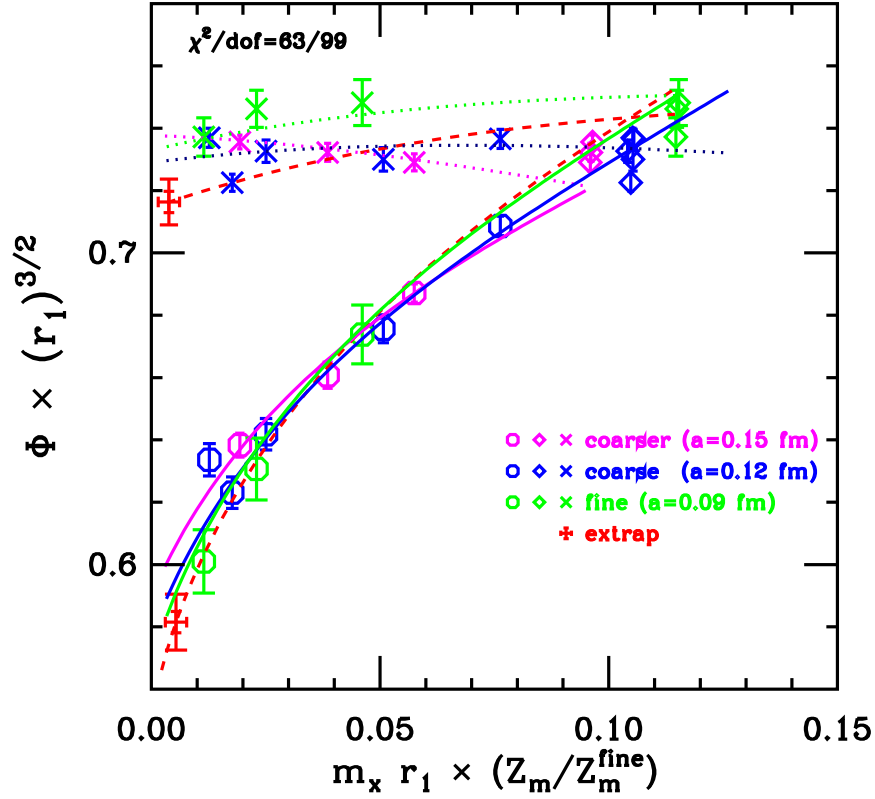}
\end{center}
\caption{Chiral extrapolation for $\Phi_D$ (octagons) and $\Phi_{D_s}$
(crosses or diamonds) by the Fermilab/MILC collaboration~\cite{Mackenzie:Lat08}.  Solid lines
are the  HM\schpt\ fit to  $\Phi_D$; dotted lines, to $\Phi_{D_s}$.   Although the full set of partially-quenched data
is included in the fit, for $\Phi_D$ the plot shows only those full
QCD points for which the light valence and sea masses are equal to $m_x$,
the mass on the abscissa.  For $\Phi_{D_s}$, only points with
the strange valence mass ($m_{sv}$) equal to the strange sea mass are shown, 
plotted either as a function of $m_l$ (crosses), or at $m_{sv}$ (diamonds).
The (red) dashed lines show the fit after removal of light-quark
discretization errors, with the fancy plus signs giving the chirally
extrapolated results with statistical errors.}
\label{fig:fD_Fermilab-MILC}
\end{figure}

\Figref{fD_Fermilab-MILC} shows the preferred  HM\schpt\ fit to data at
multiple lattice spacings for $\Phi_D$ and $\Phi_{D_s}$, which are
functions of the light valence mass, the light sea mass and the
strange sea mass.
In addition to taste-breaking discretization effects that appear as taste-splittings, taste-hairpins, and taste-violating analytic terms, there 
are ``generic'' light-quark discretization
effects, which can be thought of as changes in the physical LECs
(such as $\Phi_0$, the value of $\Phi$ in the SU(3) chiral limit) with
lattice spacing.  With the asqtad action, such effects are $\cO(\alpha_S
a^2)$. They can be (approximately) accounted for by adding additional
parameters to the HM\schpt\ fit function, with variations limited by
Bayesian priors, following \textcite{Lepage:2001ym}.  This is done in the
fit shown in \figref{fD_Fermilab-MILC}, although the effects
are quite small, and fits without the additional parameters
give almost the same results (and confidence levels), but with somewhat
smaller statistical errors.  Once the parameters of the HM\schpt\ fit are known, taste-violating
and generic discretization effects through $\cO(a^2)$ can be removed by setting $a=0$.
After taking the continuum limit, the valence and sea-quark masses are set to their physical values in order to obtain the decay constants of a $D^+$ and $D_s$ meson,
up to tiny isospin violations in the sea sector.

The HPQCD collaboration's calculation of the $B$ and
$B_s$-meson decay constants~\cite{Gamiz:2009ku} employs the NRQCD
action for the heavy $b$ quarks and the asqtad
staggered action for the light $u$, $d$, and $s$ quarks.  They use six
data points in their analysis --- four full QCD points on the coarse
ensembles and two full QCD points on the fine ensembles.  They fix
the $b$-quark mass so that the mass of a $b\bar{b}$ meson reproduces
the physical $m_\Upsilon$~\cite{Gray:2005ur}. The HPQCD computation
includes all currents of $\cO(1/m_b)$~\cite{Morningstar:1997ep}
and uses 1-loop lattice perturbation theory to match onto the
continuum~\cite{Dalgic:2003uf}.  Therefore, they include all corrections
to the heavy-light current through $\cO(\Lambda_\textrm{QCD}/m_b)$, $\cO(
\alpha_s)$, $\cO(a  \alpha_s)$, $\cO( \alpha_s/(am_b))$ and $O( \alpha_s
\Lambda_\textrm{QCD}/m_b)$.  The HPQCD collaboration uses HM\schpt\ for
the chiral extrapolations of $\Phi_B$ and $\Phi_{B_s}$ in a similar manner
to Fermilab/MILC.  They multiply the NLO expression by $[1 + c \alpha_s
a^2 +  c'a^4]$ in order to parameterize higher-order discretization
effects.  They also include an additional NNLO analytic term $\propto
(m_d - m_s)^2$ in the extrapolation of the ratio $\Phi_{B_s}/\Phi_B$.

The HPQCD collaboration's calculation of the $D$ and
$D_s$-meson decay constants~\cite{Follana:2007uv} employs the HISQ
action~\cite{Follana:2006rc} (see Sec.~\ref{sec:HISQ_ferm}) for all of
the $u$, $d$, $s$, and $c$ valence quarks.  Because they are treating
the charm quark as a light quark, the computation is similar to the
determinations of $f_\pi$ and $f_K$ described in Sec.~\ref{sec:fpi},
except for differences due to the fact that this is a mixed-action
simulation with HISQ valence quarks and asqtad sea quarks.  They use
the medium-coarse, coarse, and fine MILC lattices, and include seven
full QCD points in their analysis.  They fix the $c$-quark mass so that
the mass of the taste Goldstone $\eta_c$ meson agrees with experiment.
Because the HISQ axial current is partially-conserved, it does not need
to be renormalized.  Therefore this method avoids the use of perturbation
theory, whose truncation errors can be difficult to estimate.  The HPQCD
calculation does not use HM\schpt\ for the chiral extrapolations of
$f_D$ and $f_{D_s}$, but simply applies continuum \chpt, supplemented
by Bayesian fit parameters.
These parameters test for the expected discretization effects of the
form $\alpha_S a^2$, $a^4$, $\alpha^3_S a^2$, $\alpha^3_S a^2 \log(m_{quark})$,
and $\alpha^3_S a^2 m_{quark}$ from the asqtad action,  and the effects of
residual taste-violating interactions with HISQ valence quarks.

All of the 2+1 flavor calculations of heavy-light meson decay constants
rely upon power-counting in order to estimate the size of heavy-quark
discretization errors.  In the Fermilab method, heavy-quark discretization
errors arise due to the short-distance mismatch of higher-dimension
operators in the continuum and lattice theories.
The sizes of these mismatches are estimated using HQET as a theory of cutoff effects, as described
in \textcite{Kronfeld:2000ck} and \textcite{Harada:2001fi}.
This typically leads to errors of a few percent on the fine MILC lattices.
In simulations with NRQCD $b$ quarks, relativistic errors arise from
higher-order corrections to the NRQCD action and the heavy-light current.
Although these are not all discretization errors proportional to powers
of the lattice spacing, many are proportional to inverse powers of the
heavy-meson mass, and hence should be considered heavy-quark errors.  The
leading relativistic error comes from radiative corrections to the $\sigma
\cdot B$ term in the action, and is estimated to be of $\cO(\alpha_s
\Lambda_\textrm{QCD}/M_B) \sim 3\%$~\cite{Gamiz:2009ku}.  The HISQ action
is highly-improved, and the leading heavy-quark errors are formally of
$\cO(\alpha_s(m_c a)^2)$ and $\cO((m_c a)^4)$~\cite{Follana:2006rc}, where
$\alpha_s \sim 0.3 $ and $a m_c \sim 0.5 $ on the fine MILC lattices.
The HPQCD collaboration, however, removed errors of  $\cO(\alpha_s(m_c
a)^2)$ in the HISQ action by accounting for radiative corrections in
the coefficient of the Naik term, and also extended the traditional
Symanzik analysis to remove all $\cO((m_c a)^4)$ errors to leading
order in the charm quark's velocity.  Thus the leading charm quark
discretization errors should be of $\cO((m_c a)^4(v/c)^2) \sim 0.5\%$
or less for $D$ mesons.

\subsection{Results for masses, decay constants, and CKM matrix elements}

Although the heavy-light meson decay constants, in combination with
experimental measurements of leptonic branching fractions, can be used
to extract CKM matrix elements via the relation
\begin{equation}
\Gamma(H \to \nu \ell) = \frac{G_F^2 |V_{ab}|^2 }{8 \pi} f^2 _H m^2_\ell M_H
 \left( 1 - \frac{m^2_\ell}{ M^2_H}  \right)^2 ,
\label{eq:lept_decay}
\end{equation}
the matrix elements $|V_{cd}|$, $|V_{cs}|$, and $|V_{ub}|$ can be obtained
to better accuracy from other quantities such as neutrino scattering and
semileptonic decays~\cite{Amsler:2008zzb}.  Therefore lattice calculations
of heavy-light meson decay constants provide good tests of lattice
QCD methods, especially the treatment of heavy quarks on the lattice.
The comparison of lattice calculations with experimental measurements,
however, relies upon the assumption that, because leptonic decays occur at
tree-level in the standard model, they do not receive large corrections
from new physics.  This is generally true of most beyond-the-standard
model theories, but in a few models, such as those with leptoquarks,
this is not necessarily the case~\cite{Dobrescu:2008er}.

\begin{figure}
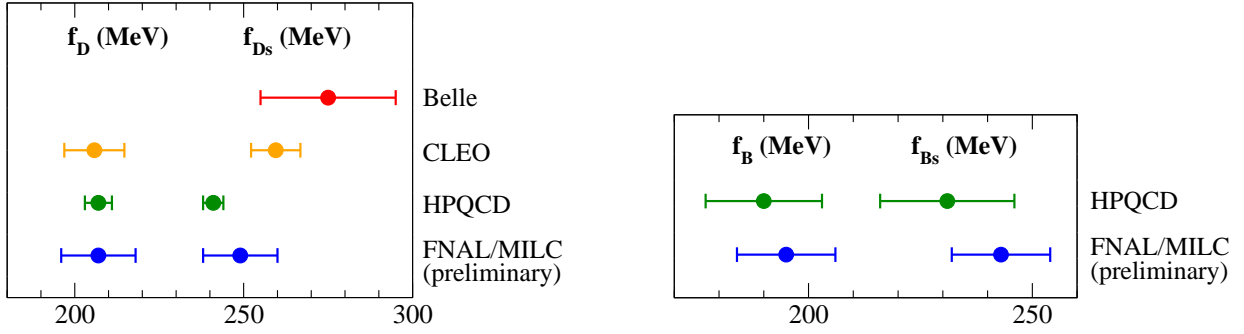

\begin{center}
\begin{tabular}{ccc}
\includegraphics[width=0.45\textwidth,angle=0]{figs/fD_RMP.eps}
& \quad \quad \quad \quad &
\includegraphics[width=0.45\textwidth,angle=0]{figs/fB_RMP.eps}
\end{tabular}
\end{center}
\caption{Comparison of lattice QCD and experimental results for $f_D$ and $f_{D_s}$ (left panel) and of lattice QCD results for $f_B$ and $f_{B_s}$ (right panel).}
\label{fig:fD_and_fB}
\end{figure}

CKM unitarity implies that
$|V_{cd}| = |V_{us}|$ and $|V_{cs}| = |V_{ud}|$ up to corrections of
$\mathcal{O}(|V_{us}|^4)$. Because both $|V_{ud}|$
and $|V_{us}|$ are known to sub-percent accuracy, experimentalists use
this relation to extract the $D$-meson decay constants from
the measured branching fractions.  The latest determinations of
$f_D$~\cite{Eisenstein:2008sq} and $f_{D_s}$~\cite{Alexander:2009ux}
from the CLEO experiment are
\begin{equation}
	f_{D^+} = 205.8  \pm 8.9 \textrm{ MeV}, \quad f_{D^+_s} = 259.5 \pm 7.3 \textrm{ MeV} \;.
\end{equation}
These results use the determination of $|V_{ud}| = 0.97418(26)$ from
superallowed $0^+ \to 0^+$ nuclear $\beta$-decay~\cite{Towner:2007np}
and of $|V_{us}| =  0.2256$ \cite{Eisenstein:2008sq}.\footnote{Although
\textcite{Eisenstein:2008sq} attribute $|V_{us}| =  0.2256$ to
FlaviaNet~\cite{Antonelli:2007mj}, \textcite{Antonelli:2007mj}
gives $|V_{us}| = 0.2246(12)$ from $K_{\ell 3}$ decays plus lattice QCD,
and $|V_{us}| = 0.2253(9)$ from $K_{\ell 2}$ and $K_{\ell 3}$ decays
plus lattice QCD.}
The Fermilab Lattice and MILC collaborations' preliminary determination of
the $D$-meson decay constants are~\cite{Mackenzie:Lat08}
 \begin{equation}
	f_D = 207(11) \textrm{ MeV}, \quad f_{D_s} = 249(11) \textrm{ MeV} \;,
 \end{equation}
where the dominant errors come from tuning the charm quark mass and from
heavy-quark discretization effects, which are each $\sim 3\%$.  Both of
these results are consistent with experiment.  The HPQCD collaboration's
determinations of the $D$-meson decay constants using HISQ fermions are
more precise~\cite{Follana:2006rc}:
\begin{equation}
	f_D  =  207(4) \textrm{ MeV}, \quad f_{D_s} = 241(3) \textrm{ MeV} \;,
\end{equation}	
with total errors each below $2\%$.  The largest contribution to the
errors comes from the uncertainty in the scale $r_1$, and is $1.4\%$
($1\%$) for $f_D$ ($f_{D_s}$).  Although HPQCD's result for $f_D$
is consistent with experiment, their value for $f_{D_s}$ is $\sim
2.5$-$\sigma$ below the CLEO measurement, where $\sigma$ is dominated
by the experimental uncertainty.  A comparison of lattice QCD and experimental results for the $D$-meson decay constants is shown in the left panel of Fig.~\ref{fig:fD_and_fB}.

Many of the statistical and systematic uncertainties that enter the
lattice calculations of $f_D$ and $f_{D_s}$ cancel in the ratio.
Therefore the quantity $f_D/f_{D_s}$ allows for a more stringent
comparison between the results of Fermilab/MILC and HPQCD.  The Fermilab
Lattice and MILC collaborations find~\cite{Mackenzie:Lat08}
\begin{equation}
	f_D/ f_{D_s} = 0.833(19) ,
\end{equation}
while the HPQCD collaboration finds~\cite{Follana:2006rc}:
\begin{equation}
	{f_{D}}/{f_{D_s}} = 0.859(8) .
\end{equation}
The lattice results for the ratio disagree slightly, but only by
$\sim 1.6$-$\sigma$.  The experimental uncertainties in $f_D$ and
$f_{D_s}$ are largely independent, and therefore add in quadrature in
the ratio~\cite{Alexander:2009ux}
\begin{equation}
	{f_{D^+}}/{f_{D^+_s}} = 0.793 \pm 0.040 .
\end{equation}
This increases the experimental errors and reduces the significance of
the discrepancy with HPQCD.

The HPQCD collaboration also uses HISQ charm quarks to compute the $D$-
and $D_s$-meson masses~\cite{Follana:2006rc}:
\begin{equation}
	M_D = 1.868(7) \textrm{ GeV}, \quad M_{D_s} = 1.962(6) \textrm{ GeV} ,
\end{equation}	
and their results agree with the experimental values $M_D =1.869
\textrm{ GeV}$ and $M_{D_s} =1.968  \textrm{ GeV}$~\cite{Amsler:2008zzb}.
This lends credibility to their calculation of $f_{D_s}$, and suggests
that both improved experimental measurements and lattice calculations
are necessary to determine whether or not this discrepancy is new
physics, a statistical fluctuation, or yet something else.  Currently,
Fermilab/MILC's determination of the $D_s$-meson decay constant lies
between the experimental measurement and the calculation of HPQCD.
Once the uncertainties in the calculation are reduced, which is expected to
occur with the addition of statistics, finer lattice spacings, and a more
sophisticated analysis, the Fermilab Lattice and MILC collaborations hope to shed light on this intriguing puzzle.

$B$-meson leptonic decays are much more difficult to observe
than $D$-meson decays because they are CKM suppressed ($\propto
|V_{ub}|^2$). In addition, $B$-decays to light leptons are suppressed
by the factor $m^2_\ell$ in Eq.~(\ref{eq:lept_decay}), and only
decays to $\tau$'s have been observed thus far. Furthermore, the
branching fraction $\Gamma(B \to \tau \nu)$ is known only to $\sim 30\%$
accuracy~\cite{Amsler:2008zzb}.  Thus there are no precise experimental
determinations of the $B$-meson decay constants, and the lattice
calculations of $f_B$ and $f_{B_s}$ should be considered predictions
that have yet to be either confirmed or refuted by experiment.

The Fermilab Lattice and MILC collaborations preliminary determinations
of $f_B$, $f_{B_s}$, and the ratio are~\cite{Mackenzie:Lat08}
 \begin{equation}
	f_B = 195(11) \textrm{ MeV}, \quad f_{B_s} = 243(11) \textrm{ MeV},
	\quad f_B/ f_{B_s} = 0.803(28) .
 \end{equation}
The largest errors in the individual decay constants are due to scale
and light-quark mass uncertainties, light-quark discretization effects,
and heavy-quark discretization effects, all of which are $\sim 2\%$. The
HPQCD collaboration's determinations are consistent and have similar
total uncertainties \cite{Gamiz:2009ku}:
 \begin{equation}
 f_B  =  190(13) \textrm{ MeV}, \quad f_{B_s}  =  231(15) \textrm{ MeV},
	\quad {f_{B}}/{f_{B_s}}  =  0.812(19) .
 \end{equation}
Their largest source of error is the $\sim 4\%$ uncertainty from 1-loop
perturbative operator matching.  A comparison of lattice QCD calculations of the $B$-meson decay constants is shown in the right panel of Fig.~\ref{fig:fD_and_fB}.

There are currently no calculations of the $B$- and $B_s$-meson masses
using the 2+1 flavor MILC lattices.  This is, in part, because the
staggered $\chi$PT expressions for heavy-light meson masses needed
to extrapolate the numerical lattice data to the physical light-quark
masses and the continuum are not known, and would require a nontrivial
extension of the continuum expressions.

%% file: RMP_sec8.tex
% File for section 8 for RMP article
%
%\section{Section 8}
\section{Semileptonic form factors}
\label{sec:semilept}

Lattice calculations of semileptonic form factors allow the extraction
of many of the CKM matrix elements from experiment.  The processes
we consider for this purpose are dominated by tree-level weak decays of
quarks at short distances, but are dressed by the strong interactions
at longer distances, such that only mesons appear on the external legs.
Given the nonperturbative form factor that parameterizes the strong
interactions of the mesons, one can extract the CKM parameters that
accompany the flavor-changing weak vertex.  With enough processes one
can over-constrain the four standard model parameters that appear in
the CKM matrix, and thus test the standard model.

\subsection{$D\rightarrow \pi \ell \nu$ and $D\rightarrow K \ell \nu$}

Semileptonic decays of $D$ mesons, $D\to K\ell\nu$ and $D\to\pi\ell\nu$,
allow determinations of the CKM matrix elements $|V_{cs}|$ and $|V_{cd}|$,
respectively.  Since these CKM matrix elements are well-determined within
the standard model by unitarity, with results for other processes, the
form factors can be obtained from experiment (assuming the standard
model), and thus serve as a strong check of lattice calculations.
Such calculations bolster confidence in similar calculations of
$B\to\pi\ell\nu$, allowing a reliable determination of $|V_{ub}|$, one
of the more important constraints on new physics in the flavor sector.
Precise calculations of semileptonic form factors for charm decays are
also interesting in their own right, given the discrepancy between the
HPQCD and experimental values for the $D_s$ leptonic decay.

The necessary hadronic amplitude $\langle P|V_\mu |D \rangle$ $(P=K,\pi)$
is parameterized in terms of form factors by
\BE
\langle P|V_\mu|D\rangle = f_+(q^2)(p_D+p_P-\Delta)_\mu + f_0(q^2)\Delta_\mu,
\EE
where $q=p_D-p_P$, $\Delta_\mu = (m_D^2-m_P^2)q_\mu/q^2$, and
$V_\mu=\overline{q}\gamma_\mu Q$.  The differential decay rate
$d\Gamma/dq^2$ is proportional to $|V_{cx}|^2|f_+(q^2)|^2$, with $x=d,s$.
The CKM matrix element $|V_{cx}|$ is determined using the experimental
decay rate and the integral over $q^2$ of the lattice determination
of $|f_+(q^2)|$.

The matrix element $\langle P|V_\mu|D\rangle$ is extracted from the
three-point function, where the $P$ meson is given a nonzero momentum
$\mathbf{p}$,
\BE\label{eq:3pt}
C_3^{D\rightarrow P}(t_x, t_y; {\mathbf p})=\sum_{{\mathbf x},{\mathbf y}}
 e^{i{\mathbf p}\cdot {\mathbf y}}\langle O_P(0)V_\mu(y) O^\dagger_D(x) \rangle,
\EE
and $O_D$ and $O_P$ are the interpolating operators for the initial and
final meson states.  
The calculation of this quantity by the Fermilab
Lattice, MILC and HPQCD Collaborations \cite{Aubin:2004ej} uses the
Fermilab action [improved through $O(\Lambda_{\textrm QCD}/m_c)$, with
$\Lambda_{\rm QCD}$ in the HQET context] for the $c$ quark and the asqtad
action for the light valence quarks.  The $D$ meson and the heavy-light
bilinears $V_\mu$ are constructed from a staggered light quark and a
Wilson-type (Fermilab) heavy quark using the procedure described in
\textcite{Wingate:2003nn} and \textcite{Bailey:2008wp}.  In order
to extract the transition amplitude $\langle P|V_\mu|D\rangle$ from
Eq.~(\ref{eq:3pt}), we need the analogous two-point correlation function,
\BE
C_2^M(t_x,{\mathbf p})=\sum_{\mathbf x}e^{i {\mathbf p} \cdot
{\mathbf x}}\langle O_M(0)O^\dagger_M(x) \rangle
~~~~~\mathrm{with} ~ M=D, P ~.
\EE
As in the case of decay constants, the renormalization factor matching
the heavy-light currents on the lattice to the continuum is
\BE
Z^{Qq}_{V_{1,4}}=\rho^{Qq}_{V_{1,4}}\sqrt{Z^{QQ}_{V_4}Z^{qq}_{V_4}},
\EE
where the factors $Z^{QQ}_{V_4}$ and $Z^{qq}_{V_4}$ are computed
nonperturbatively, and the remaining factor $\rho^{Qq}_{V_{1,4}}$
(close to 1 by construction) is determined in one-loop perturbation
theory \cite{Harada:2001fi}.
 
The quantities $f_{||}$ and $f_{\perp}$ are more natural quantities than
$f_+$ and $f_0$ in the heavy-quark effective theory, and are defined as
\BE
\langle P|V^\mu |D\rangle = \sqrt{2m_D}[v^\mu f_{||}(E)+p^{\mu}_{\perp}f_\perp(E)],
\EE
where $v=p_D/m_D$, $p_\perp=p_P-Ev$ and $E=v\cdot p_P$ is the
energy of the light meson.  The chiral extrapolation and momentum
extrapolation/interpolation are carried out in terms of these parameters,
which are then converted into $f_0$ and $f_+$.  The chiral extrapolation
in \textcite{Aubin:2004ej} was performed at fixed $E$, where $f_{||}$
and $f_{\perp}$ were fit simultaneously to the parameterization of
\textcite{Becirevic:1999kt} (BK),
\BE
f_+(q^2)=\frac{F}{(1-\tilde{q}^2)(1-\alpha \tilde{q}^2)}, \ \ \ 
f_0(q^2)=\frac{F}{1-\tilde{q}^2/\beta},
\EE
where $\tilde{q}^2=q^2/m^2_{D^*_x}$, and $F=f_+(0)$, $\alpha$ and $\beta$
are fit parameters.  The BK form contains the pole in $f_+(q^2)$ at
$q^2=m^2_{D^*_x}$.  Even so, the BK parameterization builds into the
calculation unnecessary model dependence.  The more recent calculation
of the similar semileptonic process $B\to\pi\ell\nu$ does not make use
of this assumption, as described in the next subsection.
 
\begin{figure}
\begin{center}
\includegraphics[scale=.4]{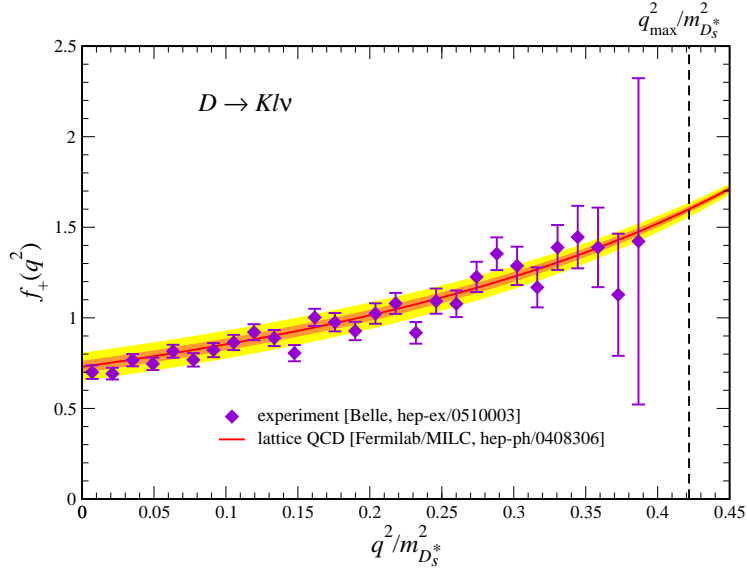}
\end{center}
\caption{Comparison of the Fermilab/MILC/HPQCD lattice prediction for
the normalized $D\rightarrow K\ell\nu$ form factor 
(bands) with the subsequent Belle results (diamonds). The orange (dark
gray) band is the 1 $\sigma$ error band from statistics, and the yellow
(light gray) band is the 1 $\sigma$ band for all errors added in
quadrature. Figure from \textcite{Kronfeld:2006sk}.}
\label{fig:Dsemilep}
\end{figure}

\textcite{Aubin:2004ej} obtain for the form factors at $q^2=0$
\BE
f_+^{D\to\pi}(0) = 0.64(3)(6), \ \ \ \  f_+^{D\to K}(0)=0.73(3)(7).
\label{eq:ff_D_to_P}
\EE
where the first error is statistical, and the second is systematic.
They also determine the shape dependence of the form factor as a
function of $q^2$.  This is shown in Fig.~\ref{fig:Dsemilep}, along
with experimental data from the Belle Collaboration \cite{Abe:2005sh}
that confirms their prediction.  Taking the most recent CLEO results
\cite{Ge:2008yi} $f^{D\to\pi}_+(0)|V_{cd}|=0.143(5)(2)$ and $f^{D\to
K}_+(0)|V_{cs}|=0.744(7)(5)$ we obtain
\BE |V_{cd}| = 0.223(8)(3)(23), \ \ \ \ |V_{cs}|=1.019(10)(7)(106),
\EE
where the first error is the (experimental) statistical error, the second
is the (experimental) systematic error, and the third is the total lattice
error.  If we use unitarity along with $|V_{ud}|$ and $|V_{us}|$, then
we can use the CLEO measurements to predict the form factors.  We then
obtain  $f^{D\to\pi}_+(0)=0.634(25)$ and $f^{D\to K}_+(0)=0.764(9)$, in
good agreement with the result in Eq.~(\ref{eq:ff_D_to_P}).  Clearly, the
lattice error still dominates the uncertainties.  The largest errors in
the lattice calculation are due to discretization errors and statistics.
Improved calculations at finer lattice spacings and higher statistics
are underway.

\subsection{$B\rightarrow\pi\ell\nu$ and $|V_{ub}|$}

Comparison between theory and experiment for $B\rightarrow\pi\ell\nu$
has been more troublesome than for other lattice calculations in CKM
physics.  Leptonic decays and $B\overline{B}$ mixing amplitudes are
described by a single parameter.  The semileptonic decays $B\rightarrow
D^{(*)}\ell\nu$ and $K\rightarrow\pi\ell\nu$ can be described to high
accuracy by a normalization and a slope.  For $B\rightarrow\pi\ell\nu$,
on the other hand, the form factors have a complicated $q^2$ dependence.
Lattice data have covered only the low momentum, high $q^2$ end of the
pion momentum spectrum, and errors are highly $q^2$-dependent and highly
correlated between $q^2$ bins in both theory and experiment.

It has long been understood that analyticity, unitarity, and crossing
symmetry can be used to constrain the possible shapes of form factors
\cite{Bourrely:1980gp,Boyd:1994tt,Boyd:1997qw,Lellouch:1995yv}.  This has
been used recently to simplify the comparison of theory and experiment
for $B\rightarrow\pi\ell\nu$.  All form factors are analytic functions
of $q^2$ except at physical poles and threshold branch points.  In the
case of the $B\to \pi l \nu$ form factors, $f(q^2)$ is analytic below
the $B\pi$ production region except at the location of the $B^*$ pole.
The fact that analytic functions can always be expressed as convergent
power series allows the form factors to be written in a particularly
useful manner.

Consider mapping the variable $q^2$ onto a new variable, $z$, in
the following way:
\begin{equation}
z(q^2, t_0) = \frac{\sqrt{1 - q^2/t_+}-\sqrt{1-t_0/t_+}}
 {\sqrt{1-q^2/t_+}+\sqrt{1-t_0/t_+}} ,
\label{eq:zee_var}
\end{equation}
where $t_+\equiv(m_B + m_\pi )^2$, $t_-\equiv(m_B - m_\pi )^2$, and
$t_0$ is a free parameter.  Although this mapping appears complicated,
it actually has a simple interpretation in terms of $q^2$;  this
transformation maps $q^2 > t_+$ (the production region) onto $|z|=1$
and maps $q^2 < t_+$ (which includes the semileptonic region) onto real
$z \in [-1,1]$.  In the case of $B\rightarrow\pi\ell\nu$, the physical
decay region is mapped into roughly $-0.3<z<0.3$.  In terms of $z$,
the form factors can be written in a simple form:
\begin{equation} f(q^2)
= \frac{1}{P(q^2) \phi(q^2,t_0)}  \sum_{k=0}^{\infty} a_k(t_0)
z(q^2,t_0)^k.
\label{eq:z_exp}
\end{equation}
Most of the $q^2$ dependence is contained in the first two, perturbatively
calculable, factors.  The Blaschke factor $P(q^2)$ is a function that
contains subthreshold poles and the outer function $\phi(q^2,t_0)$ is an
arbitrary analytic function (outside the cut from $ t_+ < q^2 < \infty$)
which is chosen to give the series coefficients $a_k$ a simple form.
See \textcite{Bailey:2008wp}, \textcite{Arnesen:2005ez},
and references therein for the explicit forms of these expressions.
With the proper choice of $\phi(q^2,t_0)$, analyticity and unitarity
require the $a_k$ to satisfy
\begin{equation}
\sum_{k=0}^{N} a_k^2 \lesssim 1.
\label{eq:a_const}
\end{equation}
The fact that  $-0.3<z<0.3$ means that according to analyticity and
unitarity, only five or six terms are required to describe the form
factors to 1\% accuracy.  (In $B\rightarrow D^{\{*\}}\ell\nu$ and
$K\rightarrow\pi\ell\nu$, $z$ is on the order of a few per cent in
the physics decay region, which is why these decays can be accurately
described by just two parameters.)  Becher and Hill have argued that
the heavy-quark expansion implies that the bound is actually much
tighter than analyticity and unitarity alone demand \cite{Becher:2005bg}.
They argue that $\sum_{k=0}^{N} a_k^2 $ should be of order $(\Lambda_{\rm
QCD}/m_b)^3$.  This would lead to the expectation that only two or three
terms will be sufficient to describe the form factors to 1\% precision.

Calculations have been performed by Fermilab Lattice and MILC
collaborations using Fermilab  $b$ quarks, and by the HPQCD
collaboration using NRQCD $b$ quarks.  Many of the details of the
Fermilab/MILC calculations are the same as those for the Fermilab/MILC
computation of heavy-light decay constants, described previously.
For the semileptonic decays, only full QCD valence masses are used,
as opposed to the partially-quenched masses used in leptonic decays.
The calculations use the $a\approx 0.12$ and $0.09$ fm gauge field
ensembles.  The HM\schpt\ continuum and chiral extrapolations are
done with the full NLO expressions plus additional NNLO analytic terms.
These formulae allow the simultaneous interpolation in pion energy along
with the continuum and chiral extrapolations, thus reducing the total
systematic uncertainty.

\begin{figure}
\includegraphics{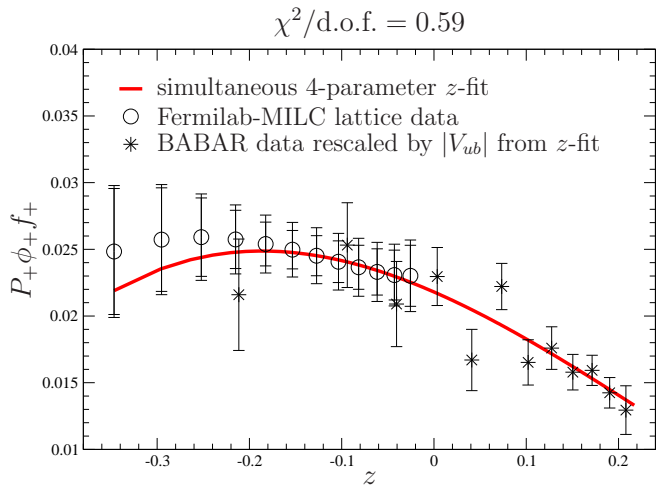}
\caption{Results for the normalized $B\rightarrow\pi\ell\nu$ form factor
$P_+\phi_+f_+$ from the Fermilab/MILC lattice calculations (circles)
and \babar\ (stars), from \textcite{Bailey:2008wp}.  The solid (red) line
is the results of a fully correlated simultaneous fit.  Requiring that
lattice and experiment have the same normalization yields $|V_{ub}|$.
}
\label{fig:BpiData}
\end{figure}

Figure~\ref{fig:BpiData} shows the result of a fully correlated
simultaneous $z$-fit to the Fermilab/MILC lattice data and the \babar\
12-bin experimental results \cite{Aubert:2006px}, with $|V_{ub}|$ being
a parameter in the fit.  The resulting $z$-fit parameters are
$a_0$ = 0.0218 $\pm$ 0.0021, 
$a_1$ = -0.0301 $\pm$ 0.0063, 
$a_2$ = -0.059 $\pm$ 0.032, 
$a_3$ = 0.079 $\pm$ 0.068, and
\begin{equation}
\label{eq:answer}
|V_{ub}| = (3.38 \pm 0.36) \times 10^{-3}
\end{equation}
\cite{Bailey:2008wp}.
The coefficients of $z^n$ are indeed of order $(\Lambda_{\rm
QCD}/m_b)^{3/2}$ as argued by \textcite{Becher:2005bg}.  Because the
$\sim 11\%$ uncertainty comes from a simultaneous fit of the lattice and
experimental data, it contains both the experimental and theoretical
errors in a way that is not simple to disentangle.  If we make the
assumption that the error in $|V_{ub}|$ is dominated by the most precisely
determined lattice point, we can estimate that the contributions are
roughly equally divided as $\sim$~6\% lattice statistical and chiral
extrapolation (combined), $\sim$~6\% lattice systematic, and $\sim$~6\%
experimental.  The largest lattice systematic uncertainties are
heavy-quark discretization, the perturbative correction, and the uncertainty
in $g_{B^*B\pi}$, all of which are about 3\%.  Our determination
is $\sim 1-2 \sigma$ lower than most inclusive determinations of
$|V_{ub}|$, where the values tend to range from $4.0-4.5\times 10^{-3}$
\cite{DiLodovico:2008}.  Our determination is, however, in good
agreement with the preferred values from the CKMfitter Collaboration
($|V_{ub}|=(3.44^{+0.22}_{-0.17})\times 10^{-3}$ \cite{Charles:2008})
and the UTfit Collaboration ($|V_{ub}|=(3.48\pm0.16)\times 10^{-3}$
\cite{Silvestrini:2008}).

\begin{figure}
\includegraphics[scale=0.5]{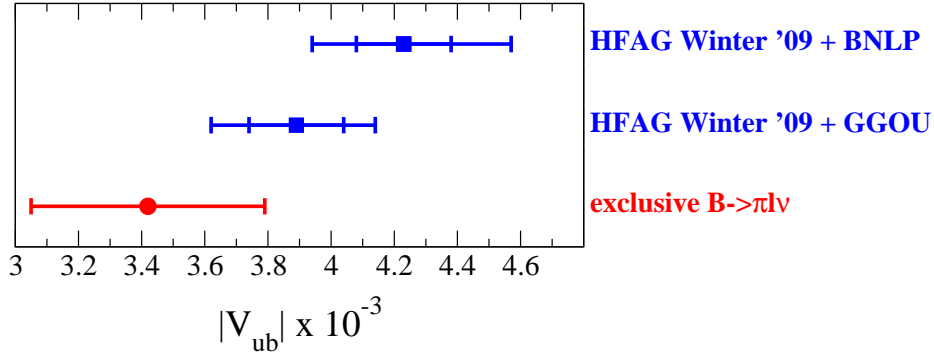}
\caption{Values of $|V_{ub}|$ obtained from averaging the exclusive
determinations compared with inclusive determinations using different
theoretical inputs.
}
\label{fig:Vub}
\end{figure}

Many of the details of the HPQCD calculation of $B\rightarrow\pi\ell\nu$
are the same as described for heavy-light decay constants in the
previous section.  They use NRQCD $b$ quarks and asqtad light quarks.
On the coarse, $a\approx 0.12$ fm ensembles, they perform the calculation
on four unquenched ensembles plus an additional two partially quenched
light quark masses on one ensemble.
They also use full QCD data on two fine, $a \approx 0.09$ fm ensembles
in order to constrain the size of discretization effects.
They use HM\schpt\ to perform the chiral extrapolations separately
for various fiducial values of $E_\pi$ after interpolating in $E_\pi$.
They also show that they obtain consistent results with simpler chiral
extrapolation methods.  They perform fits to their data using the
$z$-fit method described above, as well as several other functional forms
including the Becirevic-Kaidalov parameterization \cite{Becirevic:1999kt}
and Ball-Zwicky form \cite{Ball:2004ye}. Note that they do not use a
combined fit of experimental and lattice data using the $z$-fit method
to extract $|V_{ub}|$.  Rather, they use the various parameterizations
to integrate the form factor $f_+(q^2)$ over $q^2$, and they show that
they obtain consistent results with all methods.
Applying their results to 2008 data from Heavy Flavor Averaging Group
(HFAG) \cite{DiLodovico:2008} yields
\begin{equation}
|V_{ub}| = (3.40 \pm 0.20^{+0.59}_{-0.39}) \times 10^{-3}
\end{equation}
\cite{Dalgic:2006dt},
where the first error is experimental and the second is from the lattice
calculation.
Figure~\ref{fig:Vub} shows the comparison between an average of
the Fermilab/MILC and HPQCD results for $|V_{ub}|$ and two inclusive
determinations of the same quantity using different theoretical inputs
\cite{Lange:2005yw, Gambino:2007rp}.

\subsection{$B\rightarrow D \ell\nu$ and $B\rightarrow D^*\ell\nu$}

The CKM parameter $|V_{cb}|$ is important because it normalizes the
unitarity triangle characterizing CP-violation in the standard model,
and must be determined precisely in order to constrain new physics
in the flavor sector.  The standard model prediction for kaon mixing
contains $|V_{cb}|$ to the fourth power, for example.  It is possible
to obtain $|V_{cb}|$ from both inclusive and exclusive semileptonic
$B$ decays.  The inclusive decays \cite{Chay:1990da, Bigi:1992su,
Bigi:1992ne, Bigi:1993fe, Bigi:1997fj} make use of the heavy-quark
expansion and perturbation theory, while the exclusive decays require
the lattice calculation of the relevant form-factors.  Each of the
exclusive channels $B\to D \ell\nu$ and $B\to D^* \ell\nu$ allows
a lattice extraction of $|V_{cb}|$, and thus they provide a useful
cross-check, both of each other, and of the inclusive determination.
We have so far considered the calculations of the necessary form
factors only at zero-recoil, as this leads to considerable simplification
and reduced theoretical errors \cite{Hashimoto:2001nb}.

The differential rate for the decay $B\to D \ell \nu$ is
\begin{equation}
	\frac{d\Gamma(B\rightarrow D \ell \nu)}{dw} =
		\frac{G^2_F}{48\pi^3}m^3_{D}(m_B+m_{D})^2(w^2-1)^{3/2}
		|V_{cb}|^2 |{\cal G}(w)|^2,
\end{equation}
with
\begin{equation}
	{\cal G}(w) = h_+(w) - \frac{m_B-m_D}{m_B+m_D} h_-(w),
 \end{equation}
where $G_F$ is Fermi's constant, $h_+(w)$ and $h_-(w)$
are form factors, and $w=v'\cdot v$ is the velocity transfer from
the initial state to the final state. The differential rate for the
semileptonic decay $B\to D^*\ell\nu_\ell$ is
\begin{equation}
	\frac{d\Gamma(B\rightarrow D^* \ell \nu)}{dw} =
	\frac{G^2_F}{4\pi^3}m^3_{D^*}(m_B-m_{D^*})^2\sqrt{w^2-1}
	|V_{cb}|^2\chi(w)|{\cal F}(w)|^2,
\end{equation}
where $\chi(w)|{\cal F}(w)|^2$ contains a combination of
four form factors that must be calculated nonperturbatively.  At zero
recoil ($w=1$) we have $\chi(1)=1$, and ${\cal F}(1)$ reduces to a
single form factor, $h_{A_1}(1)$.

We compute the form factor $h_+$ at zero-recoil using the
double ratio \cite{Hashimoto:1999yp}
\begin{eqnarray}
\label{eq:doubleR_h+}
\frac{\langle D|\overline{c}\gamma_4
 b|\overline{B}\rangle \langle \overline{B}|\overline{b}\gamma_4
 c|D\rangle}{\langle D|\overline{c}\gamma_4 c|D\rangle \langle
 \overline{B}|\overline{b}\gamma_4 b|\overline{B}\rangle} =
 \left|h_{+}(1)\right|^2 ~.
\end{eqnarray}
This double ratio has the advantage that the statistical errors and many
of the systematic errors cancel.  The discretization errors are suppressed
by inverse powers of heavy-quark mass as $\alpha_s(\Lambda_{QCD}/2m_Q)^2$
and $(\Lambda_{QCD}/2m_Q)^3$ \cite{Kronfeld:2000ck}, and much of
the current renormalization cancels, leaving only a small correction
that can be computed perturbatively \cite{Harada:2001fj}.   The extra
suppression of discretization errors by a factor of $\Lambda/2m_Q$ occurs
at zero-recoil for heavy-to-heavy transitions, and is a consequence of
Luke's Theorem \cite{Luke:1990eg}.

In order to obtain $h_-$, it is necessary to consider nonzero recoil
momenta.  In this case, Luke's theorem does not apply, and the HQET
power counting leads to larger heavy-quark discretization errors.
However, this is mitigated by the small contribution of $h_-$ to the
branching fraction.  The form factor $h_-$ is determined from the double
ratio \cite{Hashimoto:1999yp}
\BE
\frac{\langle D|\overline{c}\gamma_j  b|\overline{B}\rangle \langle
D|\overline{c}\gamma_4  c|D\rangle}{\langle D|\overline{c}\gamma_4
b|\overline{B}\rangle \langle D|\overline{c}\gamma_j b|D\rangle}
 =\left[1-\frac{h_-(w)}{h_+(w)}\right]\left[1+\frac{h_-(w)}{2h_+(w)}(w-1)\right],
\EE
which is extrapolated to the zero-recoil point $w=1$.  Combining
the determinations of $h_+(1)$ and $h_-(1)$, we obtain the preliminary
result
${\cal G}(1)=1.082(18)(16)$ \cite{Okamoto:2005zg}, where the first
error is statistical and the second is the sum of all systematic errors
in quadrature, and where we have included a $0.7\%$ QED correction
\cite{Sirlin:1981ie}.  Combining this with the latest average from the
(HFAG),
${\cal G}(1)|V_{cb}|=(42.3\pm1.5)\times 10^{-3}$
\cite{DiLodovico:2008}, we obtain the preliminary result
\BE |V_{cb}|=(39.1\pm1.4\pm0.9)\times 10^{-3} ~,
\EE
where the first error is experimental, and the second is theoretical.

The form factor at zero-recoil needed for $B\to D^*\ell\nu$ is computed
using the double ratio \cite{Bernard:2008dn}
\begin{eqnarray}
\label{eq:doubleR_hA}
\frac{\langle D^*|\overline{c}\gamma_j \gamma_5
 b|\overline{B}\rangle \langle \overline{B}|\overline{b}\gamma_j
 \gamma_5 c|D^*\rangle}{\langle D^*|\overline{c}\gamma_4 c|D^*\rangle
 \langle \overline{B}|\overline{b}\gamma_4 b|\overline{B}\rangle}
 = \left|h_{A_1}(1)\right|^2 ~,
\end{eqnarray}
where again, the
discretization errors are suppressed by inverse powers of heavy-quark
mass as $\alpha_s(\Lambda_{QCD}/2m_Q)^2$ and $(\Lambda_{QCD}/2m_Q)^3$,
and much of the current renormalization cancels, leaving only a small
correction that can be computed perturbatively \cite{Harada:2001fj}.
We extrapolate to physical light quark masses using the appropriate
rHM\schpt\ \cite{Laiho:2005ue}.

\begin{figure}
\includegraphics[scale=0.5]{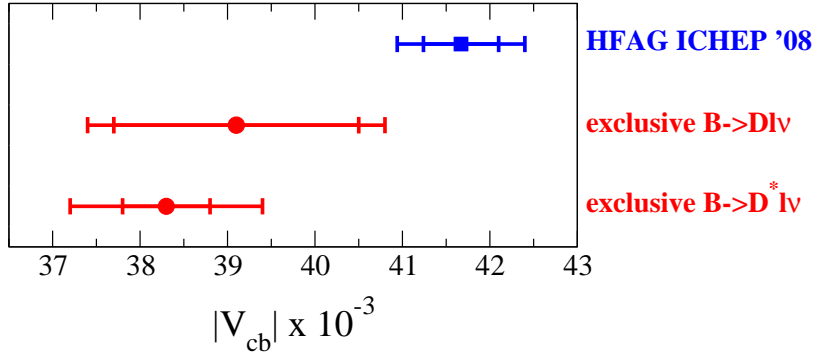}
\caption{Values of $|V_{cb}|$ from the exclusive decays $B\to D\ell\nu$,
$B\to D^*\ell\nu$, and the inclusive determination.
}
\label{fig:Vcb}
\end{figure}

Including a QED correction of $0.7\%$ \cite{Sirlin:1981ie}, we obtain
${\cal F}(1)=0.927(13)(20)$ \cite{Bernard:2008dn}, where the first
error is statistical and the second is the sum of systematic errors
in quadrature.  Taking the latest HFAG average of the experimental
determination
${\cal F}(1)|V_{cb}|=(35.49\pm0.48)\times 10^{-3}$
\cite{DiLodovico:2008}, we obtain
\BE |V_{cb}|=(38.3\pm0.5\pm1.0) \times 10^{-3} ~,
\EE
The experimental average includes all available
measurements of ${\cal F}(1)|V_{cb}|$, but we point out that the global
fit is not very consistent [$\chi^2/{\rm dof}=39/21$ (CL=0.01$\%$)].
The Particle Data Group handles this inconsistency by inflating the
experimental error by $50\%$ \cite{Amsler:2008zzb}.  The dominant
lattice errors are discretization errors and statistics, and work is
in progress to reduce these.  Note that there is some tension between
this and the inclusive determination of $|V_{cb}|=41.6(6)\times 10^{-3}$
\cite{Barberio:2007cr}, as can be seen in Fig.~\ref{fig:Vcb}.  

%

%% file: RMP_sec9.tex
% File for section 9 for RMP article
%
%\section{Section 9}
\section{Other computations using MILC lattices}
\label{sec:other_comps}

In this section, we describe a variety of additional results based on
the MILC ensembles.
Over eighty-five physicists outside the MILC
collaboration have used the MILC configurations in their
research. This includes colleagues at nearly forty institutions
throughout the world. Their research covers a very broad range of
topics including determinations of the strong coupling constant, the
quark masses, the quarkonium spectrum and decay widths, the mass
spectrum of mesons with a heavy quark and a light antiquark, the
masses of baryons with one or more heavy quarks, as well as studies of
the weak decays of mesons containing heavy quarks, the mixing of
neutral $K$ and $B$ mesons with their antiparticles, the quark and
gluon structure of hadrons, the scattering lengths of pions, kaons and
nucleons, the hadronic contributions to the muon anomalous magnetic
moment, and meson spectral functions.

\subsection{Determination of the strong coupling constant and the 
charm quark mass}
\label{sec:alpha_s}

\subsubsection{The strong coupling constant from small Wilson loops}

The HPQCD collaboration used MILC lattice ensembles to compute the
strong coupling constant $\alpha_s$ \cite{Mason:2005zx,Davies:2008sw}.
They compute nonperturbatively ({\it i.e.,} numerically on the MILC
lattices) a variety of short-distance quantities $Y$, each of which has
a perturbative expansion of the form
\begin{equation}
Y = \sum_{n=1}^\infty c_n \alpha_V^n(d/a) ~,
\label{eq:Y_expan}
\end{equation}
where $c_n$ and $d$ are dimensionless $a$-independent constants,
and $\alpha_V(d/a)$ is the running QCD coupling constant in the
so-called $V$ scheme \cite{Lepage:1992xa} for $n_f=3$ flavors of
light quarks.

The coupling $\alpha_V(d/a)$ is determined by matching the perturbative
expansion, Eq.~(\ref{eq:Y_expan}), to the nonperturbative value for $Y$.
Perturbatively converting from the $V$ to the \msbar\ scheme and running
up to the $Z$ boson mass, switching to $n_f=4$ and then $5$ at the $c$
and $b$ quark masses, gives a determination of the strong coupling
constant $\alpha_{\msbar}(M_Z, n_f=5)$.

The HPQCD collaboration considered 22 short distance quantities $Y$,
consisting of the logarithms of small Wilson loops and ratios of
small Wilson loops \cite{Davies:2008sw}. The scales $d$ in
Eq.~(\ref{eq:Y_expan}) are determined perturbatively by the method
of \textcite{Lepage:1992xa}, $c_n$ for $n=1$, 2 and 3 were computed
in lattice perturbation theory \cite{Mason:2004zt}, and higher orders,
up to $n=10$ were included in a constrained fitting procedure.
In practice, $\alpha_V(d/a)$ for all the different scales
$d/a$ used was run to a common scale of 7.5 \GeV, and $\alpha_0
\equiv\alpha_V(7.5\,\GeV)$ was used as a free fitting parameter in the
constrained fits for each of the observables.

\begin{figure}
\begin{center}
\includegraphics[width=4.0in]{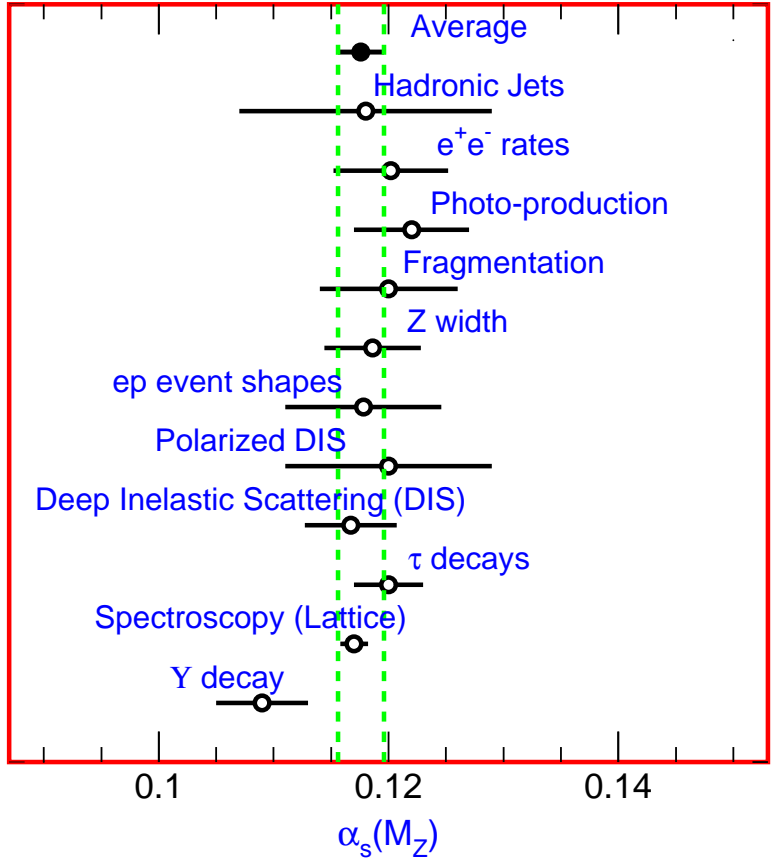}
\end{center}
\caption{Summary of determinations of the strong coupling constant
$\alpha_s(M_Z)$ from \textcite{Amsler:2008zzb}. The lattice QCD
determination is the most precise one.
}
\label{fig:alpha-s}
\end{figure}

Corrections to the perturbative form, Eq.~(\ref{eq:Y_expan}), from
condensates appearing in an operator product expansion (OPE) for
short-distance objects, were included in the constrained fitting procedure.
Other systematic errors such as finite lattice spacing effects
and scale-setting uncertainties were considered. As their final
result, the HPQCD collaboration quotes
\begin{equation}
\alpha_V(7.5\,\GeV, n_f=3) = 0.2120(28) ~~~~ \mathrm{and} ~~~~
\alpha_{\msbar}(M_Z, n_f=5) = 0.1183(8) ~.
\label{eq:alpha_V_MSbar}
\end{equation}
The lattice determination of $\alpha_{\msbar}(M_Z)$ is compared
to other determinations in Fig.~\ref{fig:alpha-s}.

In \textcite{Maltman:2008bx} a reanalysis of three of the short distance
quantities used by the HPQCD collaboration was performed with the
result
\begin{equation}
\alpha_{\msbar}(M_Z, n_f=5) = 0.1192(11) ~,
\label{eq:alpha_MSbar_II}
\end{equation}
in good agreement with other next-next-to-leading-order determinations
\cite{Bethke:2006ac}.
The two analyses differ in the way the perturbative running and
matching was done, the value of the gluon condensate used in the
OPE subtraction, the way the scale setting for each lattice ensemble
is treated and a slight difference of the value used for the scale
setting. For more details see \textcite{Maltman:2008bx}.

\subsubsection{The charm quark mass and the strong  coupling constant from
current-current correlators}

A new approach to extract $\alpha_s$ and to determine the charm
quark mass $m_c$ was used in \textcite{Allison:2008xk}. It consists
of comparing moments of charmonium current-current correlators
computed nonperturbatively on the lattice with high-order continuum
QCD perturbation theory. Vector current-current correlators have
previously been used to obtain some of the most precise determinations
of $m_c$ from the experimental $e^+e^-\rightarrow$ hadrons cross section
\cite{Kuhn:2001dm,Kuhn:2007vp}.  On the lattice, many types of correlators
are available that are not accessible to experiment.  In particular,
the pseudoscalar current-current correlator can be computed to very high
statistical accuracy, and the presence of a partially-conserved axial
vector current makes current renormalization unnecessary.

Consider the $\eta_c$ current-current correlator
\begin{equation}
G(t) = a^6 \sum_{\vec{x}} (am_{0,c})^2 \langle 0 | j_5(\vec{x},t)
 j_5(0,0) | 0 \rangle ~,
\label{eq:charm_corr}
\end{equation}
with moments
\begin{equation}
G_n = \sum_{t=-T/2}^{T/2} (t/a)^n G(t) ~.
\label{eq:charm_moms}
\end{equation}
In the continuum limit, these moments can be computed perturbatively
as
\begin{equation}
G_n(a=0) = \frac{g_n(\alpha_{\msbar}(\mu), \mu/m_c)}{(am_c(\mu))^{n-4}} ~,
\label{eq:charm_corr_pert}
\end{equation}
where $g_n$ is known to $\mathcal{O}(\alpha_s^3)$ for $n=4$, $6$ and $8$.
The approach to the continuum limit is improved by dividing by the
tree-level results, and tuning errors in $m_c$ and errors in the scale
setting are ameliorated by multiplying with factors of the lattice
$\eta_c$ mass
\begin{equation}
R_4 = G_4/G^{(0)}_4 ~~~~ \mathrm{and} ~~~~
R_n = \frac{am_{\eta_c}}{2am_{0c}} \left( G_n/G^{(0)}_n \right)^{1/(n-4)}
 ~~~~ \mathrm{for} ~~ n > 4 ~.
\label{eq:charm_rats}
\end{equation}
The ratios $R_n$ are extrapolated to the continuum limit using constrained
fits. Comparing with continuum perturbative ratios $r_4=g_4/g^{(0)}_4$
and $r_n=(g_n/g^{(0)}_n)^{1/(n-4)}$ for $n>4$, allows $\alpha_{\msbar}$
to be extracted from $R_4$ and ratios $R_n/R_{n+2}$ given the charm
quark mass, and the charm quark mass can be obtained from the $R_n$
with $n>4$, given the value of the strong coupling constant,
\begin{equation}
m_c(\mu) = \frac{m^{\mathrm{exp}}_{\eta_c}}{2}
\frac{r_n(\alpha_{\msbar}, \mu/m_c)}{R_n(a=0)} ~.
\label{eq:m_c_extract}
\end{equation}
\textcite{Allison:2008xk} used eight MILC lattice ensembles with four
different lattice spacings. The charm correlators were computed using
HISQ staggered quarks \cite{Follana:2006rc,Follana:2007uv}. They obtained
for $m_c$
\begin{equation}
m_c(3 \,\GeV, n_f=4) = 0.986(10) \, \GeV ~,~~~~ \mathrm{or} ~~
m_c(m_c, n_f=4) = 1.268(9) \, \GeV ~.
\label{eq:m_c}
\end{equation}
This is in good agreement, and about twice as precise as the best previous
determination \cite{Kuhn:2007vp}.
They obtain for $\alpha_s$
\begin{equation}
\alpha_{\msbar}(M_Z, n_f=5) = 0.1174(12)
\label{eq:alpha_s_charm}
\end{equation}
in good agreement with the lattice determination described earlier and
with other NNLO determinations \cite{Bethke:2006ac}.

\subsection{Onia and other heavy mesons}
\label{sec:onia}

Heavy quarkonia were important in the early days of QCD because
potential models could be used to approximately understand their dynamics
before first-principles calculations were possible.  The
approximate validity of potential models
helps in the selection of
operators needed in the improvement program
for quarkonia. The several
methods for formulating heavy quarks on the lattice have various
advantages and disadvantages for quarkonia.  NRQCD employs the
operators of the nonrelativistic, heavy-quark expansion.  The operator
expansion converges poorly for charmonium, and fails when
$\Lambda_{\rm QCD}/m_q$ is not small.  The Fermilab interpretation of
Wilson fermions interpolates between a nonrelativistic type of action
at $ma\gg 1$ and the usual Wilson-type action at $ma\ll 1$.  It can be
used for all $ma$, but has a more cumbersome set of operators, and has
been less highly improved than other heavy-quark actions.  The HISQ
action is a light quark action that fails when $ma\gg 1$, but has been
improved at tree level to high orders in $ma$ and works well for $ma$
close to $1$.

\subsubsection{Bottomonium with NRQCD heavy quarks}
\label{sec:bott_NRQCD}

The HPQCD and UKQCD collaborations have studied bottomonium
spectroscopy on several MILC ensembles with lattice spacings $a
\approx 0.18$, $0.12$ and $0.09$ fm \cite{Gray:2005ur}.  Even on the
finest of these ensembles, $am_b \sim 2$.  The authors have used
lattice NRQCD to formulate the $b$ quarks in the regime $am>1$
\cite{Thacker:1990bm,Lepage:1992tx,Davies:1994mp}.  The form of the
action of NRQCD was shown in Eq.~(\ref{eq:NRQCD}).  The $b$ quark is
nonrelativistic inside the bottomonium bound states, with velocity
$v^2_b \sim 0.1$.  NRQCD, as an effective field theory, can be matched
order by order to full QCD in an expansion in $v^2$ and
$\alpha_s$.  The action currently in use includes
corrections of ${\mathcal{O}}(v^2)$ beyond leading order.
Discretization errors have also been corrected to the same order in
$v^2$.

\begin{figure}[t]
\begin{center}
\begin{tabular}{c c}
\includegraphics[width=2.5in,angle=-90]{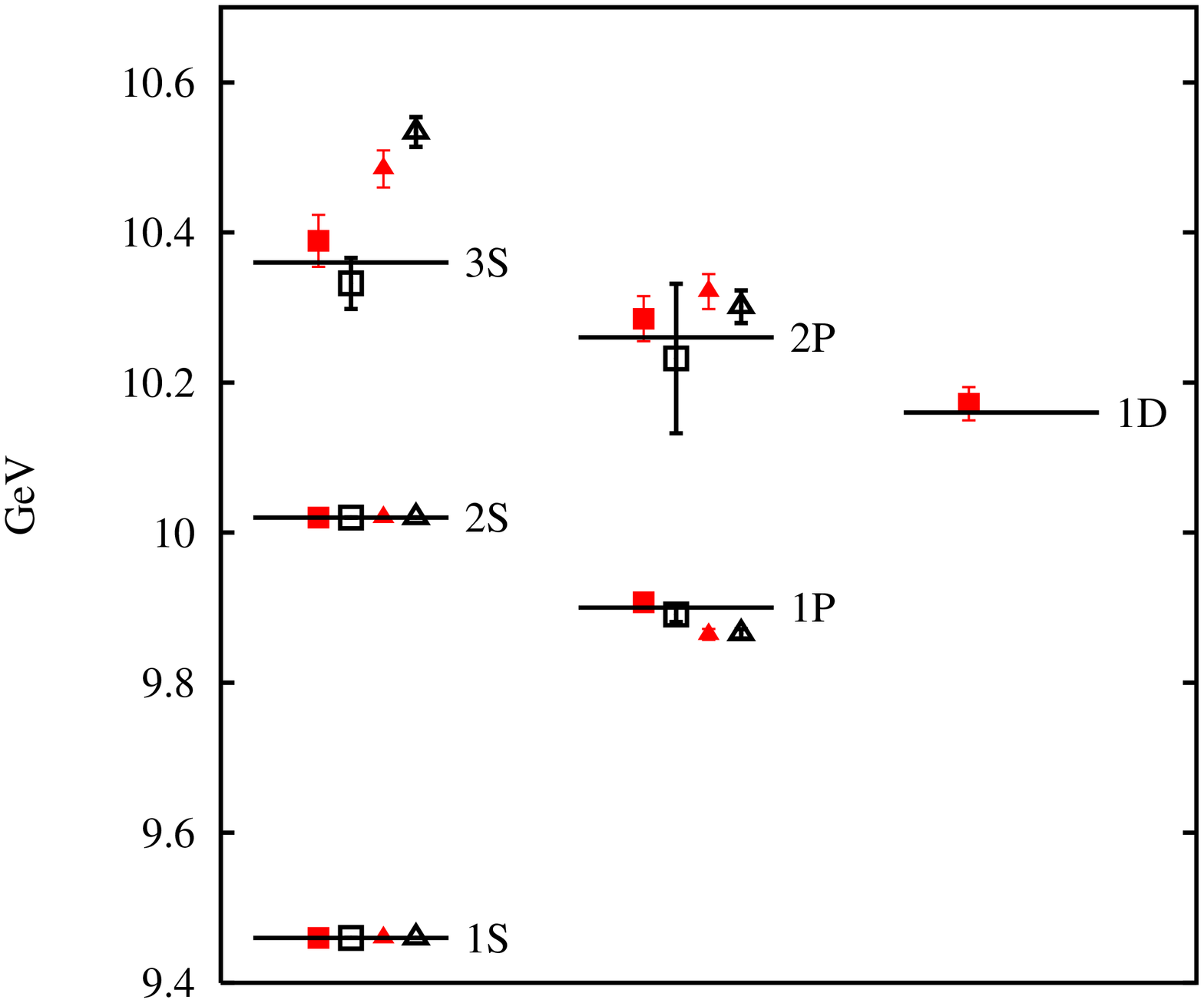}
&
\includegraphics[width=2.5in,angle=-90]{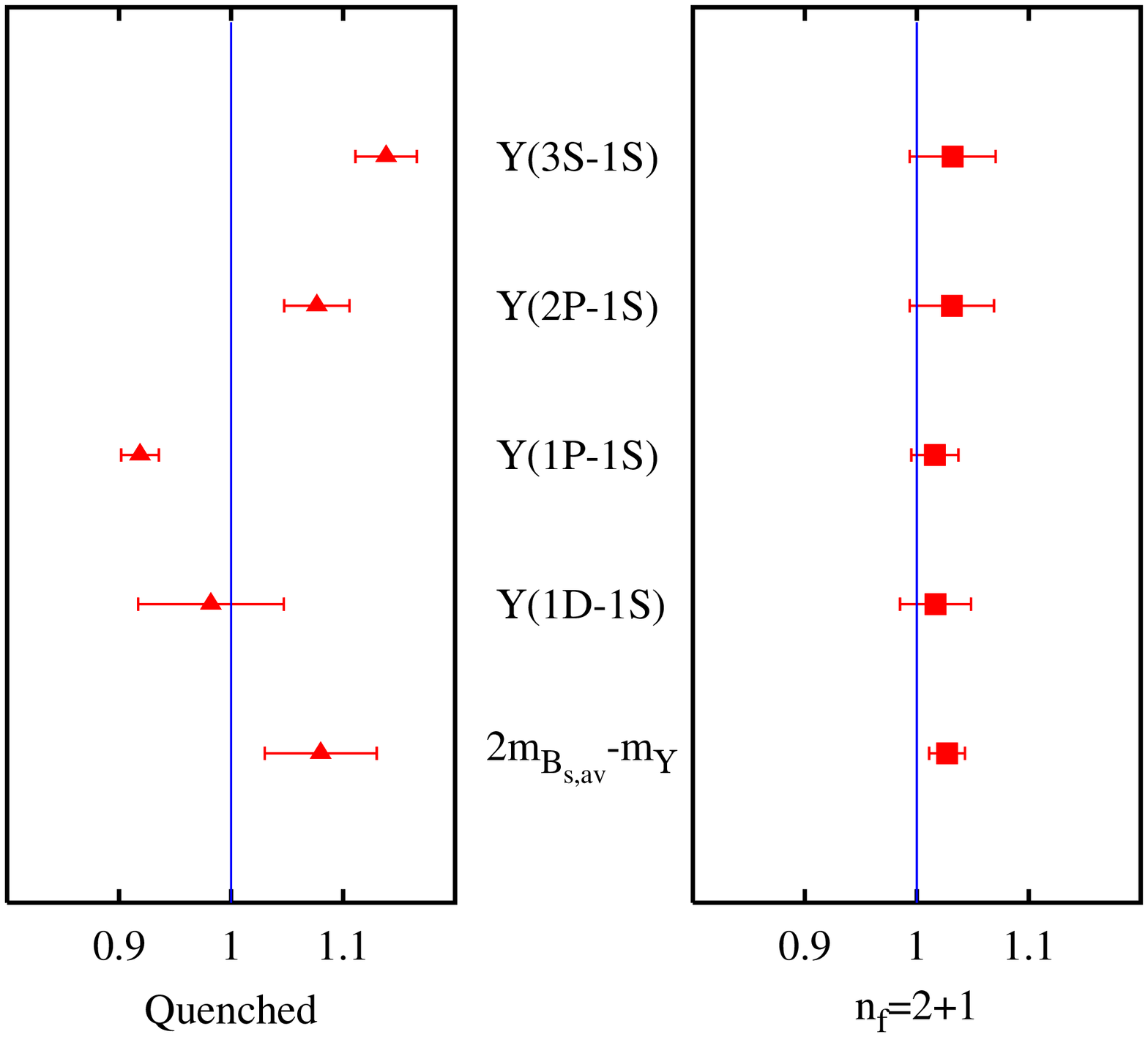}
\end{tabular}
\end{center}
\caption{Left: the $\Upsilon$ spectrum of spin-averaged radial and
orbital levels in GeV.  Closed and open symbols are from coarse and
fine lattices respectively. Squares and triangles denote unquenched and
quenched results, respectively.  Lines represent experiment.  Right:
spin-averaged mass differences from the same data divided by experiment,
in the quenched approximation (left narrow figure) and unquenched (right
narrow figure), from \textcite{Gray:2005ur}.
}
\label{fig:hpqcdSpect}
\end{figure}

The spin-averaged $\Upsilon$ mass splittings are expected to be quite
insensitive to many lattice uncertainties, such as  light sea quark
masses and normalization of
correction operators.  They are, therefore, expected to be calculable
to high accuracy on the lattice.
\textcite{Gray:2005ur}  compute spin-averaged mass splittings, $1P
- 1S$ ({\it i.e.,} $1 {}^1P_1 - 1 {}^3S_1$), $2S - 1S$ ({\it i.e.,}
$2 {}^3S_1 -1 {}^3S_1$), $2P - 1S$, and $3S - 1S$ in lattice units,
and then use the experimental splittings to determine the lattice scale,
as described in Sec.~\ref{sec:determine_a}.  Figure~\ref{fig:hpqcdSpect}
shows the results, where the lattice spacing has been set by the $2S - 1S$
splitting, and $m_b$ has been set from $M_\Upsilon$.  The left-hand figure
compares the results in GeV at two lattice spacings, for quenched and
unquenched calculations.  The right-hand figures show the  splittings
calculated on the lattice divided by experiment, in the quenched
approximation (left narrow figure) and unquenched (right narrow figure).
Clear disagreements with experiment in the quenched approximation are
removed in the unquenched calculations.

As for the $\Upsilon(1S)$ hyperfine splitting,
  \textcite{Gray:2005ur} quote $\Delta M = 61(14) \, \MeV$,
  corresponding to $r_1 \Delta M = 0.099(22)$, following an
  extrapolation to the physical point.  This result is consistent with
  the recent observation of the $\eta_b$ by the BABAR collaboration
  \cite{Babar:2008vj,Babar:2009pz} who found a splitting of 71(4) MeV
  from the $\Upsilon(1S)$.

\subsubsection{Onia with Fermilab quarks}
\label{sec:onia_clov}

The Fermilab and MILC collaborations have computed charmonium
and bottomonium masses on many of the MILC lattice ensembles with
lattice spacings from $a \approx 0.18$ fm to $a \approx 0.09$ fm
\cite{diPierro:2003bu,Gottlieb:2005me,Gottlieb:2006zz}.  For the heavy
charm and bottom quarks they use Fermilab quarks \cite{ElKhadra:1996mp}.
An updated study is underway \cite{DeTar:LAT2009}.

\begin{figure}
\begin{center}
\begin{tabular}{c c}
\includegraphics[width=2.5in]{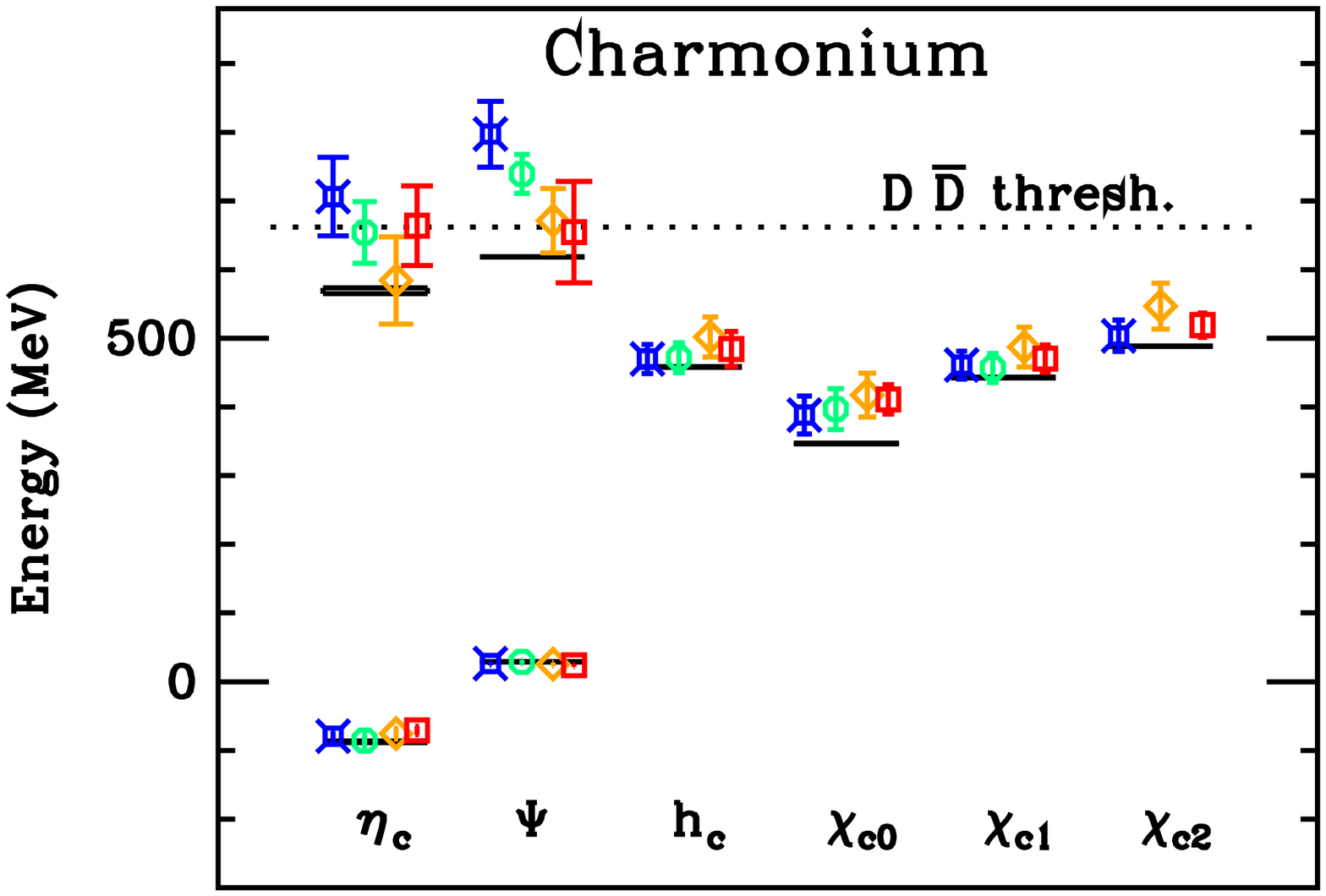}
&
\includegraphics[width=2.5in]{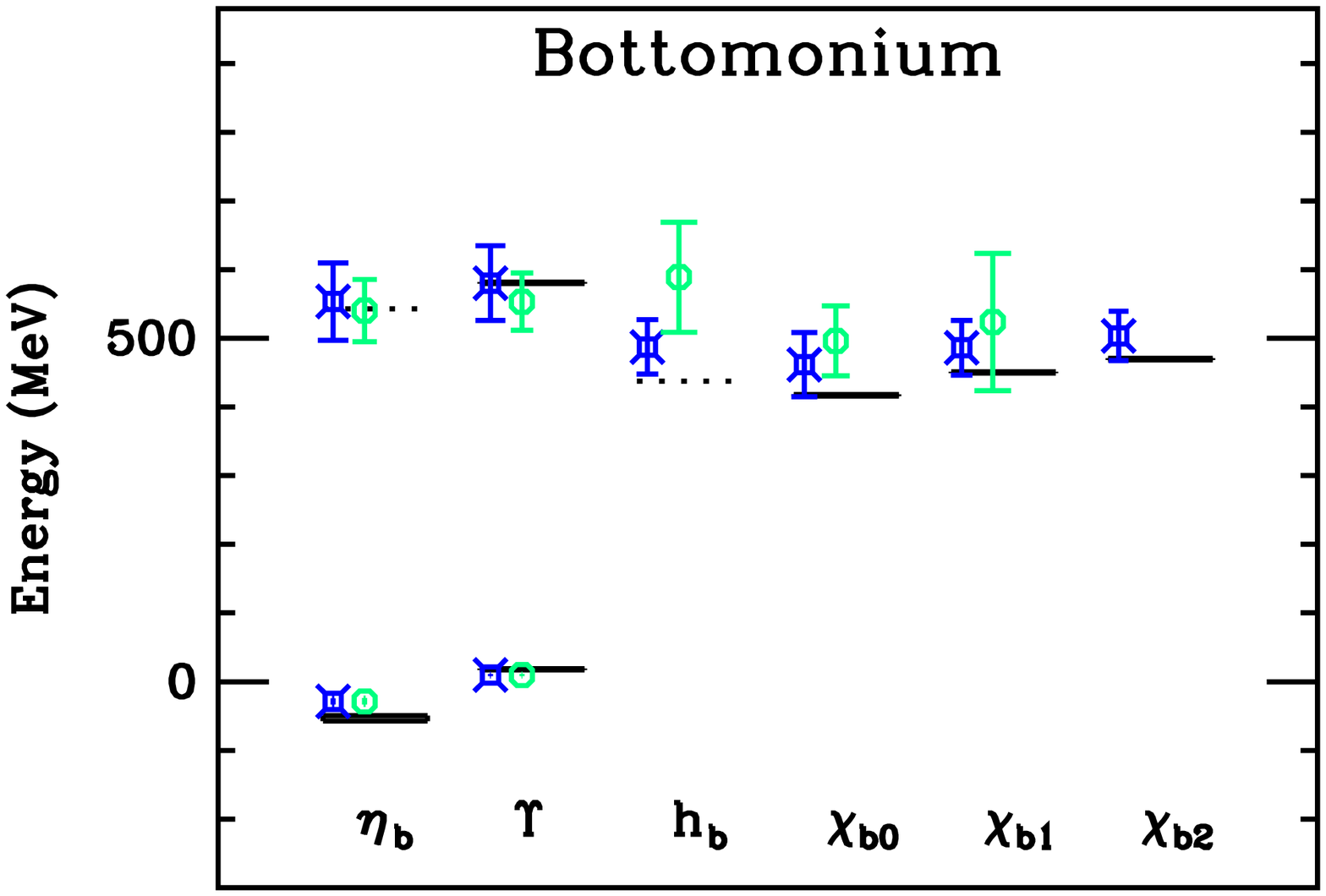}
\end{tabular}
\end{center}
\caption{Summary of the charmonium (left) and bottomonium (right)
  spectra. The fine ensemble results are in blue fancy squares, the
  coarse in green circles, the medium coarse are in orange diamonds
  and the extracoarse results are in red squares.  Included in the
  error budget is an estimated systematic uncertainty from setting the
  heavy quark masses.}
\label{fig:onia}
\end{figure}

In Fig.~\ref{fig:onia} \cite{DeTar:LAT2009} all the resulting masses
for charmonium and bottomonium are shown as splittings from the
spin-averaged $1S$ state.  Plotted are the chirally-extrapolated
values for each lattice spacing.  They are compared with the
experimental values given by solid lines, where the experimental
results are known. In the cases where they are not known and are
estimated from potential models, they are shown as dashed lines. The
charmonium spectrum shows good agreement with experiment for the
ground states, except for the $\chi_{c0}$, which
may be slightly heavier than the experimentally measured
value.  The excited $S$-wave states are also heavier than their
respective experimental results, but one has to bear in mind that
these states are difficult to determine without
careful consideration of finite-volume effects since they are
close to the $D\overline{D}$ threshold. The bottomonium summary panel
shows the general tendency of the result to approach the experimental
values as the lattice spacing decreases.

Charm
annihilation processes give a possible additional correction to the
charmonium hyperfine splitting.  \textcite{Detar:2007ni} and
\textcite{Levkova:2008qr} have started to study these quark-line
disconnected diagrams using MILC ensembles with lattice spacings $a
\approx 0.06$ and 0.09 fm. They use stochastic estimators with
unbiased subtraction \cite{Mathur:2002sf} to compute the disconnected
contribution to the $\eta_c$ propagator.
They find that annihilation processes
{\em increase} the $\eta_c$ mass a small amount (by $5.5(8) \, \MeV$
for a fine lattice and $3.4(3) \, \MeV$ for superfine), thereby
decreasing slightly the predicted hyperfine splitting
\cite{Levkova:2008qr}.

\subsubsection{Charmonium with highly improved staggered quarks}
\label{sec:stag_charm}

The HPQCD and UKQCD collaborations have studied charmonium
spectroscopy on MILC ensembles using the HISQ action for the valence
quarks.  They use MILC ensembles with lattice spacing $a \approx 0.12$
and $0.09$ fm, where $am_c = 0.66$ and $0.43$, respectively, to
demonstrate the advantages of the HISQ action, and compute the
charmonium spectrum, using the $\eta_c$ mass to tune the input value
for $am_c$.  They have corrected discretization errors in $am$ up to
order $(am)^4$, and shown that this produces a speed of light that is
independent of $p$ and equal to 1, within errors, in the equation
$E^2=p^2c^2+m^2c^4$.  The results are shown in Fig.~7 of
\textcite{Follana:2006rc}.  In particular, they find for the hyperfine
mass splitting $M_{J/\psi} - M_{\eta_c} = 109(5) \, \MeV$.
This result is the closest to the physical
value of 117(1) MeV that has yet been achieved.

\subsubsection{The $\mathbf{B}_c$ meson}
\label{sec:B_c}

The HPQCD, Fermilab Lattice and UKQCD collaborations used MILC
ensembles to predict the mass of the $B_c$ meson \cite{Allison:2004be}
before it was accurately measured. They used two different fermion
actions for the heavy bottom and charm valence quarks, choosing
the more optimal action in each case. For the bottom quark, they
used lattice NRQCD \cite{Thacker:1990bm,Lepage:1992tx,Davies:1994mp},
because it has a better treatment of the $v^4$ interactions,
where $v$ is the velocity of the heavy quark. For the charm quark, they
used the
relativistic Fermilab action \cite{ElKhadra:1996mp,Kronfeld:2000ck},
which treats higher order effects in $v^2$ better.
This is appropriate, since the velocity of the $c$ quark in $B_c$ is
not particularly small, $v^2_c \sim 0.5$.
 
\textcite{Allison:2004be} calculated mass splittings, for which many of the
systematic errors cancel, namely
\begin{equation}
\Delta_{\psi \Upsilon} = m_{B_c} - (\overline{m}_\psi + m_\Upsilon) / 2 ~,
~~~~
\Delta_{D_s B_s} = m_{B_c} - (\overline{m}_{D_s} + \overline{m}_{B_s}) ~,
\label{eq:B_c_split_defs}
\end{equation}
where $\overline{m}_\psi = (m_{\eta_c} + 3 m_{J/\psi})/4$,
$\overline{m}_{D_s} = (m_{D_s} + 3 m_{D^\ast_s})/4$, and
$\overline{m}_{B_s} = (m_{B_s} + 3 m_{B^\ast_s})/4$ are spin-averaged
masses. They found no visible lattice-spacing dependence using ensembles
with $a \approx 0.18$, $0.12$ and $0.09$ fm. Extrapolating the $a \approx
0.12$ fm results linearly in the light sea quark mass they obtain
\begin{equation}
\Delta_{\psi \Upsilon} = 39.8 \pm 3.8 \pm 11.2 \, {}^{+18}_{-0} \, \MeV ~,
~~~~
\Delta_{D_s B_s} = - [1238 \pm 30 \pm 11 \, {}^{+0}_{-37} ] \, \MeV ~.
\label{eq:B_c_split}
\end{equation}
The errors are from statistics, tuning of the heavy-quark masses, and
heavy-quark discretization effects. Since the statistical error on
the first splitting is smaller, \textcite{Allison:2004be} used that to
predict the $B_c$ mass as
\begin{equation}
m_{B_c} = 6304 \pm 4 \pm 11 \, {}^{+18}_{-0} \, \MeV ~.
\label{eq:m_B_c}
\end{equation}
Shortly after the lattice calculation was published, the CDF collaboration
announced their precise mass measurement \cite{Abulencia:2005usa}
\begin{equation}
m_{B_c} = 6287 \pm 5 \, \MeV ~,
\label{eq:m_B_c_CDF}
\end{equation}
in good agreement with the lattice prediction, {\it i.e.,}
slightly more than 1-$\sigma$ away.

\subsection{Heavy baryons}
\label{sec:heavy_baryon}

Baryons containing a heavy quark comprise a rich set of states.
For example, there are currently 17 known charmed baryons
\cite{Amsler:2008zzb}.  However, for bottom baryons, there are only a
few known states.  Thus, it is possible both to verify calculations
by comparison with known masses and to make predictions for as yet
undiscovered states.

Many of the heavy baryons contain one or more $u$ or $d$ quarks, thus
requiring a chiral extrapolation.  Although some early work on MILC
configurations \cite{Tamhankar:2002gd,Gottlieb:2003yb} used clover quarks
for $u$, $d$ and $s$, this limited how closely one could approach the
chiral limit, and recent work has used staggered light quarks instead
\cite{Na:2006qz,Na:2007pv,Na:2008hz}.  The heavy quark is dealt with as
in Sec.~\ref{sec:heavy_q}.

The pioneering lattice work on heavy baryons by the UKQCD collaboration
\cite{Bowler:1996ws} considered two operators $O_5 = \epsilon_{abc} (
\psi_1^{aT} C \gamma_5 \psi_2^b ) \Psi_H^c $ and $O_\mu = \epsilon_{abc}
( \psi_1^{aT} C \gamma_\mu \psi_2^b ) \Psi_H^c $, where $\epsilon_{abc}$
is the Levi-Civita tensor, $\psi_1$ and $\psi_2$ are light valence
quark fields for up, down, or strange quarks, $\Psi_H$ is the heavy
valence quark field for the charm or the bottom quark, $C$ is the
charge conjugation matrix, and $a$, $b$, and $c$ are color indices.
The former operator can be used to study the spin-1/2 baryons $\Lambda_h$
and $\Xi_h$.  The latter can be used, in principle, for both spin-1/2
and spin-3/2 baryons.  However, with the current formalism, for operators
with two staggered quarks, there are cancellations in the spin-3/2 sector
and $O_\mu$ can only be used for spin-1/2 baryons \cite{Na:2007pv}.
In \textcite{Gottlieb:2007ay} the taste properties of staggered di-quark
operators are considered in much the way that \textcite{Bailey:2006zn}
studied staggered baryon operators.  However, this method has not yet
been applied in calculations.  For states with two heavy quarks, both
spin-1/2 and spin-3/2 states have been studied.

Another issue when dealing with states containing heavy quarks
is the distinction between the rest and kinetic masses (see
Sec.~\ref{sec:h_l_mass}).  Calculation of kinetic masses requires looking
at states with nonzero momentum and fitting a dispersion relation.
This has not yet been done for the heavy baryons, which means that we are
restricted to reporting mass splittings.

So far, ensembles with three lattice spacings have been studied
\cite{Na:2008hz}.  With $a\approx 0.15$ fm, three ensembles with
$m_l/m_s=0.2$, 0.4 and 0.6 were used.  With $a\approx 0.12$ fm,
$m_l/m_s=0.007$, 0.01 and 0.02, and with $a\approx 0.09$ fm, only
$m_l/m_s=0.2$ and 0.4 were studied.  Seven to nine light quark masses
are used to allow for chiral extrapolation.  The charm and bottom quark
masses are as in the meson work.  Since mass splittings are desired,
ratios of hadron propagators are fit in preference to fitting each hadron
and subtracting the masses. For baryons with a heavy quark, \rschpt\
has not been worked out yet, so the chiral extrapolation is based on a
polynomial in the valence and sea masses,
\begin{equation}
\label{quad_fit}P_{\rm{quad}} = c_0 + c_1 m_l + c_2 m_l^2 + c_3 m_s + c_4 m_{\rm{sea}} ~,
\end{equation}
where $c_0$ to $c_4$ are the fitting parameters, $m_l$ is the light
valence quark mass, $m_s$ is the strange valence quark mass, and
$m_{\rm{sea}}$ is the light sea quark mass.  These fits are denoted
``quad'' in the figures.  Alternative chiral extrapolations use only the
full QCD points, \ie those in which the valence and sea light
quark masses are equal.  These are denoted ``full'' in the figures.

\begin{figure}[t]
\centering
\includegraphics[width=.47\textwidth]{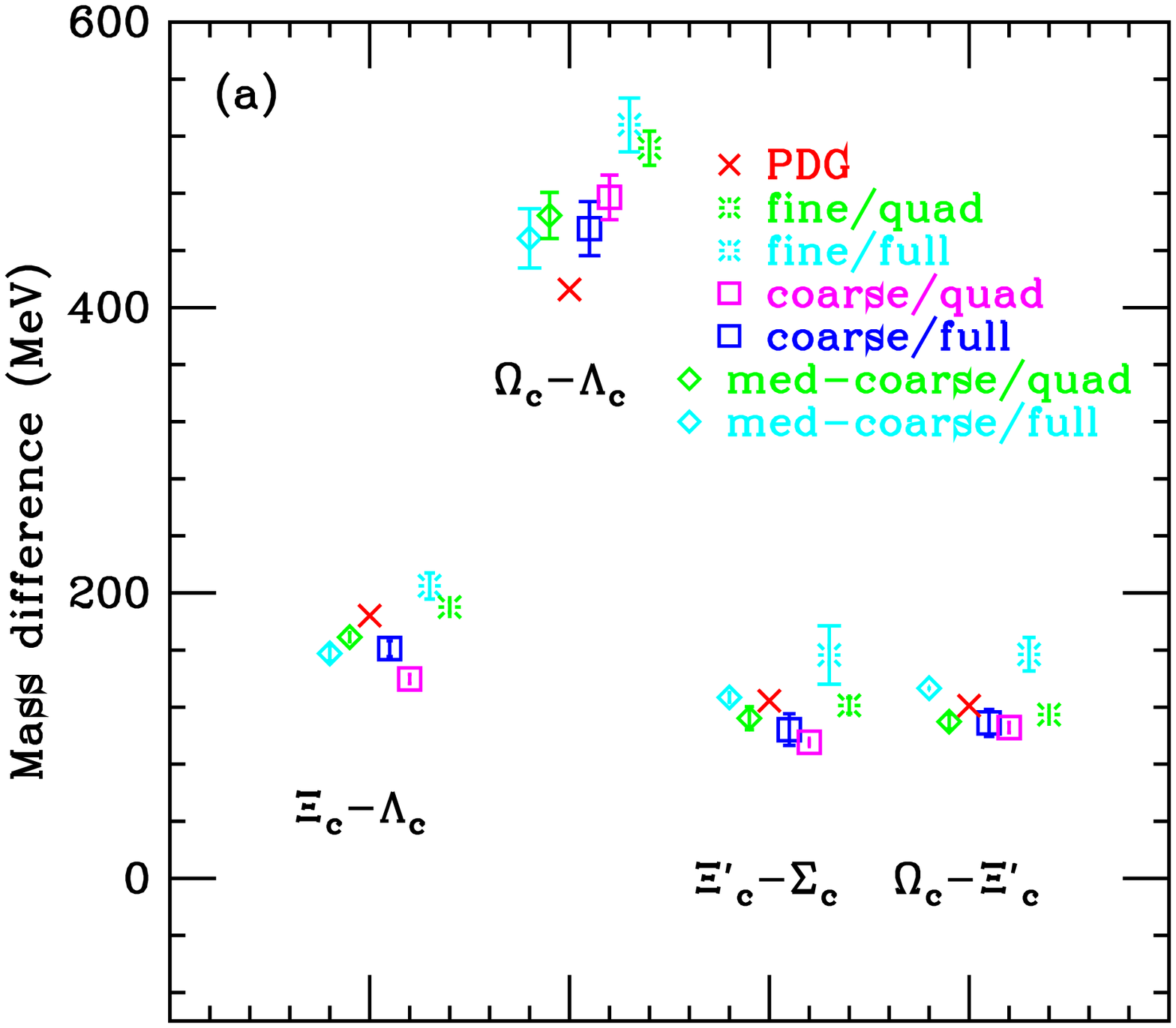}
\includegraphics[width=.47\textwidth]{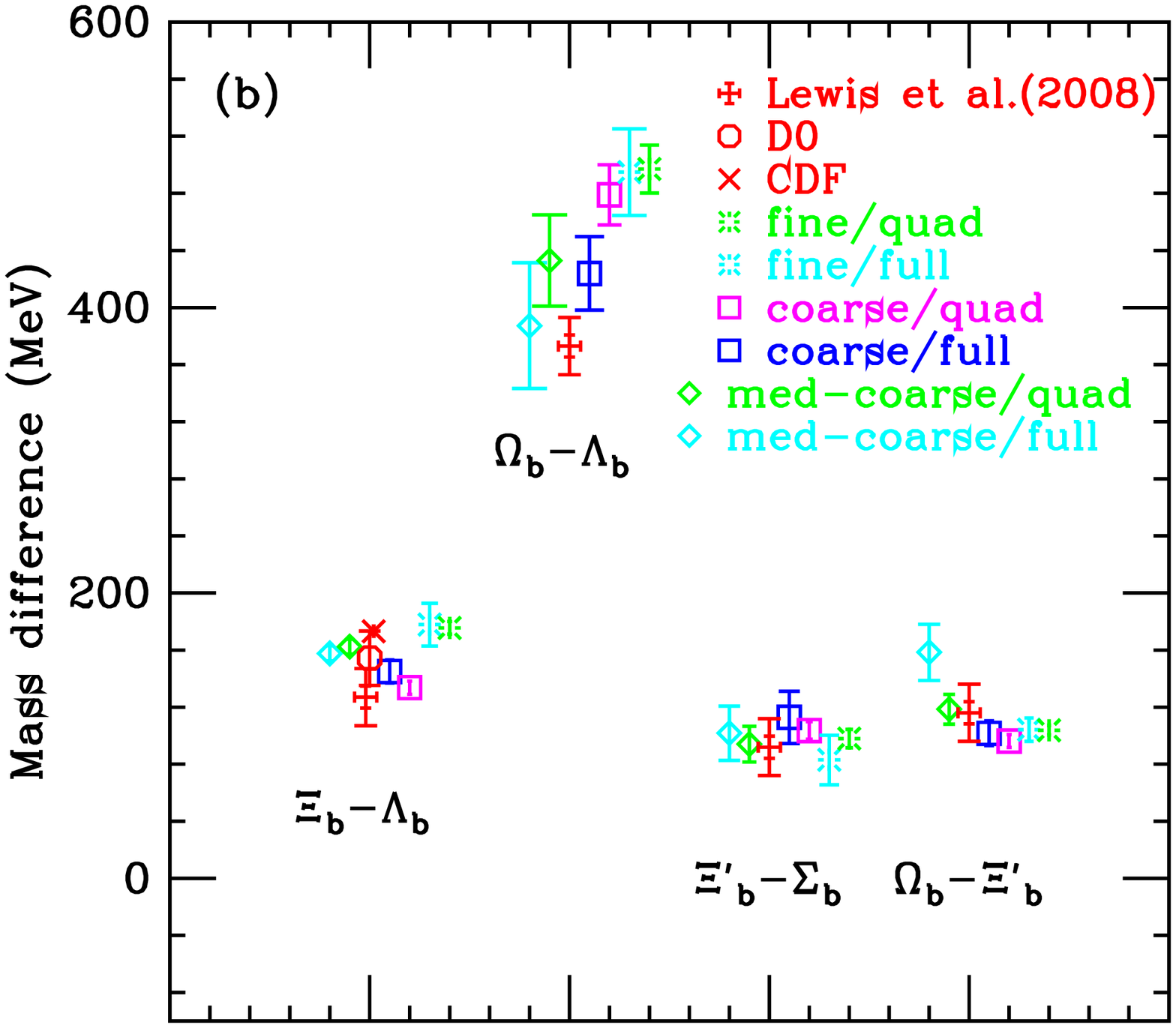}
\caption{Independent mass differences of $J^p=\frac{1}{2}^+$ 
singly charmed baryons (a), and singly bottom baryons (b). 
Figures from \textcite{Na:2008hz}.
}
\label{fig:singly_heavy}
\end{figure}

For the singly-charmed baryons in Fig.~\ref{fig:singly_heavy}(a), three
of the four differences are in good agreement with the experimental
results.  The result that is not in good agreement is one that involves
one hadron from $O_5$ and one from $O_\mu$.  The other differences
come from particles that are both determined using the same operator.
This behavior is a mystery.

In Fig.~\ref{fig:singly_heavy}(b), we consider the singly-bottom
baryons and find good agreement for the one observed difference
for $\Xi_b -\Lambda_b$.  Also shown is the comparison with a recent
lattice calculation of \textcite{Lewis:2008fu}.  The large value for
the $\Omega_b$--$\Lambda_b$ splitting is again noticeable.

\begin{figure}
\centering
\includegraphics[width=.47\textwidth]{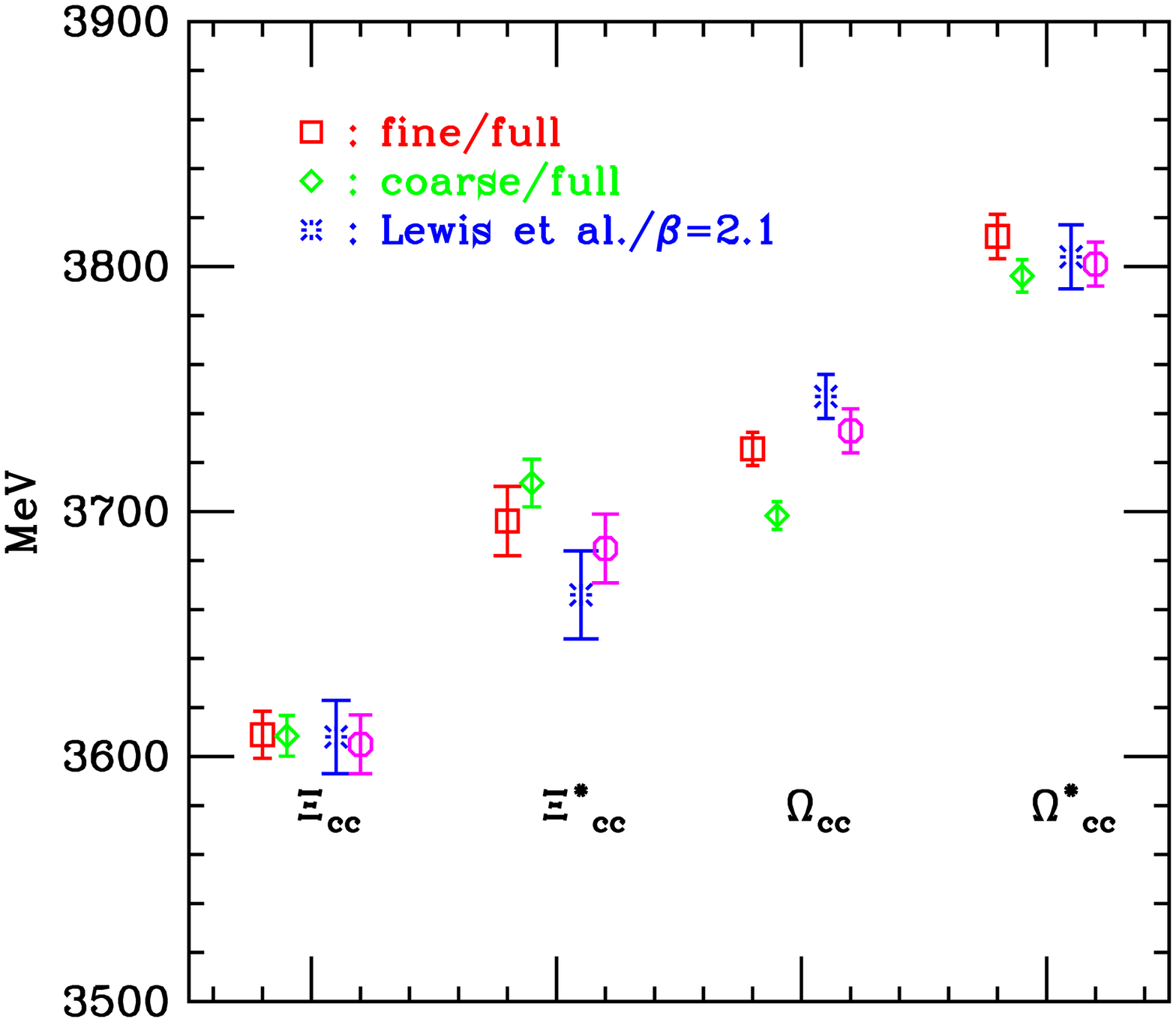}
\includegraphics[width=.47\textwidth]{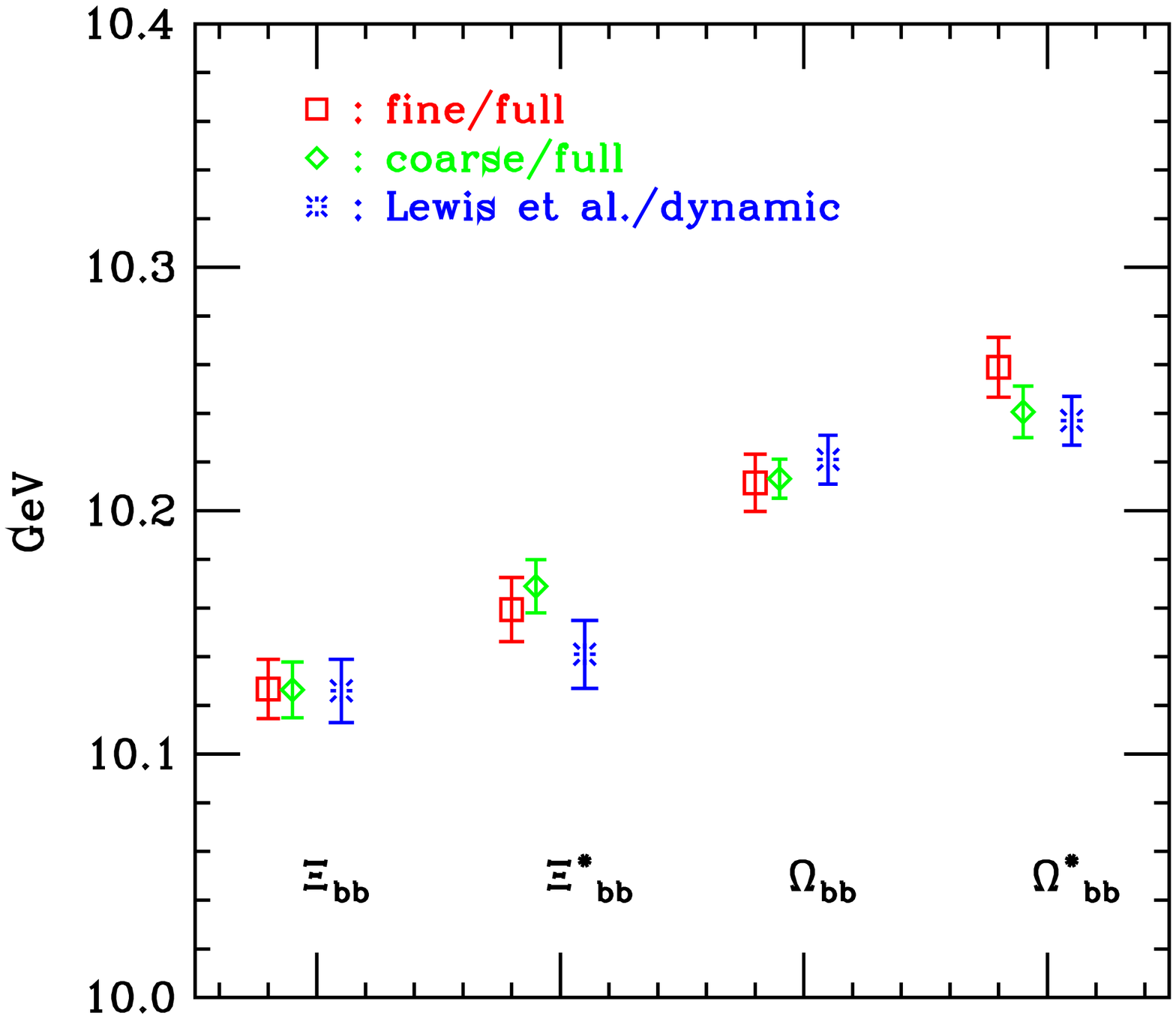}
\caption{The mass spectrum of doubly charmed and bottom baryons. 
The error bars are statistical only.
Figures from \textcite{Na:2008hz}.
}
\label{fig:doubly_heavy}
\end{figure}

In Fig.~\ref{fig:doubly_heavy}, we compare with the results of
\textcite{Lewis:2001iz} and \textcite{Lewis:2008fu} for both spin-1/2
and spin-3/2 baryons.  The earlier calculation of charmed baryons used
quenched anisotropic lattices generated with an improved gauge action.
The more recent calculation of bottom baryons uses configurations
containing the effects of dynamical quarks.  In order to compare the
two calculations, and because kinetic masses are not available in the
calculation on MILC configurations, a constant was added to the static
masses that depends on lattice spacing and whether the state contains
charm or bottom quarks, but not upon spin or light quark content.

There are a number of ways to improve upon the current work including
increasing statistics, extending the calculations to the finer ensembles,
studying the kinetic masses and studying new operators that will allow
us to explore the properties of the spin-3/2 baryons.  It is also
possible to use HISQ quarks for all of $u$, $d$, $s$ and $c$ quarks to
explore the charm sector using only staggered operators.

\subsection{$\mathbf{K^0 - \overline{K}^0}$ mixing: $\mathbf{B_K}$}
\label{sec:B_K}

Experimental measurements of the size of indirect CP-violation
in the neutral kaon system $\epsilon_K$ can be combined with
theoretical input to constrain the apex of the CKM unitarity
triangle~\cite{Buras:1998raa}. Because $\epsilon_K$ has been measured
to better than a percent accuracy~\cite{Amsler:2008zzb}, the dominant
sources of error in this procedure are the theoretical uncertainties in
the CKM matrix element $|V_{cb}|$, which enters the constraint as the
fourth power, and in the lattice determination of the nonperturbative
constant $B_K$.

The kaon bag-parameter $B_K$ encodes the hadronic contribution to
$K^0 - \overline{K}^0$ mixing~\cite{Buchalla:1995vs,Buras:1998raa}:
\begin{equation}
B_K(\mu) \equiv \frac{ \langle \overline{K}^0 | Q_{\Delta S=2}(\mu) | K^0 \rangle }
  { \frac{8}{3} \langle \overline{K}^0 | \bar{s} \gamma_0 \gamma_5 d | 0 \rangle
    \langle 0 | \bar{s} \gamma_0 \gamma_5 d | K^0 \rangle } ~,
\label{eq:B_K_def}
\end{equation}
where $Q_{\Delta S=2}$ is the effective weak four-fermion operator
\begin{equation}
Q_{\Delta S=2}(x) = [\bar{s} \gamma_\mu d]_{V-A}(x)
 [\bar{s} \gamma_\mu d]_{V-A}(x) 
\label{eq:Q_Delta_S}
\end{equation}
and $\mu$ is a renormalization scale. The dependence on $\mu$ cancels
that of a Wilson coefficient $C(\mu)$ that multiplies $B_K(\mu)$
in physical observables such as the mass difference between $K_S$
and $K_L$.  The denominator in Eq.~(\ref{eq:B_K_def}) is the value of
the matrix element with vacuum saturation of the intermediate state.
Often quoted is the value of the renormalization group invariant form
of $B_K$, $\hat{B}_K$, defined by
\begin{equation}
\hat{B}_K = C(\mu) B_K(\mu) ~.
\label{eq:B_K_hat}
\end{equation}

\textcite{Gamiz:2006sq} carried out a calculation of $B_K$ using two
MILC ensembles with lattice spacing $a \approx 0.12$ fm.  They employed
asqtad valence quarks with valence kaons made of degenerate quarks of
mass $m_s/2$.
Using one-loop matching with the coupling taken as $\alpha_V(1/a)$
they find the following value for $B_K$ in the naive dimensional regularization scheme:
\begin{equation}
B_K^{\msbar-NDR}(2 \, \GeV) = 0.618(18)(19)(30)(130) ~,
\label{eq:B_K_2GeV}
\end{equation}
where the errors are from statistics, the chiral extrapolation
\cite{VandeWater:2005uq}, discretization
errors, and the perturbative conversion to the $\msbar
- NDR$ scheme.  The value Eq.~(\ref{eq:B_K_2GeV}) corresponds to
$\hat{B}_K = 0.83 \pm 0.18$.  The error is dominated by the uncertainty
from $\mathcal{O}(\alpha_s^2)$ corrections to the perturbative
lattice-to-continuum matching.

Because the matching coefficients are
known only to one loop, the result in Eq.~(\ref{eq:B_K_2GeV}) is not competitive with
the published domain-wall fermion calculation by the RBC and UKQCD Collaborations, in which the operator renormalizatio is done nonperturbatively using the method of Rome-Southampton \cite{Martinelli:1994ty} and mixing is suppressed due
to the approximate chiral symmetry. They obtain, using a single, comparable lattice spacing, $\hat{B}_K = 0.720 \pm 0.019$ \cite{Allton:2008pn}, where the dominant uncertainty is due to discretization errors, and is estimated to be $\sim 4\%$ from the scaling behavior of quenched data.

Recently Aubin, Laiho, and Van de Water obtained the first unquenched determination of $B_K$ at two lattice spacings using domain-wall valence quarks on the MILC ensembles \cite{Aubin:2009jh}.  Because dynamical domain-wall lattice simulations are computationally expensive, this mixed-action approach is an affordable compromise that takes advantage of the best properties of both fermion formulations.
Since the MILC ensembles are available at several lattice spacings
with light pion masses and large physical volumes, this allows for
good control of the chiral extrapolation in the sea sector and the
continuum extrapolation.  Domain-wall fermions do not carry taste
quantum numbers, so there is no mixing with operators of other tastes.
Furthermore, the approximate chiral symmetry of domain-wall fermions
suppresses the mixing with wrong-chirality operators and allows the
use of nonperturbative renormalization in the same manner as in the
purely domain-wall case.  Finally, the expression for $B_K$ in mixed
action \chpt\ contains only two more parameters than in continuum \chpt\
\cite{Aubin:2006hg}, both of which are known and are, therefore, not free
parameters in the chiral and continuum extrapolation.
\textcite{Aubin:2009jh} obtain
\begin{equation}
B_K^{\msbar-NDR}(2 \, \GeV) = 0.527(6)(20),
\end{equation}
where the first error is statistical and the second is systematic.
With data on the
coarse and fine MILC lattices, \textcite{Aubin:2009jh} find that the
discretization errors in $B_K$ are small. The largest error in $B_K$
is $\sim 3\%$ and is from the renormalization factor $Z_{B_K}$, which
is computed nonperturbatively in the $\textrm{RI/MOM}$ scheme, but must
be converted to the \msbar-scheme using 1-loop continuum perturbation
theory.

\textcite{Bae:2008tb} are also computing $B_K$ with a
mixed-action approach using HYP-smeared staggered valence quarks
\cite{Hasenfratz:2001hp} on the MILC ensembles.  They have preliminary
data on the coarse, fine, and superfine MILC ensembles and are computing
$Z_{B_K}$ nonperturbatively in the $\textrm{RI/MOM}$ scheme using
the Rome-Southampton method.  When completed, their result should be
competitive with those of RBC/UKQCD and \textcite{Aubin:2009jh}.

\subsection{$\mathbf{B^0 - \bar{B}^0}$ mixing}
\label{sec:B_B}

The mass differences between the heavy and light $B^0_q$, $q=d, s$,
are given in the standard model by \cite{Buras:1990fn}
\begin{equation}
\Delta M_q^{\mathrm{theor}} = \frac{G^2_F M^2_W}{6 \pi^2}
 |V^\ast_{tq} V_{tb}|^2 \eta^B_2 S_0(x_t) M_{B_q} f^2_{B_q}
 \hat{B}_{B_q} ~,
\label{eq:Delta_M_B}
\end{equation}
where $\eta^B_2$ is a perturbative QCD correction factor and $S_0$
is the Inami-Lim function of $x_t = m^2_t/M^2_W$. $\hat{B}_{B_q}$
is the renormalization group invariant $B^0_q$ bag parameter that
can be computed in lattice QCD.

The four-fermi operators whose matrix elements between $B^0_q$ and
$\bar{B}^0_q$ are needed to study $B^0_q$ mixing in the
standard model are
\begin{eqnarray}
& OL^q \equiv [\bar{b}^a q^a]_{V-A} [\bar{b}^c q^c]_{V-A} ~,~~~~
OS^q \equiv [\bar{b}^a q^a]_{S-P} [\bar{b}^c q^c]_{S-P} ~, & \nonumber \\
& O3^q \equiv [\bar{b}^a q^c]_{S-P} [\bar{b}^c q^a]_{S-P} ~,
\label{eq:B_4ferm_op}
\end{eqnarray}
where $a, c$ are color indices.
The leading-order ${B}^0_q$-$\bar{B}^0_q$ mixing matrix element is parameterized by the product  $f^2_{B_q} B^{\msbar}_{B_q}$:
\begin{equation}
\langle \bar{B}^0_q | OL^q | B^0_q \rangle^{\msbar} (\mu) =
 \frac{8}{3} M^2_{B_q} f^2_{B_q} B^{\msbar}_{B_q}(\mu) ~,
\label{eq:B_B_def}
\end{equation}
where $B^{\msbar}_{B_q}$ is related to $\hat{B}_{B_q}$ in Eq.~(\ref{eq:Delta_M_B}) in an analogous manner to
Eq.~(\ref{eq:B_K_hat}).
Beyond tree level, the operator $OL^q$ mixes with $OS^q$, both on the
lattice and in the continuum. Including the one-loop correction, the
renormalized matrix element is given by
\begin{equation}
\frac{a^3}{2 M_{B_q}} \langle OL^q \rangle^{\msbar} (\mu) =
 \left[ 1 + \alpha_s \cdot \rho_{LL}(\mu, m_b) \right]
 \langle OL^q \rangle^{\mathrm{lat}} (a) +
 \alpha_s \cdot \rho_{LS}(\mu, m_b) \langle OS^q \rangle^{\mathrm{lat}} (a) ~.
\label{eq:B_B_op_mix}
\end{equation}
The operator $O3^q$ is only needed to compute the width difference
$\Delta \Gamma_q$ \cite{Lenz:2006hd}.

\begin{figure}
\begin{center}
\begin{tabular}{c c}
\vspace{1in}
\includegraphics{figs/ximrat.eps}
&
\vspace{1in}
\includegraphics{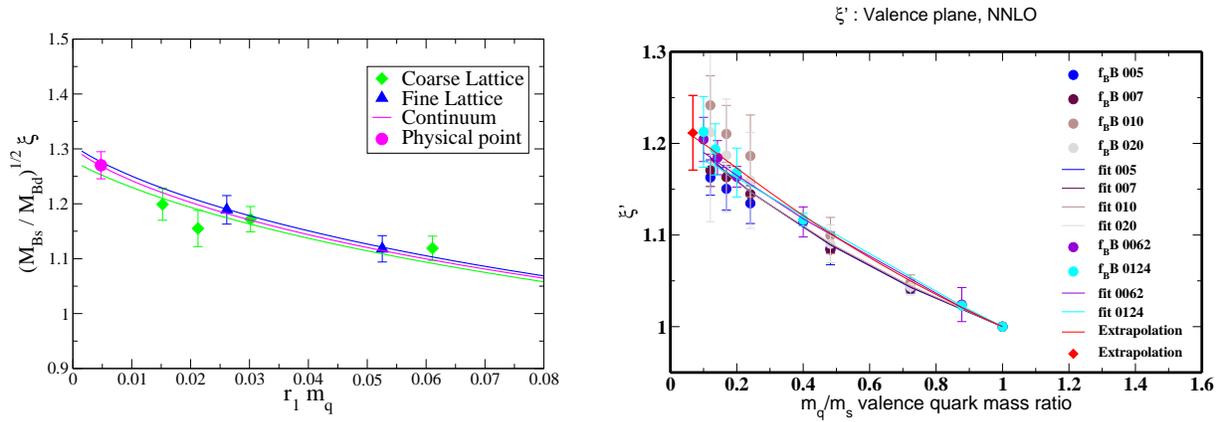}
\end{tabular}
\end{center}
\caption{The ratio $\xi^\prime = \xi \sqrt{M_{B_s} / M_{B_d} } =
f_{B_s} \sqrt{M_{B_s} B_{B_s}} / (f_{B_d} \sqrt{M_{B_d} B_{B_d}})$
as a function of the light valence quark mass together with
\rschpt\ fits and the chiral and continuum extrapolation.
The left panel is from the HPQCD collaboration \cite{Gamiz:2009ku}
and the right panel from the Fermilab/MILC collaboration
\cite{ToddEvans:2008YQ}.
}
\label{fig:f-BB-xi}
\end{figure}

The HPQCD collaboration calculated $B_{B_q}$, with $q=d,s$ on four MILC
ensembles with $a \approx 0.12$ fm and two ensembles with $a \approx 0.09$
fm, using an asqtad light valence quark and lattice NRQCD for the bottom
quark \cite{Dalgic:2006gp,Gamiz:2009ku}. With NRQCD for the heavy quark, a
dimension seven operator contributes to the relevant matrix element at
order $\mathcal{O}(\Lambda^{\mathrm{QCD}}/M_B)$, which was also taken into
account. The HPQCD collaboration finds \cite{Gamiz:2009ku}
\begin{equation}
f_{B_s} \sqrt{ \hat{B}_{B_s}} = 0.266(6)(17) \, \GeV\ ~, ~~~~~
f_{B_d} \sqrt{ \hat{B}_{B_d}} = 0.216(9)(12) \, \GeV\ ~,
\label{eq:B_B_q}
\end{equation}
and for the ratio
\begin{equation}
\xi = f_{B_s} \sqrt{ B_{B_s}} / (f_{B_d} \sqrt{ B_{B_d}})
 = 1.258(25)(21) ~,
\label{eq:xi}
\end{equation}
where the errors are from statistics plus chiral extrapolation
and from all other systematic errors added in quadrature, respectively.
The chiral and continuum extrapolation is shown in the left panel of Fig.~\ref{fig:f-BB-xi}.
Using the result in Eq.~(\ref{eq:xi}) and the experimentally measured
mass differences $\Delta M_x$, $x=s, d$, \cite{Amsler:2008zzb} they
find
\begin{equation}
\frac{|V_{td}|}{|V_{ts}|} = 0.214(1)(5) ~,
\label{eq:Vtd_Vts}
\end{equation}
where the errors are experimental and theoretical, respectively.

A similar calculation is being performed by the Fermilab Lattice and
MILC collaborations \cite{ToddEvans:2007yq,ToddEvans:2008YQ}.
They use Fermilab fermions for the heavy
quarks, and, like HPQCD, asqtad fermions for the light valence quarks.
The preliminary chiral and continuum extrapolation is shown in the right panel of Fig.~\ref{fig:f-BB-xi}.
As a preliminary result they find $\xi = 1.205(52)$, with the
statistical and systematic errors added in quadrature
\cite{ToddEvans:2008YQ}.

\subsection{Hadronic contribution to the muon anomalous magnetic moment}
\label{sec:g-2_had}

\begin{figure}
\begin{center}
\includegraphics[width=3.0in]{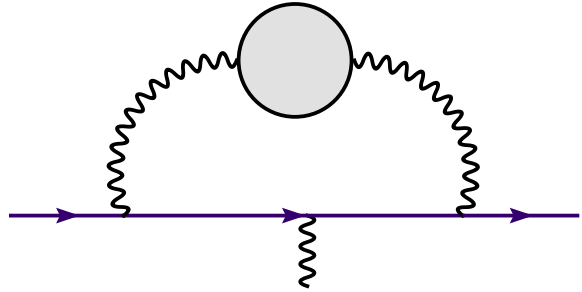}
\end{center}
\caption{The lowest-order diagram for the QCD correction to the muon
anomalous magnetic moment at $\mathcal{O}(\alpha^2)$. The bubble
represents all possible hadronic states.
Figure from \textcite{Aubin:2006xv}.
}
\label{fig:mu-hadr}
\end{figure}

One of the most precisely measured quantities, and hence an astonishingly
accurate test of QED, is the anomalous magnetic moment of the muon,
$a_\mu = (g-2)/2$. The QED contribution is known to four loops, with
the five-loop term having been estimated --- see
\textcite{Jegerlehner:2007xe,Jegerlehner:2008zz} for recent reviews.
With the experimental precision to which $a_\mu$ is known, QCD
corrections are important at leading order via the QCD contribution
to the vacuum polarization, shown in Fig.~\ref{fig:mu-hadr}.

This leading contribution can be estimated from the experimental
values of the $e^+ e^- \to$ hadrons total cross section,
$a_\mu^{\mathrm{HLO}} = (692.1 \pm 5.6) \times 10^{-10}$
\cite{Jegerlehner:2007xe,Jegerlehner:2008zz}. Using this value
the difference between experimental and theoretical value is
\begin{equation}
\delta a_\mu = a_\mu^{\mathrm{exp}} - a_\mu^{\mathrm{the}}
 = (287 \pm 91) \times 10^{-11} ~,
\label{eq:delta_a_mu}
\end{equation}
about a $3.1 \sigma$ effect and a possible hint at effects from
physics beyond the standard model.
The leading hadronic contribution can also be estimated from
$\tau \to \nu_\tau +$ hadrons, giving a result of
$10-20 \times 10^{-10}$ higher than from the $e^+ e^-$ cross section,
but this estimate is on somewhat weaker footing due to isospin-breaking
effects. A purely theoretical calculation of $a^{\mathrm{HLO}}_\mu$
is thus desirable.

The muon anomalous magnetic moment can be extracted from the full
muon--photon vertex. The first effects from QCD, at order
$\mathcal{O}(\alpha^2)$, are shown in Fig.~\ref{fig:mu-hadr}, and can
be computed from the vacuum polarization of the photons $\Pi(q^2)$ via
\cite{Blum:2002ii}
\begin{equation}
a_\mu^{\mathrm{HLO}} = \left( \frac{\alpha}{\pi} \right)^2
 \int_0^\infty dq^2 f(q^2) \Pi(q^2) ~,
\label{eq:a_mu_had}
\end{equation}
with the kernel $f(q^2)$ given in \textcite{Blum:2002ii}. The kernel
$f(q^2)$ diverges as $q^2 \to 0$. This makes a precise calculations of
$\Pi(q^2)$ at low momentum necessary, and, in particular, makes
perturbative computations unreliable.

\begin{figure}
\begin{center}
\includegraphics[width=4.0in]{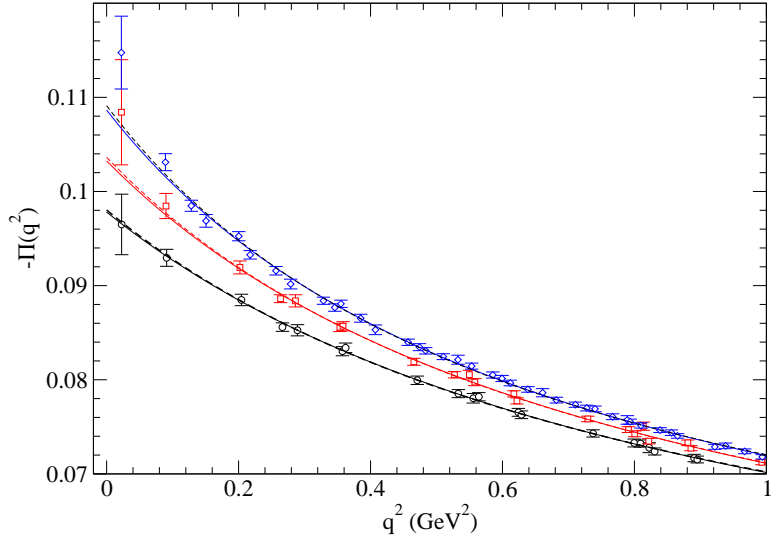}
\end{center}
\caption{Two different \rschpt\ fits to $\Pi(q^2)$ for three light
masses: $am_l=0.0031$ (diamonds), 0.0062 (squares) and 0.0124 (circles)
with $am_s=0.031$, from \textcite{Aubin:2006xv} which contains the details.
}
\label{fig:vac-pol-fits}
\end{figure}

\textcite{Aubin:2006xv} describe such a calculation based on three
MILC ensembles with lattice spacing $a \approx 0.09$ fm, and three
different light quark masses. The vacuum polarization $\Pi(q^2)$ is
computed from the correlator of the electromagnetic current in terms of
quark fields. Aubin and Blum use \rschpt\ to fit $\Pi(q^2)$ at low $q$,
(see Fig.~\ref{fig:vac-pol-fits}), and use the result in
the integral in
Eq.~(\ref{eq:a_mu_had}).

Finally, they extrapolate to the physical light quark mass, obtaining
\begin{equation}
a_\mu^{\mathrm{HLO}} = (721 \pm 15) \times 10^{-10} ~~~~ \mathrm{and} ~~
a_\mu^{\mathrm{HLO}} = (748 \pm 21) \times 10^{-10}
\label{eq:a_mu_HLO}
\end{equation}
with a linear and quadratic fit, respectively. The errors are statistical
only. Systematic errors in Eq.~(\ref{eq:a_mu_HLO}) other than due to
the quark mass extrapolation come from finite lattice spacing and
finite volume effects. Given this, the lattice result should be taken
as in broad agreement with the estimate from the $e^+ e^-$ cross
section. Further improvements need to be made before the lattice
calculation becomes competitive with other determinations.

\subsection{Quark and gluon propagators in Landau gauge}
\label{sec:q_g_prop}

Quark and gluon propagators contain perturbative and nonperturbative
information about QCD. Quark propagators play a crucial role in
hadron spectroscopy and the study of three and four-point functions
used in form factor and matrix element calculations. The propagators
are not gauge invariant, and thus have to be studied in a fixed
gauge, usually the Landau gauge. Nevertheless, they contain gauge
independent information on confinement, dynamical mass generation
and spontaneous chiral symmetry breaking. Quark and gluon propagators
can, obviously, be studied on the lattice. They are often treated
semi-analytically in the context of Dyson-Schwinger equations, see
\textcite{Roberts:2007ji} and \textcite{Fischer:2006ub} for recent
reviews.

\begin{figure}
\begin{center}
\begin{tabular}{c c}
\includegraphics[width=2.0in,angle=90]{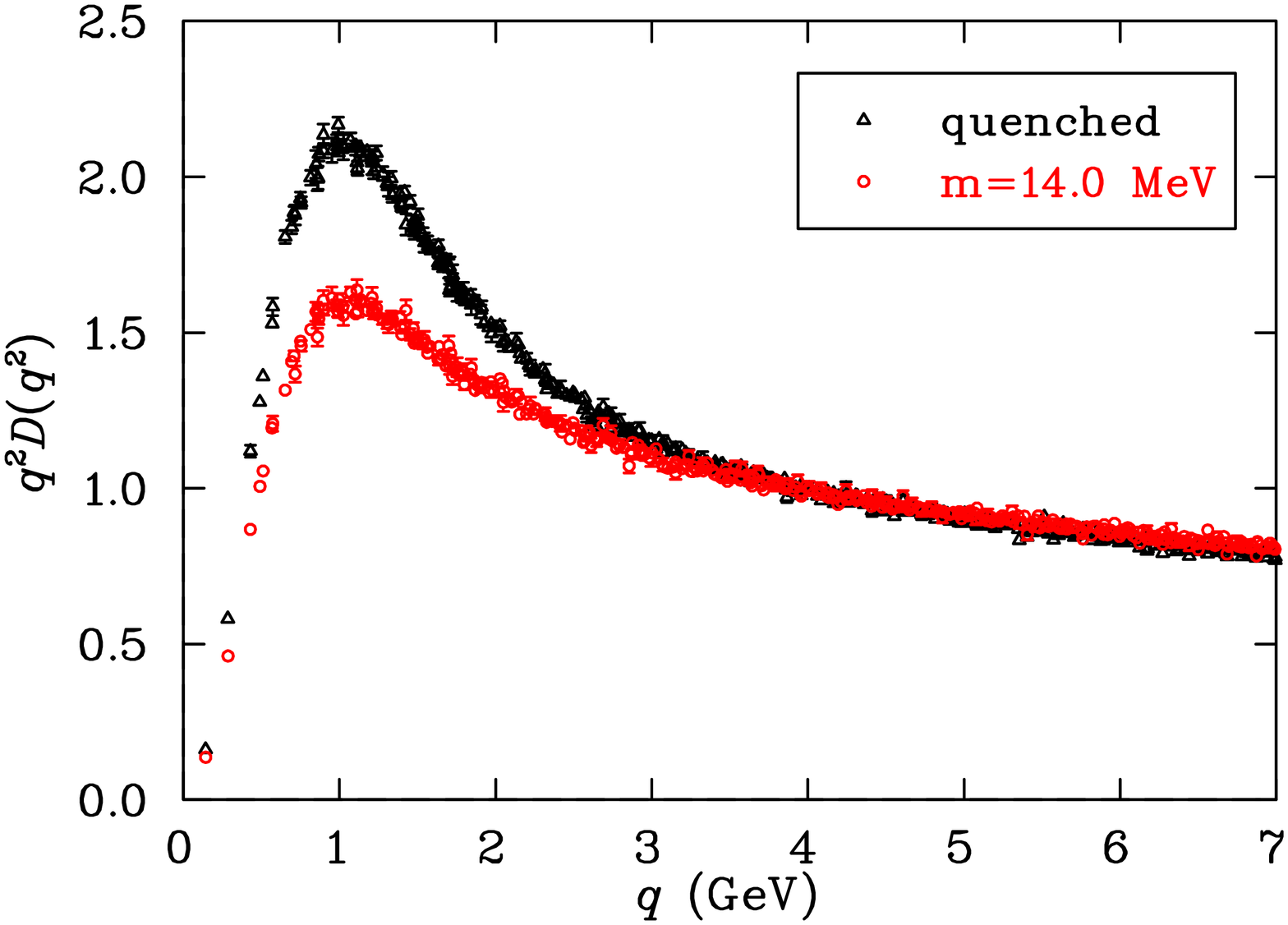}
&
\includegraphics[width=2.0in,angle=90]{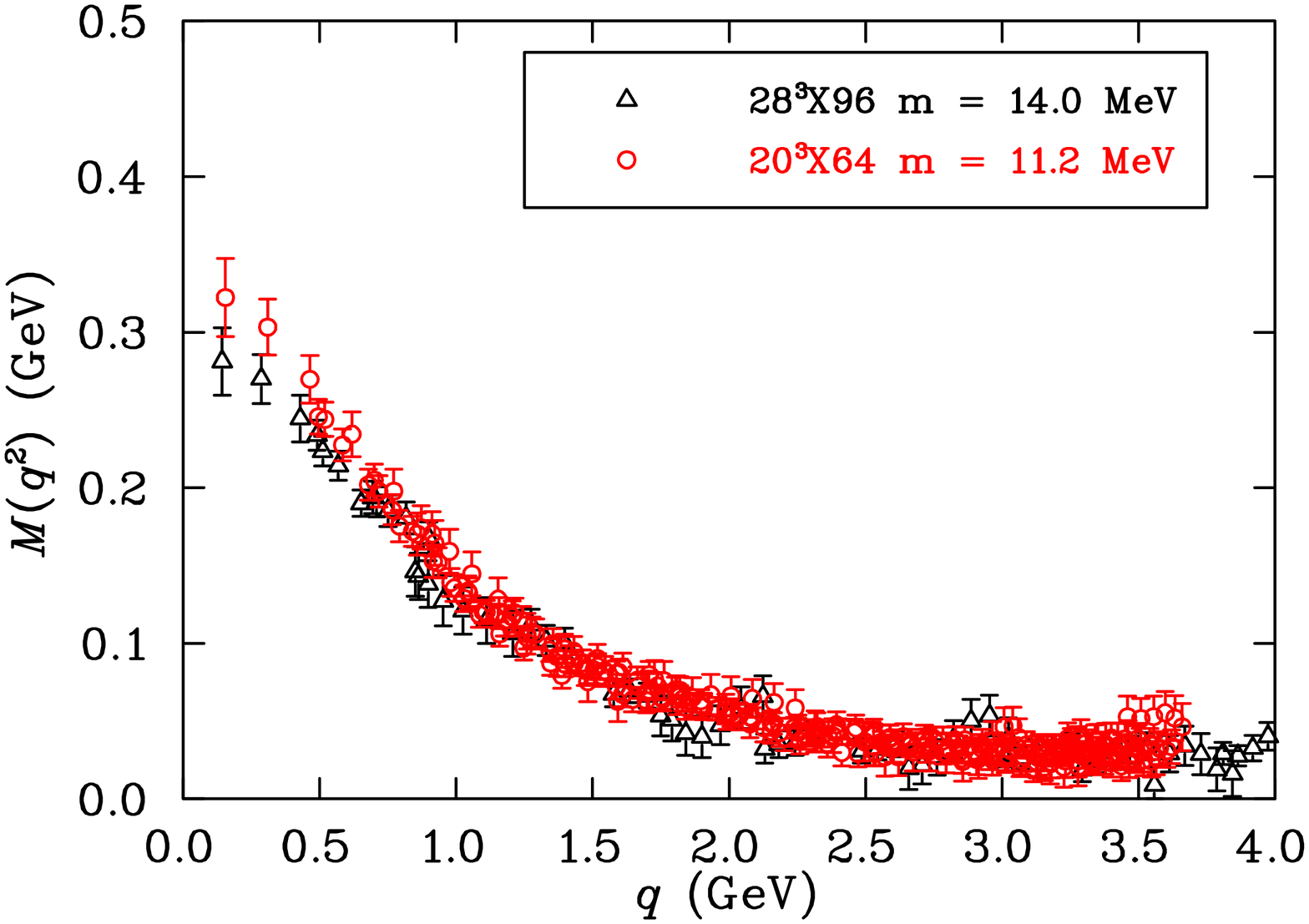}
\end{tabular}
\end{center}
\caption{The gluon dressing function $q^2 D(q^2)$ for quenched and
dynamical configurations with $a \approx 0.09$ fm, from
\textcite{Bowman:2007du} (left),
and the quark mass function for light sea quark mass in full QCD
at $a \approx 0.12$ and $0.09$ fm, from
\textcite{Parappilly:2005ei} (right).
}
\label{fig:gprop-qprop}
\end{figure}

The Landau gauge gluon propagator has been studied in full QCD
using MILC lattices by \textcite{Bowman:2004jm,Bowman:2007du}.
In the continuum, the Landau gauge gluon propagator has the tensor
structure
\begin{equation}
D^{ab}_{\mu\nu}(q) = \left( \delta_{\mu\nu} - \frac{q_\mu q_\nu}{q^2} \right)
 \delta^{ab} D(q^2) ~,
\label{eq:Land_prop}
\end{equation}
where, at tree level $D(q^2) = 1/q^2$. The bare propagator is related
to the renormalized propagator $D_R(q^2;\mu)$ by the renormalization
condition
\begin{equation}
D(q^2,a) = Z_3(a;\mu) D_R(q^2;\mu) ~,~~~~
D_R(q^2;\mu)|_{q^2 = \mu^2} = \frac{1}{\mu^2} ~.
\label{eq:renorm_gprop}
\end{equation}
The gluon propagator in full QCD is somewhat less enhanced for momenta
around $1$ \GeV\ than the quenched propagator, see
Fig.~\ref{fig:gprop-qprop} (left),
and shows good scaling behavior \cite{Bowman:2007du}.
The gluon spectral function shows clear violations of positivity in
qualitative agreement with Dyson-Schwinger equation studies
(see \textcite{Fischer:2006ub} and references therein).

The quark propagator has been studied in full QCD using MILC
lattice ensembles with lattice spacings $a \approx 0.12$ and $0.09$ fm
in \textcite{Bowman:2005vx}, \textcite{Parappilly:2005ei} and
\textcite{Furui:2006ks}.
The bare propagator can be parametrized, and related to the renormalized
propagator, by
\begin{equation}
S(p^2;a) = Z(p^2;a) [ i \gamma \cdot p + M(p^2) ]^{-1} =
 Z_2(a;\mu) S_R(p^2;\mu) ~,
\label{eq:renorm_qprop}
\end{equation}
where $Z_2(a;\mu) = Z(p^2;a)|_{p^2 = \mu^2}$, and the mass function
$M(p^2)$ is renormalization point independent.
Its asymptotic behavior as $p \to \infty$ is related via the OPE
to the RGI quark mass and the chiral condensate, see, {\it e.g.,}
\textcite{Bowman:2005zi}.

The quark mass function for light sea quark mass in full QCD simulations at
two different lattice spacings is shown in Fig.~\ref{fig:gprop-qprop}
(right).  It shows good scaling and clear indication of dynamical mass
generation (``constituent mass'') at low momenta.

\subsection{Further uses of MILC lattices}
\label{sec:further_calcs}

Besides the calculations described in the preceding subsections, the
MILC lattice ensembles have been used in other QCD calculations.
These include the study of hadronic scattering lengths and $n$-body
interactions, reviewed in \textcite{Beane:2008dv}. Furthermore,
computations of nucleon structure, moments of parton and generalized
parton distribution functions, axial nucleon couplings, electromagnetic
form factors, and nucleon transition amplitudes have been done
using MILC lattice ensembles -- see
\textcite{Orginos:2006zz}, \textcite{Hagler:2007hu} and \textcite{Zanotti:2008zm} for recent
reviews of lattice computations of these quantities.

%% file: RMP_sec10.tex
% File for section 10 for RMP article
%
%\section{Section 10}
\section{Further improvements: A look to the future}
\label{sec:future}

While the lattice QCD simulations described in this review are quite
mature, the errors of many of the observables computed can be reduced
in various ways.
Many of the calculations have omitted some of the available MILC
ensembles, in particular the more challenging ones with small lattice
spacings.
Sometimes, not all the available configurations in an ensemble
have been analyzed. Electromagnetic effects, where needed, have been taken
from nonlattice estimates (see Sec.~\ref{sec:fpi}). They can be included
directly in lattice simulations.  Discretization effects coming from the
fermion actions used can be further reduced by using improvements to the
Fermilab action for heavy quarks, and by using highly improved staggered
quarks for both valence and sea light quarks. These improvements are
briefly outlined in this section.

\subsection{Impact of new ensembles}
\label{sec:new_ensembles}

The superfine ($a\approx 0.06$~fm) and
ultrafine ($a\approx 0.045$~fm)
ensembles listed in Table~I were completed only during the past year,
as was the coarse ($a\approx 0.12$~fm) ensemble with three degenerate
light quarks.  The fine ensembles with $m_l/m_s=0.05$ and with three
degenerate light quarks are still running, but should be completed in
the near future.  In this paper, we have presented some preliminary
results from the superfine ensembles for the hadron spectrum, the
light pseudoscalar mesons and the topological susceptibility, and the
HPQCD/UKQCD collaboration has recently used some of the superfine ensembles in its
studies of charmed physics~\cite{Davies:2008hs}; however, the physics
analysis of the new ensembles is in a very early stage. When it is
completed, we expect these ensembles to have a major impact on many of
the calculations described above.

As indicated earlier, the leading finite lattice spacing artifacts
for the asqtad action are of order $a^2/\log(a)$. So these artifacts
for the superfine and
ultrafine ensembles are down from those of the fine
ensembles by factors of 2.6 and 5.2 respectively. As one can see from
Figs.~\ref{fig:toposuscfig}, \ref{fig:mnuc} and \ref{fig:decay-constants},
results obtained to date from the superfine ensembles are very close to
the \rschpt continuum extrapolations, which should significantly reduce
discretization errors in calculations that make use of them.  Furthermore,
as is illustrated in Fig.~\ref{fig:taste-split}, the decrease in taste
splitting among the pions with decreasing lattice spacing is consistent
with $a^2/\log(a)^2$, as expected. Thus, this major source of systematic
error will be significantly reduced by use of the superfine ensembles.

The $a\approx 0.045$~fm, $m_l=0.2\, m_s$ ensemble will provide an anchor
point for extrapolations to the continuum limit, and is particularly
important for calculations which use the Fermilab method for heavy valence
quarks.  For many of these quantities the discretization errors in the
heavy-quark action are the largest single source of systematic error.
Although the size of heavy-quark discretization errors can be estimated
using power-counting arguments, the precise form of the lattice spacing
dependence is not explicitly known.  It is thus important to have a range
of lattice spacings in order to study the heavy-quark discretization
effects.  The heavy-quark errors decrease as $a/\log(a)$ at the worst,
so we expect the 0.045~fm ensemble to reduce the heavy-quark errors by
a factor of two in quantities of interest involving B and D mesons,
which thus far have only been computed on ensembles with lattice
spacings $a\approx 0.09$~fm and larger. The reduction of the heavy-quark
discretization errors does not require the full set of light-quark
masses that we have calculated at coarser lattice spacings; thus, we
have generated only one ensemble at $a\approx 0.045$~fm.  By including
the superfine and
ultrafine ensembles into our work on heavy-light mesons,
in conjunction with improving the statistics,
we expect to determine the leptonic decay constants, the mixing parameters
and the corresponding semileptonic form factors to an accuracy of better
than 5\%.

The physical strange quark mass is not light enough for chiral
perturbation theory to converge rapidly in its vicinity.  To anchor chiral
fits and to test the convergence of chiral perturbation theory, it is
therefore extremely helpful to have ensembles with the strange sea quark
mass held fixed at a value well below the physical strange quark mass.
Furthermore, with three dynamical quark flavors, there are two interesting
chiral limits to be considered: the two-flavor limit, in which the $u$
and $d$ quarks become massless while the $s$ stays at its physical mass,
and the three-flavor chiral limit, where all three quarks become massless.
The difference of various quantities in these two limits is an important
probe of the nature of chiral symmetry breaking in QCD.  The extrapolation
to $m_s=0$ necessary for the three-flavor chiral limit is a long one,
with attendant large errors.  The new ensembles with three degenerate
light quarks were created to help address these issues.  We estimate
that incorporating all the superfine ensembles into the analysis, as
well as all the configurations with the strange sea quark mass held
fixed below its physical value, will allow us to reduce the systematic
errors on $f_\pi$ and $f_K$ to $2\%$ or better, and should dramatically
reduce the errors in low energy constants and quantities such as the
ratio of the two flavor to three flavor condensates, $\langle\bar
uu\rangle_{2}/\langle\bar uu\rangle_{3}$. This would be an important
milestone for lattice QCD calculations.  We also expect corresponding
improvements in other physical quantities of interest.  In particular,
our evaluation of $|V_{us}|$ should become significantly more accurate
than the current world average.

\subsection{Electromagnetic and isospin breaking effects}
\label{subsection:electromagnetic}

Most lattice calculations have not included electromagnetic or isospin
breaking effects.  However, as the precision of calculations increases,
including these effects will become increasingly important.  In fact,
we have already seen in Sec.~\ref{sec:fpi} that electromagnetic effects
are important in the determination of the $u$ and $d$ quark masses.
Another interesting challenge for lattice QCD would be to determine the
proton-neutron mass difference, which will require accounting for the
differences of both the $u$ and $d$ quark masses and their charges.

The pioneering work by \textcite*{Duncan:1996xy,Duncan:1996be}
regarding electromagnetic effects was done with quenched U(1) and
quenched SU(3) fields.  More recently, the RBC collaboration has been
pursuing such calculations but with domain-wall dynamical quarks.
In \textcite{Yamada:2005dv} and \textcite{Blum:2007cy}, electromagnetic
effects on $\pi$ and $K$ meson masses were calculated in $N_f=2$
configurations. \textcite*{Beane:2006fk} have used MILC configurations
with $a\approx 0.12$ fm to study isospin breaking for the nucleons using
domain-wall valence quarks.

Electromagnetic effects in lowest order chiral perturbation theory were
first studied some 40 years ago by \textcite{Dashen:1969eg}.  A key
result known as Dashen's Theorem is that electromagnetic splittings of
the pions and kaons are equal at this order, \ie
\begin{equation}
\Delta M_D^2 = \Delta M_K^2 - \Delta M_\pi^2 = \left( M_{K^\pm}^2
- M_{K^0}^2 \right)_{\rm em} - \left( M_{\pi^\pm}^2 - M_{\pi^0}^2
\right)_{\rm em}
\end{equation}
vanishes.

Recently, \textcite{Bijnens:2006mk} have calculated electromagnetic
corrections in partially quenched perturbation theory, which are
particularly pertinent for analysis of lattice QCD calculations.  They
have emphasized that a combination of meson masses with varying charges
and quark masses is a very close approximation
to $\Delta M_D^2$:
\begin{eqnarray}
\Delta M^2  &=& M^2(\chi_1,\chi_3,q_1,q_3)
- M^2(\chi_1,\chi_3,q_3,q_3) \nonumber \\
&-&  M^2(\chi_1,\chi_1,q_1,q_3) +
 M^2(\chi_1,\chi_1,q_3,q_3). \label{dash_eq}
\end{eqnarray}
Here $\chi_i = 2 B m_{q_i}$, where $B$ is the continuum version of
the low energy constant defined in \eq{taste-split}, and $q_i$ is the
quark charge.  In their notation, $i=1(3)$ refers to the valence $u$
($d$) quark, respectively.

\begin{figure}[t]
\begin{center}
\vspace{0.25in}
\includegraphics[width=0.54\textwidth]{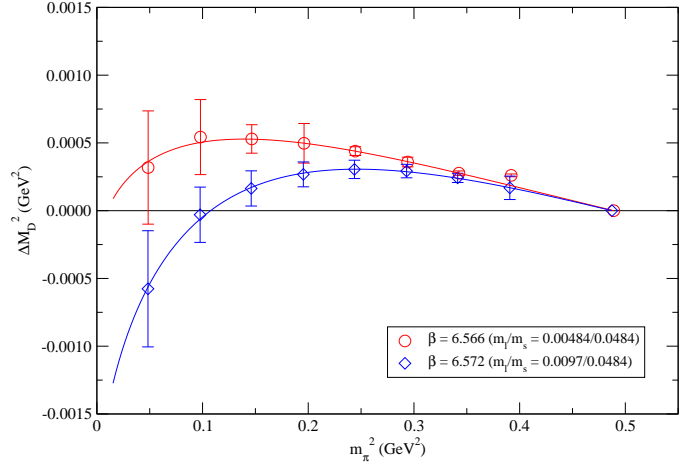}
\end{center}
\caption{Correction to Dashen's theorem, as a function of the LO $\pi$
mass squared (equivalent to the pion mass squared with $e^2=0$).
Figure from \textcite{Basak:2008na}.
}
\label{fig:delta-mD-squared}
\end{figure}

MILC has recently begun to explore electromagnetic effects on the
pseudoscalar masses \cite{Basak:2008na}, using the quenched approximation
for electromagnetism.  The initial study on $a\approx 0.15$ fm ensembles
yielded promising results.  The key result is a rough estimate of the
correction to Dashen's theorem.  In Fig.~\ref{fig:delta-mD-squared}, we
show results for two dynamical ensembles for various light valence masses.
After fitting the results and performing the chiral extrapolation, we find
that $0.7 \times 10^{-3} {\rm \GeV}^2 < \Delta M^2_D < 1.8 \times 10^{-3}
{\rm \GeV}^2 $.  A recent phenomenological estimate is $1.07\times 10^{-3}
{\rm \GeV}^2 $ \cite{Bijnens:2006mk}.

It will be very interesting to extend this work to smaller lattice
spacings and
eventually to include
dynamical electromagnetic effects.
There is also the prospect of including isospin breaking in the generation
of the configurations.

\subsection{Heavy Wilson fermion improvement program}
\label{sec:imp_FNAL}

The leading discretization errors contained in the Wilson/clover action
applied to heavy quarks have been analyzed in \textcite{Oktay:2008ex}, in
an extension to the original Fermilab formalism.  Since the heavy quarks
introduce an additional scale $1/m_Q$, they consider all the  operators
which have power counting of $\lambda^3$ ($\lambda \sim \Lambda a$
or $\Lambda/m_Q$) and $v^6$ for the heavy-light (HQET) and heavy-heavy
(NRQCD) systems, respectively.  This leads to actions containing  all
possible dimension six and some dimension seven operators.  Many of
these are redundant and  may be chosen for calculational convenience by
considering field transformations.  For example, multihop time derivative
operators (which spoil nice properties of the transfer matrix) may
be eliminated in this way.  Tree-level matching of observables in the
continuum and lattice QCD actions shows that six new operators beyond the
original Fermilab action are required at this level of improvement, four
of dimension six and two of dimension seven. In all, there are a total of
nineteen nonredundant operators at this level, and one-loop matching will
presumably introduce more of these.  One can estimate the uncertainties
due to nonzero lattice spacing by calculating the mismatch between the
lattice short-distance coefficients and their continuum counterparts.
Initial estimates show that the new lattice action reduces the errors
to the few-percent level.

\subsection{Preliminary studies of the HISQ action}
\label{sec:HISQ_study}

As discussed in \secref{Asqtad-formalism}, the HISQ action improves
taste symmetry and is well suited for future studies with dynamical
quarks.
Subtleties with dynamical HISQ simulations, in particular from
the reunitarization step, \eq{Wlinks}, which can lead to large
contributions to the force, are described in \textcite{Bazavov:2009jc}.

\begin{figure}[t]
\begin{center}
\vspace{0.25in}
\includegraphics[width=0.54\textwidth]{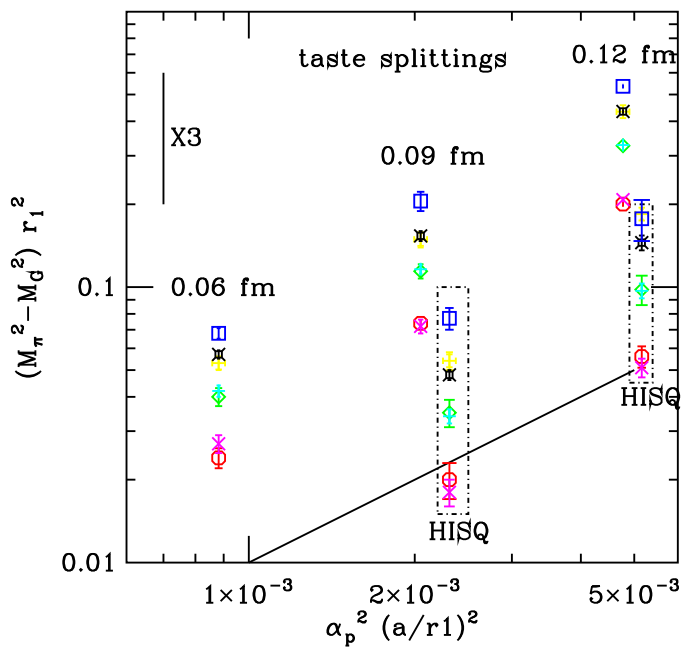}
\end{center}
\caption{
The taste splittings as function of $\alpha^2a^2$ for the asqtad and
HISQ actions (with the latter indicated by dashed boxes).}
\label{fig:taste_split_hisq}
\end{figure}

The first study of how the HISQ action reduces the splitting between
different tastes of pions was undertaken by the HPQCD and UKQCD
collaborations in \textcite{Follana:2006rc}.  They used valence HISQ on
the asqtad sea quark configurations generated by MILC. Similar findings
for HISQ sea quarks were reported in \textcite{Bazavov:2009jc}. The
results of a more recent study are
summarized in
Fig.~\ref{fig:taste_split_hisq}:
The splittings between the Goldstone and the other
pion tastes for the HISQ action are reduced by a factor of 2.5--3 compared 
to asqtad (notice a vertical line that indicates a factor of 3 in
logarithmic scale in Fig.~\ref{fig:taste_split_hisq}). Two HISQ ensembles,
with $a\approx0.09$ and 0.12 fm, are shown. 
The difference between the results presented
here and in \textcite{Bazavov:2009jc} is that
the current study uses
the
improved gauge action
with
the one-loop fermion corrections
induced by the HISQ fermions \cite{Hart:2008zi,Hart:2008sq},
and the ensembles were tuned to be close to the line of constant physics
with $m_l=0.2m_s$.

\newpage

%% file: RMP_sec11.tex
% File for section 11 for RMP article
%
\section{Summary and conclusions}
\label{sec:conclusion}

There has been a dramatic improvement in the accuracy of lattice QCD
calculations over the past decade due to a combination of developments:

\begin{itemize}
\item The use of improved actions significantly reduces finite lattice
spacing artifacts, greatly improving the accuracy of extrapolations
to the continuum limit. The asqtad improved staggered quark action
the MILC collaboration has 
used provides a particularly strong reduction in taste
symmetry breaking, the most challenging finite lattice spacing artifact
for staggered quarks.  The HISQ action
improves on asqtad in
this respect by an additional factor of three.
In general,
one finds that a HISQ ensemble has lattice artifacts approximately half
the size of an asqtad ensemble with the same lattice spacing.

\item The inclusion of up, down and strange sea quarks with realistic
masses is critical for reducing errors to the few percent level, as is
illustrated in Fig.~\ref{fig:ratio-plots}.

\item The use of partially quenched chiral perturbation theory and,
for staggered quarks, rooted staggered chiral perturbation theory have
greatly improved the accuracy of the extrapolation of lattice data to
the physical masses of the up, down and strange quarks.

\item Improved algorithms, such as RHMC, have enabled the generation of
gauge field ensembles with significantly smaller lattice spacings and
lighter quark masses than had previously been possible. These new algorithms
have changed the balance between gauge field configuration generation
and physics analysis on the configurations. Whereas the former used to
take the bulk of the computing resources, now the resources required for
an analysis project often rival those that went into the generation of
the configurations.

\item The vastly increased computing resources available to lattice gauge
theorists over the past decade have enabled us to take advantage of the
developments enumerated above. For example, between 1999 and 2008, the
total floating point operations used per year by the MILC Collaboration
increased by approximately three orders of magnitude.

\end{itemize}

The MILC collaboration has
taken advantage of these developments to generate, over the past
ten years, the ensembles of asqtad gauge field configurations 
detailed in
Table~\ref{table:runtable1}.  This is the first set of  ensembles to have
a wide enough range of small lattice spacings and light quark masses to
enable controlled extrapolations of physical quantities to the continuum
and chiral limits. These ensembles are publicly available, and we and
others are using them to calculate a wide range of physical quantities of
interest in high energy and nuclear physics. 
This work has included calculations of the strong coupling constant,
the masses of light quarks and hadrons, the properties
of light pseudoscalar mesons, the topological susceptibility, 
the masses, decays and mixings of heavy-light mesons,
the charmonium and bottomonium spectra, 
the $K^0-\bar{K^0}$ mixing parameter $B_K$, the mass of the $B_c$ meson, the
$\pi-\pi$ and $N-N$ scattering lengths, generalized parton distributions,
and hadronic contributions to the muon anomalous magnetic moment.
The errors in these quantities have typically decreased by an 
order of magnitude as the library of ensembles has grown, with further 
improvements expected as the superfine and ultrafine ensembles are fully analyzed,
and HISQ ensembles become available.

A number of quantities have
been calculated to an accuracy of a few percent, and some predictions
have been made that were later verified by experiment. The work of the
Fermilab Lattice, MILC and HPQCD/UKQCD collaborations on the decays
and mixings of heavy-light mesons and the decays of light pseudoscalar
mesons has reached a level of accuracy where it is having a significant
impact on tests of the standard model and the search for new physics.
However, high precision has been obtained only for quantities that
are most straightforward to calculate. There are many quantities, such
as scattering phase shifts, the masses and widths of hadrons that are
unstable under the strong interactions, and parton distribution functions,
which are of great interest, but continue to pose major challenges.

Because it is relatively inexpensive to simulate, the asqtad quark
action was the first to produce a set of gauge field ensembles with a
wide enough range of lattice spacings and sea quark masses to enable
controlled extrapolations to the continuum and chiral limit.  However,
such ensembles are also being produced with other quark actions, such as
Wilson-clover, twisted mass, domain wall and overlap. These ensembles are
already producing impressive results. Over the next few years one can
expect major advances on a wide variety of calculations with critical
checks coming from the use of different lattice formulations of QCD.
Finally, the techniques that have been developed for the study of QCD
can be applied to study many of the theories that have been proposed
for physics beyond the standard model. Such work is just beginning,
but appears to have a very bright future.

%% file: RMP_main.bbl
\begin{thebibliography}{384}
\expandafter\ifx\csname natexlab\endcsname\relax\def\natexlab#1{#1}\fi
\expandafter\ifx\csname bibnamefont\endcsname\relax
  \def\bibnamefont#1{#1}\fi
\expandafter\ifx\csname bibfnamefont\endcsname\relax
  \def\bibfnamefont#1{#1}\fi
\expandafter\ifx\csname citenamefont\endcsname\relax
  \def\citenamefont#1{#1}\fi
\expandafter\ifx\csname url\endcsname\relax
  \def\url#1{\texttt{#1}}\fi
\expandafter\ifx\csname urlprefix\endcsname\relax\def\urlprefix{URL }\fi
\providecommand{\bibinfo}[2]{#2}
\providecommand{\eprint}[2][]{\url{#2}}

\bibitem[{Abe \emph{et~al.}(2005)\citenamefont{Abe} \emph{et~al.}}]{Abe:2005sh}
\bibinfo{author}{\bibnamefont{Abe}, \bibfnamefont{K.}}, \emph{et~al.}
  (\bibinfo{collaboration}{BELLE}), \bibinfo{year}{2005},
  \arXiv{hep-ex/0510003}.

\bibitem[{Abulencia \emph{et~al.}(2006)\citenamefont{Abulencia}
  \emph{et~al.}}]{Abulencia:2005usa}
\bibinfo{author}{\bibnamefont{Abulencia}, \bibfnamefont{A.}}, \emph{et~al.}
  (\bibinfo{collaboration}{CDF}), \bibinfo{year}{2006}, \bibinfo{journal}{Phys.
  Rev. Lett.} \textbf{\bibinfo{volume}{96}}, \bibinfo{pages}{082002},
  \arXiv{hep-ex/0505076}.

\bibitem[{\citenamefont{Adams}(2005)}]{Adams:2004mf}
\bibinfo{author}{\bibnamefont{Adams}, \bibfnamefont{D.~H.}},
  \bibinfo{year}{2005}, \bibinfo{journal}{Phys. Rev.}
  \textbf{\bibinfo{volume}{D72}}, \bibinfo{pages}{114512},
  \arXiv{hep-lat/0411030}.

\bibitem[{\citenamefont{Adams}(2008)}]{Adams:2008db}
\bibinfo{author}{\bibnamefont{Adams}, \bibfnamefont{D.~H.}},
  \bibinfo{year}{2008}, \bibinfo{journal}{Phys. Rev.}
  \textbf{\bibinfo{volume}{D77}}, \bibinfo{pages}{105024}, \arXiv{0802.3029}.

\bibitem[{Albanese \emph{et~al.}(1987)\citenamefont{Albanese}
  \emph{et~al.}}]{Albanese:1987ds}
\bibinfo{author}{\bibnamefont{Albanese}, \bibfnamefont{M.}}, \emph{et~al.}
  (\bibinfo{collaboration}{APE}), \bibinfo{year}{1987}, \bibinfo{journal}{Phys.
  Lett.} \textbf{\bibinfo{volume}{B192}}, \bibinfo{pages}{163}.

\bibitem[{\citenamefont{Alexander}(2009)}]{Alexander:2009ux}
\bibinfo{author}{\bibnamefont{Alexander}, \bibfnamefont{.~J.~P.}}
  (\bibinfo{collaboration}{CLEO}), \bibinfo{year}{2009},
  \bibinfo{journal}{Phys. Rev.} \textbf{\bibinfo{volume}{D79}},
  \bibinfo{pages}{052001}, \arXiv{0901.1216}.

\bibitem[{\citenamefont{Alford} \emph{et~al.}(1995)\citenamefont{Alford, Dimm,
  Lepage, Hockney, and Mackenzie}}]{Alford:1995hw}
\bibinfo{author}{\bibnamefont{Alford}, \bibfnamefont{M.~G.}},
  \bibinfo{author}{\bibfnamefont{W.}~\bibnamefont{Dimm}},
  \bibinfo{author}{\bibfnamefont{G.~P.} \bibnamefont{Lepage}},
  \bibinfo{author}{\bibfnamefont{G.}~\bibnamefont{Hockney}}, and
  \bibinfo{author}{\bibfnamefont{P.~B.} \bibnamefont{Mackenzie}},
  \bibinfo{year}{1995}, \bibinfo{journal}{Phys. Lett.}
  \textbf{\bibinfo{volume}{B361}}, \bibinfo{pages}{87},
  \arXiv{hep-lat/9507010}.

\bibitem[{Allison \emph{et~al.}(2008)\citenamefont{Allison}
  \emph{et~al.}}]{Allison:2008xk}
\bibinfo{author}{\bibnamefont{Allison}, \bibfnamefont{I.}}, \emph{et~al.}
  (\bibinfo{collaboration}{HPQCD}), \bibinfo{year}{2008},
  \bibinfo{journal}{Phys. Rev.} \textbf{\bibinfo{volume}{D78}},
  \bibinfo{pages}{054513}, \arXiv{0805.2999}.

\bibitem[{Allison \emph{et~al.}(2005)\citenamefont{Allison}
  \emph{et~al.}}]{Allison:2004be}
\bibinfo{author}{\bibnamefont{Allison}, \bibfnamefont{I.~F.}}, \emph{et~al.}
  (\bibinfo{collaboration}{HPQCD}), \bibinfo{year}{2005},
  \bibinfo{journal}{Phys. Rev. Lett.} \textbf{\bibinfo{volume}{94}},
  \bibinfo{pages}{172001}, \arXiv{hep-lat/0411027}.

\bibitem[{Allton \emph{et~al.}(2008)\citenamefont{Allton}
  \emph{et~al.}}]{Allton:2008pn}
\bibinfo{author}{\bibnamefont{Allton}, \bibfnamefont{C.}}, \emph{et~al.}
  (\bibinfo{collaboration}{RBC-UKQCD}), \bibinfo{year}{2008},
  \bibinfo{journal}{Phys. Rev.} \textbf{\bibinfo{volume}{D78}},
  \bibinfo{pages}{114509}, \arXiv{0804.0473}.

\bibitem[{\citenamefont{Allton}(1996)}]{Allton:1996kr}
\bibinfo{author}{\bibnamefont{Allton}, \bibfnamefont{C.~R.}},
  \bibinfo{year}{1996}, \arXiv{hep-lat/9610016}.

\bibitem[{Ambrosino \emph{et~al.}(2006{\natexlab{a}})\citenamefont{Ambrosino}
  \emph{et~al.}}]{Ambrosino:2005fw}
\bibinfo{author}{\bibnamefont{Ambrosino}, \bibfnamefont{F.}}, \emph{et~al.}
  (\bibinfo{collaboration}{KLOE}), \bibinfo{year}{2006}{\natexlab{a}},
  \bibinfo{journal}{Phys. Lett.} \textbf{\bibinfo{volume}{B632}},
  \bibinfo{pages}{76}, \arXiv{hep-ex/0509045}.

\bibitem[{Ambrosino \emph{et~al.}(2006{\natexlab{b}})\citenamefont{Ambrosino}
  \emph{et~al.}}]{Ambrosino:2005ec}
\bibinfo{author}{\bibnamefont{Ambrosino}, \bibfnamefont{F.}}, \emph{et~al.}
  (\bibinfo{collaboration}{KLOE}), \bibinfo{year}{2006}{\natexlab{b}},
  \bibinfo{journal}{Phys. Lett.} \textbf{\bibinfo{volume}{B632}},
  \bibinfo{pages}{43}, \arXiv{hep-ex/0508027}.

\bibitem[{Amsler \emph{et~al.}(2008)\citenamefont{Amsler}
  \emph{et~al.}}]{Amsler:2008zzb}
\bibinfo{author}{\bibnamefont{Amsler}, \bibfnamefont{C.}}, \emph{et~al.}
  (\bibinfo{collaboration}{Particle Data Group}), \bibinfo{year}{2008},
  \bibinfo{journal}{Phys. Lett.} \textbf{\bibinfo{volume}{B667}},
  \bibinfo{pages}{1}.

\bibitem[{\citenamefont{Antonelli}(2007)}]{Antonelli:2007mj}
\bibinfo{author}{\bibnamefont{Antonelli}, \bibfnamefont{M.}},
  \bibinfo{year}{2007}, \arXiv{0712.0734}.

\bibitem[{Aoki \emph{et~al.}(2009)\citenamefont{Aoki}
  \emph{et~al.}}]{Aoki:2008sm}
\bibinfo{author}{\bibnamefont{Aoki}, \bibfnamefont{S.}}, \emph{et~al.}
  (\bibinfo{collaboration}{PACS-CS}), \bibinfo{year}{2009},
  \bibinfo{journal}{Phys. Rev.} \textbf{\bibinfo{volume}{D79}},
  \bibinfo{pages}{034503}, \arXiv{0807.1661}.

\bibitem[{\citenamefont{Arnesen} \emph{et~al.}(2005)\citenamefont{Arnesen,
  Grinstein, Rothstein, and Stewart}}]{Arnesen:2005ez}
\bibinfo{author}{\bibnamefont{Arnesen}, \bibfnamefont{M.~C.}},
  \bibinfo{author}{\bibfnamefont{B.}~\bibnamefont{Grinstein}},
  \bibinfo{author}{\bibfnamefont{I.~Z.} \bibnamefont{Rothstein}}, and
  \bibinfo{author}{\bibfnamefont{I.~W.} \bibnamefont{Stewart}},
  \bibinfo{year}{2005}, \bibinfo{journal}{Phys. Rev. Lett.}
  \textbf{\bibinfo{volume}{95}}, \bibinfo{pages}{071802},
  \arXiv{hep-ph/0504209}.

\bibitem[{Aubert \emph{et~al.}(2007)\citenamefont{Aubert}
  \emph{et~al.}}]{Aubert:2006px}
\bibinfo{author}{\bibnamefont{Aubert}, \bibfnamefont{B.}}, \emph{et~al.}
  (\bibinfo{collaboration}{BABAR}), \bibinfo{year}{2007},
  \bibinfo{journal}{Phys. Rev. Lett.} \textbf{\bibinfo{volume}{98}},
  \bibinfo{pages}{091801}, \arXiv{hep-ex/0612020}.

\bibitem[{Aubert \emph{et~al.}(2008)\citenamefont{Aubert}
  \emph{et~al.}}]{Babar:2008vj}
\bibinfo{author}{\bibnamefont{Aubert}, \bibfnamefont{B.}}, \emph{et~al.}
  (\bibinfo{collaboration}{BABAR}), \bibinfo{year}{2008},
  \bibinfo{journal}{Phys. Rev. Lett.} \textbf{\bibinfo{volume}{101}},
  \bibinfo{pages}{071801}, \bibinfo{note}{{Erratum-ibid.\ {\bf 102}, 029901
  (2009)}}, \arXiv{0807.1086}.

\bibitem[{Aubert \emph{et~al.}(2009)\citenamefont{Aubert}
  \emph{et~al.}}]{Babar:2009pz}
\bibinfo{author}{\bibnamefont{Aubert}, \bibfnamefont{B.}}, \emph{et~al.}
  (\bibinfo{collaboration}{BABAR}), \bibinfo{year}{2009}, \arXiv{0903.1124}.

\bibitem[{\citenamefont{Aubin and Bernard}(2003{\natexlab{a}})}]{Aubin:2003mg}
\bibinfo{author}{\bibnamefont{Aubin}, \bibfnamefont{C.}}, and
  \bibinfo{author}{\bibfnamefont{C.}~\bibnamefont{Bernard}},
  \bibinfo{year}{2003}{\natexlab{a}}, \bibinfo{journal}{Phys. Rev.}
  \textbf{\bibinfo{volume}{D68}}, \bibinfo{pages}{034014},
  \arXiv{hep-lat/0304014}.

\bibitem[{\citenamefont{Aubin and Bernard}(2003{\natexlab{b}})}]{Aubin:2003uc}
\bibinfo{author}{\bibnamefont{Aubin}, \bibfnamefont{C.}}, and
  \bibinfo{author}{\bibfnamefont{C.}~\bibnamefont{Bernard}},
  \bibinfo{year}{2003}{\natexlab{b}}, \bibinfo{journal}{Phys. Rev.}
  \textbf{\bibinfo{volume}{D68}}, \bibinfo{pages}{074011},
  \arXiv{hep-lat/0306026}.

\bibitem[{\citenamefont{Aubin and Bernard}(2004)}]{Aubin:2003rg}
\bibinfo{author}{\bibnamefont{Aubin}, \bibfnamefont{C.}}, and
  \bibinfo{author}{\bibfnamefont{C.}~\bibnamefont{Bernard}},
  \bibinfo{year}{2004}, \bibinfo{journal}{Nucl. Phys. Proc. Suppl.}
  \textbf{\bibinfo{volume}{129}}, \bibinfo{pages}{182},
  \arXiv{hep-lat/0308036}.

\bibitem[{\citenamefont{Aubin and Bernard}(2006)}]{Aubin:2005aq}
\bibinfo{author}{\bibnamefont{Aubin}, \bibfnamefont{C.}}, and
  \bibinfo{author}{\bibfnamefont{C.}~\bibnamefont{Bernard}},
  \bibinfo{year}{2006}, \bibinfo{journal}{Phys. Rev.}
  \textbf{\bibinfo{volume}{D73}}, \bibinfo{pages}{014515},
  \arXiv{hep-lat/0510088}.

\bibitem[{\citenamefont{Aubin and Bernard}(2007)}]{Aubin:2007mc}
\bibinfo{author}{\bibnamefont{Aubin}, \bibfnamefont{C.}}, and
  \bibinfo{author}{\bibfnamefont{C.}~\bibnamefont{Bernard}},
  \bibinfo{year}{2007}, \bibinfo{journal}{Phys. Rev.}
  \textbf{\bibinfo{volume}{D76}}, \bibinfo{pages}{014002}, \arXiv{0704.0795}.

\bibitem[{\citenamefont{Aubin and Blum}(2007)}]{Aubin:2006xv}
\bibinfo{author}{\bibnamefont{Aubin}, \bibfnamefont{C.}}, and
  \bibinfo{author}{\bibfnamefont{T.}~\bibnamefont{Blum}}, \bibinfo{year}{2007},
  \bibinfo{journal}{Phys. Rev.} \textbf{\bibinfo{volume}{D75}},
  \bibinfo{pages}{114502}, \arXiv{hep-lat/0608011}.

\bibitem[{\citenamefont{Aubin}
  \emph{et~al.}(2007{\natexlab{a}})\citenamefont{Aubin, Laiho, and Van~de
  Water}}]{Aubin:2007pt}
\bibinfo{author}{\bibnamefont{Aubin}, \bibfnamefont{C.}},
  \bibinfo{author}{\bibfnamefont{J.}~\bibnamefont{Laiho}}, and
  \bibinfo{author}{\bibfnamefont{R.~S.} \bibnamefont{Van~de Water}},
  \bibinfo{year}{2007}{\natexlab{a}}, \bibinfo{journal}{PoS}
  \textbf{\bibinfo{volume}{LAT2007}}, \bibinfo{pages}{375}, \arXiv{0710.1121}.

\bibitem[{\citenamefont{Aubin}
  \emph{et~al.}(2007{\natexlab{b}})\citenamefont{Aubin, Laiho, and Van~de
  Water}}]{Aubin:2006hg}
\bibinfo{author}{\bibnamefont{Aubin}, \bibfnamefont{C.}},
  \bibinfo{author}{\bibfnamefont{J.}~\bibnamefont{Laiho}}, and
  \bibinfo{author}{\bibfnamefont{R.~S.} \bibnamefont{Van~de Water}},
  \bibinfo{year}{2007}{\natexlab{b}}, \bibinfo{journal}{Phys. Rev.}
  \textbf{\bibinfo{volume}{D75}}, \bibinfo{pages}{034502},
  \arXiv{hep-lat/0609009}.

\bibitem[{\citenamefont{Aubin} \emph{et~al.}(2008)\citenamefont{Aubin, Laiho,
  and Van~de Water}}]{Aubin:2008wk}
\bibinfo{author}{\bibnamefont{Aubin}, \bibfnamefont{C.}},
  \bibinfo{author}{\bibfnamefont{J.}~\bibnamefont{Laiho}}, and
  \bibinfo{author}{\bibfnamefont{R.~S.} \bibnamefont{Van~de Water}},
  \bibinfo{year}{2008}, \bibinfo{journal}{Phys. Rev.}
  \textbf{\bibinfo{volume}{D77}}, \bibinfo{pages}{114501}, \arXiv{0803.0129}.

\bibitem[{\citenamefont{Aubin}
  \emph{et~al.}(2009{\natexlab{a}})\citenamefont{Aubin, Laiho, and Van~de
  Water}}]{Aubin:2008ie}
\bibinfo{author}{\bibnamefont{Aubin}, \bibfnamefont{C.}},
  \bibinfo{author}{\bibfnamefont{J.}~\bibnamefont{Laiho}}, and
  \bibinfo{author}{\bibfnamefont{R.~S.} \bibnamefont{Van~de Water}},
  \bibinfo{year}{2009}{\natexlab{a}}, \bibinfo{journal}{PoS}
  \textbf{\bibinfo{volume}{LAT2008}}, \bibinfo{pages}{105}, \arXiv{0810.4328}.

\bibitem[{\citenamefont{Aubin}
  \emph{et~al.}(2009{\natexlab{b}})\citenamefont{Aubin, Laiho, and Van~de
  Water}}]{Aubin:2009jh}
\bibinfo{author}{\bibnamefont{Aubin}, \bibfnamefont{C.}},
  \bibinfo{author}{\bibfnamefont{J.}~\bibnamefont{Laiho}}, and
  \bibinfo{author}{\bibfnamefont{R.~S.} \bibnamefont{Van~de Water}},
  \bibinfo{year}{2009}{\natexlab{b}}, \arXiv{0905.3947}.

\bibitem[{Aubin \emph{et~al.}(2004{\natexlab{a}})\citenamefont{Aubin}
  \emph{et~al.}}]{Aubin:2004wf}
\bibinfo{author}{\bibnamefont{Aubin}, \bibfnamefont{C.}}, \emph{et~al.},
  \bibinfo{year}{2004}{\natexlab{a}}, \bibinfo{journal}{Phys. Rev.}
  \textbf{\bibinfo{volume}{D70}}, \bibinfo{pages}{094505},
  \arXiv{hep-lat/0402030}.

\bibitem[{Aubin \emph{et~al.}(2004{\natexlab{b}})\citenamefont{Aubin}
  \emph{et~al.}}]{Aubin:2004fs}
\bibinfo{author}{\bibnamefont{Aubin}, \bibfnamefont{C.}}, \emph{et~al.}
  (\bibinfo{collaboration}{MILC}), \bibinfo{year}{2004}{\natexlab{b}},
  \bibinfo{journal}{Phys. Rev.} \textbf{\bibinfo{volume}{D70}},
  \bibinfo{pages}{114501}, \arXiv{hep-lat/0407028}.

\bibitem[{Aubin \emph{et~al.}(2005{\natexlab{a}})\citenamefont{Aubin}
  \emph{et~al.}}]{Aubin:2005ar}
\bibinfo{author}{\bibnamefont{Aubin}, \bibfnamefont{C.}}, \emph{et~al.},
  \bibinfo{year}{2005}{\natexlab{a}}, \bibinfo{journal}{Phys. Rev. Lett.}
  \textbf{\bibinfo{volume}{95}}, \bibinfo{pages}{122002},
  \arXiv{hep-lat/0506030}.

\bibitem[{Aubin \emph{et~al.}(2005{\natexlab{b}})\citenamefont{Aubin}
  \emph{et~al.}}]{Aubin:2004ej}
\bibinfo{author}{\bibnamefont{Aubin}, \bibfnamefont{C.}}, \emph{et~al.}
  (\bibinfo{collaboration}{Fermilab Lattice}),
  \bibinfo{year}{2005}{\natexlab{b}}, \bibinfo{journal}{Phys. Rev. Lett.}
  \textbf{\bibinfo{volume}{94}}, \bibinfo{pages}{011601},
  \arXiv{hep-ph/0408306}.

\bibitem[{Bae \emph{et~al.}(2009)\citenamefont{Bae} \emph{et~al.}}]{Bae:2008tb}
\bibinfo{author}{\bibnamefont{Bae}, \bibfnamefont{T.}}, \emph{et~al.},
  \bibinfo{year}{2009}, \bibinfo{journal}{PoS}
  \textbf{\bibinfo{volume}{LAT2008}}, \bibinfo{pages}{275}, \arXiv{0809.1220}.

\bibitem[{Bailey \emph{et~al.}(2009)\citenamefont{Bailey}
  \emph{et~al.}}]{Bailey:2008wp}
\bibinfo{author}{\bibnamefont{Bailey}, \bibfnamefont{J.}}, \emph{et~al.},
  \bibinfo{year}{2009}, \bibinfo{journal}{Phys. Rev.}
  \textbf{\bibinfo{volume}{D79}}, \bibinfo{pages}{054507}, \arXiv{0811.3640}.

\bibitem[{\citenamefont{Bailey}(2007)}]{Bailey:2006zn}
\bibinfo{author}{\bibnamefont{Bailey}, \bibfnamefont{J.~A.}},
  \bibinfo{year}{2007}, \bibinfo{journal}{Phys. Rev.}
  \textbf{\bibinfo{volume}{D75}}, \bibinfo{pages}{114505},
  \arXiv{hep-lat/0611023}.

\bibitem[{\citenamefont{Ball and Zwicky}(2005)}]{Ball:2004ye}
\bibinfo{author}{\bibnamefont{Ball}, \bibfnamefont{P.}}, and
  \bibinfo{author}{\bibfnamefont{R.}~\bibnamefont{Zwicky}},
  \bibinfo{year}{2005}, \bibinfo{journal}{Phys. Rev.}
  \textbf{\bibinfo{volume}{D71}}, \bibinfo{pages}{014015},
  \arXiv{hep-ph/0406232}.

\bibitem[{\citenamefont{Banks} \emph{et~al.}(1976)\citenamefont{Banks,
  Susskind, and Kogut}}]{Banks:1975gq}
\bibinfo{author}{\bibnamefont{Banks}, \bibfnamefont{T.}},
  \bibinfo{author}{\bibfnamefont{L.}~\bibnamefont{Susskind}}, and
  \bibinfo{author}{\bibfnamefont{J.~B.} \bibnamefont{Kogut}},
  \bibinfo{year}{1976}, \bibinfo{journal}{Phys. Rev.}
  \textbf{\bibinfo{volume}{D13}}, \bibinfo{pages}{1043}.

\bibitem[{Banks \emph{et~al.}(1977)\citenamefont{Banks}
  \emph{et~al.}}]{Banks:1976ia}
\bibinfo{author}{\bibnamefont{Banks}, \bibfnamefont{T.}}, \emph{et~al.}
  (\bibinfo{collaboration}{Cornell-Oxford-Tel Aviv-Yeshiva}),
  \bibinfo{year}{1977}, \bibinfo{journal}{Phys. Rev.}
  \textbf{\bibinfo{volume}{D15}}, \bibinfo{pages}{1111}.

\bibitem[{\citenamefont{B{\"a}r} \emph{et~al.}(2005)\citenamefont{B{\"a}r,
  Bernard, Rupak, and Shoresh}}]{Bar:2005tu}
\bibinfo{author}{\bibnamefont{B{\"a}r}, \bibfnamefont{O.}},
  \bibinfo{author}{\bibfnamefont{C.}~\bibnamefont{Bernard}},
  \bibinfo{author}{\bibfnamefont{G.}~\bibnamefont{Rupak}}, and
  \bibinfo{author}{\bibfnamefont{N.}~\bibnamefont{Shoresh}},
  \bibinfo{year}{2005}, \bibinfo{journal}{Phys. Rev.}
  \textbf{\bibinfo{volume}{D72}}, \bibinfo{pages}{054502},
  \arXiv{hep-lat/0503009}.

\bibitem[{\citenamefont{B{\"a}r} \emph{et~al.}(2003)\citenamefont{B{\"a}r,
  Rupak, and Shoresh}}]{Bar:2002nr}
\bibinfo{author}{\bibnamefont{B{\"a}r}, \bibfnamefont{O.}},
  \bibinfo{author}{\bibfnamefont{G.}~\bibnamefont{Rupak}}, and
  \bibinfo{author}{\bibfnamefont{N.}~\bibnamefont{Shoresh}},
  \bibinfo{year}{2003}, \bibinfo{journal}{Phys. Rev.}
  \textbf{\bibinfo{volume}{D67}}, \bibinfo{pages}{114505},
  \arXiv{hep-lat/0210050}.

\bibitem[{\citenamefont{B{\"a}r} \emph{et~al.}(2004)\citenamefont{B{\"a}r,
  Rupak, and Shoresh}}]{Bar:2003mh}
\bibinfo{author}{\bibnamefont{B{\"a}r}, \bibfnamefont{O.}},
  \bibinfo{author}{\bibfnamefont{G.}~\bibnamefont{Rupak}}, and
  \bibinfo{author}{\bibfnamefont{N.}~\bibnamefont{Shoresh}},
  \bibinfo{year}{2004}, \bibinfo{journal}{Phys. Rev.}
  \textbf{\bibinfo{volume}{D70}}, \bibinfo{pages}{034508},
  \arXiv{hep-lat/0306021}.

\bibitem[{Barberio \emph{et~al.}(2007)\citenamefont{Barberio}
  \emph{et~al.}}]{Barberio:2007cr}
\bibinfo{author}{\bibnamefont{Barberio}, \bibfnamefont{E.}}, \emph{et~al.}
  (\bibinfo{collaboration}{Heavy Flavor Averaging Group (HFAG)}),
  \bibinfo{year}{2007}, \arXiv{0704.3575}.

\bibitem[{Basak \emph{et~al.}(2009)\citenamefont{Basak}
  \emph{et~al.}}]{Basak:2008na}
\bibinfo{author}{\bibnamefont{Basak}, \bibfnamefont{S.}}, \emph{et~al.},
  \bibinfo{year}{2009}, \bibinfo{journal}{PoS}
  \textbf{\bibinfo{volume}{LAT2008}}, \bibinfo{pages}{127}, \arXiv{0812.4486}.

\bibitem[{Bazavov \emph{et~al.}(2009)\citenamefont{Bazavov}
  \emph{et~al.}}]{Bazavov:2009jc}
\bibinfo{author}{\bibnamefont{Bazavov}, \bibfnamefont{A.}}, \emph{et~al.},
  \bibinfo{year}{2009}, \bibinfo{journal}{PoS}
  \textbf{\bibinfo{volume}{LAT2008}}, \bibinfo{pages}{033}, \arXiv{0903.0874}.

\bibitem[{\citenamefont{Beane}
  \emph{et~al.}(2007{\natexlab{a}})\citenamefont{Beane, Bedaque, Orginos, and
  Savage}}]{Beane:2006kx}
\bibinfo{author}{\bibnamefont{Beane}, \bibfnamefont{S.~R.}},
  \bibinfo{author}{\bibfnamefont{P.~F.} \bibnamefont{Bedaque}},
  \bibinfo{author}{\bibfnamefont{K.}~\bibnamefont{Orginos}}, and
  \bibinfo{author}{\bibfnamefont{M.~J.} \bibnamefont{Savage}},
  \bibinfo{year}{2007}{\natexlab{a}}, \bibinfo{journal}{Phys. Rev.}
  \textbf{\bibinfo{volume}{D75}}, \bibinfo{pages}{094501},
  \arXiv{hep-lat/0606023}.

\bibitem[{\citenamefont{Beane}
  \emph{et~al.}(2007{\natexlab{b}})\citenamefont{Beane, Orginos, and
  Savage}}]{Beane:2006fk}
\bibinfo{author}{\bibnamefont{Beane}, \bibfnamefont{S.~R.}},
  \bibinfo{author}{\bibfnamefont{K.}~\bibnamefont{Orginos}}, and
  \bibinfo{author}{\bibfnamefont{M.~J.} \bibnamefont{Savage}},
  \bibinfo{year}{2007}{\natexlab{b}}, \bibinfo{journal}{Nucl. Phys.}
  \textbf{\bibinfo{volume}{B768}}, \bibinfo{pages}{38},
  \arXiv{hep-lat/0605014}.

\bibitem[{\citenamefont{Beane}
  \emph{et~al.}(2008{\natexlab{a}})\citenamefont{Beane, Orginos,
  and Savage}}]{Beane:2008dv}
\bibinfo{author}{\bibnamefont{Beane}, \bibfnamefont{S.~R.}},
  \bibinfo{author}{\bibfnamefont{K.}~\bibnamefont{Orginos}}, and
  \bibinfo{author}{\bibfnamefont{M.~J.} \bibnamefont{Savage}},
  \bibinfo{year}{2008}{\natexlab{a}}, \bibinfo{journal}{Int. J. Mod. Phys.}
  \textbf{\bibinfo{volume}{E17}}, \bibinfo{pages}{1157}, \arXiv{0805.4629}.

\bibitem[{Beane \emph{et~al.}(2007{\natexlab{c}})\citenamefont{Beane}
  \emph{et~al.}}]{Beane:2006gf}
\bibinfo{author}{\bibnamefont{Beane}, \bibfnamefont{S.~R.}}, \emph{et~al.}
  (\bibinfo{collaboration}{NPLQCD}), \bibinfo{year}{2007}{\natexlab{c}},
  \bibinfo{journal}{Nucl. Phys.} \textbf{\bibinfo{volume}{A794}},
  \bibinfo{pages}{62}, \arXiv{hep-lat/0612026}.

\bibitem[{Beane \emph{et~al.}(2008{\natexlab{b}})\citenamefont{Beane}
  \emph{et~al.}}]{Beane:2007es}
\bibinfo{author}{\bibnamefont{Beane}, \bibfnamefont{S.~R.}}, \emph{et~al.},
  \bibinfo{year}{2008}{\natexlab{b}}, \bibinfo{journal}{Phys. Rev. Lett.}
  \textbf{\bibinfo{volume}{100}}, \bibinfo{pages}{082004}, \arXiv{0710.1827}.

\bibitem[{Beane \emph{et~al.}(2008{\natexlab{c}})\citenamefont{Beane}
  \emph{et~al.}}]{Beane:2007xs}
\bibinfo{author}{\bibnamefont{Beane}, \bibfnamefont{S.~R.}}, \emph{et~al.},
  \bibinfo{year}{2008}{\natexlab{c}}, \bibinfo{journal}{Phys. Rev.}
  \textbf{\bibinfo{volume}{D77}}, \bibinfo{pages}{014505}, \arXiv{0706.3026}.

\bibitem[{Beane \emph{et~al.}(2008{\natexlab{d}})\citenamefont{Beane}
  \emph{et~al.}}]{Beane:2007uh}
\bibinfo{author}{\bibnamefont{Beane}, \bibfnamefont{S.~R.}}, \emph{et~al.}
  (\bibinfo{collaboration}{NPLQCD}), \bibinfo{year}{2008}{\natexlab{d}},
  \bibinfo{journal}{Phys. Rev.} \textbf{\bibinfo{volume}{D77}},
  \bibinfo{pages}{094507}, \arXiv{0709.1169}.

\bibitem[{\citenamefont{Becher and Joos}(1982)}]{Becher:1982ud}
\bibinfo{author}{\bibnamefont{Becher}, \bibfnamefont{P.}}, and
  \bibinfo{author}{\bibfnamefont{H.}~\bibnamefont{Joos}}, \bibinfo{year}{1982},
  \bibinfo{journal}{Zeit. Phys.} \textbf{\bibinfo{volume}{C15}},
  \bibinfo{pages}{343}.

\bibitem[{\citenamefont{Becher and Hill}(2006)}]{Becher:2005bg}
\bibinfo{author}{\bibnamefont{Becher}, \bibfnamefont{T.}}, and
  \bibinfo{author}{\bibfnamefont{R.~J.} \bibnamefont{Hill}},
  \bibinfo{year}{2006}, \bibinfo{journal}{Phys. Lett.}
  \textbf{\bibinfo{volume}{B633}}, \bibinfo{pages}{61}, \arXiv{hep-ph/0509090}.

\bibitem[{\citenamefont{Becirevic and Kaidalov}(2000)}]{Becirevic:1999kt}
\bibinfo{author}{\bibnamefont{Becirevic}, \bibfnamefont{D.}}, and
  \bibinfo{author}{\bibfnamefont{A.~B.} \bibnamefont{Kaidalov}},
  \bibinfo{year}{2000}, \bibinfo{journal}{Phys. Lett.}
  \textbf{\bibinfo{volume}{B478}}, \bibinfo{pages}{417},
  \arXiv{hep-ph/9904490}.

\bibitem[{\citenamefont{Bernard}(2002)}]{Bernard:2001yj}
\bibinfo{author}{\bibnamefont{Bernard}, \bibfnamefont{C.}}
  (\bibinfo{collaboration}{MILC}), \bibinfo{year}{2002},
  \bibinfo{journal}{Phys. Rev.} \textbf{\bibinfo{volume}{D65}},
  \bibinfo{pages}{054031}, \arXiv{hep-lat/0111051}.

\bibitem[{\citenamefont{Bernard}(2005)}]{Bernard:2004ab}
\bibinfo{author}{\bibnamefont{Bernard}, \bibfnamefont{C.}},
  \bibinfo{year}{2005}, \bibinfo{journal}{Phys. Rev.}
  \textbf{\bibinfo{volume}{D71}}, \bibinfo{pages}{094020},
  \arXiv{hep-lat/0412030}.

\bibitem[{\citenamefont{Bernard}(2006)}]{Bernard:2006zw}
\bibinfo{author}{\bibnamefont{Bernard}, \bibfnamefont{C.}},
  \bibinfo{year}{2006}, \bibinfo{journal}{Phys. Rev.}
  \textbf{\bibinfo{volume}{D73}}, \bibinfo{pages}{114503},
  \arXiv{hep-lat/0603011}.

\bibitem[{\citenamefont{Bernard}
  \emph{et~al.}(2006{\natexlab{a}})\citenamefont{Bernard, DeTar, Fu, and
  Prelovsek}}]{Bernard:2006gj}
\bibinfo{author}{\bibnamefont{Bernard}, \bibfnamefont{C.}},
  \bibinfo{author}{\bibfnamefont{C.~E.} \bibnamefont{DeTar}},
  \bibinfo{author}{\bibfnamefont{Z.}~\bibnamefont{Fu}}, and
  \bibinfo{author}{\bibfnamefont{S.}~\bibnamefont{Prelovsek}},
  \bibinfo{year}{2006}{\natexlab{a}}, \bibinfo{journal}{PoS}
  \textbf{\bibinfo{volume}{LAT2006}}, \bibinfo{pages}{173},
  \arXiv{hep-lat/0610031}.

\bibitem[{\citenamefont{Bernard}
  \emph{et~al.}(2007{\natexlab{a}})\citenamefont{Bernard, DeTar, Fu, and
  Prelovsek}}]{Bernard:2007qf}
\bibinfo{author}{\bibnamefont{Bernard}, \bibfnamefont{C.}},
  \bibinfo{author}{\bibfnamefont{C.~E.} \bibnamefont{DeTar}},
  \bibinfo{author}{\bibfnamefont{Z.}~\bibnamefont{Fu}}, and
  \bibinfo{author}{\bibfnamefont{S.}~\bibnamefont{Prelovsek}},
  \bibinfo{year}{2007}{\natexlab{a}}, \bibinfo{journal}{Phys. Rev.}
  \textbf{\bibinfo{volume}{D76}}, \bibinfo{pages}{094504}, \arXiv{0707.2402}.

\bibitem[{\citenamefont{Bernard}
  \emph{et~al.}(2006{\natexlab{b}})\citenamefont{Bernard, Golterman, and
  Shamir}}]{Bernard:2006ee}
\bibinfo{author}{\bibnamefont{Bernard}, \bibfnamefont{C.}},
  \bibinfo{author}{\bibfnamefont{M.}~\bibnamefont{Golterman}}, and
  \bibinfo{author}{\bibfnamefont{Y.}~\bibnamefont{Shamir}},
  \bibinfo{year}{2006}{\natexlab{b}}, \bibinfo{journal}{Phys. Rev.}
  \textbf{\bibinfo{volume}{D73}}, \bibinfo{pages}{114511},
  \arXiv{hep-lat/0604017}.

\bibitem[{\citenamefont{Bernard}
  \emph{et~al.}(2008{\natexlab{a}})\citenamefont{Bernard, Golterman, and
  Shamir}}]{Bernard:2007ma}
\bibinfo{author}{\bibnamefont{Bernard}, \bibfnamefont{C.}},
  \bibinfo{author}{\bibfnamefont{M.}~\bibnamefont{Golterman}}, and
  \bibinfo{author}{\bibfnamefont{Y.}~\bibnamefont{Shamir}},
  \bibinfo{year}{2008}{\natexlab{a}}, \bibinfo{journal}{Phys. Rev.}
  \textbf{\bibinfo{volume}{D77}}, \bibinfo{pages}{074505}, \arXiv{0712.2560}.

\bibitem[{\citenamefont{Bernard}
  \emph{et~al.}(2007{\natexlab{b}})\citenamefont{Bernard, Golterman, Shamir,
  and Sharpe}}]{Bernard:2006vv}
\bibinfo{author}{\bibnamefont{Bernard}, \bibfnamefont{C.}},
  \bibinfo{author}{\bibfnamefont{M.}~\bibnamefont{Golterman}},
  \bibinfo{author}{\bibfnamefont{Y.}~\bibnamefont{Shamir}}, and
  \bibinfo{author}{\bibfnamefont{S.~R.} \bibnamefont{Sharpe}},
  \bibinfo{year}{2007}{\natexlab{b}}, \bibinfo{journal}{Phys. Lett.}
  \textbf{\bibinfo{volume}{B649}}, \bibinfo{pages}{235},
  \arXiv{hep-lat/0603027}.

\bibitem[{\citenamefont{Bernard}
  \emph{et~al.}(2008{\natexlab{b}})\citenamefont{Bernard, Golterman, Shamir,
  and Sharpe}}]{Bernard:2008gr}
\bibinfo{author}{\bibnamefont{Bernard}, \bibfnamefont{C.}},
  \bibinfo{author}{\bibfnamefont{M.}~\bibnamefont{Golterman}},
  \bibinfo{author}{\bibfnamefont{Y.}~\bibnamefont{Shamir}}, and
  \bibinfo{author}{\bibfnamefont{S.~R.} \bibnamefont{Sharpe}},
  \bibinfo{year}{2008}{\natexlab{b}}, \bibinfo{journal}{Phys. Rev.}
  \textbf{\bibinfo{volume}{D78}}, \bibinfo{pages}{078502}, \arXiv{0808.2056}.

\bibitem[{\citenamefont{Bernard}
  \emph{et~al.}(2008{\natexlab{c}})\citenamefont{Bernard, Golterman, Shamir,
  and Sharpe}}]{Bernard:2007eh}
\bibinfo{author}{\bibnamefont{Bernard}, \bibfnamefont{C.}},
  \bibinfo{author}{\bibfnamefont{M.}~\bibnamefont{Golterman}},
  \bibinfo{author}{\bibfnamefont{Y.}~\bibnamefont{Shamir}}, and
  \bibinfo{author}{\bibfnamefont{S.~R.} \bibnamefont{Sharpe}},
  \bibinfo{year}{2008}{\natexlab{c}}, \bibinfo{journal}{Phys. Rev.}
  \textbf{\bibinfo{volume}{D77}}, \bibinfo{pages}{114504}, \arXiv{0711.0696}.

\bibitem[{\citenamefont{Bernard and Golterman}(1994)}]{Bernard:1993sv}
\bibinfo{author}{\bibnamefont{Bernard}, \bibfnamefont{C.}}, and
  \bibinfo{author}{\bibfnamefont{M.~F.~L.} \bibnamefont{Golterman}},
  \bibinfo{year}{1994}, \bibinfo{journal}{Phys. Rev.}
  \textbf{\bibinfo{volume}{D49}}, \bibinfo{pages}{486},
  \arXiv{hep-lat/9306005}.

\bibitem[{\citenamefont{Bernard and Golterman}(2009)}]{BGinprogress}
\bibinfo{author}{\bibnamefont{Bernard}, \bibfnamefont{C.}}, and
  \bibinfo{author}{\bibfnamefont{M.~F.~L.} \bibnamefont{Golterman}},
  \bibinfo{year}{2009}, \bibinfo{note}{work in progress}.

\bibitem[{Bernard \emph{et~al.}(1998)\citenamefont{Bernard}
  \emph{et~al.}}]{Bernard:1997mz}
\bibinfo{author}{\bibnamefont{Bernard}, \bibfnamefont{C.}}, \emph{et~al.}
  (\bibinfo{collaboration}{MILC}), \bibinfo{year}{1998},
  \bibinfo{journal}{Phys. Rev.} \textbf{\bibinfo{volume}{D58}},
  \bibinfo{pages}{014503}, \arXiv{hep-lat/9712010}.

\bibitem[{Bernard \emph{et~al.}(2000{\natexlab{a}})\citenamefont{Bernard}
  \emph{et~al.}}]{Bernard:1999xx}
\bibinfo{author}{\bibnamefont{Bernard}, \bibfnamefont{C.}}, \emph{et~al.}
  (\bibinfo{collaboration}{MILC}), \bibinfo{year}{2000}{\natexlab{a}},
  \bibinfo{journal}{Phys. Rev.} \textbf{\bibinfo{volume}{D61}},
  \bibinfo{pages}{111502}, \arXiv{hep-lat/9912018}.

\bibitem[{Bernard \emph{et~al.}(2000{\natexlab{b}})\citenamefont{Bernard}
  \emph{et~al.}}]{Bernard:2000gd}
\bibinfo{author}{\bibnamefont{Bernard}, \bibfnamefont{C.}}, \emph{et~al.},
  \bibinfo{year}{2000}{\natexlab{b}}, \bibinfo{journal}{Phys. Rev.}
  \textbf{\bibinfo{volume}{D62}}, \bibinfo{pages}{034503},
  \arXiv{hep-lat/0002028}.

\bibitem[{Bernard \emph{et~al.}(2001)\citenamefont{Bernard}
  \emph{et~al.}}]{Bernard:2001av}
\bibinfo{author}{\bibnamefont{Bernard}, \bibfnamefont{C.}}, \emph{et~al.},
  \bibinfo{year}{2001}, \bibinfo{journal}{Phys. Rev.}
  \textbf{\bibinfo{volume}{D64}}, \bibinfo{pages}{054506},
  \arXiv{hep-lat/0104002}.

\bibitem[{Bernard \emph{et~al.}(2003{\natexlab{a}})\citenamefont{Bernard}
  \emph{et~al.}}]{Bernard:2003jq}
\bibinfo{author}{\bibnamefont{Bernard}, \bibfnamefont{C.}}, \emph{et~al.}
  (\bibinfo{collaboration}{MILC}), \bibinfo{year}{2003}{\natexlab{a}},
  \bibinfo{journal}{Nucl. Phys. Proc. Suppl.} \textbf{\bibinfo{volume}{119}},
  \bibinfo{pages}{769}.

\bibitem[{Bernard \emph{et~al.}(2003{\natexlab{b}})\citenamefont{Bernard}
  \emph{et~al.}}]{Bernard:2002rz}
\bibinfo{author}{\bibnamefont{Bernard}, \bibfnamefont{C.}}, \emph{et~al.},
  \bibinfo{year}{2003}{\natexlab{b}}, \bibinfo{journal}{Nucl. Phys. Proc.
  Suppl.} \textbf{\bibinfo{volume}{119}}, \bibinfo{pages}{260},
  \arXiv{hep-lat/0209097}.

\bibitem[{Bernard \emph{et~al.}(2003{\natexlab{c}})\citenamefont{Bernard}
  \emph{et~al.}}]{Bernard:2003jd}
\bibinfo{author}{\bibnamefont{Bernard}, \bibfnamefont{C.}}, \emph{et~al.},
  \bibinfo{year}{2003}{\natexlab{c}}, \bibinfo{journal}{Phys. Rev.}
  \textbf{\bibinfo{volume}{D68}}, \bibinfo{pages}{074505},
  \arXiv{hep-lat/0301024}.

\bibitem[{Bernard \emph{et~al.}(2003{\natexlab{d}})\citenamefont{Bernard}
  \emph{et~al.}}]{Bernard:2003gq}
\bibinfo{author}{\bibnamefont{Bernard}, \bibfnamefont{C.}}, \emph{et~al.},
  \bibinfo{year}{2003}{\natexlab{d}}, \bibinfo{journal}{Phys. Rev.}
  \textbf{\bibinfo{volume}{D68}}, \bibinfo{pages}{114501},
  \arXiv{hep-lat/0308019}.

\bibitem[{Bernard \emph{et~al.}(2005)\citenamefont{Bernard}
  \emph{et~al.}}]{Bernard:2004je}
\bibinfo{author}{\bibnamefont{Bernard}, \bibfnamefont{C.}}, \emph{et~al.}
  (\bibinfo{collaboration}{MILC}), \bibinfo{year}{2005},
  \bibinfo{journal}{Phys. Rev.} \textbf{\bibinfo{volume}{D71}},
  \bibinfo{pages}{034504}, \arXiv{hep-lat/0405029}.

\bibitem[{Bernard \emph{et~al.}(2006{\natexlab{c}})\citenamefont{Bernard}
  \emph{et~al.}}]{Bernard:2005gf}
\bibinfo{author}{\bibnamefont{Bernard}, \bibfnamefont{C.}}, \emph{et~al.},
  \bibinfo{year}{2006}{\natexlab{c}}, \bibinfo{journal}{PoS}
  \textbf{\bibinfo{volume}{LAT2005}}, \bibinfo{pages}{114},
  \arXiv{hep-lat/0509176}.

\bibitem[{Bernard \emph{et~al.}(2006{\natexlab{d}})\citenamefont{Bernard}
  \emph{et~al.}}]{Bernard:2005ei}
\bibinfo{author}{\bibnamefont{Bernard}, \bibfnamefont{C.}}, \emph{et~al.}
  (\bibinfo{collaboration}{MILC}), \bibinfo{year}{2006}{\natexlab{d}},
  \bibinfo{journal}{PoS} \textbf{\bibinfo{volume}{LAT2005}},
  \bibinfo{pages}{025}, \arXiv{hep-lat/0509137}.

\bibitem[{Bernard \emph{et~al.}(2006{\natexlab{e}})\citenamefont{Bernard}
  \emph{et~al.}}]{Bernard:2006wx}
\bibinfo{author}{\bibnamefont{Bernard}, \bibfnamefont{C.}}, \emph{et~al.}
  (\bibinfo{collaboration}{MILC}), \bibinfo{year}{2006}{\natexlab{e}},
  \bibinfo{journal}{PoS} \textbf{\bibinfo{volume}{LAT2006}},
  \bibinfo{pages}{163}, \arXiv{hep-lat/0609053}.

\bibitem[{Bernard \emph{et~al.}(2007{\natexlab{c}})\citenamefont{Bernard}
  \emph{et~al.}}]{Bernard:2007ux}
\bibinfo{author}{\bibnamefont{Bernard}, \bibfnamefont{C.}}, \emph{et~al.}
  (\bibinfo{collaboration}{MILC}), \bibinfo{year}{2007}{\natexlab{c}},
  \bibinfo{journal}{PoS} \textbf{\bibinfo{volume}{LAT2007}},
  \bibinfo{pages}{137}, \arXiv{0711.0021}.

\bibitem[{Bernard \emph{et~al.}(2007{\natexlab{d}})\citenamefont{Bernard}
  \emph{et~al.}}]{Bernard:2006nj}
\bibinfo{author}{\bibnamefont{Bernard}, \bibfnamefont{C.}}, \emph{et~al.},
  \bibinfo{year}{2007}{\natexlab{d}}, \bibinfo{journal}{Phys. Rev.}
  \textbf{\bibinfo{volume}{D75}}, \bibinfo{pages}{094505},
  \arXiv{hep-lat/0611031}.

\bibitem[{Bernard \emph{et~al.}(2007{\natexlab{e}})\citenamefont{Bernard}
  \emph{et~al.}}]{Bernard:2007ps}
\bibinfo{author}{\bibnamefont{Bernard}, \bibfnamefont{C.}}, \emph{et~al.},
  \bibinfo{year}{2007}{\natexlab{e}}, \bibinfo{journal}{PoS}
  \textbf{\bibinfo{volume}{LAT2007}}, \bibinfo{pages}{090}, \arXiv{0710.1118}.

\bibitem[{Bernard \emph{et~al.}(2007{\natexlab{f}})\citenamefont{Bernard}
  \emph{et~al.}}]{Bernard:2007ez}
\bibinfo{author}{\bibnamefont{Bernard}, \bibfnamefont{C.}}, \emph{et~al.},
  \bibinfo{year}{2007}{\natexlab{f}}, \bibinfo{journal}{PoS}
  \textbf{\bibinfo{volume}{LAT2007}}, \bibinfo{pages}{310}, \arXiv{0710.3124}.

\bibitem[{Bernard \emph{et~al.}(2009{\natexlab{a}})\citenamefont{Bernard}
  \emph{et~al.}}]{Bernard:2008dn}
\bibinfo{author}{\bibnamefont{Bernard}, \bibfnamefont{C.}}, \emph{et~al.},
  \bibinfo{year}{2009}{\natexlab{a}}, \bibinfo{journal}{Phys. Rev.}
  \textbf{\bibinfo{volume}{D79}}, \bibinfo{pages}{014506}, \arXiv{0808.2519}.

\bibitem[{Bernard \emph{et~al.}(2009{\natexlab{b}})\citenamefont{Bernard}
  \emph{et~al.}}]{Mackenzie:Lat08}
\bibinfo{author}{\bibnamefont{Bernard}, \bibfnamefont{C.}}, \emph{et~al.}
  (\bibinfo{collaboration}{Fermilab Lattice and MILC}),
  \bibinfo{year}{2009}{\natexlab{b}}, \bibinfo{journal}{PoS}
  \textbf{\bibinfo{volume}{LAT2008}}, \bibinfo{pages}{278}.

\bibitem[{\citenamefont{Bernard} \emph{et~al.}(1993)\citenamefont{Bernard,
  Kaiser, and Meissner}}]{Bernard:1993nj}
\bibinfo{author}{\bibnamefont{Bernard}, \bibfnamefont{V.}},
  \bibinfo{author}{\bibfnamefont{N.}~\bibnamefont{Kaiser}}, and
  \bibinfo{author}{\bibfnamefont{U.~G.} \bibnamefont{Meissner}},
  \bibinfo{year}{1993}, \bibinfo{journal}{Z. Phys.}
  \textbf{\bibinfo{volume}{C60}}, \bibinfo{pages}{111}, \arXiv{hep-ph/9303311}.

\bibitem[{\citenamefont{Bethke}(2007)}]{Bethke:2006ac}
\bibinfo{author}{\bibnamefont{Bethke}, \bibfnamefont{S.}},
  \bibinfo{year}{2007}, \bibinfo{journal}{Prog. Part. Nucl. Phys.}
  \textbf{\bibinfo{volume}{58}}, \bibinfo{pages}{351}, \arXiv{hep-ex/0606035}.

\bibitem[{\citenamefont{Bigi}
  \emph{et~al.}(1992{\natexlab{a}})\citenamefont{Bigi, Blok, Shifman, Uraltsev,
  and Vainshtein}}]{Bigi:1992ne}
\bibinfo{author}{\bibnamefont{Bigi}, \bibfnamefont{I.~I.~Y.}},
  \bibinfo{author}{\bibfnamefont{B.}~\bibnamefont{Blok}},
  \bibinfo{author}{\bibfnamefont{M.~A.} \bibnamefont{Shifman}},
  \bibinfo{author}{\bibfnamefont{N.~G.} \bibnamefont{Uraltsev}}, and
  \bibinfo{author}{\bibfnamefont{A.~I.} \bibnamefont{Vainshtein}},
  \bibinfo{year}{1992}{\natexlab{a}}, \arXiv{hep-ph/9212227}.

\bibitem[{\citenamefont{Bigi} \emph{et~al.}(1997)\citenamefont{Bigi, Shifman,
  and Uraltsev}}]{Bigi:1997fj}
\bibinfo{author}{\bibnamefont{Bigi}, \bibfnamefont{I.~I.~Y.}},
  \bibinfo{author}{\bibfnamefont{M.~A.} \bibnamefont{Shifman}}, and
  \bibinfo{author}{\bibfnamefont{N.}~\bibnamefont{Uraltsev}},
  \bibinfo{year}{1997}, \bibinfo{journal}{Ann. Rev. Nucl. Part. Sci.}
  \textbf{\bibinfo{volume}{47}}, \bibinfo{pages}{591}, \arXiv{hep-ph/9703290}.

\bibitem[{\citenamefont{Bigi} \emph{et~al.}(1993)\citenamefont{Bigi, Shifman,
  Uraltsev, and Vainshtein}}]{Bigi:1993fe}
\bibinfo{author}{\bibnamefont{Bigi}, \bibfnamefont{I.~I.~Y.}},
  \bibinfo{author}{\bibfnamefont{M.~A.} \bibnamefont{Shifman}},
  \bibinfo{author}{\bibfnamefont{N.~G.} \bibnamefont{Uraltsev}}, and
  \bibinfo{author}{\bibfnamefont{A.~I.} \bibnamefont{Vainshtein}},
  \bibinfo{year}{1993}, \bibinfo{journal}{Phys. Rev. Lett.}
  \textbf{\bibinfo{volume}{71}}, \bibinfo{pages}{496}, \arXiv{hep-ph/9304225}.

\bibitem[{\citenamefont{Bigi}
  \emph{et~al.}(1992{\natexlab{b}})\citenamefont{Bigi, Uraltsev, and
  Vainshtein}}]{Bigi:1992su}
\bibinfo{author}{\bibnamefont{Bigi}, \bibfnamefont{I.~I.~Y.}},
  \bibinfo{author}{\bibfnamefont{N.~G.} \bibnamefont{Uraltsev}}, and
  \bibinfo{author}{\bibfnamefont{A.~I.} \bibnamefont{Vainshtein}},
  \bibinfo{year}{1992}{\natexlab{b}}, \bibinfo{journal}{Phys. Lett.}
  \textbf{\bibinfo{volume}{B293}}, \bibinfo{pages}{430},
  \bibinfo{note}{{Erratum-ibid.\ {\bf B297}, 477 (1993)}},
  \arXiv{hep-ph/9207214}.

\bibitem[{\citenamefont{Bijnens}(2007)}]{Bijnens:2007yd}
\bibinfo{author}{\bibnamefont{Bijnens}, \bibfnamefont{J.}},
  \bibinfo{year}{2007}, \bibinfo{journal}{PoS}
  \textbf{\bibinfo{volume}{LAT2007}}, \bibinfo{pages}{004}, \arXiv{0708.1377}.

\bibitem[{\citenamefont{Bijnens and Danielsson}(2007)}]{Bijnens:2006mk}
\bibinfo{author}{\bibnamefont{Bijnens}, \bibfnamefont{J.}}, and
  \bibinfo{author}{\bibfnamefont{N.}~\bibnamefont{Danielsson}},
  \bibinfo{year}{2007}, \bibinfo{journal}{Phys. Rev.}
  \textbf{\bibinfo{volume}{D75}}, \bibinfo{pages}{014505},
  \arXiv{hep-lat/0610127}.

\bibitem[{\citenamefont{Bijnens} \emph{et~al.}(2004)\citenamefont{Bijnens,
  Danielsson, and Lahde}}]{Bijnens:2004hk}
\bibinfo{author}{\bibnamefont{Bijnens}, \bibfnamefont{J.}},
  \bibinfo{author}{\bibfnamefont{N.}~\bibnamefont{Danielsson}}, and
  \bibinfo{author}{\bibfnamefont{T.~A.} \bibnamefont{Lahde}},
  \bibinfo{year}{2004}, \bibinfo{journal}{Phys. Rev.}
  \textbf{\bibinfo{volume}{D70}}, \bibinfo{pages}{111503},
  \arXiv{hep-lat/0406017}.

\bibitem[{\citenamefont{Bijnens} \emph{et~al.}(2006)\citenamefont{Bijnens,
  Danielsson, and Lahde}}]{Bijnens:2006jv}
\bibinfo{author}{\bibnamefont{Bijnens}, \bibfnamefont{J.}},
  \bibinfo{author}{\bibfnamefont{N.}~\bibnamefont{Danielsson}}, and
  \bibinfo{author}{\bibfnamefont{T.~A.} \bibnamefont{Lahde}},
  \bibinfo{year}{2006}, \bibinfo{journal}{Phys. Rev.}
  \textbf{\bibinfo{volume}{D73}}, \bibinfo{pages}{074509},
  \arXiv{hep-lat/0602003}.

\bibitem[{\citenamefont{Bijnens and Lahde}(2005)}]{Bijnens:2005ae}
\bibinfo{author}{\bibnamefont{Bijnens}, \bibfnamefont{J.}}, and
  \bibinfo{author}{\bibfnamefont{T.~A.} \bibnamefont{Lahde}},
  \bibinfo{year}{2005}, \bibinfo{journal}{Phys. Rev.}
  \textbf{\bibinfo{volume}{D71}}, \bibinfo{pages}{094502},
  \arXiv{hep-lat/0501014}.

\bibitem[{\citenamefont{Billeter} \emph{et~al.}(2004)\citenamefont{Billeter,
  DeTar, and Osborn}}]{Billeter:2004wx}
\bibinfo{author}{\bibnamefont{Billeter}, \bibfnamefont{B.}},
  \bibinfo{author}{\bibfnamefont{C.~E.} \bibnamefont{DeTar}}, and
  \bibinfo{author}{\bibfnamefont{J.}~\bibnamefont{Osborn}},
  \bibinfo{year}{2004}, \bibinfo{journal}{Phys. Rev.}
  \textbf{\bibinfo{volume}{D70}}, \bibinfo{pages}{077502},
  \arXiv{hep-lat/0406032}.

\bibitem[{\citenamefont{Blum}(2003)}]{Blum:2002ii}
\bibinfo{author}{\bibnamefont{Blum}, \bibfnamefont{T.}}, \bibinfo{year}{2003},
  \bibinfo{journal}{Phys. Rev. Lett.} \textbf{\bibinfo{volume}{91}},
  \bibinfo{pages}{052001}, \arXiv{hep-lat/0212018}.

\bibitem[{\citenamefont{Blum} \emph{et~al.}(2007)\citenamefont{Blum, Doi,
  Hayakawa, Izubuchi, and Yamada}}]{Blum:2007cy}
\bibinfo{author}{\bibnamefont{Blum}, \bibfnamefont{T.}},
  \bibinfo{author}{\bibfnamefont{T.}~\bibnamefont{Doi}},
  \bibinfo{author}{\bibfnamefont{M.}~\bibnamefont{Hayakawa}},
  \bibinfo{author}{\bibfnamefont{T.}~\bibnamefont{Izubuchi}}, and
  \bibinfo{author}{\bibfnamefont{N.}~\bibnamefont{Yamada}},
  \bibinfo{year}{2007}, \bibinfo{journal}{Phys. Rev.}
  \textbf{\bibinfo{volume}{D76}}, \bibinfo{pages}{114508}, \arXiv{0708.0484}.

\bibitem[{Blum \emph{et~al.}(1997)\citenamefont{Blum}
  \emph{et~al.}}]{Blum:1996uf}
\bibinfo{author}{\bibnamefont{Blum}, \bibfnamefont{T.}}, \emph{et~al.},
  \bibinfo{year}{1997}, \bibinfo{journal}{Phys. Rev.}
  \textbf{\bibinfo{volume}{D55}}, \bibinfo{pages}{1133},
  \arXiv{hep-lat/9609036}.

\bibitem[{Booth \emph{et~al.}(1992)\citenamefont{Booth}
  \emph{et~al.}}]{Booth:1992bm}
\bibinfo{author}{\bibnamefont{Booth}, \bibfnamefont{S.~P.}}, \emph{et~al.}
  (\bibinfo{collaboration}{UKQCD}), \bibinfo{year}{1992},
  \bibinfo{journal}{Phys. Lett.} \textbf{\bibinfo{volume}{B294}},
  \bibinfo{pages}{385}, \arXiv{hep-lat/9209008}.

\bibitem[{\citenamefont{Borici}(1999)}]{Borici:1998mr}
\bibinfo{author}{\bibnamefont{Borici}, \bibfnamefont{A.}},
  \bibinfo{year}{1999}, \bibinfo{journal}{Phys. Lett.}
  \textbf{\bibinfo{volume}{B453}}, \bibinfo{pages}{46},
  \arXiv{hep-lat/9810064}.

\bibitem[{\citenamefont{Borici}(2000)}]{Borici:1999da}
\bibinfo{author}{\bibnamefont{Borici}, \bibfnamefont{A.}},
  \bibinfo{year}{2000}, \bibinfo{note}{in: Lattice Fermions and Structure of
  the Vacuum, eds. V. Mitrjushkin and G. Schierholz (Kluwer Academic
  Publishers, 2000) p. 41}, \arXiv{hep-lat/9912040}.

\bibitem[{\citenamefont{Bourrely} \emph{et~al.}(1981)\citenamefont{Bourrely,
  Machet, and de~Rafael}}]{Bourrely:1980gp}
\bibinfo{author}{\bibnamefont{Bourrely}, \bibfnamefont{C.}},
  \bibinfo{author}{\bibfnamefont{B.}~\bibnamefont{Machet}}, and
  \bibinfo{author}{\bibfnamefont{E.}~\bibnamefont{de~Rafael}},
  \bibinfo{year}{1981}, \bibinfo{journal}{Nucl. Phys.}
  \textbf{\bibinfo{volume}{B189}}, \bibinfo{pages}{157}.

\bibitem[{\citenamefont{Bowler} \emph{et~al.}(1987)\citenamefont{Bowler,
  Chalmers, Kenway, Pawley, and Roweth}}]{Bowler:1986fw}
\bibinfo{author}{\bibnamefont{Bowler}, \bibfnamefont{K.~C.}},
  \bibinfo{author}{\bibfnamefont{C.~B.} \bibnamefont{Chalmers}},
  \bibinfo{author}{\bibfnamefont{R.~D.} \bibnamefont{Kenway}},
  \bibinfo{author}{\bibfnamefont{G.~S.} \bibnamefont{Pawley}}, and
  \bibinfo{author}{\bibfnamefont{D.}~\bibnamefont{Roweth}},
  \bibinfo{year}{1987}, \bibinfo{journal}{Nucl. Phys.}
  \textbf{\bibinfo{volume}{B284}}, \bibinfo{pages}{299}.

\bibitem[{Bowler \emph{et~al.}(1996)\citenamefont{Bowler}
  \emph{et~al.}}]{Bowler:1996ws}
\bibinfo{author}{\bibnamefont{Bowler}, \bibfnamefont{K.~C.}}, \emph{et~al.}
  (\bibinfo{collaboration}{UKQCD}), \bibinfo{year}{1996},
  \bibinfo{journal}{Phys. Rev.} \textbf{\bibinfo{volume}{D54}},
  \bibinfo{pages}{3619}, \arXiv{hep-lat/9601022}.

\bibitem[{Bowler \emph{et~al.}(2000)\citenamefont{Bowler}
  \emph{et~al.}}]{Bowler:1999ae}
\bibinfo{author}{\bibnamefont{Bowler}, \bibfnamefont{K.~C.}}, \emph{et~al.}
  (\bibinfo{collaboration}{UKQCD}), \bibinfo{year}{2000},
  \bibinfo{journal}{Phys. Rev.} \textbf{\bibinfo{volume}{D62}},
  \bibinfo{pages}{054506}, \arXiv{hep-lat/9910022}.

\bibitem[{\citenamefont{Bowman} \emph{et~al.}(2004)\citenamefont{Bowman,
  Heller, Leinweber, Parappilly, and Williams}}]{Bowman:2004jm}
\bibinfo{author}{\bibnamefont{Bowman}, \bibfnamefont{P.~O.}},
  \bibinfo{author}{\bibfnamefont{U.~M.} \bibnamefont{Heller}},
  \bibinfo{author}{\bibfnamefont{D.~B.} \bibnamefont{Leinweber}},
  \bibinfo{author}{\bibfnamefont{M.~B.} \bibnamefont{Parappilly}}, and
  \bibinfo{author}{\bibfnamefont{A.~G.} \bibnamefont{Williams}},
  \bibinfo{year}{2004}, \bibinfo{journal}{Phys. Rev.}
  \textbf{\bibinfo{volume}{D70}}, \bibinfo{pages}{034509},
  \arXiv{hep-lat/0402032}.

\bibitem[{\citenamefont{Bowman}
  \emph{et~al.}(2005{\natexlab{a}})\citenamefont{Bowman,
  Heller, Leinweber, Williams, and Zhang}}]{Bowman:2005zi}
\bibinfo{author}{\bibnamefont{Bowman}, \bibfnamefont{P.~O.}},
  \bibinfo{author}{\bibfnamefont{U.~M.} \bibnamefont{Heller}},
  \bibinfo{author}{\bibfnamefont{D.~B.} \bibnamefont{Leinweber}},
  \bibinfo{author}{\bibfnamefont{A.~G.} \bibnamefont{Williams}}, and
  \bibinfo{author}{\bibfnamefont{J.~B.} \bibnamefont{Zhang}},
  \bibinfo{year}{2005}{\natexlab{a}}, \bibinfo{journal}{Lect. Notes Phys.}
  \textbf{\bibinfo{volume}{663}}, \bibinfo{pages}{17}.

\bibitem[{Bowman \emph{et~al.}(2005{\natexlab{b}})\citenamefont{Bowman}
  \emph{et~al.}}]{Bowman:2005vx}
\bibinfo{author}{\bibnamefont{Bowman}, \bibfnamefont{P.~O.}}, \emph{et~al.},
  \bibinfo{year}{2005}{\natexlab{b}}, \bibinfo{journal}{Phys. Rev.}
  \textbf{\bibinfo{volume}{D71}}, \bibinfo{pages}{054507},
  \arXiv{hep-lat/0501019}.

\bibitem[{Bowman \emph{et~al.}(2007)\citenamefont{Bowman}
  \emph{et~al.}}]{Bowman:2007du}
\bibinfo{author}{\bibnamefont{Bowman}, \bibfnamefont{P.~O.}}, \emph{et~al.},
  \bibinfo{year}{2007}, \bibinfo{journal}{Phys. Rev.}
  \textbf{\bibinfo{volume}{D76}}, \bibinfo{pages}{094505},
  \arXiv{hep-lat/0703022}.

\bibitem[{\citenamefont{Boyd} \emph{et~al.}(1995)\citenamefont{Boyd, Grinstein,
  and Lebed}}]{Boyd:1994tt}
\bibinfo{author}{\bibnamefont{Boyd}, \bibfnamefont{C.~G.}},
  \bibinfo{author}{\bibfnamefont{B.}~\bibnamefont{Grinstein}}, and
  \bibinfo{author}{\bibfnamefont{R.~F.} \bibnamefont{Lebed}},
  \bibinfo{year}{1995}, \bibinfo{journal}{Phys. Rev. Lett.}
  \textbf{\bibinfo{volume}{74}}, \bibinfo{pages}{4603}, \arXiv{hep-ph/9412324}.

\bibitem[{\citenamefont{Boyd and Savage}(1997)}]{Boyd:1997qw}
\bibinfo{author}{\bibnamefont{Boyd}, \bibfnamefont{C.~G.}}, and
  \bibinfo{author}{\bibfnamefont{M.~J.} \bibnamefont{Savage}},
  \bibinfo{year}{1997}, \bibinfo{journal}{Phys. Rev.}
  \textbf{\bibinfo{volume}{D56}}, \bibinfo{pages}{303}, \arXiv{hep-ph/9702300}.

\bibitem[{Bratt \emph{et~al.}(2009)\citenamefont{Bratt}
  \emph{et~al.}}]{Bratt:2008uf}
\bibinfo{author}{\bibnamefont{Bratt}, \bibfnamefont{J.~D.}}, \emph{et~al.}
  (\bibinfo{collaboration}{LHP}), \bibinfo{year}{2009}, \bibinfo{journal}{PoS}
  \textbf{\bibinfo{volume}{LAT2008}}, \bibinfo{pages}{141}, \arXiv{0810.1933}.

\bibitem[{\citenamefont{Buchalla} \emph{et~al.}(1996)\citenamefont{Buchalla,
  Buras, and Lautenbacher}}]{Buchalla:1995vs}
\bibinfo{author}{\bibnamefont{Buchalla}, \bibfnamefont{G.}},
  \bibinfo{author}{\bibfnamefont{A.~J.} \bibnamefont{Buras}}, and
  \bibinfo{author}{\bibfnamefont{M.~E.} \bibnamefont{Lautenbacher}},
  \bibinfo{year}{1996}, \bibinfo{journal}{Rev. Mod. Phys.}
  \textbf{\bibinfo{volume}{68}}, \bibinfo{pages}{1125}, \arXiv{hep-ph/9512380}.

\bibitem[{\citenamefont{Buras}(1998)}]{Buras:1998raa}
\bibinfo{author}{\bibnamefont{Buras}, \bibfnamefont{A.~J.}},
  \bibinfo{year}{1998}, \bibinfo{note}{{Les Houches Lectures, ``Probing the
  Standard Model of Particle Interactions'', F. David and R. Gupta, eds,
  Elsevier Science B.V.}}, \arXiv{hep-ph/9806471}.

\bibitem[{\citenamefont{Buras} \emph{et~al.}(1990)\citenamefont{Buras, Jamin,
  and Weisz}}]{Buras:1990fn}
\bibinfo{author}{\bibnamefont{Buras}, \bibfnamefont{A.~J.}},
  \bibinfo{author}{\bibfnamefont{M.}~\bibnamefont{Jamin}}, and
  \bibinfo{author}{\bibfnamefont{P.~H.} \bibnamefont{Weisz}},
  \bibinfo{year}{1990}, \bibinfo{journal}{Nucl. Phys.}
  \textbf{\bibinfo{volume}{B347}}, \bibinfo{pages}{491}.

\bibitem[{\citenamefont{Burch} \emph{et~al.}(2001)\citenamefont{Burch, Orginos,
  and Toussaint}}]{Burch:2001tr}
\bibinfo{author}{\bibnamefont{Burch}, \bibfnamefont{T.}},
  \bibinfo{author}{\bibfnamefont{K.}~\bibnamefont{Orginos}}, and
  \bibinfo{author}{\bibfnamefont{D.}~\bibnamefont{Toussaint}},
  \bibinfo{year}{2001}, \bibinfo{journal}{Phys. Rev.}
  \textbf{\bibinfo{volume}{D64}}, \bibinfo{pages}{074505},
  \arXiv{hep-lat/0103025}.

\bibitem[{\citenamefont{Burch} \emph{et~al.}(2002)\citenamefont{Burch, Orginos,
  and Toussaint}}]{Burch:2001nk}
\bibinfo{author}{\bibnamefont{Burch}, \bibfnamefont{T.}},
  \bibinfo{author}{\bibfnamefont{K.}~\bibnamefont{Orginos}}, and
  \bibinfo{author}{\bibfnamefont{D.}~\bibnamefont{Toussaint}},
  \bibinfo{year}{2002}, \bibinfo{journal}{Nucl. Phys. Proc. Suppl.}
  \textbf{\bibinfo{volume}{106}}, \bibinfo{pages}{382},
  \arXiv{hep-lat/0110001}.

\bibitem[{\citenamefont{Burch and Toussaint}(2003)}]{Burch:2003zf}
\bibinfo{author}{\bibnamefont{Burch}, \bibfnamefont{T.}}, and
  \bibinfo{author}{\bibfnamefont{D.}~\bibnamefont{Toussaint}}
  (\bibinfo{collaboration}{MILC}), \bibinfo{year}{2003},
  \bibinfo{journal}{Phys. Rev.} \textbf{\bibinfo{volume}{D68}},
  \bibinfo{pages}{094504}, \arXiv{hep-lat/0305008}.

\bibitem[{\citenamefont{Callaway and Rahman}(1982)}]{Callaway:1982eb}
\bibinfo{author}{\bibnamefont{Callaway}, \bibfnamefont{D.~J.~E.}}, and
  \bibinfo{author}{\bibfnamefont{A.}~\bibnamefont{Rahman}},
  \bibinfo{year}{1982}, \bibinfo{journal}{Phys. Rev. Lett.}
  \textbf{\bibinfo{volume}{49}}, \bibinfo{pages}{613}.

\bibitem[{\citenamefont{Callaway and Rahman}(1983)}]{Callaway:1983ee}
\bibinfo{author}{\bibnamefont{Callaway}, \bibfnamefont{D.~J.~E.}}, and
  \bibinfo{author}{\bibfnamefont{A.}~\bibnamefont{Rahman}},
  \bibinfo{year}{1983}, \bibinfo{journal}{Phys. Rev.}
  \textbf{\bibinfo{volume}{D28}}, \bibinfo{pages}{1506}.

\bibitem[{\citenamefont{Caswell and Lepage}(1986)}]{Caswell:1985ui}
\bibinfo{author}{\bibnamefont{Caswell}, \bibfnamefont{W.~E.}}, and
  \bibinfo{author}{\bibfnamefont{G.~P.} \bibnamefont{Lepage}},
  \bibinfo{year}{1986}, \bibinfo{journal}{Phys. Lett.}
  \textbf{\bibinfo{volume}{B167}}, \bibinfo{pages}{437}.

\bibitem[{Charles \emph{et~al.}(2008)\citenamefont{Charles}
  \emph{et~al.}}]{Charles:2008}
\bibinfo{author}{\bibnamefont{Charles}, \bibfnamefont{J.}}, \emph{et~al.},
  \bibinfo{year}{2008}, \bibinfo{note}{preliminary results for Summer 2008, \\
  http://ckmfitter.in2p3.fr/plots$\_$Summer2008/ckmEval$\_$results.html}.

\bibitem[{\citenamefont{Chay} \emph{et~al.}(1990)\citenamefont{Chay, Georgi,
  and Grinstein}}]{Chay:1990da}
\bibinfo{author}{\bibnamefont{Chay}, \bibfnamefont{J.}},
  \bibinfo{author}{\bibfnamefont{H.}~\bibnamefont{Georgi}}, and
  \bibinfo{author}{\bibfnamefont{B.}~\bibnamefont{Grinstein}},
  \bibinfo{year}{1990}, \bibinfo{journal}{Phys. Lett.}
  \textbf{\bibinfo{volume}{B247}}, \bibinfo{pages}{399}.

\bibitem[{\citenamefont{Chen}
  \emph{et~al.}(2009{\natexlab{a}})\citenamefont{Chen, Golterman, O'Connell,
  and Walker-Loud}}]{Chen:2009su}
\bibinfo{author}{\bibnamefont{Chen}, \bibfnamefont{J.-W.}},
  \bibinfo{author}{\bibfnamefont{M.}~\bibnamefont{Golterman}},
  \bibinfo{author}{\bibfnamefont{D.}~\bibnamefont{O'Connell}}, and
  \bibinfo{author}{\bibfnamefont{A.}~\bibnamefont{Walker-Loud}},
  \bibinfo{year}{2009}{\natexlab{a}}, \bibinfo{journal}{Phys. Rev.}
  \textbf{\bibinfo{volume}{D79}}, \bibinfo{pages}{117502}, \arXiv{0905.2566}.

\bibitem[{\citenamefont{Chen} \emph{et~al.}(2007)\citenamefont{Chen, O'Connell,
  and Walker-Loud}}]{Chen:2006wf}
\bibinfo{author}{\bibnamefont{Chen}, \bibfnamefont{J.-W.}},
  \bibinfo{author}{\bibfnamefont{D.}~\bibnamefont{O'Connell}}, and
  \bibinfo{author}{\bibfnamefont{A.}~\bibnamefont{Walker-Loud}},
  \bibinfo{year}{2007}, \bibinfo{journal}{Phys. Rev.}
  \textbf{\bibinfo{volume}{D75}}, \bibinfo{pages}{054501},
  \arXiv{hep-lat/0611003}.

\bibitem[{\citenamefont{Chen}
  \emph{et~al.}(2009{\natexlab{b}})\citenamefont{Chen, O'Connell, and
  Walker-Loud}}]{Chen:2007ug}
\bibinfo{author}{\bibnamefont{Chen}, \bibfnamefont{J.-W.}},
  \bibinfo{author}{\bibfnamefont{D.}~\bibnamefont{O'Connell}}, and
  \bibinfo{author}{\bibfnamefont{A.~A.} \bibnamefont{Walker-Loud}},
  \bibinfo{year}{2009}{\natexlab{b}}, \bibinfo{journal}{JHEP}
  \textbf{\bibinfo{volume}{04}}, \bibinfo{pages}{090}, \arXiv{0706.0035}.

\bibitem[{\citenamefont{Chen} \emph{et~al.}(2006)\citenamefont{Chen, O'Connell,
  Van~de Water, and Walker-Loud}}]{Chen:2005ab}
\bibinfo{author}{\bibnamefont{Chen}, \bibfnamefont{J.-W.}},
  \bibinfo{author}{\bibfnamefont{D.}~\bibnamefont{O'Connell}},
  \bibinfo{author}{\bibfnamefont{R.~S.} \bibnamefont{Van~de Water}}, and
  \bibinfo{author}{\bibfnamefont{A.}~\bibnamefont{Walker-Loud}},
  \bibinfo{year}{2006}, \bibinfo{journal}{Phys. Rev.}
  \textbf{\bibinfo{volume}{D73}}, \bibinfo{pages}{074510},
  \arXiv{hep-lat/0510024}.

\bibitem[{\citenamefont{Clark and Kennedy}(2004)}]{Clark:2003na}
\bibinfo{author}{\bibnamefont{Clark}, \bibfnamefont{M.~A.}}, and
  \bibinfo{author}{\bibfnamefont{A.~D.} \bibnamefont{Kennedy}},
  \bibinfo{year}{2004}, \bibinfo{journal}{Nucl. Phys. Proc. Suppl.}
  \textbf{\bibinfo{volume}{129}}, \bibinfo{pages}{850},
  \arXiv{hep-lat/0309084}.

\bibitem[{\citenamefont{Clark and Kennedy}(2005)}]{Clark:2004cq}
\bibinfo{author}{\bibnamefont{Clark}, \bibfnamefont{M.~A.}}, and
  \bibinfo{author}{\bibfnamefont{A.~D.} \bibnamefont{Kennedy}},
  \bibinfo{year}{2005}, \bibinfo{journal}{Nucl. Phys. Proc. Suppl.}
  \textbf{\bibinfo{volume}{140}}, \bibinfo{pages}{838},
  \arXiv{hep-lat/0409134}.

\bibitem[{\citenamefont{Colangelo} \emph{et~al.}(2005)\citenamefont{Colangelo,
  D{\"u}rr, and Haefeli}}]{Colangelo:2005gd}
\bibinfo{author}{\bibnamefont{Colangelo}, \bibfnamefont{G.}},
  \bibinfo{author}{\bibfnamefont{S.}~\bibnamefont{D{\"u}rr}}, and
  \bibinfo{author}{\bibfnamefont{C.}~\bibnamefont{Haefeli}},
  \bibinfo{year}{2005}, \bibinfo{journal}{Nucl. Phys.}
  \textbf{\bibinfo{volume}{B721}}, \bibinfo{pages}{136},
  \arXiv{hep-lat/0503014}.

\bibitem[{\citenamefont{Collins} \emph{et~al.}(1997)\citenamefont{Collins,
  Edwards, Heller, and Sloan}}]{Collins:1996zc}
\bibinfo{author}{\bibnamefont{Collins}, \bibfnamefont{S.}},
  \bibinfo{author}{\bibfnamefont{R.~G.} \bibnamefont{Edwards}},
  \bibinfo{author}{\bibfnamefont{U.~M.} \bibnamefont{Heller}}, and
  \bibinfo{author}{\bibfnamefont{J.~H.} \bibnamefont{Sloan}},
  \bibinfo{year}{1997}, \bibinfo{journal}{Nucl. Phys. Proc. Suppl.}
  \textbf{\bibinfo{volume}{53}}, \bibinfo{pages}{877}, \arXiv{hep-lat/9608021}.

\bibitem[{\citenamefont{Creutz}(1980)}]{Creutz:1980zw}
\bibinfo{author}{\bibnamefont{Creutz}, \bibfnamefont{M.}},
  \bibinfo{year}{1980}, \bibinfo{journal}{Phys. Rev.}
  \textbf{\bibinfo{volume}{D21}}, \bibinfo{pages}{2308}.

\bibitem[{\citenamefont{Creutz}(2006{\natexlab{a}})}]{Creutz:2006wv}
\bibinfo{author}{\bibnamefont{Creutz}, \bibfnamefont{M.}},
  \bibinfo{year}{2006}{\natexlab{a}}, \bibinfo{journal}{PoS}
  \textbf{\bibinfo{volume}{LAT2006}}, \bibinfo{pages}{208},
  \arXiv{hep-lat/0608020}.

\bibitem[{\citenamefont{Creutz}(2006{\natexlab{b}})}]{Creutz:2006ys}
\bibinfo{author}{\bibnamefont{Creutz}, \bibfnamefont{M.}},
  \bibinfo{year}{2006}{\natexlab{b}}, \arXiv{hep-lat/0603020}.

\bibitem[{\citenamefont{Creutz}(2007{\natexlab{a}})}]{Creutz:2007pr}
\bibinfo{author}{\bibnamefont{Creutz}, \bibfnamefont{M.}},
  \bibinfo{year}{2007}{\natexlab{a}}, \bibinfo{journal}{Phys. Lett.}
  \textbf{\bibinfo{volume}{B649}}, \bibinfo{pages}{241}, \arXiv{0704.2016}.

\bibitem[{\citenamefont{Creutz}(2007{\natexlab{b}})}]{Creutz:2007yg}
\bibinfo{author}{\bibnamefont{Creutz}, \bibfnamefont{M.}},
  \bibinfo{year}{2007}{\natexlab{b}}, \bibinfo{journal}{Phys. Lett.}
  \textbf{\bibinfo{volume}{B649}}, \bibinfo{pages}{230},
  \arXiv{hep-lat/0701018}.

\bibitem[{\citenamefont{Creutz}(2007{\natexlab{c}})}]{Creutz:2007rk}
\bibinfo{author}{\bibnamefont{Creutz}, \bibfnamefont{M.}},
  \bibinfo{year}{2007}{\natexlab{c}}, \bibinfo{journal}{PoS}
  \textbf{\bibinfo{volume}{LAT2007}}, \bibinfo{pages}{007}, \arXiv{0708.1295}.

\bibitem[{\citenamefont{Creutz}(2008{\natexlab{a}})}]{Creutz:2008kb}
\bibinfo{author}{\bibnamefont{Creutz}, \bibfnamefont{M.}},
  \bibinfo{year}{2008}{\natexlab{a}}, \bibinfo{journal}{Phys. Rev.}
  \textbf{\bibinfo{volume}{D78}}, \bibinfo{pages}{078501}, \arXiv{0805.1350}.

\bibitem[{\citenamefont{Creutz}(2008{\natexlab{b}})}]{Creutz:2008nk}
\bibinfo{author}{\bibnamefont{Creutz}, \bibfnamefont{M.}},
  \bibinfo{year}{2008}{\natexlab{b}}, \bibinfo{journal}{PoS}
  \textbf{\bibinfo{volume}{CONFINEMENT8}}, \bibinfo{pages}{016},
  \arXiv{0810.4526}.

\bibitem[{\citenamefont{Dalgic} \emph{et~al.}(2004)\citenamefont{Dalgic,
  Shigemitsu, and Wingate}}]{Dalgic:2003uf}
\bibinfo{author}{\bibnamefont{Dalgic}, \bibfnamefont{E.}},
  \bibinfo{author}{\bibfnamefont{J.}~\bibnamefont{Shigemitsu}}, and
  \bibinfo{author}{\bibfnamefont{M.}~\bibnamefont{Wingate}},
  \bibinfo{year}{2004}, \bibinfo{journal}{Phys. Rev.}
  \textbf{\bibinfo{volume}{D69}}, \bibinfo{pages}{074501},
  \arXiv{hep-lat/0312017}.

\bibitem[{Dalgic \emph{et~al.}(2006)\citenamefont{Dalgic}
  \emph{et~al.}}]{Dalgic:2006dt}
\bibinfo{author}{\bibnamefont{Dalgic}, \bibfnamefont{E.}}, \emph{et~al.},
  \bibinfo{year}{2006}, \bibinfo{journal}{Phys. Rev.}
  \textbf{\bibinfo{volume}{D73}}, \bibinfo{pages}{074502},
  \bibinfo{note}{{Erratum-ibid.\ {\bf D75}, 119906 (2007)}},
  \arXiv{hep-lat/0601021}.

\bibitem[{Dalgic \emph{et~al.}(2007)\citenamefont{Dalgic}
  \emph{et~al.}}]{Dalgic:2006gp}
\bibinfo{author}{\bibnamefont{Dalgic}, \bibfnamefont{E.}}, \emph{et~al.},
  \bibinfo{year}{2007}, \bibinfo{journal}{Phys. Rev.}
  \textbf{\bibinfo{volume}{D76}}, \bibinfo{pages}{011501},
  \arXiv{hep-lat/0610104}.

\bibitem[{\citenamefont{Dashen}(1969)}]{Dashen:1969eg}
\bibinfo{author}{\bibnamefont{Dashen}, \bibfnamefont{R.~F.}},
  \bibinfo{year}{1969}, \bibinfo{journal}{Phys. Rev.}
  \textbf{\bibinfo{volume}{183}}, \bibinfo{pages}{1245}.

\bibitem[{\citenamefont{Dashen}(1971)}]{Dashen:1970et}
\bibinfo{author}{\bibnamefont{Dashen}, \bibfnamefont{R.~F.}},
  \bibinfo{year}{1971}, \bibinfo{journal}{Phys. Rev.}
  \textbf{\bibinfo{volume}{D3}}, \bibinfo{pages}{1879}.

\bibitem[{\citenamefont{Davies}(2008)}]{Davies:2008hs}
\bibinfo{author}{\bibnamefont{Davies}, \bibfnamefont{C.~T.~H.}}
  (\bibinfo{collaboration}{HPQCD}), \bibinfo{year}{2008}, \arXiv{0810.3309}.

\bibitem[{Davies \emph{et~al.}(1994)\citenamefont{Davies}
  \emph{et~al.}}]{Davies:1994mp}
\bibinfo{author}{\bibnamefont{Davies}, \bibfnamefont{C.~T.~H.}}, \emph{et~al.},
  \bibinfo{year}{1994}, \bibinfo{journal}{Phys. Rev.}
  \textbf{\bibinfo{volume}{D50}}, \bibinfo{pages}{6963},
  \arXiv{hep-lat/9406017}.

\bibitem[{Davies \emph{et~al.}(2004)\citenamefont{Davies}
  \emph{et~al.}}]{Davies:2003ik}
\bibinfo{author}{\bibnamefont{Davies}, \bibfnamefont{C.~T.~H.}}, \emph{et~al.}
  (\bibinfo{collaboration}{HPQCD}), \bibinfo{year}{2004},
  \bibinfo{journal}{Phys. Rev. Lett.} \textbf{\bibinfo{volume}{92}},
  \bibinfo{pages}{022001}, \arXiv{hep-lat/0304004}.

\bibitem[{Davies \emph{et~al.}(2008)\citenamefont{Davies}
  \emph{et~al.}}]{Davies:2008sw}
\bibinfo{author}{\bibnamefont{Davies}, \bibfnamefont{C.~T.~H.}}, \emph{et~al.}
  (\bibinfo{collaboration}{HPQCD}), \bibinfo{year}{2008},
  \bibinfo{journal}{Phys. Rev.} \textbf{\bibinfo{volume}{D78}},
  \bibinfo{pages}{114507}, \arXiv{0807.1687}.

\bibitem[{Davies \emph{et~al.}(2009)\citenamefont{Davies}
  \emph{et~al.}}]{Davies:2008nq}
\bibinfo{author}{\bibnamefont{Davies}, \bibfnamefont{C.~T.~H.}}, \emph{et~al.}
  (\bibinfo{collaboration}{HPQCD}), \bibinfo{year}{2009},
  \bibinfo{journal}{PoS} \textbf{\bibinfo{volume}{LAT2008}},
  \bibinfo{pages}{118}, \arXiv{0810.3548}.

\bibitem[{\citenamefont{DeGrand} \emph{et~al.}(1997)\citenamefont{DeGrand,
  Hasenfratz, and Kovacs}}]{DeGrand:1997gu}
\bibinfo{author}{\bibnamefont{DeGrand}, \bibfnamefont{T.~A.}},
  \bibinfo{author}{\bibfnamefont{A.}~\bibnamefont{Hasenfratz}}, and
  \bibinfo{author}{\bibfnamefont{T.~G.} \bibnamefont{Kovacs}},
  \bibinfo{year}{1997}, \bibinfo{journal}{Nucl. Phys.}
  \textbf{\bibinfo{volume}{B505}}, \bibinfo{pages}{417},
  \arXiv{hep-lat/9705009}.

\bibitem[{\citenamefont{DeGrand and Heller}(2002)}]{DeGrand:2002gm}
\bibinfo{author}{\bibnamefont{DeGrand}, \bibfnamefont{T.~A.}}, and
  \bibinfo{author}{\bibfnamefont{U.~M.} \bibnamefont{Heller}}
  (\bibinfo{collaboration}{MILC}), \bibinfo{year}{2002},
  \bibinfo{journal}{Phys. Rev.} \textbf{\bibinfo{volume}{D65}},
  \bibinfo{pages}{114501}, \arXiv{hep-lat/0202001}.

\bibitem[{\citenamefont{DeTar and Heller}(2009)}]{DeTar:2009ef}
\bibinfo{author}{\bibnamefont{DeTar}, \bibfnamefont{C.}}, and
  \bibinfo{author}{\bibfnamefont{U.~M.} \bibnamefont{Heller}},
  \bibinfo{year}{2009}, \arXiv{0905.2949}.

\bibitem[{DeTar \emph{et~al.}(2009)\citenamefont{DeTar}
  \emph{et~al.}}]{DeTar:LAT2009}
\bibinfo{author}{\bibnamefont{DeTar}, \bibfnamefont{C.}}, \emph{et~al.}
  (\bibinfo{collaboration}{Fermilab Lattice and MILC}), \bibinfo{year}{2009},
  \bibinfo{journal}{PoS} \textbf{\bibinfo{volume}{LAT2009}}, \bibinfo{note}{to
  apprear}.

\bibitem[{\citenamefont{DeTar and Levkova}(2007)}]{Detar:2007ni}
\bibinfo{author}{\bibnamefont{DeTar}, \bibfnamefont{C.~E.}}, and
  \bibinfo{author}{\bibfnamefont{L.}~\bibnamefont{Levkova}}
  (\bibinfo{collaboration}{Fermilab Lattice}), \bibinfo{year}{2007},
  \bibinfo{journal}{PoS} \textbf{\bibinfo{volume}{LAT2007}},
  \bibinfo{pages}{116}, \arXiv{0710.1322}.

\bibitem[{Detmold \emph{et~al.}(2008{\natexlab{a}})\citenamefont{Detmold}
  \emph{et~al.}}]{Detmold:2008yn}
\bibinfo{author}{\bibnamefont{Detmold}, \bibfnamefont{W.}}, \emph{et~al.},
  \bibinfo{year}{2008}{\natexlab{a}}, \bibinfo{journal}{Phys. Rev.}
  \textbf{\bibinfo{volume}{D78}}, \bibinfo{pages}{054514}, \arXiv{0807.1856}.

\bibitem[{Detmold \emph{et~al.}(2008{\natexlab{b}})\citenamefont{Detmold}
  \emph{et~al.}}]{Detmold:2008fn}
\bibinfo{author}{\bibnamefont{Detmold}, \bibfnamefont{W.}}, \emph{et~al.},
  \bibinfo{year}{2008}{\natexlab{b}}, \bibinfo{journal}{Phys. Rev.}
  \textbf{\bibinfo{volume}{D78}}, \bibinfo{pages}{014507}, \arXiv{0803.2728}.

\bibitem[{\citenamefont{Di~Lodovico}(2008)}]{DiLodovico:2008}
\bibinfo{author}{\bibnamefont{Di~Lodovico}, \bibfnamefont{F.}},
  \bibinfo{year}{2008}, \bibinfo{note}{update presented at ICHEP 2008, \\
  http://www.slac.stanford. edu/xorg/hfag/semi/ichep08/home.shtml}.

\bibitem[{\citenamefont{Dobrescu and Kronfeld}(2008)}]{Dobrescu:2008er}
\bibinfo{author}{\bibnamefont{Dobrescu}, \bibfnamefont{B.~A.}}, and
  \bibinfo{author}{\bibfnamefont{A.~S.} \bibnamefont{Kronfeld}},
  \bibinfo{year}{2008}, \bibinfo{journal}{Phys. Rev. Lett.}
  \textbf{\bibinfo{volume}{100}}, \bibinfo{pages}{241802}, \arXiv{0803.0512}.

\bibitem[{\citenamefont{van~den Doel and Smit}(1983)}]{vandenDoel:1983mf}
\bibinfo{author}{\bibnamefont{van~den Doel}, \bibfnamefont{C.}}, and
  \bibinfo{author}{\bibfnamefont{J.}~\bibnamefont{Smit}}, \bibinfo{year}{1983},
  \bibinfo{journal}{Nucl. Phys.} \textbf{\bibinfo{volume}{B228}},
  \bibinfo{pages}{122}.

\bibitem[{\citenamefont{Dong and Liu}(1994)}]{Dong:1993pk}
\bibinfo{author}{\bibnamefont{Dong}, \bibfnamefont{S.-J.}}, and
  \bibinfo{author}{\bibfnamefont{K.-F.} \bibnamefont{Liu}},
  \bibinfo{year}{1994}, \bibinfo{journal}{Phys. Lett.}
  \textbf{\bibinfo{volume}{B328}}, \bibinfo{pages}{130},
  \arXiv{hep-lat/9308015}.

\bibitem[{\citenamefont{Duane} \emph{et~al.}(1987)\citenamefont{Duane, Kennedy,
  Pendleton, and Roweth}}]{Duane:1987de}
\bibinfo{author}{\bibnamefont{Duane}, \bibfnamefont{S.}},
  \bibinfo{author}{\bibfnamefont{A.~D.} \bibnamefont{Kennedy}},
  \bibinfo{author}{\bibfnamefont{B.~J.} \bibnamefont{Pendleton}}, and
  \bibinfo{author}{\bibfnamefont{D.}~\bibnamefont{Roweth}},
  \bibinfo{year}{1987}, \bibinfo{journal}{Phys. Lett.}
  \textbf{\bibinfo{volume}{B195}}, \bibinfo{pages}{216}.

\bibitem[{\citenamefont{Duane and Kogut}(1985)}]{Duane:1985hz}
\bibinfo{author}{\bibnamefont{Duane}, \bibfnamefont{S.}}, and
  \bibinfo{author}{\bibfnamefont{J.~B.} \bibnamefont{Kogut}},
  \bibinfo{year}{1985}, \bibinfo{journal}{Phys. Rev. Lett.}
  \textbf{\bibinfo{volume}{55}}, \bibinfo{pages}{2774}.

\bibitem[{\citenamefont{Duane and Kogut}(1986)}]{Duane:1986iw}
\bibinfo{author}{\bibnamefont{Duane}, \bibfnamefont{S.}}, and
  \bibinfo{author}{\bibfnamefont{J.~B.} \bibnamefont{Kogut}},
  \bibinfo{year}{1986}, \bibinfo{journal}{Nucl. Phys.}
  \textbf{\bibinfo{volume}{B275}}, \bibinfo{pages}{398}.

\bibitem[{\citenamefont{Duncan} \emph{et~al.}(1996)\citenamefont{Duncan,
  Eichten, and Thacker}}]{Duncan:1996xy}
\bibinfo{author}{\bibnamefont{Duncan}, \bibfnamefont{A.}},
  \bibinfo{author}{\bibfnamefont{E.}~\bibnamefont{Eichten}}, and
  \bibinfo{author}{\bibfnamefont{H.}~\bibnamefont{Thacker}},
  \bibinfo{year}{1996}, \bibinfo{journal}{Phys. Rev. Lett.}
  \textbf{\bibinfo{volume}{76}}, \bibinfo{pages}{3894},
  \arXiv{hep-lat/9602005}.

\bibitem[{\citenamefont{Duncan} \emph{et~al.}(1997)\citenamefont{Duncan,
  Eichten, and Thacker}}]{Duncan:1996be}
\bibinfo{author}{\bibnamefont{Duncan}, \bibfnamefont{A.}},
  \bibinfo{author}{\bibfnamefont{E.}~\bibnamefont{Eichten}}, and
  \bibinfo{author}{\bibfnamefont{H.}~\bibnamefont{Thacker}},
  \bibinfo{year}{1997}, \bibinfo{journal}{Phys. Lett.}
  \textbf{\bibinfo{volume}{B409}}, \bibinfo{pages}{387},
  \arXiv{hep-lat/9607032}.

\bibitem[{\citenamefont{Duncan} \emph{et~al.}(1982)\citenamefont{Duncan,
  Roskies, and Vaidya}}]{Duncan:1982xe}
\bibinfo{author}{\bibnamefont{Duncan}, \bibfnamefont{A.}},
  \bibinfo{author}{\bibfnamefont{R.}~\bibnamefont{Roskies}}, and
  \bibinfo{author}{\bibfnamefont{H.}~\bibnamefont{Vaidya}},
  \bibinfo{year}{1982}, \bibinfo{journal}{Phys. Lett.}
  \textbf{\bibinfo{volume}{B114}}, \bibinfo{pages}{439}.

\bibitem[{\citenamefont{D{\"u}rr and Hoelbling}(2005)}]{Durr:2004ta}
\bibinfo{author}{\bibnamefont{D{\"u}rr}, \bibfnamefont{S.}}, and
  \bibinfo{author}{\bibfnamefont{C.}~\bibnamefont{Hoelbling}},
  \bibinfo{year}{2005}, \bibinfo{journal}{Phys. Rev.}
  \textbf{\bibinfo{volume}{D71}}, \bibinfo{pages}{054501},
  \arXiv{hep-lat/0411022}.

\bibitem[{\citenamefont{D{\"u}rr and Hoelbling}(2006)}]{Durr:2006ze}
\bibinfo{author}{\bibnamefont{D{\"u}rr}, \bibfnamefont{S.}}, and
  \bibinfo{author}{\bibfnamefont{C.}~\bibnamefont{Hoelbling}},
  \bibinfo{year}{2006}, \bibinfo{journal}{Phys. Rev.}
  \textbf{\bibinfo{volume}{D74}}, \bibinfo{pages}{014513},
  \arXiv{hep-lat/0604005}.

\bibitem[{\citenamefont{D{\"u}rr} \emph{et~al.}(2004)\citenamefont{D{\"u}rr,
  Hoelbling, and Wenger}}]{Durr:2004as}
\bibinfo{author}{\bibnamefont{D{\"u}rr}, \bibfnamefont{S.}},
  \bibinfo{author}{\bibfnamefont{C.}~\bibnamefont{Hoelbling}}, and
  \bibinfo{author}{\bibfnamefont{U.}~\bibnamefont{Wenger}},
  \bibinfo{year}{2004}, \bibinfo{journal}{Phys. Rev.}
  \textbf{\bibinfo{volume}{D70}}, \bibinfo{pages}{094502},
  \arXiv{hep-lat/0406027}.

\bibitem[{D{\"u}rr \emph{et~al.}(2008)\citenamefont{D{\"u}rr}
  \emph{et~al.}}]{Durr:2008zz}
\bibinfo{author}{\bibnamefont{D{\"u}rr}, \bibfnamefont{S.}}, \emph{et~al.},
  \bibinfo{year}{2008}, \bibinfo{journal}{Science}
  \textbf{\bibinfo{volume}{322}}, \bibinfo{pages}{1224}.

\bibitem[{D{\"u}rr \emph{et~al.}(2009)\citenamefont{D{\"u}rr}
  \emph{et~al.}}]{Durr:2008rw}
\bibinfo{author}{\bibnamefont{D{\"u}rr}, \bibfnamefont{S.}}, \emph{et~al.},
  \bibinfo{year}{2009}, \bibinfo{journal}{Phys. Rev.}
  \textbf{\bibinfo{volume}{D79}}, \bibinfo{pages}{014501}, \arXiv{0802.2706}.

\bibitem[{\citenamefont{Edwards and Heller}(2001)}]{Edwards:2000qv}
\bibinfo{author}{\bibnamefont{Edwards}, \bibfnamefont{R.~G.}}, and
  \bibinfo{author}{\bibfnamefont{U.~M.} \bibnamefont{Heller}},
  \bibinfo{year}{2001}, \bibinfo{journal}{Phys. Rev.}
  \textbf{\bibinfo{volume}{D63}}, \bibinfo{pages}{094505},
  \arXiv{hep-lat/0005002}.

\bibitem[{\citenamefont{Edwards} \emph{et~al.}(1999)\citenamefont{Edwards,
  Heller, and Narayanan}}]{Edwards:1998yw}
\bibinfo{author}{\bibnamefont{Edwards}, \bibfnamefont{R.~G.}},
  \bibinfo{author}{\bibfnamefont{U.~M.} \bibnamefont{Heller}}, and
  \bibinfo{author}{\bibfnamefont{R.}~\bibnamefont{Narayanan}},
  \bibinfo{year}{1999}, \bibinfo{journal}{Nucl. Phys.}
  \textbf{\bibinfo{volume}{B540}}, \bibinfo{pages}{457},
  \arXiv{hep-lat/9807017}.

\bibitem[{Edwards \emph{et~al.}(2006{\natexlab{a}})\citenamefont{Edwards}
  \emph{et~al.}}]{Edwards:2006zza}
\bibinfo{author}{\bibnamefont{Edwards}, \bibfnamefont{R.~G.}}, \emph{et~al.}
  (\bibinfo{collaboration}{LHPC}), \bibinfo{year}{2006}{\natexlab{a}},
  \bibinfo{journal}{PoS} \textbf{\bibinfo{volume}{LAT2006}},
  \bibinfo{pages}{195}.

\bibitem[{Edwards \emph{et~al.}(2006{\natexlab{b}})\citenamefont{Edwards}
  \emph{et~al.}}]{Edwards:2005ym}
\bibinfo{author}{\bibnamefont{Edwards}, \bibfnamefont{R.~G.}}, \emph{et~al.}
  (\bibinfo{collaboration}{LHPC}), \bibinfo{year}{2006}{\natexlab{b}},
  \bibinfo{journal}{Phys. Rev. Lett.} \textbf{\bibinfo{volume}{96}},
  \bibinfo{pages}{052001}, \arXiv{hep-lat/0510062}.

\bibitem[{\citenamefont{Eichten and Hill}(1990)}]{Eichten:1989zv}
\bibinfo{author}{\bibnamefont{Eichten}, \bibfnamefont{E.}}, and
  \bibinfo{author}{\bibfnamefont{B.~R.} \bibnamefont{Hill}},
  \bibinfo{year}{1990}, \bibinfo{journal}{Phys. Lett.}
  \textbf{\bibinfo{volume}{B234}}, \bibinfo{pages}{511}.

\bibitem[{Eisenstein \emph{et~al.}(2008)\citenamefont{Eisenstein}
  \emph{et~al.}}]{Eisenstein:2008sq}
\bibinfo{author}{\bibnamefont{Eisenstein}, \bibfnamefont{B.~I.}}, \emph{et~al.}
  (\bibinfo{collaboration}{CLEO}), \bibinfo{year}{2008},
  \bibinfo{journal}{Phys. Rev.} \textbf{\bibinfo{volume}{D78}},
  \bibinfo{pages}{052003}, \arXiv{0806.2112}.

\bibitem[{\citenamefont{El-Khadra} \emph{et~al.}(2007)\citenamefont{El-Khadra,
  Gamiz, Kronfeld, and Nobes}}]{ElKhadra:2007qe}
\bibinfo{author}{\bibnamefont{El-Khadra}, \bibfnamefont{A.~X.}},
  \bibinfo{author}{\bibfnamefont{E.}~\bibnamefont{Gamiz}},
  \bibinfo{author}{\bibfnamefont{A.~S.} \bibnamefont{Kronfeld}}, and
  \bibinfo{author}{\bibfnamefont{M.~A.} \bibnamefont{Nobes}},
  \bibinfo{year}{2007}, \bibinfo{journal}{PoS}
  \textbf{\bibinfo{volume}{LAT2007}}, \bibinfo{pages}{242}, \arXiv{0710.1437}.

\bibitem[{\citenamefont{El-Khadra} \emph{et~al.}(1997)\citenamefont{El-Khadra,
  Kronfeld, and Mackenzie}}]{ElKhadra:1996mp}
\bibinfo{author}{\bibnamefont{El-Khadra}, \bibfnamefont{A.~X.}},
  \bibinfo{author}{\bibfnamefont{A.~S.} \bibnamefont{Kronfeld}}, and
  \bibinfo{author}{\bibfnamefont{P.~B.} \bibnamefont{Mackenzie}},
  \bibinfo{year}{1997}, \bibinfo{journal}{Phys. Rev.}
  \textbf{\bibinfo{volume}{D55}}, \bibinfo{pages}{3933},
  \arXiv{hep-lat/9604004}.

\bibitem[{\citenamefont{van~den Eshof} \emph{et~al.}(2002)\citenamefont{van~den
  Eshof, Frommer, Lippert, Schilling, and van~der Vorst}}]{vandenEshof:2002ms}
\bibinfo{author}{\bibnamefont{van~den Eshof}, \bibfnamefont{J.}},
  \bibinfo{author}{\bibfnamefont{A.}~\bibnamefont{Frommer}},
  \bibinfo{author}{\bibfnamefont{T.}~\bibnamefont{Lippert}},
  \bibinfo{author}{\bibfnamefont{K.}~\bibnamefont{Schilling}}, and
  \bibinfo{author}{\bibfnamefont{H.~A.} \bibnamefont{van~der Vorst}},
  \bibinfo{year}{2002}, \bibinfo{journal}{Comput. Phys. Commun.}
  \textbf{\bibinfo{volume}{146}}, \bibinfo{pages}{203},
  \arXiv{hep-lat/0202025}.

\bibitem[{\citenamefont{Evans} \emph{et~al.}(2009)\citenamefont{Evans,
  El-Khadra, and Gamiz}}]{ToddEvans:2008YQ}
\bibinfo{author}{\bibnamefont{Evans}, \bibfnamefont{R.~T.}},
  \bibinfo{author}{\bibfnamefont{A.~X.} \bibnamefont{El-Khadra}}, and
  \bibinfo{author}{\bibfnamefont{E.}~\bibnamefont{Gamiz}},
  \bibinfo{year}{2009}, \bibinfo{journal}{PoS}
  \textbf{\bibinfo{volume}{LAT2008}}, \bibinfo{pages}{052}.

\bibitem[{\citenamefont{Evans} \emph{et~al.}(2007)\citenamefont{Evans, Gamiz,
  El-Khadra, and Di~Pierro}}]{ToddEvans:2007yq}
\bibinfo{author}{\bibnamefont{Evans}, \bibfnamefont{R.~T.}},
  \bibinfo{author}{\bibfnamefont{E.}~\bibnamefont{Gamiz}},
  \bibinfo{author}{\bibfnamefont{A.~X.} \bibnamefont{El-Khadra}}, and
  \bibinfo{author}{\bibfnamefont{M.}~\bibnamefont{Di~Pierro}},
  \bibinfo{year}{2007}, \bibinfo{journal}{PoS}
  \textbf{\bibinfo{volume}{LAT2007}}, \bibinfo{pages}{354}, \arXiv{0710.2880}.

\bibitem[{\citenamefont{Fischer}(2006)}]{Fischer:2006ub}
\bibinfo{author}{\bibnamefont{Fischer}, \bibfnamefont{C.~S.}},
  \bibinfo{year}{2006}, \bibinfo{journal}{J. Phys.}
  \textbf{\bibinfo{volume}{G32}}, \bibinfo{pages}{R253},
  \arXiv{hep-ph/0605173}.

\bibitem[{Foley \emph{et~al.}(2005)\citenamefont{Foley}
  \emph{et~al.}}]{Foley:2005ac}
\bibinfo{author}{\bibnamefont{Foley}, \bibfnamefont{J.}}, \emph{et~al.},
  \bibinfo{year}{2005}, \bibinfo{journal}{Comput. Phys. Commun.}
  \textbf{\bibinfo{volume}{172}}, \bibinfo{pages}{145},
  \arXiv{hep-lat/0505023}.

\bibitem[{\citenamefont{Follana} \emph{et~al.}(2008)\citenamefont{Follana,
  Davies, Lepage, and Shigemitsu}}]{Follana:2007uv}
\bibinfo{author}{\bibnamefont{Follana}, \bibfnamefont{E.}},
  \bibinfo{author}{\bibfnamefont{C.~T.~H.} \bibnamefont{Davies}},
  \bibinfo{author}{\bibfnamefont{G.~P.} \bibnamefont{Lepage}}, and
  \bibinfo{author}{\bibfnamefont{J.}~\bibnamefont{Shigemitsu}}
  (\bibinfo{collaboration}{HPQCD}), \bibinfo{year}{2008},
  \bibinfo{journal}{Phys. Rev. Lett.} \textbf{\bibinfo{volume}{100}},
  \bibinfo{pages}{062002}, \arXiv{0706.1726}.

\bibitem[{\citenamefont{Follana} \emph{et~al.}(2004)\citenamefont{Follana,
  Hart, and Davies}}]{Follana:2004sz}
\bibinfo{author}{\bibnamefont{Follana}, \bibfnamefont{E.}},
  \bibinfo{author}{\bibfnamefont{A.}~\bibnamefont{Hart}}, and
  \bibinfo{author}{\bibfnamefont{C.~T.~H.} \bibnamefont{Davies}}
  (\bibinfo{collaboration}{HPQCD}), \bibinfo{year}{2004},
  \bibinfo{journal}{Phys. Rev. Lett.} \textbf{\bibinfo{volume}{93}},
  \bibinfo{pages}{241601}, \arXiv{hep-lat/0406010}.

\bibitem[{Follana \emph{et~al.}(2007)\citenamefont{Follana}
  \emph{et~al.}}]{Follana:2006rc}
\bibinfo{author}{\bibnamefont{Follana}, \bibfnamefont{E.}}, \emph{et~al.}
  (\bibinfo{collaboration}{HPQCD}), \bibinfo{year}{2007},
  \bibinfo{journal}{Phys. Rev.} \textbf{\bibinfo{volume}{D75}},
  \bibinfo{pages}{054502}, \arXiv{hep-lat/0610092}.

\bibitem[{\citenamefont{de~Forcrand}
  \emph{et~al.}(1997)\citenamefont{de~Forcrand, Garcia~Perez, and
  Stamatescu}}]{deForcrand:1997sq}
\bibinfo{author}{\bibnamefont{de~Forcrand}, \bibfnamefont{P.}},
  \bibinfo{author}{\bibfnamefont{M.}~\bibnamefont{Garcia~Perez}}, and
  \bibinfo{author}{\bibfnamefont{I.-O.} \bibnamefont{Stamatescu}},
  \bibinfo{year}{1997}, \bibinfo{journal}{Nucl. Phys.}
  \textbf{\bibinfo{volume}{B499}}, \bibinfo{pages}{409},
  \arXiv{hep-lat/9701012}.

\bibitem[{\citenamefont{Frezzotti} \emph{et~al.}(2000)\citenamefont{Frezzotti,
  Grassi, Sint, and Weisz}}]{Frezzotti:1999vv}
\bibinfo{author}{\bibnamefont{Frezzotti}, \bibfnamefont{R.}},
  \bibinfo{author}{\bibfnamefont{P.~A.} \bibnamefont{Grassi}},
  \bibinfo{author}{\bibfnamefont{S.}~\bibnamefont{Sint}}, and
  \bibinfo{author}{\bibfnamefont{P.}~\bibnamefont{Weisz}},
  \bibinfo{year}{2000}, \bibinfo{journal}{Nucl. Phys. Proc. Suppl.}
  \textbf{\bibinfo{volume}{83}}, \bibinfo{pages}{941}, \arXiv{hep-lat/9909003}.

\bibitem[{\citenamefont{Frezzotti} \emph{et~al.}(2001)\citenamefont{Frezzotti,
  Grassi, Sint, and Weisz}}]{Frezzotti:2000nk}
\bibinfo{author}{\bibnamefont{Frezzotti}, \bibfnamefont{R.}},
  \bibinfo{author}{\bibfnamefont{P.~A.} \bibnamefont{Grassi}},
  \bibinfo{author}{\bibfnamefont{S.}~\bibnamefont{Sint}}, and
  \bibinfo{author}{\bibfnamefont{P.}~\bibnamefont{Weisz}}
  (\bibinfo{collaboration}{Alpha}), \bibinfo{year}{2001},
  \bibinfo{journal}{JHEP} \textbf{\bibinfo{volume}{08}}, \bibinfo{pages}{058},
  \arXiv{hep-lat/0101001}.

\bibitem[{\citenamefont{Frezzotti and Rossi}(2004)}]{Frezzotti:2003ni}
\bibinfo{author}{\bibnamefont{Frezzotti}, \bibfnamefont{R.}}, and
  \bibinfo{author}{\bibfnamefont{G.~C.} \bibnamefont{Rossi}},
  \bibinfo{year}{2004}, \bibinfo{journal}{JHEP} \textbf{\bibinfo{volume}{08}},
  \bibinfo{pages}{007}, \arXiv{hep-lat/0306014}.

\bibitem[{\citenamefont{Frommer} \emph{et~al.}(1995)\citenamefont{Frommer,
  N{\"o}ckel, G{\"u}sken, Lippert, and Schilling}}]{Frommer:1995ik}
\bibinfo{author}{\bibnamefont{Frommer}, \bibfnamefont{A.}},
  \bibinfo{author}{\bibfnamefont{B.}~\bibnamefont{N{\"o}ckel}},
  \bibinfo{author}{\bibfnamefont{S.}~\bibnamefont{G{\"u}sken}},
  \bibinfo{author}{\bibfnamefont{T.}~\bibnamefont{Lippert}}, and
  \bibinfo{author}{\bibfnamefont{K.}~\bibnamefont{Schilling}},
  \bibinfo{year}{1995}, \bibinfo{journal}{Int. J. Mod. Phys.}
  \textbf{\bibinfo{volume}{C6}}, \bibinfo{pages}{627}, \arXiv{hep-lat/9504020}.

\bibitem[{\citenamefont{Fukugita} \emph{et~al.}(1993)\citenamefont{Fukugita,
  Ishizuka, Mino, Okawa, and Ukawa}}]{Fukugita:1992hr}
\bibinfo{author}{\bibnamefont{Fukugita}, \bibfnamefont{M.}},
  \bibinfo{author}{\bibfnamefont{N.}~\bibnamefont{Ishizuka}},
  \bibinfo{author}{\bibfnamefont{H.}~\bibnamefont{Mino}},
  \bibinfo{author}{\bibfnamefont{M.}~\bibnamefont{Okawa}}, and
  \bibinfo{author}{\bibfnamefont{A.}~\bibnamefont{Ukawa}},
  \bibinfo{year}{1993}, \bibinfo{journal}{Phys. Rev.}
  \textbf{\bibinfo{volume}{D47}}, \bibinfo{pages}{4739}.

\bibitem[{\citenamefont{Furman and Shamir}(1995)}]{Furman:1994ky}
\bibinfo{author}{\bibnamefont{Furman}, \bibfnamefont{V.}}, and
  \bibinfo{author}{\bibfnamefont{Y.}~\bibnamefont{Shamir}},
  \bibinfo{year}{1995}, \bibinfo{journal}{Nucl. Phys.}
  \textbf{\bibinfo{volume}{B439}}, \bibinfo{pages}{54},
  \arXiv{hep-lat/9405004}.

\bibitem[{\citenamefont{Furui and Nakajima}(2006)}]{Furui:2006ks}
\bibinfo{author}{\bibnamefont{Furui}, \bibfnamefont{S.}}, and
  \bibinfo{author}{\bibfnamefont{H.}~\bibnamefont{Nakajima}},
  \bibinfo{year}{2006}, \bibinfo{journal}{Phys. Rev.}
  \textbf{\bibinfo{volume}{D73}}, \bibinfo{pages}{074503}.

\bibitem[{\citenamefont{Gambino} \emph{et~al.}(2007)\citenamefont{Gambino,
  Giordano, Ossola, and Uraltsev}}]{Gambino:2007rp}
\bibinfo{author}{\bibnamefont{Gambino}, \bibfnamefont{P.}},
  \bibinfo{author}{\bibfnamefont{P.}~\bibnamefont{Giordano}},
  \bibinfo{author}{\bibfnamefont{G.}~\bibnamefont{Ossola}}, and
  \bibinfo{author}{\bibfnamefont{N.}~\bibnamefont{Uraltsev}},
  \bibinfo{year}{2007}, \bibinfo{journal}{JHEP} \textbf{\bibinfo{volume}{10}},
  \bibinfo{pages}{058}, \arXiv{0707.2493}.

\bibitem[{\citenamefont{Gamiz} \emph{et~al.}(2009)\citenamefont{Gamiz, Davies,
  Lepage, Shigemitsu, and Wingate}}]{Gamiz:2009ku}
\bibinfo{author}{\bibnamefont{Gamiz}, \bibfnamefont{E.}},
  \bibinfo{author}{\bibfnamefont{C.~T.~H.} \bibnamefont{Davies}},
  \bibinfo{author}{\bibfnamefont{G.~P.} \bibnamefont{Lepage}},
  \bibinfo{author}{\bibfnamefont{J.}~\bibnamefont{Shigemitsu}}, and
  \bibinfo{author}{\bibfnamefont{M.}~\bibnamefont{Wingate}}
  (\bibinfo{collaboration}{HPQCD}), \bibinfo{year}{2009},
  \bibinfo{journal}{Phys. Rev.} \textbf{\bibinfo{volume}{D80}},
  \bibinfo{pages}{014503}, \arXiv{0902.1815}.

\bibitem[{Gamiz \emph{et~al.}(2006)\citenamefont{Gamiz}
  \emph{et~al.}}]{Gamiz:2006sq}
\bibinfo{author}{\bibnamefont{Gamiz}, \bibfnamefont{E.}}, \emph{et~al.}
  (\bibinfo{collaboration}{HPQCD}), \bibinfo{year}{2006},
  \bibinfo{journal}{Phys. Rev.} \textbf{\bibinfo{volume}{D73}},
  \bibinfo{pages}{114502}, \arXiv{hep-lat/0603023}.

\bibitem[{\citenamefont{Gasser and Leutwyler}(1984)}]{Gasser:1983yg}
\bibinfo{author}{\bibnamefont{Gasser}, \bibfnamefont{J.}}, and
  \bibinfo{author}{\bibfnamefont{H.}~\bibnamefont{Leutwyler}},
  \bibinfo{year}{1984}, \bibinfo{journal}{Ann. Phys.}
  \textbf{\bibinfo{volume}{158}}, \bibinfo{pages}{142}.

\bibitem[{\citenamefont{Gasser and Leutwyler}(1985)}]{Gasser:1984gg}
\bibinfo{author}{\bibnamefont{Gasser}, \bibfnamefont{J.}}, and
  \bibinfo{author}{\bibfnamefont{H.}~\bibnamefont{Leutwyler}},
  \bibinfo{year}{1985}, \bibinfo{journal}{Nucl. Phys.}
  \textbf{\bibinfo{volume}{B250}}, \bibinfo{pages}{465}.

\bibitem[{\citenamefont{Gattringer}(2001)}]{Gattringer:2000js}
\bibinfo{author}{\bibnamefont{Gattringer}, \bibfnamefont{C.}},
  \bibinfo{year}{2001}, \bibinfo{journal}{Phys. Rev.}
  \textbf{\bibinfo{volume}{D63}}, \bibinfo{pages}{114501},
  \arXiv{hep-lat/0003005}.

\bibitem[{Ge \emph{et~al.}(2009)\citenamefont{Ge} \emph{et~al.}}]{Ge:2008yi}
\bibinfo{author}{\bibnamefont{Ge}, \bibfnamefont{J.}}, \emph{et~al.}
  (\bibinfo{collaboration}{CLEO}), \bibinfo{year}{2009},
  \bibinfo{journal}{Phys. Rev.} \textbf{\bibinfo{volume}{D79}},
  \bibinfo{pages}{052010}, \arXiv{0810.3878}.

\bibitem[{\citenamefont{Ginsparg and Wilson}(1982)}]{Ginsparg:1981bj}
\bibinfo{author}{\bibnamefont{Ginsparg}, \bibfnamefont{P.~H.}}, and
  \bibinfo{author}{\bibfnamefont{K.~G.} \bibnamefont{Wilson}},
  \bibinfo{year}{1982}, \bibinfo{journal}{Phys. Rev.}
  \textbf{\bibinfo{volume}{D25}}, \bibinfo{pages}{2649}.

\bibitem[{\citenamefont{Gliozzi}(1982)}]{Gliozzi:1982ib}
\bibinfo{author}{\bibnamefont{Gliozzi}, \bibfnamefont{F.}},
  \bibinfo{year}{1982}, \bibinfo{journal}{Nucl. Phys.}
  \textbf{\bibinfo{volume}{B204}}, \bibinfo{pages}{419}.

\bibitem[{\citenamefont{Gockeler}(1984)}]{Gockeler:1984rq}
\bibinfo{author}{\bibnamefont{Gockeler}, \bibfnamefont{M.}},
  \bibinfo{year}{1984}, \bibinfo{journal}{Phys. Lett.}
  \textbf{\bibinfo{volume}{B142}}, \bibinfo{pages}{197}.

\bibitem[{\citenamefont{Golterman}(2008)}]{Golterman:2008gt}
\bibinfo{author}{\bibnamefont{Golterman}, \bibfnamefont{M.}},
  \bibinfo{year}{2008}, \bibinfo{journal}{PoS}
  \textbf{\bibinfo{volume}{CONFINEMENT8}}, \bibinfo{pages}{014},
  \arXiv{0812.3110}.

\bibitem[{\citenamefont{Golterman} \emph{et~al.}(2005)\citenamefont{Golterman,
  Izubuchi, and Shamir}}]{Golterman:2005xa}
\bibinfo{author}{\bibnamefont{Golterman}, \bibfnamefont{M.}},
  \bibinfo{author}{\bibfnamefont{T.}~\bibnamefont{Izubuchi}}, and
  \bibinfo{author}{\bibfnamefont{Y.}~\bibnamefont{Shamir}},
  \bibinfo{year}{2005}, \bibinfo{journal}{Phys. Rev.}
  \textbf{\bibinfo{volume}{D71}}, \bibinfo{pages}{114508},
  \arXiv{hep-lat/0504013}.

\bibitem[{\citenamefont{Golterman} \emph{et~al.}(2006)\citenamefont{Golterman,
  Shamir, and Svetitsky}}]{Golterman:2006rw}
\bibinfo{author}{\bibnamefont{Golterman}, \bibfnamefont{M.}},
  \bibinfo{author}{\bibfnamefont{Y.}~\bibnamefont{Shamir}}, and
  \bibinfo{author}{\bibfnamefont{B.}~\bibnamefont{Svetitsky}},
  \bibinfo{year}{2006}, \bibinfo{journal}{Phys. Rev.}
  \textbf{\bibinfo{volume}{D74}}, \bibinfo{pages}{071501},
  \arXiv{hep-lat/0602026}.

\bibitem[{\citenamefont{Golterman}(1986{\natexlab{a}})}]{Golterman:1986jf}
\bibinfo{author}{\bibnamefont{Golterman}, \bibfnamefont{M.~F.~L.}},
  \bibinfo{year}{1986}{\natexlab{a}}, \bibinfo{journal}{Nucl. Phys.}
  \textbf{\bibinfo{volume}{B278}}, \bibinfo{pages}{417}.

\bibitem[{\citenamefont{Golterman}(1986{\natexlab{b}})}]{Golterman:1985dz}
\bibinfo{author}{\bibnamefont{Golterman}, \bibfnamefont{M.~F.~L.}},
  \bibinfo{year}{1986}{\natexlab{b}}, \bibinfo{journal}{Nucl. Phys.}
  \textbf{\bibinfo{volume}{B273}}, \bibinfo{pages}{663}.

\bibitem[{\citenamefont{Golterman and Smit}(1984)}]{Golterman:1984cy}
\bibinfo{author}{\bibnamefont{Golterman}, \bibfnamefont{M.~F.~L.}}, and
  \bibinfo{author}{\bibfnamefont{J.}~\bibnamefont{Smit}}, \bibinfo{year}{1984},
  \bibinfo{journal}{Nucl. Phys.} \textbf{\bibinfo{volume}{B245}},
  \bibinfo{pages}{61}.

\bibitem[{\citenamefont{Golterman and Smit}(1985)}]{Golterman:1984dn}
\bibinfo{author}{\bibnamefont{Golterman}, \bibfnamefont{M.~F.~L.}}, and
  \bibinfo{author}{\bibfnamefont{J.}~\bibnamefont{Smit}}, \bibinfo{year}{1985},
  \bibinfo{journal}{Nucl. Phys.} \textbf{\bibinfo{volume}{B255}},
  \bibinfo{pages}{328}.

\bibitem[{\citenamefont{Gottlieb} \emph{et~al.}(2008)\citenamefont{Gottlieb,
  Na, and Nagata}}]{Gottlieb:2007ay}
\bibinfo{author}{\bibnamefont{Gottlieb}, \bibfnamefont{S.}},
  \bibinfo{author}{\bibfnamefont{H.}~\bibnamefont{Na}}, and
  \bibinfo{author}{\bibfnamefont{K.}~\bibnamefont{Nagata}},
  \bibinfo{year}{2008}, \bibinfo{journal}{Phys. Rev.}
  \textbf{\bibinfo{volume}{D77}}, \bibinfo{pages}{017505}, \arXiv{0707.3537}.

\bibitem[{Gottlieb \emph{et~al.}(2006{\natexlab{a}})\citenamefont{Gottlieb}
  \emph{et~al.}}]{Gottlieb:2005me}
\bibinfo{author}{\bibnamefont{Gottlieb}, \bibfnamefont{S.}}, \emph{et~al.},
  \bibinfo{year}{2006}{\natexlab{a}}, \bibinfo{journal}{PoS}
  \textbf{\bibinfo{volume}{LAT2005}}, \bibinfo{pages}{203},
  \arXiv{hep-lat/0510072}.

\bibitem[{\citenamefont{Gottlieb} \emph{et~al.}(1987)\citenamefont{Gottlieb,
  Liu, Toussaint, Renken, and Sugar}}]{Gottlieb:1987mq}
\bibinfo{author}{\bibnamefont{Gottlieb}, \bibfnamefont{S.~A.}},
  \bibinfo{author}{\bibfnamefont{W.}~\bibnamefont{Liu}},
  \bibinfo{author}{\bibfnamefont{D.}~\bibnamefont{Toussaint}},
  \bibinfo{author}{\bibfnamefont{R.~L.} \bibnamefont{Renken}}, and
  \bibinfo{author}{\bibfnamefont{R.~L.} \bibnamefont{Sugar}},
  \bibinfo{year}{1987}, \bibinfo{journal}{Phys. Rev.}
  \textbf{\bibinfo{volume}{D35}}, \bibinfo{pages}{2531}.

\bibitem[{\citenamefont{Gottlieb and Tamhankar}(2003)}]{Gottlieb:2003yb}
\bibinfo{author}{\bibnamefont{Gottlieb}, \bibfnamefont{S.~A.}}, and
  \bibinfo{author}{\bibfnamefont{S.}~\bibnamefont{Tamhankar}},
  \bibinfo{year}{2003}, \bibinfo{journal}{Nucl. Phys. Proc. Suppl.}
  \textbf{\bibinfo{volume}{119}}, \bibinfo{pages}{644},
  \arXiv{hep-lat/0301022}.

\bibitem[{Gottlieb \emph{et~al.}(2006{\natexlab{b}})\citenamefont{Gottlieb}
  \emph{et~al.}}]{Gottlieb:2006zz}
\bibinfo{author}{\bibnamefont{Gottlieb}, \bibfnamefont{S.~A.}}, \emph{et~al.}
  (\bibinfo{collaboration}{Fermilab Lattice and MILC}),
  \bibinfo{year}{2006}{\natexlab{b}}, \bibinfo{journal}{PoS}
  \textbf{\bibinfo{volume}{LAT2006}}, \bibinfo{pages}{175}.

\bibitem[{Gray \emph{et~al.}(2003)\citenamefont{Gray}
  \emph{et~al.}}]{Gray:2002vk}
\bibinfo{author}{\bibnamefont{Gray}, \bibfnamefont{A.}}, \emph{et~al.}
  (\bibinfo{collaboration}{HPQCD}), \bibinfo{year}{2003},
  \bibinfo{journal}{Nucl. Phys. Proc. Suppl.} \textbf{\bibinfo{volume}{119}},
  \bibinfo{pages}{592}, \arXiv{hep-lat/0209022}.

\bibitem[{Gray \emph{et~al.}(2005)\citenamefont{Gray}
  \emph{et~al.}}]{Gray:2005ur}
\bibinfo{author}{\bibnamefont{Gray}, \bibfnamefont{A.}}, \emph{et~al.},
  \bibinfo{year}{2005}, \bibinfo{journal}{Phys. Rev.}
  \textbf{\bibinfo{volume}{D72}}, \bibinfo{pages}{094507},
  \arXiv{hep-lat/0507013}.

\bibitem[{\citenamefont{Gregory} \emph{et~al.}(2007)\citenamefont{Gregory,
  Irving, Richards, McNeile, and Hart}}]{Gregory:2007ce}
\bibinfo{author}{\bibnamefont{Gregory}, \bibfnamefont{E.~B.}},
  \bibinfo{author}{\bibfnamefont{A.}~\bibnamefont{Irving}},
  \bibinfo{author}{\bibfnamefont{C.~M.} \bibnamefont{Richards}},
  \bibinfo{author}{\bibfnamefont{C.}~\bibnamefont{McNeile}}, and
  \bibinfo{author}{\bibfnamefont{A.}~\bibnamefont{Hart}}, \bibinfo{year}{2007},
  \bibinfo{journal}{PoS} \textbf{\bibinfo{volume}{LAT2007}},
  \bibinfo{pages}{099}, \arXiv{0710.1725}.

\bibitem[{\citenamefont{Gregory} \emph{et~al.}(2006)\citenamefont{Gregory,
  Irving, McNeile, Miller, and Sroczynski}}]{Gregory:2005yr}
\bibinfo{author}{\bibnamefont{Gregory}, \bibfnamefont{E.~B.}},
  \bibinfo{author}{\bibfnamefont{A.~C.} \bibnamefont{Irving}},
  \bibinfo{author}{\bibfnamefont{C.~C.} \bibnamefont{McNeile}},
  \bibinfo{author}{\bibfnamefont{S.}~\bibnamefont{Miller}}, and
  \bibinfo{author}{\bibfnamefont{Z.}~\bibnamefont{Sroczynski}},
  \bibinfo{year}{2006}, \bibinfo{journal}{PoS}
  \textbf{\bibinfo{volume}{LAT2005}}, \bibinfo{pages}{027},
  \arXiv{hep-lat/0510066}.

\bibitem[{\citenamefont{Gregory} \emph{et~al.}(2008)\citenamefont{Gregory,
  Irving, Richards, and McNeile}}]{Gregory:2007ev}
\bibinfo{author}{\bibnamefont{Gregory}, \bibfnamefont{E.~B.}},
  \bibinfo{author}{\bibfnamefont{A.~C.} \bibnamefont{Irving}},
  \bibinfo{author}{\bibfnamefont{C.~M.} \bibnamefont{Richards}}, and
  \bibinfo{author}{\bibfnamefont{C.}~\bibnamefont{McNeile}},
  \bibinfo{year}{2008}, \bibinfo{journal}{Phys. Rev.}
  \textbf{\bibinfo{volume}{D77}}, \bibinfo{pages}{065019}, \arXiv{0709.4224}.

\bibitem[{\citenamefont{Gregory} \emph{et~al.}(2009)\citenamefont{Gregory,
  McNeile, Irving, and Richards}}]{Gregory:2008mn}
\bibinfo{author}{\bibnamefont{Gregory}, \bibfnamefont{E.~B.}},
  \bibinfo{author}{\bibfnamefont{C.}~\bibnamefont{McNeile}},
  \bibinfo{author}{\bibfnamefont{A.~C.} \bibnamefont{Irving}}, and
  \bibinfo{author}{\bibfnamefont{C.}~\bibnamefont{Richards}}
  (\bibinfo{collaboration}{UKQCD}), \bibinfo{year}{2009},
  \bibinfo{journal}{PoS} \textbf{\bibinfo{volume}{LAT2008}},
  \bibinfo{pages}{286}, \arXiv{0810.0136}.

\bibitem[{\citenamefont{Gupta} \emph{et~al.}(1991)\citenamefont{Gupta,
  Guralnik, Kilcup, and Sharpe}}]{Gupta:1990mr}
\bibinfo{author}{\bibnamefont{Gupta}, \bibfnamefont{R.}},
  \bibinfo{author}{\bibfnamefont{G.}~\bibnamefont{Guralnik}},
  \bibinfo{author}{\bibfnamefont{G.~W.} \bibnamefont{Kilcup}}, and
  \bibinfo{author}{\bibfnamefont{S.~R.} \bibnamefont{Sharpe}},
  \bibinfo{year}{1991}, \bibinfo{journal}{Phys. Rev.}
  \textbf{\bibinfo{volume}{D43}}, \bibinfo{pages}{2003}.

\bibitem[{\citenamefont{H{\"a}gler}(2007)}]{Hagler:2007hu}
\bibinfo{author}{\bibnamefont{H{\"a}gler}, \bibfnamefont{P.}},
  \bibinfo{year}{2007}, \bibinfo{journal}{PoS}
  \textbf{\bibinfo{volume}{LAT2007}}, \bibinfo{pages}{013}, \arXiv{0711.0819}.

\bibitem[{H{\"a}gler \emph{et~al.}(2008)\citenamefont{H{\"a}gler}
  \emph{et~al.}}]{Hagler:2007xi}
\bibinfo{author}{\bibnamefont{H{\"a}gler}, \bibfnamefont{P.}}, \emph{et~al.}
  (\bibinfo{collaboration}{LHPC}), \bibinfo{year}{2008},
  \bibinfo{journal}{Phys. Rev.} \textbf{\bibinfo{volume}{D77}},
  \bibinfo{pages}{094502}, \arXiv{0705.4295}.

\bibitem[{\citenamefont{Hao} \emph{et~al.}(2007)\citenamefont{Hao, von Hippel,
  Horgan, Mason, and Trottier}}]{Hao:2007iz}
\bibinfo{author}{\bibnamefont{Hao}, \bibfnamefont{Z.}},
  \bibinfo{author}{\bibfnamefont{G.~M.} \bibnamefont{von Hippel}},
  \bibinfo{author}{\bibfnamefont{R.~R.} \bibnamefont{Horgan}},
  \bibinfo{author}{\bibfnamefont{Q.~J.} \bibnamefont{Mason}}, and
  \bibinfo{author}{\bibfnamefont{H.~D.} \bibnamefont{Trottier}},
  \bibinfo{year}{2007}, \bibinfo{journal}{Phys. Rev.}
  \textbf{\bibinfo{volume}{D76}}, \bibinfo{pages}{034507}, \arXiv{0705.4660}.

\bibitem[{\citenamefont{Harada}
  \emph{et~al.}(2002{\natexlab{a}})\citenamefont{Harada,
  Hashimoto, Kronfeld, and Onogi}}]{Harada:2001fj}
\bibinfo{author}{\bibnamefont{Harada}, \bibfnamefont{J.}},
  \bibinfo{author}{\bibfnamefont{S.}~\bibnamefont{Hashimoto}},
  \bibinfo{author}{\bibfnamefont{A.~S.} \bibnamefont{Kronfeld}}, and
  \bibinfo{author}{\bibfnamefont{T.}~\bibnamefont{Onogi}},
  \bibinfo{year}{2002}{\natexlab{a}}, \bibinfo{journal}{Phys. Rev.}
  \textbf{\bibinfo{volume}{D65}}, \bibinfo{pages}{094514},
  \arXiv{hep-lat/0112045}.

\bibitem[{Harada \emph{et~al.}(2002{\natexlab{b}})\citenamefont{Harada}
  \emph{et~al.}}]{Harada:2001fi}
\bibinfo{author}{\bibnamefont{Harada}, \bibfnamefont{J.}}, \emph{et~al.},
  \bibinfo{year}{2002}{\natexlab{b}}, \bibinfo{journal}{Phys. Rev.}
  \textbf{\bibinfo{volume}{D65}}, \bibinfo{pages}{094513},
  \bibinfo{note}{{Erratum-ibid.\ {\bf D71}, 019903 (2005)}},
  \arXiv{hep-lat/0112044}.

\bibitem[{\citenamefont{Hart}
  \emph{et~al.}(2009{\natexlab{a}})\citenamefont{Hart, von Hippel, and
  Horgan}}]{Hart:2008zi}
\bibinfo{author}{\bibnamefont{Hart}, \bibfnamefont{A.}},
  \bibinfo{author}{\bibfnamefont{G.~M.} \bibnamefont{von Hippel}}, and
  \bibinfo{author}{\bibfnamefont{R.~R.} \bibnamefont{Horgan}},
  \bibinfo{year}{2009}{\natexlab{a}}, \bibinfo{journal}{PoS}
  \textbf{\bibinfo{volume}{LAT2008}}, \bibinfo{pages}{046}, \arXiv{0808.1791}.

\bibitem[{\citenamefont{Hart}
  \emph{et~al.}(2009{\natexlab{b}})\citenamefont{Hart, von Hippel, and
  Horgan}}]{Hart:2008sq}
\bibinfo{author}{\bibnamefont{Hart}, \bibfnamefont{A.}},
  \bibinfo{author}{\bibfnamefont{G.~M.} \bibnamefont{von Hippel}}, and
  \bibinfo{author}{\bibfnamefont{R.~R.} \bibnamefont{Horgan}}
  (\bibinfo{collaboration}{HPQCD}), \bibinfo{year}{2009}{\natexlab{b}},
  \bibinfo{journal}{Phys. Rev.} \textbf{\bibinfo{volume}{D79}},
  \bibinfo{pages}{074008}, \arXiv{0812.0503}.

\bibitem[{\citenamefont{Hasenbusch}(2001)}]{Hasenbusch:2001ne}
\bibinfo{author}{\bibnamefont{Hasenbusch}, \bibfnamefont{M.}},
  \bibinfo{year}{2001}, \bibinfo{journal}{Phys. Lett.}
  \textbf{\bibinfo{volume}{B519}}, \bibinfo{pages}{177},
  \arXiv{hep-lat/0107019}.

\bibitem[{\citenamefont{Hasenbusch and Jansen}(2003)}]{Hasenbusch:2002ai}
\bibinfo{author}{\bibnamefont{Hasenbusch}, \bibfnamefont{M.}}, and
  \bibinfo{author}{\bibfnamefont{K.}~\bibnamefont{Jansen}},
  \bibinfo{year}{2003}, \bibinfo{journal}{Nucl. Phys.}
  \textbf{\bibinfo{volume}{B659}}, \bibinfo{pages}{299},
  \arXiv{hep-lat/0211042}.

\bibitem[{\citenamefont{Hasenfratz}
  \emph{et~al.}(2007)\citenamefont{Hasenfratz, Hoffmann, and
  Schaefer}}]{Hasenfratz:2007rf}
\bibinfo{author}{\bibnamefont{Hasenfratz}, \bibfnamefont{A.}},
  \bibinfo{author}{\bibfnamefont{R.}~\bibnamefont{Hoffmann}}, and
  \bibinfo{author}{\bibfnamefont{S.}~\bibnamefont{Schaefer}},
  \bibinfo{year}{2007}, \bibinfo{journal}{JHEP} \textbf{\bibinfo{volume}{05}},
  \bibinfo{pages}{029}, \arXiv{hep-lat/0702028}.

\bibitem[{\citenamefont{Hasenfratz and Knechtli}(2001)}]{Hasenfratz:2001hp}
\bibinfo{author}{\bibnamefont{Hasenfratz}, \bibfnamefont{A.}}, and
  \bibinfo{author}{\bibfnamefont{F.}~\bibnamefont{Knechtli}},
  \bibinfo{year}{2001}, \bibinfo{journal}{Phys. Rev.}
  \textbf{\bibinfo{volume}{D64}}, \bibinfo{pages}{034504},
  \arXiv{hep-lat/0103029}.

\bibitem[{\citenamefont{Hasenfratz}(1998)}]{Hasenfratz:1997ft}
\bibinfo{author}{\bibnamefont{Hasenfratz}, \bibfnamefont{P.}},
  \bibinfo{year}{1998}, \bibinfo{journal}{Nucl. Phys. Proc. Suppl.}
  \textbf{\bibinfo{volume}{63}}, \bibinfo{pages}{53}, \arXiv{hep-lat/9709110}.

\bibitem[{\citenamefont{Hasenfratz}
  \emph{et~al.}(1998)\citenamefont{Hasenfratz, Laliena, and
  Niedermayer}}]{Hasenfratz:1998ri}
\bibinfo{author}{\bibnamefont{Hasenfratz}, \bibfnamefont{P.}},
  \bibinfo{author}{\bibfnamefont{V.}~\bibnamefont{Laliena}}, and
  \bibinfo{author}{\bibfnamefont{F.}~\bibnamefont{Niedermayer}},
  \bibinfo{year}{1998}, \bibinfo{journal}{Phys. Lett.}
  \textbf{\bibinfo{volume}{B427}}, \bibinfo{pages}{125},
  \arXiv{hep-lat/9801021}.

\bibitem[{\citenamefont{Hashimoto} \emph{et~al.}(2002)\citenamefont{Hashimoto,
  Kronfeld, Mackenzie, Ryan, and Simone}}]{Hashimoto:2001nb}
\bibinfo{author}{\bibnamefont{Hashimoto}, \bibfnamefont{S.}},
  \bibinfo{author}{\bibfnamefont{A.~S.} \bibnamefont{Kronfeld}},
  \bibinfo{author}{\bibfnamefont{P.~B.} \bibnamefont{Mackenzie}},
  \bibinfo{author}{\bibfnamefont{S.~M.} \bibnamefont{Ryan}}, and
  \bibinfo{author}{\bibfnamefont{J.~N.} \bibnamefont{Simone}},
  \bibinfo{year}{2002}, \bibinfo{journal}{Phys. Rev.}
  \textbf{\bibinfo{volume}{D66}}, \bibinfo{pages}{014503},
  \arXiv{hep-ph/0110253}.

\bibitem[{Hashimoto \emph{et~al.}(1999)\citenamefont{Hashimoto}
  \emph{et~al.}}]{Hashimoto:1999yp}
\bibinfo{author}{\bibnamefont{Hashimoto}, \bibfnamefont{S.}}, \emph{et~al.},
  \bibinfo{year}{1999}, \bibinfo{journal}{Phys. Rev.}
  \textbf{\bibinfo{volume}{D61}}, \bibinfo{pages}{014502},
  \arXiv{hep-ph/9906376}.

\bibitem[{\citenamefont{Heller} \emph{et~al.}(1999)\citenamefont{Heller,
  Karsch, and Sturm}}]{Heller:1999xz}
\bibinfo{author}{\bibnamefont{Heller}, \bibfnamefont{U.~M.}},
  \bibinfo{author}{\bibfnamefont{F.}~\bibnamefont{Karsch}}, and
  \bibinfo{author}{\bibfnamefont{B.}~\bibnamefont{Sturm}},
  \bibinfo{year}{1999}, \bibinfo{journal}{Phys. Rev.}
  \textbf{\bibinfo{volume}{D60}}, \bibinfo{pages}{114502},
  \arXiv{hep-lat/9901010}.

\bibitem[{\citenamefont{Isgur and Wise}(1992)}]{Isgur:HQET}
\bibinfo{author}{\bibnamefont{Isgur}, \bibfnamefont{N.}}, and
  \bibinfo{author}{\bibfnamefont{M.~B.} \bibnamefont{Wise}},
  \bibinfo{year}{1992}, \emph{\bibinfo{title}{B {D}ecays}}
  (\bibinfo{publisher}{World Scientific}), p. \bibinfo{pages}{489}.

\bibitem[{\citenamefont{Ishizuka} \emph{et~al.}(1994)\citenamefont{Ishizuka,
  Fukugita, Mino, Okawa, and Ukawa}}]{Ishizuka:1993mt}
\bibinfo{author}{\bibnamefont{Ishizuka}, \bibfnamefont{N.}},
  \bibinfo{author}{\bibfnamefont{M.}~\bibnamefont{Fukugita}},
  \bibinfo{author}{\bibfnamefont{H.}~\bibnamefont{Mino}},
  \bibinfo{author}{\bibfnamefont{M.}~\bibnamefont{Okawa}}, and
  \bibinfo{author}{\bibfnamefont{A.}~\bibnamefont{Ukawa}},
  \bibinfo{year}{1994}, \bibinfo{journal}{Nucl. Phys.}
  \textbf{\bibinfo{volume}{B411}}, \bibinfo{pages}{875}.

\bibitem[{\citenamefont{Jegerlehner}(1996)}]{Jegerlehner:1996pm}
\bibinfo{author}{\bibnamefont{Jegerlehner}, \bibfnamefont{B.}},
  \bibinfo{year}{1996}, \arXiv{hep-lat/9612014}.

\bibitem[{\citenamefont{Jegerlehner}(1998)}]{Jegerlehner:1997rn}
\bibinfo{author}{\bibnamefont{Jegerlehner}, \bibfnamefont{B.}},
  \bibinfo{year}{1998}, \bibinfo{journal}{Nucl. Phys. Proc. Suppl.}
  \textbf{\bibinfo{volume}{63}}, \bibinfo{pages}{958}, \arXiv{hep-lat/9708029}.

\bibitem[{\citenamefont{Jegerlehner}(2007)}]{Jegerlehner:2007xe}
\bibinfo{author}{\bibnamefont{Jegerlehner}, \bibfnamefont{F.}},
  \bibinfo{year}{2007}, \bibinfo{journal}{Acta Phys. Polon.}
  \textbf{\bibinfo{volume}{B38}}, \bibinfo{pages}{3021},
  \arXiv{hep-ph/0703125}.

\bibitem[{\citenamefont{Jegerlehner}(2008)}]{Jegerlehner:2008zz}
\bibinfo{author}{\bibnamefont{Jegerlehner}, \bibfnamefont{F.}},
  \bibinfo{year}{2008}, \bibinfo{note}{in: Springer, Berlin, Germany (2008) 426
  p}.

\bibitem[{\citenamefont{Jenkins}(1992)}]{Jenkins:1991ts}
\bibinfo{author}{\bibnamefont{Jenkins}, \bibfnamefont{E.~E.}},
  \bibinfo{year}{1992}, \bibinfo{journal}{Nucl. Phys.}
  \textbf{\bibinfo{volume}{B368}}, \bibinfo{pages}{190}.

\bibitem[{\citenamefont{Kaplan}(1992)}]{Kaplan:1992bt}
\bibinfo{author}{\bibnamefont{Kaplan}, \bibfnamefont{D.~B.}},
  \bibinfo{year}{1992}, \bibinfo{journal}{Phys. Lett.}
  \textbf{\bibinfo{volume}{B288}}, \bibinfo{pages}{342},
  \arXiv{hep-lat/9206013}.

\bibitem[{\citenamefont{Karsten and Smit}(1981)}]{Karsten:1980wd}
\bibinfo{author}{\bibnamefont{Karsten}, \bibfnamefont{L.~H.}}, and
  \bibinfo{author}{\bibfnamefont{J.}~\bibnamefont{Smit}}, \bibinfo{year}{1981},
  \bibinfo{journal}{Nucl. Phys.} \textbf{\bibinfo{volume}{B183}},
  \bibinfo{pages}{103}.

\bibitem[{\citenamefont{Kawamoto and Smit}(1981)}]{Kawamoto:1981hw}
\bibinfo{author}{\bibnamefont{Kawamoto}, \bibfnamefont{N.}}, and
  \bibinfo{author}{\bibfnamefont{J.}~\bibnamefont{Smit}}, \bibinfo{year}{1981},
  \bibinfo{journal}{Nucl. Phys.} \textbf{\bibinfo{volume}{B192}},
  \bibinfo{pages}{100}.

\bibitem[{\citenamefont{Kennedy and Pendleton}(1985)}]{Kennedy:1985nu}
\bibinfo{author}{\bibnamefont{Kennedy}, \bibfnamefont{A.~D.}}, and
  \bibinfo{author}{\bibfnamefont{B.~J.} \bibnamefont{Pendleton}},
  \bibinfo{year}{1985}, \bibinfo{journal}{Phys. Lett.}
  \textbf{\bibinfo{volume}{B156}}, \bibinfo{pages}{393}.

\bibitem[{\citenamefont{Kikukawa and Noguchi}(1999)}]{Kikukawa:1999sy}
\bibinfo{author}{\bibnamefont{Kikukawa}, \bibfnamefont{Y.}}, and
  \bibinfo{author}{\bibfnamefont{T.}~\bibnamefont{Noguchi}},
  \bibinfo{year}{1999}, \arXiv{hep-lat/9902022}.

\bibitem[{\citenamefont{Kilcup and Sharpe}(1987)}]{Kilcup:1986dg}
\bibinfo{author}{\bibnamefont{Kilcup}, \bibfnamefont{G.~W.}}, and
  \bibinfo{author}{\bibfnamefont{S.~R.} \bibnamefont{Sharpe}},
  \bibinfo{year}{1987}, \bibinfo{journal}{Nucl. Phys.}
  \textbf{\bibinfo{volume}{B283}}, \bibinfo{pages}{493}.

\bibitem[{\citenamefont{Kluberg-Stern}
  \emph{et~al.}(1981)\citenamefont{Kluberg-Stern, Morel, Napoly, and
  Petersson}}]{KlubergStern:1981wz}
\bibinfo{author}{\bibnamefont{Kluberg-Stern}, \bibfnamefont{H.}},
  \bibinfo{author}{\bibfnamefont{A.}~\bibnamefont{Morel}},
  \bibinfo{author}{\bibfnamefont{O.}~\bibnamefont{Napoly}}, and
  \bibinfo{author}{\bibfnamefont{B.}~\bibnamefont{Petersson}},
  \bibinfo{year}{1981}, \bibinfo{journal}{Nucl. Phys.}
  \textbf{\bibinfo{volume}{B190}}, \bibinfo{pages}{504}.

\bibitem[{\citenamefont{Kluberg-Stern}
  \emph{et~al.}(1983{\natexlab{a}})\citenamefont{Kluberg-Stern, Morel, Napoly,
  and Petersson}}]{KlubergStern:1983dg}
\bibinfo{author}{\bibnamefont{Kluberg-Stern}, \bibfnamefont{H.}},
  \bibinfo{author}{\bibfnamefont{A.}~\bibnamefont{Morel}},
  \bibinfo{author}{\bibfnamefont{O.}~\bibnamefont{Napoly}}, and
  \bibinfo{author}{\bibfnamefont{B.}~\bibnamefont{Petersson}},
  \bibinfo{year}{1983}{\natexlab{a}}, \bibinfo{journal}{Nucl. Phys.}
  \textbf{\bibinfo{volume}{B220}}, \bibinfo{pages}{447}.

\bibitem[{\citenamefont{Kluberg-Stern}
  \emph{et~al.}(1983{\natexlab{b}})\citenamefont{Kluberg-Stern, Morel, and
  Petersson}}]{KlubergStern:1982bs}
\bibinfo{author}{\bibnamefont{Kluberg-Stern}, \bibfnamefont{H.}},
  \bibinfo{author}{\bibfnamefont{A.}~\bibnamefont{Morel}}, and
  \bibinfo{author}{\bibfnamefont{B.}~\bibnamefont{Petersson}},
  \bibinfo{year}{1983}{\natexlab{b}}, \bibinfo{journal}{Nucl. Phys.}
  \textbf{\bibinfo{volume}{B215}}, \bibinfo{pages}{527}.

\bibitem[{\citenamefont{Kogut and Susskind}(1975)}]{Kogut:1974ag}
\bibinfo{author}{\bibnamefont{Kogut}, \bibfnamefont{J.~B.}}, and
  \bibinfo{author}{\bibfnamefont{L.}~\bibnamefont{Susskind}},
  \bibinfo{year}{1975}, \bibinfo{journal}{Phys. Rev.}
  \textbf{\bibinfo{volume}{D11}}, \bibinfo{pages}{395}.

\bibitem[{\citenamefont{Kronfeld}(2000)}]{Kronfeld:2000ck}
\bibinfo{author}{\bibnamefont{Kronfeld}, \bibfnamefont{A.~S.}},
  \bibinfo{year}{2000}, \bibinfo{journal}{Phys. Rev.}
  \textbf{\bibinfo{volume}{D62}}, \bibinfo{pages}{014505},
  \arXiv{hep-lat/0002008}.

\bibitem[{\citenamefont{Kronfeld}(2004)}]{Kronfeld:2003sd}
\bibinfo{author}{\bibnamefont{Kronfeld}, \bibfnamefont{A.~S.}},
  \bibinfo{year}{2004}, \bibinfo{journal}{Nucl. Phys. Proc. Suppl.}
  \textbf{\bibinfo{volume}{129}}, \bibinfo{pages}{46}, \arXiv{hep-lat/0310063}.

\bibitem[{\citenamefont{Kronfeld}(2006)}]{Kronfeld:2006sk}
\bibinfo{author}{\bibnamefont{Kronfeld}, \bibfnamefont{A.~S.}}
  (\bibinfo{collaboration}{Fermilab Lattice}), \bibinfo{year}{2006},
  \bibinfo{journal}{J. Phys. Conf. Ser.} \textbf{\bibinfo{volume}{46}},
  \bibinfo{pages}{147}, \arXiv{hep-lat/0607011}.

\bibitem[{\citenamefont{Kronfeld}(2007)}]{Kronfeld:2007ek}
\bibinfo{author}{\bibnamefont{Kronfeld}, \bibfnamefont{A.~S.}},
  \bibinfo{year}{2007}, \bibinfo{journal}{PoS}
  \textbf{\bibinfo{volume}{LAT2007}}, \bibinfo{pages}{016}, \arXiv{0711.0699}.

\bibitem[{\citenamefont{K{\"u}hn and Steinhauser}(2001)}]{Kuhn:2001dm}
\bibinfo{author}{\bibnamefont{K{\"u}hn}, \bibfnamefont{J.~H.}}, and
  \bibinfo{author}{\bibfnamefont{M.}~\bibnamefont{Steinhauser}},
  \bibinfo{year}{2001}, \bibinfo{journal}{Nucl. Phys.}
  \textbf{\bibinfo{volume}{B619}}, \bibinfo{pages}{588},
  \bibinfo{note}{{Erratum-ibid.\ {\bf B640}, 415 (2002)}},
  \arXiv{hep-ph/0109084}.

\bibitem[{\citenamefont{K{\"u}hn} \emph{et~al.}(2007)\citenamefont{K{\"u}hn,
  Steinhauser, and Sturm}}]{Kuhn:2007vp}
\bibinfo{author}{\bibnamefont{K{\"u}hn}, \bibfnamefont{J.~H.}},
  \bibinfo{author}{\bibfnamefont{M.}~\bibnamefont{Steinhauser}}, and
  \bibinfo{author}{\bibfnamefont{C.}~\bibnamefont{Sturm}},
  \bibinfo{year}{2007}, \bibinfo{journal}{Nucl. Phys.}
  \textbf{\bibinfo{volume}{B778}}, \bibinfo{pages}{192},
  \arXiv{hep-ph/0702103}.

\bibitem[{\citenamefont{Kuramashi} \emph{et~al.}(1994)\citenamefont{Kuramashi,
  Fukugita, Mino, Okawa, and Ukawa}}]{Kuramashi:1994aj}
\bibinfo{author}{\bibnamefont{Kuramashi}, \bibfnamefont{Y.}},
  \bibinfo{author}{\bibfnamefont{M.}~\bibnamefont{Fukugita}},
  \bibinfo{author}{\bibfnamefont{H.}~\bibnamefont{Mino}},
  \bibinfo{author}{\bibfnamefont{M.}~\bibnamefont{Okawa}}, and
  \bibinfo{author}{\bibfnamefont{A.}~\bibnamefont{Ukawa}},
  \bibinfo{year}{1994}, \bibinfo{journal}{Phys. Rev. Lett.}
  \textbf{\bibinfo{volume}{72}}, \bibinfo{pages}{3448}.

\bibitem[{\citenamefont{Lagae and Sinclair}(1999)}]{Lagae:1998pe}
\bibinfo{author}{\bibnamefont{Lagae}, \bibfnamefont{J.~F.}}, and
  \bibinfo{author}{\bibfnamefont{D.~K.} \bibnamefont{Sinclair}},
  \bibinfo{year}{1999}, \bibinfo{journal}{Phys. Rev.}
  \textbf{\bibinfo{volume}{D59}}, \bibinfo{pages}{014511},
  \arXiv{hep-lat/9806014}.

\bibitem[{\citenamefont{Laiho and Van~de Water}(2006)}]{Laiho:2005ue}
\bibinfo{author}{\bibnamefont{Laiho}, \bibfnamefont{J.}}, and
  \bibinfo{author}{\bibfnamefont{R.~S.} \bibnamefont{Van~de Water}},
  \bibinfo{year}{2006}, \bibinfo{journal}{Phys. Rev.}
  \textbf{\bibinfo{volume}{D73}}, \bibinfo{pages}{054501},
  \arXiv{hep-lat/0512007}.

\bibitem[{\citenamefont{Lange} \emph{et~al.}(2005)\citenamefont{Lange, Neubert,
  and Paz}}]{Lange:2005yw}
\bibinfo{author}{\bibnamefont{Lange}, \bibfnamefont{B.~O.}},
  \bibinfo{author}{\bibfnamefont{M.}~\bibnamefont{Neubert}}, and
  \bibinfo{author}{\bibfnamefont{G.}~\bibnamefont{Paz}}, \bibinfo{year}{2005},
  \bibinfo{journal}{Phys. Rev.} \textbf{\bibinfo{volume}{D72}},
  \bibinfo{pages}{073006}, \arXiv{hep-ph/0504071}.

\bibitem[{\citenamefont{Lee and Sharpe}(1999)}]{Lee:1999zxa}
\bibinfo{author}{\bibnamefont{Lee}, \bibfnamefont{W.-J.}}, and
  \bibinfo{author}{\bibfnamefont{S.~R.} \bibnamefont{Sharpe}},
  \bibinfo{year}{1999}, \bibinfo{journal}{Phys. Rev.}
  \textbf{\bibinfo{volume}{D60}}, \bibinfo{pages}{114503},
  \arXiv{hep-lat/9905023}.

\bibitem[{\citenamefont{Lellouch}(1996)}]{Lellouch:1995yv}
\bibinfo{author}{\bibnamefont{Lellouch}, \bibfnamefont{L.}},
  \bibinfo{year}{1996}, \bibinfo{journal}{Nucl. Phys.}
  \textbf{\bibinfo{volume}{B479}}, \bibinfo{pages}{353},
  \arXiv{hep-ph/9509358}.

\bibitem[{\citenamefont{Lenz and Nierste}(2007)}]{Lenz:2006hd}
\bibinfo{author}{\bibnamefont{Lenz}, \bibfnamefont{A.}}, and
  \bibinfo{author}{\bibfnamefont{U.}~\bibnamefont{Nierste}},
  \bibinfo{year}{2007}, \bibinfo{journal}{JHEP} \textbf{\bibinfo{volume}{06}},
  \bibinfo{pages}{072}, \arXiv{hep-ph/0612167}.

\bibitem[{\citenamefont{Lepage}(1990)}]{Lepage:1989hd}
\bibinfo{author}{\bibnamefont{Lepage}, \bibfnamefont{G.~P.}},
  \bibinfo{year}{1990}, \bibinfo{note}{in: From Actions to Answers: Proceedings
  of the 1989 Theoretical Advanced Study Institute in Elementary Particle
  Physics, eds. T. DeGrand and D. Toussaint (World Scientific, Singapore, 1990)
  p. 197}.

\bibitem[{\citenamefont{Lepage}(1999)}]{Lepage:1998vj}
\bibinfo{author}{\bibnamefont{Lepage}, \bibfnamefont{G.~P.}},
  \bibinfo{year}{1999}, \bibinfo{journal}{Phys. Rev.}
  \textbf{\bibinfo{volume}{D59}}, \bibinfo{pages}{074502},
  \arXiv{hep-lat/9809157}.

\bibitem[{\citenamefont{Lepage and Mackenzie}(1993)}]{Lepage:1992xa}
\bibinfo{author}{\bibnamefont{Lepage}, \bibfnamefont{G.~P.}}, and
  \bibinfo{author}{\bibfnamefont{P.~B.} \bibnamefont{Mackenzie}},
  \bibinfo{year}{1993}, \bibinfo{journal}{Phys. Rev.}
  \textbf{\bibinfo{volume}{D48}}, \bibinfo{pages}{2250},
  \arXiv{hep-lat/9209022}.

\bibitem[{\citenamefont{Lepage} \emph{et~al.}(1992)\citenamefont{Lepage,
  Magnea, Nakhleh, Magnea, and Hornbostel}}]{Lepage:1992tx}
\bibinfo{author}{\bibnamefont{Lepage}, \bibfnamefont{G.~P.}},
  \bibinfo{author}{\bibfnamefont{L.}~\bibnamefont{Magnea}},
  \bibinfo{author}{\bibfnamefont{C.}~\bibnamefont{Nakhleh}},
  \bibinfo{author}{\bibfnamefont{U.}~\bibnamefont{Magnea}}, and
  \bibinfo{author}{\bibfnamefont{K.}~\bibnamefont{Hornbostel}},
  \bibinfo{year}{1992}, \bibinfo{journal}{Phys. Rev.}
  \textbf{\bibinfo{volume}{D46}}, \bibinfo{pages}{4052},
  \arXiv{hep-lat/9205007}.

\bibitem[{Lepage \emph{et~al.}(2002)\citenamefont{Lepage}
  \emph{et~al.}}]{Lepage:2001ym}
\bibinfo{author}{\bibnamefont{Lepage}, \bibfnamefont{G.~P.}}, \emph{et~al.},
  \bibinfo{year}{2002}, \bibinfo{journal}{Nucl. Phys. Proc. Suppl.}
  \textbf{\bibinfo{volume}{106}}, \bibinfo{pages}{12}, \arXiv{hep-lat/0110175}.

\bibitem[{\citenamefont{Lepage}(1998)}]{Lepage:1997id}
\bibinfo{author}{\bibnamefont{Lepage}, \bibfnamefont{P.}},
  \bibinfo{year}{1998}, \bibinfo{journal}{Nucl. Phys. Proc. Suppl.}
  \textbf{\bibinfo{volume}{60A}}, \bibinfo{pages}{267},
  \arXiv{hep-lat/9707026}.

\bibitem[{\citenamefont{Leutwyler}(1994)}]{Leutwyler:1993iq}
\bibinfo{author}{\bibnamefont{Leutwyler}, \bibfnamefont{H.}},
  \bibinfo{year}{1994}, \bibinfo{journal}{Ann. Phys.}
  \textbf{\bibinfo{volume}{235}}, \bibinfo{pages}{165}, \arXiv{hep-ph/9311274}.

\bibitem[{\citenamefont{Leutwyler}(2006)}]{Leutwyler:2006qq}
\bibinfo{author}{\bibnamefont{Leutwyler}, \bibfnamefont{H.}},
  \bibinfo{year}{2006}, \arXiv{hep-ph/0612112}.

\bibitem[{\citenamefont{Leutwyler and Smilga}(1992)}]{Leutwyler:1992yt}
\bibinfo{author}{\bibnamefont{Leutwyler}, \bibfnamefont{H.}}, and
  \bibinfo{author}{\bibfnamefont{A.~V.} \bibnamefont{Smilga}},
  \bibinfo{year}{1992}, \bibinfo{journal}{Phys. Rev.}
  \textbf{\bibinfo{volume}{D46}}, \bibinfo{pages}{5607}.

\bibitem[{\citenamefont{Levkova and DeTar}(2009)}]{Levkova:2008qr}
\bibinfo{author}{\bibnamefont{Levkova}, \bibfnamefont{L.}}, and
  \bibinfo{author}{\bibfnamefont{C.}~\bibnamefont{DeTar}},
  \bibinfo{year}{2009}, \bibinfo{journal}{PoS}
  \textbf{\bibinfo{volume}{LAT2008}}, \bibinfo{pages}{133}, \arXiv{0809.5086}.

\bibitem[{\citenamefont{Lewis} \emph{et~al.}(2001)\citenamefont{Lewis, Mathur,
  and Woloshyn}}]{Lewis:2001iz}
\bibinfo{author}{\bibnamefont{Lewis}, \bibfnamefont{R.}},
  \bibinfo{author}{\bibfnamefont{N.}~\bibnamefont{Mathur}}, and
  \bibinfo{author}{\bibfnamefont{R.~M.} \bibnamefont{Woloshyn}},
  \bibinfo{year}{2001}, \bibinfo{journal}{Phys. Rev.}
  \textbf{\bibinfo{volume}{D64}}, \bibinfo{pages}{094509},
  \arXiv{hep-ph/0107037}.

\bibitem[{\citenamefont{Lewis and Woloshyn}(2009)}]{Lewis:2008fu}
\bibinfo{author}{\bibnamefont{Lewis}, \bibfnamefont{R.}}, and
  \bibinfo{author}{\bibfnamefont{R.~M.} \bibnamefont{Woloshyn}},
  \bibinfo{year}{2009}, \bibinfo{journal}{Phys. Rev.}
  \textbf{\bibinfo{volume}{D79}}, \bibinfo{pages}{014502}, \arXiv{0806.4783}.

\bibitem[{\citenamefont{Luke}(1990)}]{Luke:1990eg}
\bibinfo{author}{\bibnamefont{Luke}, \bibfnamefont{M.~E.}},
  \bibinfo{year}{1990}, \bibinfo{journal}{Phys. Lett.}
  \textbf{\bibinfo{volume}{B252}}, \bibinfo{pages}{447}.

\bibitem[{\citenamefont{Luo}(1997)}]{Luo:1996vt}
\bibinfo{author}{\bibnamefont{Luo}, \bibfnamefont{Y.-b.}},
  \bibinfo{year}{1997}, \bibinfo{journal}{Phys. Rev.}
  \textbf{\bibinfo{volume}{D55}}, \bibinfo{pages}{353},
  \arXiv{hep-lat/9604025}.

\bibitem[{\citenamefont{L{\"u}scher}(1998)}]{Luscher:1998pqa}
\bibinfo{author}{\bibnamefont{L{\"u}scher}, \bibfnamefont{M.}},
  \bibinfo{year}{1998}, \bibinfo{journal}{Phys. Lett.}
  \textbf{\bibinfo{volume}{B428}}, \bibinfo{pages}{342},
  \arXiv{hep-lat/9802011}.

\bibitem[{\citenamefont{L{\"u}scher}
  \emph{et~al.}(1996)\citenamefont{L{\"u}scher, Sint, Sommer, and
  Weisz}}]{Luscher:1996sc}
\bibinfo{author}{\bibnamefont{L{\"u}scher}, \bibfnamefont{M.}},
  \bibinfo{author}{\bibfnamefont{S.}~\bibnamefont{Sint}},
  \bibinfo{author}{\bibfnamefont{R.}~\bibnamefont{Sommer}}, and
  \bibinfo{author}{\bibfnamefont{P.}~\bibnamefont{Weisz}},
  \bibinfo{year}{1996}, \bibinfo{journal}{Nucl. Phys.}
  \textbf{\bibinfo{volume}{B478}}, \bibinfo{pages}{365},
  \arXiv{hep-lat/9605038}.

\bibitem[{\citenamefont{L{\"u}scher}
  \emph{et~al.}(1997)\citenamefont{L{\"u}scher, Sint, Sommer, Weisz, and
  Wolff}}]{Luscher:1996ug}
\bibinfo{author}{\bibnamefont{L{\"u}scher}, \bibfnamefont{M.}},
  \bibinfo{author}{\bibfnamefont{S.}~\bibnamefont{Sint}},
  \bibinfo{author}{\bibfnamefont{R.}~\bibnamefont{Sommer}},
  \bibinfo{author}{\bibfnamefont{P.}~\bibnamefont{Weisz}}, and
  \bibinfo{author}{\bibfnamefont{U.}~\bibnamefont{Wolff}},
  \bibinfo{year}{1997}, \bibinfo{journal}{Nucl. Phys.}
  \textbf{\bibinfo{volume}{B491}}, \bibinfo{pages}{323},
  \arXiv{hep-lat/9609035}.

\bibitem[{\citenamefont{L{\"u}scher and
  Weisz}(1985{\natexlab{a}})}]{Luscher:1985zq}
\bibinfo{author}{\bibnamefont{L{\"u}scher}, \bibfnamefont{M.}}, and
  \bibinfo{author}{\bibfnamefont{P.}~\bibnamefont{Weisz}},
  \bibinfo{year}{1985}{\natexlab{a}}, \bibinfo{journal}{Phys. Lett.}
  \textbf{\bibinfo{volume}{B158}}, \bibinfo{pages}{250}.

\bibitem[{\citenamefont{L{\"u}scher and
  Weisz}(1985{\natexlab{b}})}]{Luscher:1984xn}
\bibinfo{author}{\bibnamefont{L{\"u}scher}, \bibfnamefont{M.}}, and
  \bibinfo{author}{\bibfnamefont{P.}~\bibnamefont{Weisz}},
  \bibinfo{year}{1985}{\natexlab{b}}, \bibinfo{journal}{Commun. Math. Phys.}
  \textbf{\bibinfo{volume}{97}}, \bibinfo{pages}{59}.

\bibitem[{\citenamefont{L{\"u}scher and Weisz}(1996)}]{Luscher:1996vw}
\bibinfo{author}{\bibnamefont{L{\"u}scher}, \bibfnamefont{M.}}, and
  \bibinfo{author}{\bibfnamefont{P.}~\bibnamefont{Weisz}},
  \bibinfo{year}{1996}, \bibinfo{journal}{Nucl. Phys.}
  \textbf{\bibinfo{volume}{B479}}, \bibinfo{pages}{429},
  \arXiv{hep-lat/9606016}.

\bibitem[{\citenamefont{Maltman} \emph{et~al.}(2008)\citenamefont{Maltman,
  Leinweber, Moran, and Sternbeck}}]{Maltman:2008bx}
\bibinfo{author}{\bibnamefont{Maltman}, \bibfnamefont{K.}},
  \bibinfo{author}{\bibfnamefont{D.}~\bibnamefont{Leinweber}},
  \bibinfo{author}{\bibfnamefont{P.}~\bibnamefont{Moran}}, and
  \bibinfo{author}{\bibfnamefont{A.}~\bibnamefont{Sternbeck}},
  \bibinfo{year}{2008}, \bibinfo{journal}{Phys. Rev.}
  \textbf{\bibinfo{volume}{D78}}, \bibinfo{pages}{114504}, \arXiv{0807.2020}.

\bibitem[{\citenamefont{Manohar and Wise}(2000)}]{Manohar:2000dt}
\bibinfo{author}{\bibnamefont{Manohar}, \bibfnamefont{A.~V.}}, and
  \bibinfo{author}{\bibfnamefont{M.~B.} \bibnamefont{Wise}},
  \bibinfo{year}{2000}, \bibinfo{journal}{Camb. Monogr. Part. Phys. Nucl. Phys.
  Cosmol.} \textbf{\bibinfo{volume}{10}}, \bibinfo{pages}{1},
  \bibinfo{note}{and references therein.}

\bibitem[{\citenamefont{Marciano}(2004)}]{Marciano:2004uf}
\bibinfo{author}{\bibnamefont{Marciano}, \bibfnamefont{W.~J.}},
  \bibinfo{year}{2004}, \bibinfo{journal}{Phys. Rev. Lett.}
  \textbf{\bibinfo{volume}{93}}, \bibinfo{pages}{231803},
  \arXiv{hep-ph/0402299}.

\bibitem[{\citenamefont{Marinari}
  \emph{et~al.}(1981{\natexlab{a}})\citenamefont{Marinari, Parisi, and
  Rebbi}}]{Marinari:1981nu}
\bibinfo{author}{\bibnamefont{Marinari}, \bibfnamefont{E.}},
  \bibinfo{author}{\bibfnamefont{G.}~\bibnamefont{Parisi}}, and
  \bibinfo{author}{\bibfnamefont{C.}~\bibnamefont{Rebbi}},
  \bibinfo{year}{1981}{\natexlab{a}}, \bibinfo{journal}{Phys. Rev. Lett.}
  \textbf{\bibinfo{volume}{47}}, \bibinfo{pages}{1795}.

\bibitem[{\citenamefont{Marinari}
  \emph{et~al.}(1981{\natexlab{b}})\citenamefont{Marinari, Parisi, and
  Rebbi}}]{Marinari:1981qf}
\bibinfo{author}{\bibnamefont{Marinari}, \bibfnamefont{E.}},
  \bibinfo{author}{\bibfnamefont{G.}~\bibnamefont{Parisi}}, and
  \bibinfo{author}{\bibfnamefont{C.}~\bibnamefont{Rebbi}},
  \bibinfo{year}{1981}{\natexlab{b}}, \bibinfo{journal}{Nucl. Phys.}
  \textbf{\bibinfo{volume}{B190}}, \bibinfo{pages}{734}.

\bibitem[{\citenamefont{Martinelli}
  \emph{et~al.}(1995)\citenamefont{Martinelli, Pittori, Sachrajda, Testa, and
  Vladikas}}]{Martinelli:1994ty}
\bibinfo{author}{\bibnamefont{Martinelli}, \bibfnamefont{G.}},
  \bibinfo{author}{\bibfnamefont{C.}~\bibnamefont{Pittori}},
  \bibinfo{author}{\bibfnamefont{C.~T.} \bibnamefont{Sachrajda}},
  \bibinfo{author}{\bibfnamefont{M.}~\bibnamefont{Testa}}, and
  \bibinfo{author}{\bibfnamefont{A.}~\bibnamefont{Vladikas}},
  \bibinfo{year}{1995}, \bibinfo{journal}{Nucl. Phys.}
  \textbf{\bibinfo{volume}{B445}}, \bibinfo{pages}{81},
  \arXiv{hep-lat/9411010}.

\bibitem[{\citenamefont{Mason} \emph{et~al.}(2006)\citenamefont{Mason,
  Trottier, Horgan, Davies, and Lepage}}]{Mason:2005bj}
\bibinfo{author}{\bibnamefont{Mason}, \bibfnamefont{Q.}},
  \bibinfo{author}{\bibfnamefont{H.~D.} \bibnamefont{Trottier}},
  \bibinfo{author}{\bibfnamefont{R.}~\bibnamefont{Horgan}},
  \bibinfo{author}{\bibfnamefont{C.~T.~H.} \bibnamefont{Davies}}, and
  \bibinfo{author}{\bibfnamefont{G.~P.} \bibnamefont{Lepage}}
  (\bibinfo{collaboration}{HPQCD}), \bibinfo{year}{2006},
  \bibinfo{journal}{Phys. Rev.} \textbf{\bibinfo{volume}{D73}},
  \bibinfo{pages}{114501}, \arXiv{hep-ph/0511160}.

\bibitem[{Mason \emph{et~al.}(2005)\citenamefont{Mason}
  \emph{et~al.}}]{Mason:2005zx}
\bibinfo{author}{\bibnamefont{Mason}, \bibfnamefont{Q.}}, \emph{et~al.}
  (\bibinfo{collaboration}{HPQCD}), \bibinfo{year}{2005},
  \bibinfo{journal}{Phys. Rev. Lett.} \textbf{\bibinfo{volume}{95}},
  \bibinfo{pages}{052002}, \arXiv{hep-lat/0503005}.

\bibitem[{\citenamefont{Mason}(2004)}]{Mason:2004zt}
\bibinfo{author}{\bibnamefont{Mason}, \bibfnamefont{Q.~J.}},
  \bibinfo{year}{2004}, \bibinfo{note}{{Cornell University Ph.D. thesis,
  UMI-31-14569}}.

\bibitem[{\citenamefont{Mathur and Dong}(2003)}]{Mathur:2002sf}
\bibinfo{author}{\bibnamefont{Mathur}, \bibfnamefont{N.}}, and
  \bibinfo{author}{\bibfnamefont{S.-J.} \bibnamefont{Dong}},
  \bibinfo{year}{2003}, \bibinfo{journal}{Nucl. Phys. Proc. Suppl.}
  \textbf{\bibinfo{volume}{119}}, \bibinfo{pages}{401},
  \arXiv{hep-lat/0209055}.

\bibitem[{\citenamefont{McNeile and Michael}(2001)}]{McNeile:2000xx}
\bibinfo{author}{\bibnamefont{McNeile}, \bibfnamefont{C.}}, and
  \bibinfo{author}{\bibfnamefont{C.}~\bibnamefont{Michael}}
  (\bibinfo{collaboration}{UKQCD}), \bibinfo{year}{2001},
  \bibinfo{journal}{Phys. Rev.} \textbf{\bibinfo{volume}{D63}},
  \bibinfo{pages}{114503}, \arXiv{hep-lat/0010019}.

\bibitem[{\citenamefont{Metropolis}
  \emph{et~al.}(1953)\citenamefont{Metropolis, Rosenbluth, Rosenbluth, Teller,
  and Teller}}]{Metropolis:1953am}
\bibinfo{author}{\bibnamefont{Metropolis}, \bibfnamefont{N.}},
  \bibinfo{author}{\bibfnamefont{A.~W.} \bibnamefont{Rosenbluth}},
  \bibinfo{author}{\bibfnamefont{M.~N.} \bibnamefont{Rosenbluth}},
  \bibinfo{author}{\bibfnamefont{A.~H.} \bibnamefont{Teller}}, and
  \bibinfo{author}{\bibfnamefont{E.}~\bibnamefont{Teller}},
  \bibinfo{year}{1953}, \bibinfo{journal}{J. Chem. Phys.}
  \textbf{\bibinfo{volume}{21}}, \bibinfo{pages}{1087}.

\bibitem[{\citenamefont{Michael}(1994)}]{Michael:1993yj}
\bibinfo{author}{\bibnamefont{Michael}, \bibfnamefont{C.}},
  \bibinfo{year}{1994}, \bibinfo{journal}{Phys. Rev.}
  \textbf{\bibinfo{volume}{D49}}, \bibinfo{pages}{2616},
  \arXiv{hep-lat/9310026}.

\bibitem[{\citenamefont{Mitra and Weisz}(1983)}]{Mitra:1983bi}
\bibinfo{author}{\bibnamefont{Mitra}, \bibfnamefont{P.}}, and
  \bibinfo{author}{\bibfnamefont{P.}~\bibnamefont{Weisz}},
  \bibinfo{year}{1983}, \bibinfo{journal}{Phys. Lett.}
  \textbf{\bibinfo{volume}{B126}}, \bibinfo{pages}{355}.

\bibitem[{\citenamefont{Morningstar and Peardon}(2004)}]{Morningstar:2003gk}
\bibinfo{author}{\bibnamefont{Morningstar}, \bibfnamefont{C.}}, and
  \bibinfo{author}{\bibfnamefont{M.~J.} \bibnamefont{Peardon}},
  \bibinfo{year}{2004}, \bibinfo{journal}{Phys. Rev.}
  \textbf{\bibinfo{volume}{D69}}, \bibinfo{pages}{054501},
  \arXiv{hep-lat/0311018}.

\bibitem[{\citenamefont{Morningstar and Shigemitsu}(1998)}]{Morningstar:1997ep}
\bibinfo{author}{\bibnamefont{Morningstar}, \bibfnamefont{C.~J.}}, and
  \bibinfo{author}{\bibfnamefont{J.}~\bibnamefont{Shigemitsu}},
  \bibinfo{year}{1998}, \bibinfo{journal}{Phys. Rev.}
  \textbf{\bibinfo{volume}{D57}}, \bibinfo{pages}{6741},
  \arXiv{hep-lat/9712016}.

\bibitem[{\citenamefont{Na and Gottlieb}(2006)}]{Na:2006qz}
\bibinfo{author}{\bibnamefont{Na}, \bibfnamefont{H.}}, and
  \bibinfo{author}{\bibfnamefont{S.}~\bibnamefont{Gottlieb}},
  \bibinfo{year}{2006}, \bibinfo{journal}{PoS}
  \textbf{\bibinfo{volume}{LAT2006}}, \bibinfo{pages}{191},
  \arXiv{hep-lat/0610009}.

\bibitem[{\citenamefont{Na and Gottlieb}(2009)}]{Na:2008hz}
\bibinfo{author}{\bibnamefont{Na}, \bibfnamefont{H.}}, and
  \bibinfo{author}{\bibfnamefont{S.}~\bibnamefont{Gottlieb}},
  \bibinfo{year}{2009}, \bibinfo{journal}{PoS}
  \textbf{\bibinfo{volume}{LAT2008}}, \bibinfo{pages}{119}, \arXiv{0812.1235}.

\bibitem[{\citenamefont{Na and Gottlieb}(2007)}]{Na:2007pv}
\bibinfo{author}{\bibnamefont{Na}, \bibfnamefont{H.}}, and
  \bibinfo{author}{\bibfnamefont{S.~A.} \bibnamefont{Gottlieb}},
  \bibinfo{year}{2007}, \bibinfo{journal}{PoS}
  \textbf{\bibinfo{volume}{LAT2007}}, \bibinfo{pages}{124}, \arXiv{0710.1422}.

\bibitem[{\citenamefont{Naik}(1989)}]{Naik:1986bn}
\bibinfo{author}{\bibnamefont{Naik}, \bibfnamefont{S.}}, \bibinfo{year}{1989},
  \bibinfo{journal}{Nucl. Phys.} \textbf{\bibinfo{volume}{B316}},
  \bibinfo{pages}{238}.

\bibitem[{\citenamefont{Narayanan and Neuberger}(1995)}]{Narayanan:1994gw}
\bibinfo{author}{\bibnamefont{Narayanan}, \bibfnamefont{R.}}, and
  \bibinfo{author}{\bibfnamefont{H.}~\bibnamefont{Neuberger}},
  \bibinfo{year}{1995}, \bibinfo{journal}{Nucl. Phys.}
  \textbf{\bibinfo{volume}{B443}}, \bibinfo{pages}{305},
  \arXiv{hep-th/9411108}.

\bibitem[{\citenamefont{Neuberger}(1998{\natexlab{a}})}]{Neuberger:1998my}
\bibinfo{author}{\bibnamefont{Neuberger}, \bibfnamefont{H.}},
  \bibinfo{year}{1998}{\natexlab{a}}, \bibinfo{journal}{Phys. Rev. Lett.}
  \textbf{\bibinfo{volume}{81}}, \bibinfo{pages}{4060},
  \arXiv{hep-lat/9806025}.

\bibitem[{\citenamefont{Neuberger}(1998{\natexlab{b}})}]{Neuberger:1997fp}
\bibinfo{author}{\bibnamefont{Neuberger}, \bibfnamefont{H.}},
  \bibinfo{year}{1998}{\natexlab{b}}, \bibinfo{journal}{Phys. Lett.}
  \textbf{\bibinfo{volume}{B417}}, \bibinfo{pages}{141},
  \arXiv{hep-lat/9707022}.

\bibitem[{\citenamefont{Neuberger}(1998{\natexlab{c}})}]{Neuberger:1997bg}
\bibinfo{author}{\bibnamefont{Neuberger}, \bibfnamefont{H.}},
  \bibinfo{year}{1998}{\natexlab{c}}, \bibinfo{journal}{Phys. Rev.}
  \textbf{\bibinfo{volume}{D57}}, \bibinfo{pages}{5417},
  \arXiv{hep-lat/9710089}.

\bibitem[{\citenamefont{Neubert}(1994)}]{Neubert:1993mb}
\bibinfo{author}{\bibnamefont{Neubert}, \bibfnamefont{M.}},
  \bibinfo{year}{1994}, \bibinfo{journal}{Phys. Rept.}
  \textbf{\bibinfo{volume}{245}}, \bibinfo{pages}{259}, \arXiv{hep-ph/9306320}.

\bibitem[{\citenamefont{Nielsen and Ninomiya}(1981)}]{Nielsen:1981hk}
\bibinfo{author}{\bibnamefont{Nielsen}, \bibfnamefont{H.~B.}}, and
  \bibinfo{author}{\bibfnamefont{M.}~\bibnamefont{Ninomiya}},
  \bibinfo{year}{1981}, \bibinfo{journal}{Phys. Lett.}
  \textbf{\bibinfo{volume}{B105}}, \bibinfo{pages}{219}.

\bibitem[{\citenamefont{Okamoto}(2006)}]{Okamoto:2005zg}
\bibinfo{author}{\bibnamefont{Okamoto}, \bibfnamefont{M.}},
  \bibinfo{year}{2006}, \bibinfo{journal}{PoS}
  \textbf{\bibinfo{volume}{LAT2005}}, \bibinfo{pages}{013},
  \arXiv{hep-lat/0510113}.

\bibitem[{\citenamefont{Oktay and Kronfeld}(2008)}]{Oktay:2008ex}
\bibinfo{author}{\bibnamefont{Oktay}, \bibfnamefont{M.~B.}}, and
  \bibinfo{author}{\bibfnamefont{A.~S.} \bibnamefont{Kronfeld}},
  \bibinfo{year}{2008}, \bibinfo{journal}{Phys. Rev.}
  \textbf{\bibinfo{volume}{D78}}, \bibinfo{pages}{014504}, \arXiv{0803.0523}.

\bibitem[{\citenamefont{Omelyan}
  \emph{et~al.}(2002{\natexlab{a}})\citenamefont{Omelyan, Mryglod, and
  Folk}}]{Omelyan:2002E2}
\bibinfo{author}{\bibnamefont{Omelyan}, \bibfnamefont{I.~P.}},
  \bibinfo{author}{\bibfnamefont{I.~M.} \bibnamefont{Mryglod}}, and
  \bibinfo{author}{\bibfnamefont{R.}~\bibnamefont{Folk}},
  \bibinfo{year}{2002}{\natexlab{a}}, \bibinfo{journal}{Phys. Rev.}
  \textbf{\bibinfo{volume}{E66}}, \bibinfo{pages}{026701}.

\bibitem[{\citenamefont{Omelyan}
  \emph{et~al.}(2002{\natexlab{b}})\citenamefont{Omelyan, Mryglod, and
  Folk}}]{Omelyan:2002E1}
\bibinfo{author}{\bibnamefont{Omelyan}, \bibfnamefont{I.~P.}},
  \bibinfo{author}{\bibfnamefont{I.~M.} \bibnamefont{Mryglod}}, and
  \bibinfo{author}{\bibfnamefont{R.}~\bibnamefont{Folk}},
  \bibinfo{year}{2002}{\natexlab{b}}, \bibinfo{journal}{Phys. Rev.}
  \textbf{\bibinfo{volume}{E65}}, \bibinfo{pages}{056706}.

\bibitem[{\citenamefont{Omelyan} \emph{et~al.}(2003)\citenamefont{Omelyan,
  Mryglod, and Folk}}]{Omelyan:2003CC}
\bibinfo{author}{\bibnamefont{Omelyan}, \bibfnamefont{I.~P.}},
  \bibinfo{author}{\bibfnamefont{I.~M.} \bibnamefont{Mryglod}}, and
  \bibinfo{author}{\bibfnamefont{R.}~\bibnamefont{Folk}}, \bibinfo{year}{2003},
  \bibinfo{journal}{Comput. Phys. Commun.} \textbf{\bibinfo{volume}{151}},
  \bibinfo{pages}{272}.

\bibitem[{\citenamefont{Orginos}(2006)}]{Orginos:2006zz}
\bibinfo{author}{\bibnamefont{Orginos}, \bibfnamefont{K.}},
  \bibinfo{year}{2006}, \bibinfo{journal}{PoS}
  \textbf{\bibinfo{volume}{LAT2006}}, \bibinfo{pages}{018}.

\bibitem[{\citenamefont{Orginos} \emph{et~al.}(2000)\citenamefont{Orginos,
  Sugar, and Toussaint}}]{Orginos:1999kg}
\bibinfo{author}{\bibnamefont{Orginos}, \bibfnamefont{K.}},
  \bibinfo{author}{\bibfnamefont{R.}~\bibnamefont{Sugar}}, and
  \bibinfo{author}{\bibfnamefont{D.}~\bibnamefont{Toussaint}},
  \bibinfo{year}{2000}, \bibinfo{journal}{Nucl. Phys. Proc. Suppl.}
  \textbf{\bibinfo{volume}{83}}, \bibinfo{pages}{878}, \arXiv{hep-lat/9909087}.

\bibitem[{\citenamefont{Orginos and Toussaint}(1999)}]{Orginos:1998ue}
\bibinfo{author}{\bibnamefont{Orginos}, \bibfnamefont{K.}}, and
  \bibinfo{author}{\bibfnamefont{D.}~\bibnamefont{Toussaint}}
  (\bibinfo{collaboration}{MILC}), \bibinfo{year}{1999},
  \bibinfo{journal}{Phys. Rev.} \textbf{\bibinfo{volume}{D59}},
  \bibinfo{pages}{014501}, \arXiv{hep-lat/9805009}.

\bibitem[{\citenamefont{Orginos} \emph{et~al.}(1999)\citenamefont{Orginos,
  Toussaint, and Sugar}}]{Orginos:1999cr}
\bibinfo{author}{\bibnamefont{Orginos}, \bibfnamefont{K.}},
  \bibinfo{author}{\bibfnamefont{D.}~\bibnamefont{Toussaint}}, and
  \bibinfo{author}{\bibfnamefont{R.~L.} \bibnamefont{Sugar}}
  (\bibinfo{collaboration}{MILC}), \bibinfo{year}{1999},
  \bibinfo{journal}{Phys. Rev.} \textbf{\bibinfo{volume}{D60}},
  \bibinfo{pages}{054503}, \arXiv{hep-lat/9903032}.

\bibitem[{\citenamefont{Orginos and Walker-Loud}(2008)}]{Orginos:2007tw}
\bibinfo{author}{\bibnamefont{Orginos}, \bibfnamefont{K.}}, and
  \bibinfo{author}{\bibfnamefont{A.}~\bibnamefont{Walker-Loud}},
  \bibinfo{year}{2008}, \bibinfo{journal}{Phys. Rev.}
  \textbf{\bibinfo{volume}{D77}}, \bibinfo{pages}{094505}, \arXiv{0705.0572}.

\bibitem[{Parappilly \emph{et~al.}(2006)\citenamefont{Parappilly}
  \emph{et~al.}}]{Parappilly:2005ei}
\bibinfo{author}{\bibnamefont{Parappilly}, \bibfnamefont{M.~B.}},
  \emph{et~al.}, \bibinfo{year}{2006}, \bibinfo{journal}{Phys. Rev.}
  \textbf{\bibinfo{volume}{D73}}, \bibinfo{pages}{054504},
  \arXiv{hep-lat/0511007}.

\bibitem[{di~Pierro \emph{et~al.}(2004)\citenamefont{di~Pierro}
  \emph{et~al.}}]{diPierro:2003bu}
\bibinfo{author}{\bibnamefont{di~Pierro}, \bibfnamefont{M.}}, \emph{et~al.},
  \bibinfo{year}{2004}, \bibinfo{journal}{Nucl. Phys. Proc. Suppl.}
  \textbf{\bibinfo{volume}{129}}, \bibinfo{pages}{340},
  \arXiv{hep-lat/0310042}.

\bibitem[{\citenamefont{Prelovsek}(2006{\natexlab{a}})}]{Prelovsek:2005qc}
\bibinfo{author}{\bibnamefont{Prelovsek}, \bibfnamefont{S.}},
  \bibinfo{year}{2006}{\natexlab{a}}, \bibinfo{journal}{PoS}
  \textbf{\bibinfo{volume}{LAT2005}}, \bibinfo{pages}{085},
  \arXiv{hep-lat/0509083}.

\bibitem[{\citenamefont{Prelovsek}(2006{\natexlab{b}})}]{Prelovsek:2005rf}
\bibinfo{author}{\bibnamefont{Prelovsek}, \bibfnamefont{S.}},
  \bibinfo{year}{2006}{\natexlab{b}}, \bibinfo{journal}{Phys. Rev.}
  \textbf{\bibinfo{volume}{D73}}, \bibinfo{pages}{014506},
  \arXiv{hep-lat/0510080}.

\bibitem[{Renner \emph{et~al.}(2005)\citenamefont{Renner}
  \emph{et~al.}}]{Renner:2004ck}
\bibinfo{author}{\bibnamefont{Renner}, \bibfnamefont{D.~B.}}, \emph{et~al.}
  (\bibinfo{collaboration}{LHP}), \bibinfo{year}{2005}, \bibinfo{journal}{Nucl.
  Phys. Proc. Suppl.} \textbf{\bibinfo{volume}{140}}, \bibinfo{pages}{255},
  \arXiv{hep-lat/0409130}.

\bibitem[{Renner \emph{et~al.}(2007)\citenamefont{Renner}
  \emph{et~al.}}]{Renner:2007pb}
\bibinfo{author}{\bibnamefont{Renner}, \bibfnamefont{D.~B.}}, \emph{et~al.}
  (\bibinfo{collaboration}{LHPC}), \bibinfo{year}{2007}, \bibinfo{journal}{PoS}
  \textbf{\bibinfo{volume}{LAT2007}}, \bibinfo{pages}{160}, \arXiv{0710.1373}.

\bibitem[{\citenamefont{Roberts}(2008)}]{Roberts:2007ji}
\bibinfo{author}{\bibnamefont{Roberts}, \bibfnamefont{C.~D.}},
  \bibinfo{year}{2008}, \bibinfo{journal}{Prog. Part. Nucl. Phys.}
  \textbf{\bibinfo{volume}{61}}, \bibinfo{pages}{50}, \arXiv{0712.0633}.

\bibitem[{\citenamefont{Schaefer} \emph{et~al.}(2007)\citenamefont{Schaefer,
  Hasenfratz, and Hoffmann}}]{Schaefer:2007dc}
\bibinfo{author}{\bibnamefont{Schaefer}, \bibfnamefont{S.}},
  \bibinfo{author}{\bibfnamefont{A.}~\bibnamefont{Hasenfratz}}, and
  \bibinfo{author}{\bibfnamefont{R.}~\bibnamefont{Hoffmann}},
  \bibinfo{year}{2007}, \bibinfo{journal}{PoS}
  \textbf{\bibinfo{volume}{LAT2007}}, \bibinfo{pages}{132}, \arXiv{0709.4130}.

\bibitem[{\citenamefont{Sexton and Weingarten}(1992)}]{Sexton:1992nu}
\bibinfo{author}{\bibnamefont{Sexton}, \bibfnamefont{J.~C.}}, and
  \bibinfo{author}{\bibfnamefont{D.~H.} \bibnamefont{Weingarten}},
  \bibinfo{year}{1992}, \bibinfo{journal}{Nucl. Phys.}
  \textbf{\bibinfo{volume}{B380}}, \bibinfo{pages}{665}.

\bibitem[{\citenamefont{Shamir}(1993)}]{Shamir:1993zy}
\bibinfo{author}{\bibnamefont{Shamir}, \bibfnamefont{Y.}},
  \bibinfo{year}{1993}, \bibinfo{journal}{Nucl. Phys.}
  \textbf{\bibinfo{volume}{B406}}, \bibinfo{pages}{90},
  \arXiv{hep-lat/9303005}.

\bibitem[{\citenamefont{Shamir}(2005)}]{Shamir:2004zc}
\bibinfo{author}{\bibnamefont{Shamir}, \bibfnamefont{Y.}},
  \bibinfo{year}{2005}, \bibinfo{journal}{Phys. Rev.}
  \textbf{\bibinfo{volume}{D71}}, \bibinfo{pages}{034509},
  \arXiv{hep-lat/0412014}.

\bibitem[{\citenamefont{Shamir}(2007)}]{Shamir:2006nj}
\bibinfo{author}{\bibnamefont{Shamir}, \bibfnamefont{Y.}},
  \bibinfo{year}{2007}, \bibinfo{journal}{Phys. Rev.}
  \textbf{\bibinfo{volume}{D75}}, \bibinfo{pages}{054503},
  \arXiv{hep-lat/0607007}.

\bibitem[{\citenamefont{Sharatchandra}
  \emph{et~al.}(1981)\citenamefont{Sharatchandra, Thun, and
  Weisz}}]{Sharatchandra:1981si}
\bibinfo{author}{\bibnamefont{Sharatchandra}, \bibfnamefont{H.~S.}},
  \bibinfo{author}{\bibfnamefont{H.~J.} \bibnamefont{Thun}}, and
  \bibinfo{author}{\bibfnamefont{P.}~\bibnamefont{Weisz}},
  \bibinfo{year}{1981}, \bibinfo{journal}{Nucl. Phys.}
  \textbf{\bibinfo{volume}{B192}}, \bibinfo{pages}{205}.

\bibitem[{\citenamefont{Sharpe}(1990)}]{Sharpe:1989bb}
\bibinfo{author}{\bibnamefont{Sharpe}, \bibfnamefont{S.~R.}},
  \bibinfo{year}{1990}, \bibinfo{journal}{Nucl. Phys. Proc. Suppl.}
  \textbf{\bibinfo{volume}{17}}, \bibinfo{pages}{146}.

\bibitem[{\citenamefont{Sharpe}(1992)}]{Sharpe:1992ft}
\bibinfo{author}{\bibnamefont{Sharpe}, \bibfnamefont{S.~R.}},
  \bibinfo{year}{1992}, \bibinfo{journal}{Phys. Rev.}
  \textbf{\bibinfo{volume}{D46}}, \bibinfo{pages}{3146},
  \arXiv{hep-lat/9205020}.

\bibitem[{\citenamefont{Sharpe}(2006{\natexlab{a}})}]{Sharpe:2006pu}
\bibinfo{author}{\bibnamefont{Sharpe}, \bibfnamefont{S.~R.}},
  \bibinfo{year}{2006}{\natexlab{a}}, \arXiv{hep-lat/0607016}.

\bibitem[{\citenamefont{Sharpe}(2006{\natexlab{b}})}]{Sharpe:2006re}
\bibinfo{author}{\bibnamefont{Sharpe}, \bibfnamefont{S.~R.}},
  \bibinfo{year}{2006}{\natexlab{b}}, \bibinfo{journal}{PoS}
  \textbf{\bibinfo{volume}{LAT2006}}, \bibinfo{pages}{022},
  \arXiv{hep-lat/0610094}.

\bibitem[{\citenamefont{Sharpe and Patel}(1994)}]{Sharpe:1993ur}
\bibinfo{author}{\bibnamefont{Sharpe}, \bibfnamefont{S.~R.}}, and
  \bibinfo{author}{\bibfnamefont{A.}~\bibnamefont{Patel}},
  \bibinfo{year}{1994}, \bibinfo{journal}{Nucl. Phys.}
  \textbf{\bibinfo{volume}{B417}}, \bibinfo{pages}{307},
  \arXiv{hep-lat/9310004}.

\bibitem[{\citenamefont{Sharpe and Shoresh}(2001)}]{Sharpe:2001fh}
\bibinfo{author}{\bibnamefont{Sharpe}, \bibfnamefont{S.~R.}}, and
  \bibinfo{author}{\bibfnamefont{N.}~\bibnamefont{Shoresh}},
  \bibinfo{year}{2001}, \bibinfo{journal}{Phys. Rev.}
  \textbf{\bibinfo{volume}{D64}}, \bibinfo{pages}{114510},
  \arXiv{hep-lat/0108003}.

\bibitem[{\citenamefont{Sharpe and Van~de Water}(2005)}]{Sharpe:2004is}
\bibinfo{author}{\bibnamefont{Sharpe}, \bibfnamefont{S.~R.}}, and
  \bibinfo{author}{\bibfnamefont{R.~S.} \bibnamefont{Van~de Water}},
  \bibinfo{year}{2005}, \bibinfo{journal}{Phys. Rev.}
  \textbf{\bibinfo{volume}{D71}}, \bibinfo{pages}{114505},
  \arXiv{hep-lat/0409018}.

\bibitem[{\citenamefont{Sheikholeslami and
  Wohlert}(1985)}]{Sheikholeslami:1985ij}
\bibinfo{author}{\bibnamefont{Sheikholeslami}, \bibfnamefont{B.}}, and
  \bibinfo{author}{\bibfnamefont{R.}~\bibnamefont{Wohlert}},
  \bibinfo{year}{1985}, \bibinfo{journal}{Nucl. Phys.}
  \textbf{\bibinfo{volume}{B259}}, \bibinfo{pages}{572}.

\bibitem[{\citenamefont{Silvestrini}(2008)}]{Silvestrini:2008}
\bibinfo{author}{\bibnamefont{Silvestrini}, \bibfnamefont{L.}},
  \bibinfo{year}{2008}, \bibinfo{note}{preliminary result presented at Lattice
  2008, http://www. utfit.org/}.

\bibitem[{\citenamefont{Sirlin}(1982)}]{Sirlin:1981ie}
\bibinfo{author}{\bibnamefont{Sirlin}, \bibfnamefont{A.}},
  \bibinfo{year}{1982}, \bibinfo{journal}{Nucl. Phys.}
  \textbf{\bibinfo{volume}{B196}}, \bibinfo{pages}{83}.

\bibitem[{\citenamefont{Smit and Vink}(1987)}]{Smit:1986fn}
\bibinfo{author}{\bibnamefont{Smit}, \bibfnamefont{J.}}, and
  \bibinfo{author}{\bibfnamefont{J.~C.} \bibnamefont{Vink}},
  \bibinfo{year}{1987}, \bibinfo{journal}{Nucl. Phys.}
  \textbf{\bibinfo{volume}{B286}}, \bibinfo{pages}{485}.

\bibitem[{\citenamefont{Sommer}(1994)}]{Sommer:1993ce}
\bibinfo{author}{\bibnamefont{Sommer}, \bibfnamefont{R.}},
  \bibinfo{year}{1994}, \bibinfo{journal}{Nucl. Phys.}
  \textbf{\bibinfo{volume}{B411}}, \bibinfo{pages}{839},
  \arXiv{hep-lat/9310022}.

\bibitem[{Struckmann \emph{et~al.}(2001)\citenamefont{Struckmann}
  \emph{et~al.}}]{Struckmann:2000bt}
\bibinfo{author}{\bibnamefont{Struckmann}, \bibfnamefont{T.}}, \emph{et~al.}
  (\bibinfo{collaboration}{TXL}), \bibinfo{year}{2001}, \bibinfo{journal}{Phys.
  Rev.} \textbf{\bibinfo{volume}{D63}}, \bibinfo{pages}{074503},
  \arXiv{hep-lat/0010005}.

\bibitem[{\citenamefont{Susskind}(1977)}]{Susskind:1976jm}
\bibinfo{author}{\bibnamefont{Susskind}, \bibfnamefont{L.}},
  \bibinfo{year}{1977}, \bibinfo{journal}{Phys. Rev.}
  \textbf{\bibinfo{volume}{D16}}, \bibinfo{pages}{3031}.

\bibitem[{\citenamefont{Symanzik}(1980)}]{Symanzik:1980UH}
\bibinfo{author}{\bibnamefont{Symanzik}, \bibfnamefont{K.}},
  \bibinfo{year}{1980}, \bibinfo{note}{in: Recent Developments in Gauge
  Theories, edited by G. 't Hooft et al. (Plenum Press, New York, 1980), p.
  313}.

\bibitem[{\citenamefont{Symanzik}(1983)}]{Symanzik:1983dc}
\bibinfo{author}{\bibnamefont{Symanzik}, \bibfnamefont{K.}},
  \bibinfo{year}{1983}, \bibinfo{journal}{Nucl. Phys.}
  \textbf{\bibinfo{volume}{B226}}, \bibinfo{pages}{187}.

\bibitem[{\citenamefont{Takaishi and de~Forcrand}(2006)}]{Takaishi:2005E1}
\bibinfo{author}{\bibnamefont{Takaishi}, \bibfnamefont{T.}}, and
  \bibinfo{author}{\bibfnamefont{P.}~\bibnamefont{de~Forcrand}},
  \bibinfo{year}{2006}, \bibinfo{journal}{Phys. Rev.}
  \textbf{\bibinfo{volume}{E73}}, \bibinfo{pages}{036706},
  \arXiv{hep-lat/0505020}.

\bibitem[{\citenamefont{Tamhankar}(2002)}]{Tamhankar:2002gd}
\bibinfo{author}{\bibnamefont{Tamhankar}, \bibfnamefont{S.~S.}},
  \bibinfo{year}{2002}, \bibinfo{note}{uMI-30-75964}.

\bibitem[{\citenamefont{Thacker and Lepage}(1991)}]{Thacker:1990bm}
\bibinfo{author}{\bibnamefont{Thacker}, \bibfnamefont{B.~A.}}, and
  \bibinfo{author}{\bibfnamefont{G.~P.} \bibnamefont{Lepage}},
  \bibinfo{year}{1991}, \bibinfo{journal}{Phys. Rev.}
  \textbf{\bibinfo{volume}{D43}}, \bibinfo{pages}{196}.

\bibitem[{\citenamefont{Toussaint and Davies}(2005)}]{Toussaint:2004cj}
\bibinfo{author}{\bibnamefont{Toussaint}, \bibfnamefont{D.}}, and
  \bibinfo{author}{\bibfnamefont{C.~T.~H.} \bibnamefont{Davies}},
  \bibinfo{year}{2005}, \bibinfo{journal}{Nucl. Phys. Proc. Suppl.}
  \textbf{\bibinfo{volume}{140}}, \bibinfo{pages}{234},
  \arXiv{hep-lat/0409129}.

\bibitem[{\citenamefont{Toussaint and Freeman}(2008)}]{Toussaint:2008ke}
\bibinfo{author}{\bibnamefont{Toussaint}, \bibfnamefont{D.}}, and
  \bibinfo{author}{\bibfnamefont{W.}~\bibnamefont{Freeman}},
  \bibinfo{year}{2008}, \arXiv{0808.2211}.

\bibitem[{\citenamefont{Towner and Hardy}(2008)}]{Towner:2007np}
\bibinfo{author}{\bibnamefont{Towner}, \bibfnamefont{I.~S.}}, and
  \bibinfo{author}{\bibfnamefont{J.~C.} \bibnamefont{Hardy}},
  \bibinfo{year}{2008}, \bibinfo{journal}{Phys. Rev.}
  \textbf{\bibinfo{volume}{C77}}, \bibinfo{pages}{025501}, \arXiv{0710.3181}.

\bibitem[{Ukita \emph{et~al.}(2007)\citenamefont{Ukita}
  \emph{et~al.}}]{Ukita:2007cu}
\bibinfo{author}{\bibnamefont{Ukita}, \bibfnamefont{N.}}, \emph{et~al.}
  (\bibinfo{collaboration}{PACS-CS}), \bibinfo{year}{2007},
  \bibinfo{journal}{PoS} \textbf{\bibinfo{volume}{LAT2007}},
  \bibinfo{pages}{138}, \arXiv{0710.3462}.

\bibitem[{Ukita \emph{et~al.}(2009)\citenamefont{Ukita}
  \emph{et~al.}}]{Ukita:2008mq}
\bibinfo{author}{\bibnamefont{Ukita}, \bibfnamefont{N.}}, \emph{et~al.}
  (\bibinfo{collaboration}{PACS-CS}), \bibinfo{year}{2009},
  \bibinfo{journal}{PoS} \textbf{\bibinfo{volume}{LAT2008}},
  \bibinfo{pages}{097}, \arXiv{0810.0563}.

\bibitem[{\citenamefont{Veneziano}(1979)}]{Veneziano:1979ec}
\bibinfo{author}{\bibnamefont{Veneziano}, \bibfnamefont{G.}},
  \bibinfo{year}{1979}, \bibinfo{journal}{Nucl. Phys.}
  \textbf{\bibinfo{volume}{B159}}, \bibinfo{pages}{213}.

\bibitem[{\citenamefont{Venkataraman and Kilcup}(1997)}]{Venkataraman:1997xi}
\bibinfo{author}{\bibnamefont{Venkataraman}, \bibfnamefont{L.}}, and
  \bibinfo{author}{\bibfnamefont{G.}~\bibnamefont{Kilcup}},
  \bibinfo{year}{1997}, \arXiv{hep-lat/9711006}.

\bibitem[{\citenamefont{Venkataraman and Kilcup}(1998)}]{Venkataraman:1997yj}
\bibinfo{author}{\bibnamefont{Venkataraman}, \bibfnamefont{L.}}, and
  \bibinfo{author}{\bibfnamefont{G.}~\bibnamefont{Kilcup}},
  \bibinfo{year}{1998}, \bibinfo{journal}{Nucl. Phys. Proc. Suppl.}
  \textbf{\bibinfo{volume}{63}}, \bibinfo{pages}{826}, \arXiv{hep-lat/9710086}.

\bibitem[{Walker-Loud \emph{et~al.}(2009)\citenamefont{Walker-Loud}
  \emph{et~al.}}]{WalkerLoud:2008bp}
\bibinfo{author}{\bibnamefont{Walker-Loud}, \bibfnamefont{A.}}, \emph{et~al.},
  \bibinfo{year}{2009}, \bibinfo{journal}{Phys. Rev.}
  \textbf{\bibinfo{volume}{D79}}, \bibinfo{pages}{054502}, \arXiv{0806.4549}.

\bibitem[{\citenamefont{Van~de Water and Sharpe}(2006)}]{VandeWater:2005uq}
\bibinfo{author}{\bibnamefont{Van~de Water}, \bibfnamefont{R.~S.}}, and
  \bibinfo{author}{\bibfnamefont{S.~R.} \bibnamefont{Sharpe}},
  \bibinfo{year}{2006}, \bibinfo{journal}{Phys. Rev.}
  \textbf{\bibinfo{volume}{D73}}, \bibinfo{pages}{014003},
  \arXiv{hep-lat/0507012}.

\bibitem[{\citenamefont{Weinberg}(1979)}]{Weinberg:1978kz}
\bibinfo{author}{\bibnamefont{Weinberg}, \bibfnamefont{S.}},
  \bibinfo{year}{1979}, \bibinfo{journal}{Physica}
  \textbf{\bibinfo{volume}{A96}}, \bibinfo{pages}{327}.

\bibitem[{\citenamefont{Wilcox}(1999)}]{Wilcox:1999ab}
\bibinfo{author}{\bibnamefont{Wilcox}, \bibfnamefont{W.}},
  \bibinfo{year}{1999}, \arXiv{hep-lat/9911013}.

\bibitem[{\citenamefont{Wilson}(1974)}]{Wilson:1974sk}
\bibinfo{author}{\bibnamefont{Wilson}, \bibfnamefont{K.~G.}},
  \bibinfo{year}{1974}, \bibinfo{journal}{Phys. Rev.}
  \textbf{\bibinfo{volume}{D10}}, \bibinfo{pages}{2445}.

\bibitem[{\citenamefont{Wilson}(1975)}]{Wilson:1975hf}
\bibinfo{author}{\bibnamefont{Wilson}, \bibfnamefont{K.~G.}},
  \bibinfo{year}{1975}, \bibinfo{note}{in: New Phenomena In Subnuclear Physics,
  ed., A. Zichichi (Erice, 1975), (Plenum Press, New York 1977) p. 13-32}.

\bibitem[{\citenamefont{Wingate} \emph{et~al.}(2004)\citenamefont{Wingate,
  Davies, Gray, Lepage, and Shigemitsu}}]{Wingate:2003gm}
\bibinfo{author}{\bibnamefont{Wingate}, \bibfnamefont{M.}},
  \bibinfo{author}{\bibfnamefont{C.~T.~H.} \bibnamefont{Davies}},
  \bibinfo{author}{\bibfnamefont{A.}~\bibnamefont{Gray}},
  \bibinfo{author}{\bibfnamefont{G.~P.} \bibnamefont{Lepage}}, and
  \bibinfo{author}{\bibfnamefont{J.}~\bibnamefont{Shigemitsu}},
  \bibinfo{year}{2004}, \bibinfo{journal}{Phys. Rev. Lett.}
  \textbf{\bibinfo{volume}{92}}, \bibinfo{pages}{162001},
  \arXiv{hep-ph/0311130}.

\bibitem[{\citenamefont{Wingate} \emph{et~al.}(2003)\citenamefont{Wingate,
  Shigemitsu, Davies, Lepage, and Trottier}}]{Wingate:2003nn}
\bibinfo{author}{\bibnamefont{Wingate}, \bibfnamefont{M.}},
  \bibinfo{author}{\bibfnamefont{J.}~\bibnamefont{Shigemitsu}},
  \bibinfo{author}{\bibfnamefont{C.~T.} \bibnamefont{Davies}},
  \bibinfo{author}{\bibfnamefont{G.~P.} \bibnamefont{Lepage}}, and
  \bibinfo{author}{\bibfnamefont{H.~D.} \bibnamefont{Trottier}},
  \bibinfo{year}{2003}, \bibinfo{journal}{Phys. Rev.}
  \textbf{\bibinfo{volume}{D67}}, \bibinfo{pages}{054505},
  \arXiv{hep-lat/0211014}.

\bibitem[{\citenamefont{Witten}(1979)}]{Witten:1979vv}
\bibinfo{author}{\bibnamefont{Witten}, \bibfnamefont{E.}},
  \bibinfo{year}{1979}, \bibinfo{journal}{Nucl. Phys.}
  \textbf{\bibinfo{volume}{B156}}, \bibinfo{pages}{269}.

\bibitem[{\citenamefont{Witten}(1980)}]{Witten:1980sp}
\bibinfo{author}{\bibnamefont{Witten}, \bibfnamefont{E.}},
  \bibinfo{year}{1980}, \bibinfo{journal}{Ann. Phys.}
  \textbf{\bibinfo{volume}{128}}, \bibinfo{pages}{363}.

\bibitem[{\citenamefont{Wohlert}(1987)}]{Wohlert:1987rf}
\bibinfo{author}{\bibnamefont{Wohlert}, \bibfnamefont{R.}},
  \bibinfo{year}{1987}, \bibinfo{note}{{DESY 87/069}}.

\bibitem[{\citenamefont{Yamada} \emph{et~al.}(2006)\citenamefont{Yamada, Blum,
  Hayakawa, and Izubuchi}}]{Yamada:2005dv}
\bibinfo{author}{\bibnamefont{Yamada}, \bibfnamefont{N.}},
  \bibinfo{author}{\bibfnamefont{T.}~\bibnamefont{Blum}},
  \bibinfo{author}{\bibfnamefont{M.}~\bibnamefont{Hayakawa}}, and
  \bibinfo{author}{\bibfnamefont{T.}~\bibnamefont{Izubuchi}}
  (\bibinfo{collaboration}{RBC}), \bibinfo{year}{2006}, \bibinfo{journal}{PoS}
  \textbf{\bibinfo{volume}{LAT2005}}, \bibinfo{pages}{092},
  \arXiv{hep-lat/0509124}.

\bibitem[{\citenamefont{Zanotti}(2009)}]{Zanotti:2008zm}
\bibinfo{author}{\bibnamefont{Zanotti}, \bibfnamefont{J.~M.}},
  \bibinfo{year}{2009}, \bibinfo{journal}{PoS}
  \textbf{\bibinfo{volume}{LAT2008}}, \bibinfo{pages}{007}, \arXiv{0812.3845}.

\bibitem[{Zanotti \emph{et~al.}(2002)\citenamefont{Zanotti}
  \emph{et~al.}}]{Zanotti:2001yb}
\bibinfo{author}{\bibnamefont{Zanotti}, \bibfnamefont{J.~M.}}, \emph{et~al.}
  (\bibinfo{collaboration}{CSSM Lattice}), \bibinfo{year}{2002},
  \bibinfo{journal}{Phys. Rev.} \textbf{\bibinfo{volume}{D65}},
  \bibinfo{pages}{074507}, \arXiv{hep-lat/0110216}.

\end{thebibliography}
